\documentclass[draftclsnofoot,onecolumn, 10pt]{IEEEtran}
\newcounter{todocounter}

\usepackage{amssymb}
\usepackage{wrapfig}

\usepackage{bigdelim}
\usepackage[rgb,x11names]{xcolor}
\usepackage{import}
\usepackage[subpreambles=true]{standalone}
\usepackage{graphicx}
\usepackage{amsfonts,amsmath,amssymb}
\usepackage[tight]{subfigure}
\usepackage{booktabs}
\usepackage[flushleft]{threeparttable}
\usepackage{multirow,bigstrut}
\usepackage{tabu}
\usepackage{algorithm}
\usepackage{algorithmicx}
\usepackage{algpseudocode}
\usepackage[algo2e]{algorithm2e}
\usepackage{pgfplotstable}
\usepackage{hhline}
\usepackage{setspace}
\usepackage{xspace}
\usepackage{pgfplots}
\usepackage{filecontents}
\usepackage{paralist}
\usepackage{amsthm}

\usepackage{enumitem}
\usepackage{tikz}
\usetikzlibrary{arrows,decorations.markings,backgrounds,positioning,fit,plotmarks,shadows,intersections,patterns,matrix,arrows.meta}
\usepackage{tikz-3dplot}
\usepackage[latin1]{inputenc}
\usepackage{tikz}
\usetikzlibrary{shapes,arrows,tikzmark}
\usepackage{enumitem}
\usepackage{booktabs}
\usepackage{array}
\usepackage{mdframed}%
\usepackage{xcolor}%
\usepackage{filecontents}
\usepackage{algorithm}
\usepackage[textwidth=0.2\textwidth,textsize=scriptsize]{todonotes}
\usetikzlibrary{matrix}
\newlength{\iwidth}
\newlength{\iheight}

\newlength{\colorbarlength}
\newlength{\rowspacing}

\usepackage{breqn}

\def \figDir{Figures}

\usepgfplotslibrary{statistics} %
\usetikzlibrary[pgfplots.statistics] %
\usepackage[square,sort,comma,numbers]{natbib}
\makeatletter
\def\BState{\State\hskip-\ALG@thistlm}
\makeatother

\def\eg{e.g.\@\xspace}
\def\ie{i.e.\@\xspace}
\def\etal{et~al.\@\xspace}

\tikzstyle{fancylabel} = [rounded corners, fill=Blue4, draw=black, thick, text=white, inner sep=0pt, minimum size=15pt, yshift=0pt]

\tikzstyle{mylabel} = [text=orange, ultra thick, inner sep=1pt, minimum size=15pt, yshift=-9pt, xshift=9pt] %

\usepackage{booktabs}

\emergencystretch=\hsize
\usepackage{verbatim}

\usepackage{siunitx}

\DeclareMathOperator{\erf}{erf}

\usepackage{xr}
\usepackage{hyperref}

\begin{document}
	
	\title{
		Probing Tissue Microarchitecture of the Baby Brain via Spherical Mean Spectrum Imaging
	}
	
	\author{
		Khoi Minh Huynh$^{\dagger}$,
		Tiantian Xu$^{\dagger}$,
		Ye Wu$^{\dagger}$,
		Xifeng Wang,
		Geng Chen,
		Kim-Han Thung,
		Haiyong Wu,
		Weili Lin,
		Dinggang Shen, and
		Pew-Thian Yap$^{*}$
		\thanks{\Copyright}
		\thanks{\Funding}
		\thanks{\RADBRIC}
		\thanks{\Equal}
		\thanks{\CorrAuthor}
	}
	
	\def\Copyright{Copyright (c) 2020 IEEE. Personal use of this material is permitted.  Permission from IEEE must be obtained for all other uses, in any current or future media, including reprinting/republishing this material for advertising or promotional purposes, creating new collective works, for resale or redistribution to servers or lists, or reuse of any copyrighted component of this work in other works.}
	
	\def\RADBRIC{K.M.~Huynh, T.~Xu, Y.~Wu, G.~Chen, W.~Lin, D.~Shen, and P.-T.~Yap are with the Department of Radiology and Biomedical Research Imaging Center (BRIC), University of North Carolina at Chapel Hill, NC, U.S.A. 
		K.M.~Huynh, D.~Shen, and P.-T.~Yap are also with the Department of Biomedical Engineering, University of North Carolina at Chapel Hill, NC, U.S.A.
		H.~Wu is with School of Information Engineering, Xiaozhuang University, Nanjing, China.
		X.~Wang is with Department of Biostatistics, University of North Carolina at Chapel Hill, North Carolina, U.S.A.}
	
	\def\Equal{$^{\dagger}$Khoi Minh Huynh, Tiantian Xu, and Ye Wu contributed equally to this work.}
	\def\CorrAuthor{Corresponding author: Pew-Thian Yap (\texttt{ptyap@med.unc.edu})}
	
	\def\Funding{This work was supported in part by NIH grants (NS093842, EB022880, MH104324, and MH110274) and the efforts of the UNC/UMN Baby Connectome Project Consortium.}
	
	\maketitle

	\begin{abstract}
		During the first years of life, the human brain undergoes dynamic spatially-heterogeneous changes, involving differentiation of neuronal types, dendritic arborization, axonal ingrowth, outgrowth and retraction, synaptogenesis, and myelination. To better quantify these changes, this article presents a method for probing tissue microarchitecture by characterizing water diffusion in a \emph{spectrum} of length scales, factoring out the effects of intra-voxel orientation heterogeneity. Our method is based on the spherical means of the diffusion signal, computed over gradient directions for a set of diffusion weightings (i.e., $b$-values). We decompose the spherical mean profile at each voxel into a spherical mean spectrum (SMS), which essentially encodes the fractions of spin packets undergoing fine- to coarse-scale diffusion processes, characterizing restricted and hindered diffusion stemming respectively from intra- and extra-cellular water compartments. From the SMS, multiple orientation distribution invariant indices can be computed, allowing for example the quantification of neurite density, microscopic fractional anisotropy ($\mu$FA), per-axon axial/radial diffusivity, and free/restricted isotropic diffusivity.
		We show that these indices can be computed for the developing brain for greater sensitivity and specificity to development related changes in tissue microstructure.
		Also, we demonstrate that our method, called spherical mean spectrum imaging (SMSI), is fast, accurate, and can overcome the biases associated with other state-of-the-art microstructure models.
	\end{abstract}


	\clearpage
	\section{Introduction}
	
	\IEEEPARstart{B}{iophysical} diffusion models play a vital role in characterizing complex changes in tissue microstructure, such as dendrites, axons, and glial cells, in the developing brain, giving important insights into the structural basis of the human brain. Microstructural analysis of the human brain has revealed important information on the maturational processes that occur in newborns \citep{kunz2014assessing}.

	Diffusion tensor imaging (DTI) is commonly used to assess microstructural changes in the human brain. DTI indices such as mean, radial, and axial diffusivities (MD, RD, AD), and fractional anisotropy (FA) can be used as quantitative indicators of brain developmental changes. However, DTI does not differentiate between white matter intra- and extra-axonal compartments. Moreover, FA can only measure voxel-level anisotropy, which mingles the effects of neurite microscopic-level anisotropy and orientation dispersion  \citep{lampinen2017neurite}. 
	
	Considerable efforts have been dedicated to deriving suitable diffusion indices to probe tissue microstructural properties.
	Assaf and Basser \citep{assaf2005composite} introduced the composite hindered and restricted model of diffusion (CHARMED) to address the deficiencies of DTI.
	This framework was later extended in \citep{assaf2008axcaliber} using a model called AxCaliber to estimate the axon diameter distribution.
	Alexander~\etal introduced orientationally invariant indices of axon diameter using a four-compartment tissue model combined with an optimized multi-shell acquisition scheme \citep{alexander2010orientationally}. 
	Using diffusion kurtosis imaging (DKI), Fieremans~\etal \citep{fieremans2011white} probed restricted water diffusion using two non-exchanging compartments representing intra- and extra-axonal spaces.
	Taking a step forward, Zhang~\etal \citep{zhang2012noddi} introduced NODDI to quantify neurite orientation density and dispersion.
	Daducci~\etal \citep{daducci2015accelerated} presented AMICO to significantly decrease NODDI computation time by linearizing the fitting problem.
	White~\etal \citep{white2013probing} demonstrated how restriction spectrum imaging (RSI), which involves a straightforward extension of the linear spherical deconvolution (SD) model \citep{tournier2004direct,tournier2007robust}, can be used to probe tissue orientation structures over a spectrum of length scales with minimal assumptions on the underlying microarchitecture. Kaden~\etal \citep{kaden2016quantitative} presented the spherical mean technique (SMT) method for estimating per-axon microscopic features, not confounded by the effects of fiber crossing and dispersion. SMT was extended in \citep{kaden2016multi} to take into consideration the presence of multiple compartments (MC-SMT). 
	DIAMOND \citep{scherrer2015characterizing} is based on a tridimensional extension of the statistical model of the apparent diffusion coefficient \citep{yablonskiy03statistical} and characterizes microstructural diffusivity with consideration of intra-voxel heterogeneities. 
	Diffusion basis spectrum imaging (DBSI) \citep{wang2011quantification} characterizes water diffusion by considering the diffusion signal as a linear combination of multiple anisotropic tensors and a spectrum of isotropic diffusion tensors.

	The infant brain develops rapidly in terms of total brain volume and tissue microarchitecture.
	The MR signal reflects the effects of various biological factors associated with maturation-related changes \citep{paus2001maturation}. To quantify these changes, existing studies mostly focus on the grey-white matter contrast given by T1- and T2-weighted images, diffusion indices given by DTI (e.g., FA and apparent diffusion coefficient (ADC)) \citep{partridge2004diffusion}, and major fascicles in the infant brain \citep{dubois2006assessment}. DTI has been used to show white matter changes in preterm infants \citep{anjari2007diffusion,kersbergen2014microstructural,partridge2004diffusion,rose2014brain} and for investigating brain-behavior relationships and maturation in infant white matter bundles 
	\citep{dubois2006assessment,dubois2008asynchrony,sullivan2006diffusion}.%
	
	With the advanced microstructural analysis methods described previously, distinct properties, such as neurite density, axon diameter, and orientation dispersion of the developing brain can be quantified. 
	Kunz~\etal \citep{kunz2014assessing} applied CHARMED and NODDI to study the maturation processes of newborn brains. Jelescu~\etal \citep{jelescu2015one} studied the microstructural changes in the infant brain using DKI and NODDI. Both models reveal a non-linear increase in intra-axonal water fraction and in tortuosity of the extra-axonal space as a function of age in the genu and splenium of the corpus callosum and the posterior limb of the internal capsule. Neurite density estimated using NODDI combined with myelin content information can be used to obtain the myelin $g$-ratio, which is a reliable measure of axonal myelination defined as the ratio of the inner axonal diameter to the total outer diameter \citep{dean2016mapping}.

	The aforementioned approaches are limited in that they
	\begin{inparaenum}[(i)]
		\item assume a predefined number of compartments (\eg, CHARMED, MC-SMT, SMT, NODDI),
		\item fix the diffusivity of one or more compartments (\eg, NODDI, RSI), or
		\item model only a portion of the diffusion spectrum (\eg, DBSI, RSI).
	\end{inparaenum}
	Given the complex tissue microstructure \citep{jones2010diffusion} and its dynamic developmental changes \citep{huppi1998microstructural,dubois2014early}, such assumptions are not necessarily ideal for accurate characterization of microstructural properties. 
	To better quantify the changes in the developing brain by tackling the mentioned problems, we present in this article a method for probing tissue microarchitecture by characterizing water diffusion 
	with a full \emph{spectrum} of diffusion shapes and scales and at the same time  factoring out the effects of intra-voxel orientation dispersion. 
	Our method is based on the spherical means of the diffusion signal, computed over gradient directions for a set of diffusion weightings \citep{huynh2019probing,kaden2016quantitative,kaden2016multi}. We decompose the spherical mean profile at each voxel into a spherical mean spectrum (SMS), which encodes the fractions of spin packets undergoing fine- to coarse-scale diffusion processes.
	From the SMS, multiple rotation invariant indices can be computed for greater sensitivity and specificity to changes in tissue microstructure.

	\section{Method}
	
	In this section, we will first provide a brief summary of SMT \citep{kaden2016quantitative,kaden2016multi} and then describe our method, called spherical mean spectrum imaging (SMSI), the implementation details, and the associated diffusion indices.
	
	\subsection{Spherical Mean Technique (SMT) }
	
	Spherical mean technique (SMT) \citep{kaden2016quantitative} estimates per-axon parallel and perpendicular diffusivities by factoring out the effects of orientation dispersion. It is based on the observation that 
	the spherical mean of the diffusion-attenuated signal over the gradient directions $g$, \ie,
	\begin{equation}
	\bar{S}_b = 
	\frac{1}{4\pi} 
	\int_{g\in\mathbb{S}^2}S_b(g)dg
	\end{equation}
	does not depend on the fiber orientation distribution. 
	Assuming that the signal
	can be represented as the spherical convolution of a fiber orientation distribution function (fODF) $p(\omega)$ ($p(\omega)\geq 0$, $\int_{\mathbb{S}^2}p(\omega)d\omega = 1$, $p(\omega) = p(-\omega)$, $\omega\in \mathbb{S}^2$)  with an axial and antipodal symmetric kernel $h_b(g |\omega) = h_b(\omega | g) \equiv h_{b}\left(| \langle g,\omega \rangle | \right)$, i.e.,
	\begin{equation}\label{eq:fODFConvoKernel}
	S_b(g) = S_{0}\int_{\omega\in\mathbb{S}^2}h_b(g | \omega)p(\omega)d\omega,
	\end{equation}
	it can be shown that 
	\begin{equation}\label{eq:SMTEquation}
	\bar{S}_b = S_{0}\bar{h}_b,
	\end{equation}
	where $\bar{h}_b$ is the kernel spherical mean.
	
	Setting the kernel as an axial symmetric diffusion tensor \citep{anderson2005measurement}, which is parameterized by orientation $\omega$, parallel diffusivity 
	$\lambda_{\parallel}
	$, and perpendicular diffusivity $\lambda_{\perp}
	$, i.e.,
	\begin{equation}
	\begin{split}
	h_{b}(g | \omega, \lambda_{\parallel}, \lambda_{\perp}) &= \underbrace{\exp\left(-b\langle g,\omega\rangle^{2}\lambda_{\parallel}\right)}_{\text{longitudinal}} \underbrace{\exp\left(-b\left(  1 - \langle g,\omega\rangle^{2}\right)\lambda_{\perp}\right)}_{\text{transverse}}\\
	&= \exp\left(-b\lambda_{\perp}\right)\exp\left(-b ( \lambda_{\parallel} - \lambda_{\perp}) \langle g,\omega\rangle^{2}\right),
	\end{split}
	\end{equation}
	it is straightforward, by noting
	
	\begin{align}
	&\bar{h}_b = \int_{g\in\mathbb{S}^2}h_{b}(g | \omega)dg =\int_{0}^{1} h_{b}(x)dx, \quad x \equiv \langle g,\omega\rangle \\
	&\erf(x) =\frac{2}{\sqrt{\pi}}\int_{0}^{x}\exp(-t^{2})dt,
	\end{align}
	
	to show that 
	\begin{align}
	\bar{h}_b(\lambda_{\parallel}, \lambda_{\perp}) &= 
	\frac{1}{4\pi} \int_{g\in\mathbb{S}^2} h_{b}(g | \omega, \lambda_{\parallel}, \lambda_{\perp}) dg \\
	&=
	\exp\left(-b \lambda_{\bot}\right) \frac{\sqrt{\pi}\erf\left(\sqrt{b(\lambda_{\parallel}-\lambda_{\bot})}\right)}{2 \sqrt{b (\lambda_{\parallel}-\lambda_{\bot})}}\label{2}.
	\end{align} 
	
	Note that $\bar{h}_b$ is not dependent on $\omega$. In SMT, the above equation is substituted in \eqref{eq:SMTEquation} to solve for $\lambda_{\parallel}$ and $\lambda_{\perp}$:
	\begin{equation}\label{eq:SMTcost}
	\frac{ \bar{S}_b}{S_{0}}=
	\begin{cases}
	\exp\left(-b \lambda_{\bot}\right),& \lambda_{\bot}=\lambda_{\parallel},\\
	\exp\left(-b \lambda_{\bot}\right) \frac{\sqrt{\pi}\erf\left(\sqrt{b(\lambda_{\parallel}-\lambda_{\bot})}\right)}{2 \sqrt{b (\lambda_{\parallel}-\lambda_{\bot})}}, & \lambda_{\bot}<\lambda_{\parallel}.\\
	\end{cases}
	\end{equation}

	\subsection{Spherical Mean Spectrum Imaging (SMSI)}
	
	\subsubsection{Ensemble of Spin Packets}
\begin{figure}
		\hspace{10pt}
		\subimport{Figures/SpinPackets/}{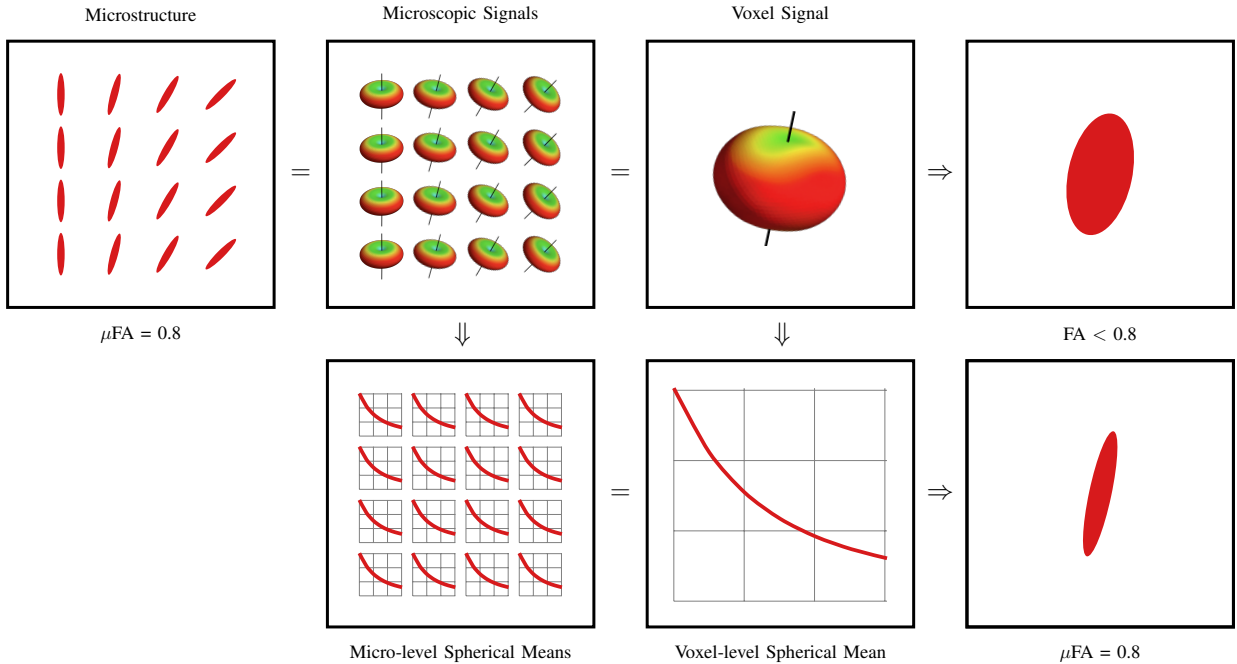}
		\caption{\textbf{Spherical Mean \& Microstructure.} The spherical mean can be used to quantify the diffusion patterns of spin packets in microenvironments, unconfounded by the orientation distribution. Unlike microscopic FA ($\mu$FA), voxel-level DTI-FA underestimates the anisotropy due to orientation dispersion.}
		\label{fig:spinpackets}
	\end{figure}

	We assume the signal measurements at each voxel to be a collective outcome of an ensemble of homogeneous spin packets originating from different positions within the voxel, each undergoing local anisotropic or isotropic diffusion represented by an axial-symmetric diffusion tensor model and contributes to the signal for gradient direction $g$ by $h_{b}(g |\omega,\lambda_{\parallel},\lambda_{\perp})$ 
	\citep{yablonskiy03statistical}. 
	Bigger heterogeneous spin packets, such as those assumed in \citep{scherrer2015characterizing}, can be decomposed into smaller homogeneous ones.
	The diffusion patterns of the spin packets are determined by microstructural barriers in intra- and extra-cellular spaces.
	Encoding the fractions of the spin packets using probability distribution $p(\omega,\lambda_{\parallel},\lambda_{\perp})$, the diffusion-attenuated signal $S$ can be written as
	\begin{equation}\label{eq:SMSIRepresentation}
	S_{b}(g) = S_{0}\int_{\omega,\lambda_{\parallel}, \lambda_{\perp}}p(\omega,\lambda_{\parallel},\lambda_{\perp})
	h_{b}(g | \omega,\lambda_{\parallel},\lambda_{\perp})
	d\omega d\lambda_{\parallel} d\lambda_{\perp}.
	\end{equation}
	Computing the spherical mean of the signal results in 
	\begin{equation}
	\bar{S}_{b} = S_{0}\int_{\omega,\lambda_{\parallel}, \lambda_{\perp}}p(\omega,\lambda_{\parallel},\lambda_{\perp})
	\bar{h}_{b}(\lambda_{\parallel},\lambda_{\perp})
	d\omega d\lambda_{\parallel} d\lambda_{\perp}.
	\end{equation}
	The variable $\omega$ can be marginalized out, giving
	\begin{equation}\label{eq:SMSContinuous}
	\bar{S}_{b} = S_{0}\int_{\lambda_{\parallel}, \lambda_{\perp}}p(\lambda_{\parallel},\lambda_{\perp})
	\bar{h}_{b}(\lambda_{\parallel},\lambda_{\perp})
	d\lambda_{\parallel} d\lambda_{\perp}.
	\end{equation}
	The spherical mean signal of each voxel can thus be seen as the weighted combination of the spherical mean signals of the spin packets. 
	Note that in the derivation, the antipodal symmetry assumption of the fiber orientation distributions is not needed.
	If the spin packets can be represented by a single set of diffusivities $(\lambda_{\parallel}^{*},\lambda_{\perp}^{*})$, $p(\lambda_{\parallel},\lambda_{\perp})$ can be defined using the delta function, i.e., $p(\lambda_{\parallel},\lambda_{\perp})=\delta(\lambda_{\parallel}-\lambda_{\parallel}^{*})\delta(\lambda_{\perp}-\lambda_{\perp}^{*})$, giving $\bar{S}_{b} = S_{0}\bar{h}_{b}(\lambda_{\parallel}^{*},\lambda_{\perp}^{*})$, which is identical to \eqref{eq:SMTEquation} of SMT. 
	Fig.~\ref{fig:spinpackets} illustrates how the spherical mean can be used to quantify microstructural properties. We name $p(\lambda_{\parallel},\lambda_{\perp})$ the spherical mean spectrum (SMS) because it encodes the probability of diffusivity pairs $(\lambda_{\parallel},\lambda_{\perp})$ according to the spherical mean profile.
	
\begin{figure}[htp]
	
	\definecolor{color1}{RGB}{128,245,152}
	\definecolor{color2}{RGB}{255,177,20}
	
	\centering
	\resizebox{0.7\linewidth}{!}
	{
	\begin{tikzpicture}[scale=0.80]
	
	\begin{scope}
	
	\clip (0cm,0cm) -- (8cm,8cm) -- (0cm,8cm) -- cycle;
	
	\foreach \x in {0,0.01,...,0.99} 
	{
		\pgfmathparse{(1-0.33* \x)}
		\definecolor{currentcolor}{hsb}{\pgfmathresult , 1, 1}
		\draw[draw=none, fill=currentcolor]
		(0cm,0cm) -- (90 - 45*\x:11.5cm) arc (90 - 45*\x:89.55 - 45*\x:11.5cm)	-- cycle;
	}
	
	\foreach \y in {1,2,...,12} {
		\draw[ultra thick,draw=black!30!white, opacity=0.5] (0cm,2*\y) -- (\y,0cm);
	}
	
	\end{scope}

	\tikzset{>=latex}
	\draw[|->|,line width = 2.5pt] (0cm,8.60cm) node[above=4pt]{$0$} -- node[above=4pt]{$\lambda_{\perp}$} (8cm,8.60cm) node[above=4pt]{{$\lambda_{\text{FW}}$}};
	\draw[|->|,line width = 2.5pt] (-0.60cm,0cm) node[left=4pt]{$0$} -- node[left=4pt]{$\lambda_{\parallel}$} (-0.60cm,8cm) node[left=4pt]{{$\lambda_{\text{FW}}$}};

	\draw[->,line width = 2.5pt,draw=white] (45:2.5cm) arc (45:90:2.5cm) node[midway,above=6pt]{\textcolor{white}{$\boldsymbol\mu$\textbf{FA}}};
	
	\draw[->,line width = 2.5pt,draw=white] (2cm,4.8cm) -- node[below=8pt]{\textcolor{white}{$\boldsymbol\mu$\textbf{MD}}}+(26.56505:2cm);
	
	\tikzstyle{ellipsecolor} = [color=color1,draw=black,circular drop shadow={opacity=0.7,fill=black}]

	\fill[ellipsecolor] (1.5,1.5) ellipse [x radius=0.15cm,y radius=0.15cm];
	\fill[ellipsecolor] (3.8,3.8) ellipse [x radius=0.38cm,y radius=0.38cm];
	\fill[ellipsecolor] (6.5,6.5) ellipse [x radius=0.65cm,y radius=0.65cm];

  \fill[ellipsecolor] (0,1.5) ellipse [x radius=0.045cm,y radius=0.15cm,rotate=-30];
  \fill[ellipsecolor] (0,3.8) ellipse [x radius=0.045cm,y radius=0.38cm,rotate=-30];
  \fill[ellipsecolor] (0,6.5) ellipse [x radius=0.045cm,y radius=0.65cm,rotate=-30];

  \fill[ellipsecolor] (0.6,1.5) ellipse [x radius=0.06cm,y radius=0.15cm,rotate=-30];
  \fill[ellipsecolor] (1.44,3.8) ellipse [x radius=0.144cm,y radius=0.38cm,rotate=-30];
  \fill[ellipsecolor] (2.4,6.5) ellipse [x radius=0.24cm,y radius=0.65cm,rotate=-30];
		
	\end{tikzpicture}
}
	\caption{\textbf{Spherical Mean Spectrum (SMS).} The SMS map with constraint $0< \lambda_{\perp} < \lambda_{\parallel}< \lambda_{\text{FW}}$. $\mu$FA ranges from 0 at the blue extreme to 1 at the red extreme. $\mu$MD increases perpendicular to the gray lines, on which $\mu$MD is constant.\label{fig:SMS}}
	\vspace{-1pt}	
\end{figure}
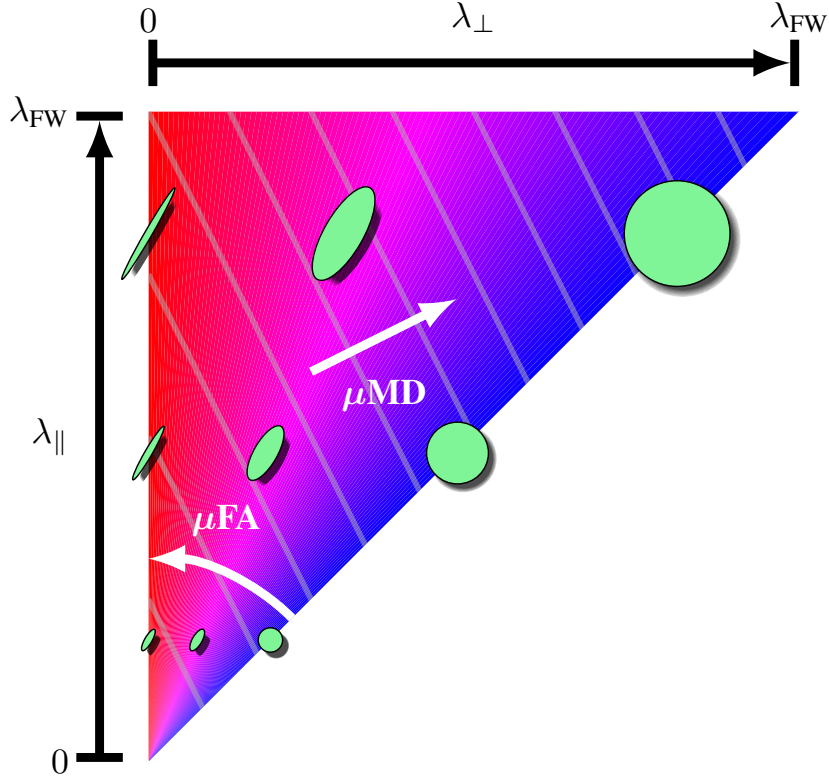
	
	\subsubsection{Spherical Mean Spectrum (SMS)}
	We relax the assumption of SMT and introduce a method to estimate the SMS, $p(\lambda_{\parallel},\lambda_{\perp})$, directly without imposing any constraints that restrict its shape.
	By studying the SMS (see Fig.~\ref{fig:SMS}), we can for example examine the fractions of spin packets undergoing isotropic ($\lambda_{\parallel}=\lambda_{\perp}$) or anisotropic ($\lambda_{\parallel}>\lambda_{\perp}$) diffusion and separate anisotropic diffusion into restricted (small $\lambda_{\perp}$)  and hindered (larger $\lambda_{\perp}$) diffusion. Similar to RSI \citep{white2013probing}, the SMS allows us to probe tissue microarchitecture using a spectrum of diffusion scales.
	Dissimilar to RSI, the SMS is invariant to the fODF and is therefore well suited for 
	regions with complex axonal geometries such as fanning and bending \citep{wu2018multi}.

	For the sake of feasibility, we discretize \eqref{eq:SMSContinuous} by defining 
	\begin{equation}
	p(\lambda_{\parallel},\lambda_{\perp}) = \sum_{i}\nu[i]\delta(\lambda_{\parallel}-\lambda_{\parallel}[i])\delta(\lambda_{\perp}-\lambda_{\perp}[i])
	\end{equation}
	to obtain
	\begin{equation}\label{eq:SMSDiscrete}
	\bar{S}_{b} = S_{0}\sum_{i} \nu[i] \bar{h}_{b}(\lambda_{\parallel}[i], \lambda_{\perp}[i])
	\end{equation}
	with volume fractions $\{\nu[1],\nu[2],\hdots\}$. 
	The ranges of $\lambda_{\parallel}[i]$ and $\lambda_{\parallel}[i]$ are set according to constraint $0< \lambda_{\perp}[i] < \lambda_{\parallel}[i] < \lambda_{\text{FW}}$, $\forall i$, where $\lambda_{\text{FW}}$ is the diffusivity of free water (see Fig.~\ref{fig:SMS}).
	Note that since 
	$
	\int_{\lambda_{\parallel}, \lambda_{\perp}}p(\lambda_{\parallel},\lambda_{\perp})
	d\lambda_{\parallel} d\lambda_{\perp} = 1
	$, 
	we have $\sum_{i}\nu[i] = 1$.

	Solving for $\nu$ using \eqref{eq:SMSDiscrete} is an ill-posed inverse problem since there are typically more unknowns than observations. With dictionary $A = \left[ \bar{h}_{b}(\lambda_{\parallel}[1], \lambda_{\perp}[1]), \bar{h}_{b}(\lambda_{\parallel}[2], \lambda_{\perp}[2]), \hdots \right] \in \mathbb{R}^{n\times p}$, where $n$ is the number of $b$-shells and $p$ is the number of dictionary atoms, we propose a solution based on elastic net \citep{zou2005regularization}:
	\begin{equation}\label{eq:L21Problem}
	\nu=\underset{\nu \succeq 0}{\mathrm{arg} \,\mathrm{min}}{\| A\nu-\bar{S} \| _2^2 + \gamma _{1} \| \text{diag}(w)\nu \|_{1} + \gamma _{2}\| \nu \|_2^2},
	\end{equation}
	where the first term ensures data fidelity, and $\gamma_{1}$ and $\gamma_{1}$ control the lasso ($\ell_{1}$-norm) penalty and ridge ($\ell_{2}$-norm) penalty, respectively. $\bar{S}$ is a vector containing the spherical means $\{ \bar{S}_{b} \}$ for different $b$-shells. $w$ is a weight vector. The reasons for elastic net are as follows:
	\begin{enumerate}
		\item Sparsity --- Ridge penalization keeps all atoms in the model and is hence not parsimonious. Lasso penalization promotes sparse solutions and hence improves interpretability.
		\item Stability --- If the atoms are highly correlated, lasso tends to select only one of them indiscriminately. Elastic net has the ability to select `grouped' predictors, a property that is not shared by lasso.
		\item Super-resolution --- Lasso selects at most $n$ atoms before it saturates. Elastic net can be seen a stabilized version of lasso and can be written as an augmented problem \citep{zou2005regularization}:
		\begin{equation}\label{eq:augLasso}
		\nu=\underset{\nu\succeq 0}{\mathrm{arg} \,\mathrm{min}}{\left\|
			\begin{pmatrix}
			A\\
			\sqrt{\gamma_{2}}I
			\end{pmatrix}
			\nu
			-
			\begin{pmatrix}
			\bar{S}\\
			0
			\end{pmatrix}
			\right\|_2^2
		}
		+\gamma _{1}\|\text{diag}(w)\nu\|_{1},
		\end{equation}
		allowing it to potentially select all $p$ atoms in all situations. This property was also used in \citep{tournier2007robust} to improve estimation of fiber orientation distributions.
	\end{enumerate}
	\vspace{-1.0pt}
	Fig.~\ref{fig:overview} illustrates how SMSI determines the microstructural compartments.
	
	\input{Figures/description/overview.tex}
	
	\subsubsection{Diffusion Indices}
	We divide the SMS into three compartments: isotropic, hindered, and restricted. Note that this is based on compartments commonly used in the literature, but is not the only way to divide the SMS. Also, each compartment can be represented by multiple, instead of one, diffusion kernels.
	It is assumed that on the timescale of a typical diffusion experiment, the effects of tissue membrane permeability is negligible. This is necessary in order to enforce strict signal compartmentalization.
	The isotropic diffusion compartment is represented by atoms with $\lambda_{\parallel} = \lambda_{\bot}$ 
	and a spectrum of diffusivity ranging from 0 to $3\,\text{mm}^{2}/\text{s}$, similar to \citep{wang2011quantification}.
	The hindered and restricted compartments are anisotropic with $\lambda_{\parallel} > \lambda_{\bot}$. 
	We define the restricted compartment with $\frac{\lambda_{\parallel}}{\lambda_{\bot}} \geq \tau ^2$ and the hindered compartment with $\frac{\lambda_{\parallel}}{\lambda_{\bot}} < \tau ^2$, where $\tau$ is the geometric tortuosity \citep{white2013probing}.
	Bihan suggested a value of $\frac{\pi}{2} \approx 1.57$ for $\tau$ \citep{bihan1995molecular}.
	The perpendicular diffusivity of the restricted compartment is 0 in \citep{kaden2016multi}, corresponding to $\tau \to \infty$.
	We determine $\tau$ automatically via grid search based on the voxels in the body of the corpus callosum, where  fiber dispersion and isotropic diffusion contamination are low, by exploring all possible values of $\tau$ estimated via MC-SMT.
	We found that $\tau$ is typically 2.6 for the Human Connectome Project (HCP) and Baby Connectome Project (BCP) datasets.
	In practice, $\lambda_{\bot}$ reflects the combined effects of nonvanishing permeability and possibly water in the extra-cellular space experiencing cylindrical diffusion symmetry \cite{jespersen2007modeling}.
	Intra-axonal diffusion is restricted and extra-axonal diffusion is hindered  \citep{assaf2004new,kaden2016multi}. 
	
	\noindent\emph{Microscopic Anisotropy ---}
	We present here a new measure of microscopic anisotropy for multi-compartmental models.
	We note that the orientations of the tensors used to represent the spin-packets in the microenvironments are between totally coherent with no dispersion and totally incoherent with full dispersion in all directions.
	For full dispersion, we have
	$
	p(\omega,\lambda_{\parallel},\lambda_{\perp}) = p(\omega)p(\lambda_{\parallel},\lambda_{\perp})=\frac{1}{4\pi}p(\lambda_{\parallel},\lambda_{\perp})$.
	Therefore, it is straightforward to show from \eqref{eq:SMSIRepresentation} that the signal resulting from this configuration is actually the spherical mean $\bar{S}_{b}$:
	\begin{equation}\label{eq:fulldispersion}
	\begin{split}
	S_{b}(g) &= S_{0}\int_{\omega,\lambda_{\parallel}, \lambda_{\perp}}
	\frac{1}{4\pi}p(\lambda_{\parallel},\lambda_{\perp})
	h_{b}(g | \omega,\lambda_{\parallel},\lambda_{\perp})
	d\omega d\lambda_{\parallel} d\lambda_{\perp}\\
	&= S_{0}\int_{\lambda_{\parallel}, \lambda_{\perp}}
	p(\lambda_{\parallel},\lambda_{\perp})
	\bar{h}_{b}(\lambda_{\parallel},\lambda_{\perp})
	d\lambda_{\parallel} d\lambda_{\perp}\\
	& \approx S_{0}\sum_{i} \nu[i] \bar{h}_b(\lambda_{\parallel}[i], \lambda_{\perp}[i])
	\\
	&
	\approx \bar{S}_{b}.
	\end{split}
	\end{equation}
	For no dispersion, the signal $S_{b}^{\uparrow}(g)$ is given by aligning the spin-packet tensors, \ie, $p(\omega,\lambda_{\parallel},\lambda_{\perp})=\delta(\omega - \omega_{0})p(\lambda_{\parallel},\lambda_{\perp})$ for an arbitrary $\omega_{0}$:
	\begin{equation}\label{eq:aligntensors}
	\begin{split}
	S_{b}(g) &= S_{0}\int_{\lambda_{\parallel}, \lambda_{\perp}}
	p(\lambda_{\parallel},\lambda_{\perp}) h_{b}(g | \omega_{0},\lambda_{\parallel},\lambda_{\perp})
	d\lambda_{\parallel} d\lambda_{\perp}\\
	& \approx S_{0}\sum_{i} \nu[i] h_b(g | \omega_{0}, \lambda_{\parallel}[i], \lambda_{\perp}[i])
	\\
	&
	=S_{b}^{\uparrow}(g).
	\end{split}
	\end{equation}
	A measure of anisotropy of the spin-packets can be defined as
	\begin{equation}\label{eq:AnisotropyUnnormalized}
	\frac{1}{4\pi}\sum_{b}\int_{\mathbb{S}^{2}}[S_{b}^{\uparrow}(g) - \bar{S}_b]^2 dg.
	\end{equation}
	We normalize \eqref{eq:AnisotropyUnnormalized} with the maximum anisotropy, which happens when we set for all anisotropic terms $\lambda_{\perp}[i]=0$. Denoting the signal and mean in this case respectively as $S_{b}^{\uparrow,*}(g)$ and $\bar{S}_{b}^{*}$, the microscopic anisotropy index (MAI) is defined as
	\begin{equation}
	\text{MAI} = 
	\sqrt
	{
		\frac
		{ \sum_{b}\int_{\mathcal{S}^{2}}[S_{b}^{\uparrow}(g) - \bar{S}_{b}]^2 dg }
		{ \sum_{b}\int_{\mathcal{S}^{2}}[S_{b}^{\uparrow,*}(g) - \bar{S}_{b}^{*}]^2 dg }
	}.
	\end{equation}%
	Similar to FA, MAI ranges from 0 to 1.
	Note that MAI is free from the influence of dispersion and can be used for multi-compartmental models, including SMSI, SMT, MC-SMT, and NODDI, provided that the diffusivities and volume fractions of the compartments are known.  
	
	\smallskip
	
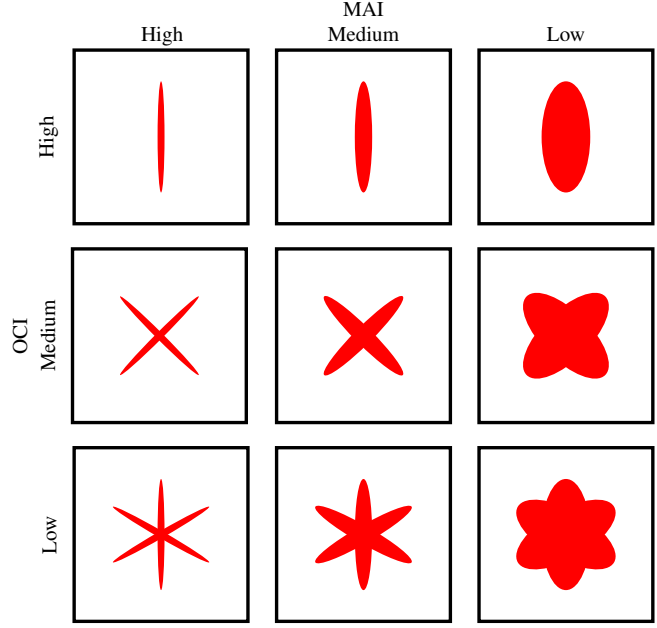
\begin{figure}[t]
		\newcommand\columnText[1]{
		\begin{tikzpicture}[baseline,trim left]
		\node[rotate=90] at (0.5\iwidth, 0.5\iheight) {#1};
		\end{tikzpicture}
	}
	\centering
	\resizebox{0.55\linewidth}{!}
	{
	\begin{tabular}{cccc}
		& MAI \\
		\qquad \quad High & Medium & Low\\
		\begin{tikzpicture}[scale=0.58]
		\tikzstyle{ellipsestyle} = [x radius=.07cm,y radius=0.4cm]
		\tikzstyle{ellipsecolor} = [color=red]
		\draw[ultra thick] (0,0) rectangle (5,5);
		\fill[ellipsecolor] (2.5,2.5) ellipse [x radius=.1cm,y radius=1.6cm,rotate=0];
		\node at (-0.5,2.5) {\columnText{High}};
		\node at (-1.5,2.5) {\columnText{\textcolor{white}{~}}};
		\end{tikzpicture}
		&
		\begin{tikzpicture}[scale=0.58]
		\tikzstyle{ellipsestyle} = [x radius=.07cm,y radius=0.4cm]
		\tikzstyle{ellipsecolor} = [color=red]
		\draw[ultra thick] (0,0) rectangle (5,5);
		\fill[ellipsecolor] (2.5,2.5) ellipse [x radius=.25cm,y radius=1.6cm,rotate=0];
		
		\end{tikzpicture}
		&
		\begin{tikzpicture}[scale=0.58]
		\tikzstyle{ellipsestyle} = [x radius=.07cm,y radius=.4cm]
		\tikzstyle{ellipsecolor} = [color=red]
		\draw[ultra thick] (0,0) rectangle (5,5);
		
		\fill[ellipsecolor] (2.5,2.5) ellipse [x radius=0.7cm,y radius=1.6cm,rotate=0];
		
		\end{tikzpicture}
	
		\\[1.5ex]
		\begin{tikzpicture}[scale=0.58]
		\draw[ultra thick] (0,0) rectangle (5,5);
		\tikzstyle{ellipsecolor} = [color=red]
		\fill[ellipsecolor] (2.5,2.5) ellipse [x radius=.1cm,y radius=1.6cm,rotate=45];
		\fill[ellipsecolor] (2.5,2.5) ellipse [x radius=.1cm,y radius=1.6cm,rotate=-45];
		\node at (-0.5,2.5) {\columnText{Medium}};
		\node at (-1.3,2.5) {\columnText{OCI}};
		\end{tikzpicture}
		&
		\begin{tikzpicture}[scale=0.58]
		\draw[ultra thick] (0,0) rectangle (5,5);
		\tikzstyle{ellipsecolor} = [color=red]
		\fill[ellipsecolor] (2.5,2.5) ellipse [x radius=.250cm,y radius=1.6cm,rotate=45];
		\fill[ellipsecolor] (2.5,2.5) ellipse [x radius=.250cm,y radius=1.6cm,rotate=-45];
		\end{tikzpicture}
		&
		\begin{tikzpicture}[scale=0.58]
		\draw[ultra thick] (0,0) rectangle (5,5);
		\tikzstyle{ellipsecolor} = [color=red]
		\fill[ellipsecolor] (2.5,2.5) ellipse [x radius=0.7cm,y radius=1.6cm,rotate=45];
		\fill[ellipsecolor] (2.5,2.5) ellipse [x radius=0.7cm,y radius=1.6cm,rotate=-45];
		\end{tikzpicture}
		
		\\[1.5ex]
		\begin{tikzpicture}[scale=0.58]
		\draw[ultra thick] (0,0) rectangle (5,5);
		\tikzstyle{ellipsecolor} = [color=red]
		\fill[ellipsecolor] (2.5,2.5) ellipse [x radius=0.1cm,y radius=1.6cm,rotate=0];
		\fill[ellipsecolor] (2.5,2.5) ellipse [x radius=.1cm,y radius=1.6cm,rotate=60];
		\fill[ellipsecolor] (2.5,2.5) ellipse [x radius=.1cm,y radius=1.6cm,rotate=-60];
		\node at (-0.5,2.5) {\columnText{Low}};
		\node at (-1.5,2.5) {\columnText{\textcolor{white}{~}}};
		\end{tikzpicture}
		&
		\begin{tikzpicture}[scale=0.58]
		\draw[ultra thick] (0,0) rectangle (5,5);
		\tikzstyle{ellipsecolor} = [color=red]
		\fill[ellipsecolor] (2.5,2.5) ellipse [x radius=.250cm,y radius=1.6cm,rotate=0];
		\fill[ellipsecolor] (2.5,2.5) ellipse [x radius=.250cm,y radius=1.6cm,rotate=60];
		\fill[ellipsecolor] (2.5,2.5) ellipse [x radius=.250cm,y radius=1.6cm,rotate=-60];
		\end{tikzpicture}
		&
		\begin{tikzpicture}[scale=0.58]
		\draw[ultra thick] (0,0) rectangle (5,5);
		\tikzstyle{ellipsecolor} = [color=red]
		\fill[ellipsecolor] (2.5,2.5) ellipse [x radius=0.7cm,y radius=1.6cm,rotate=0];
		\fill[ellipsecolor] (2.5,2.5) ellipse [x radius=0.7cm,y radius=1.6cm,rotate=60];
		\fill[ellipsecolor] (2.5,2.5) ellipse [x radius=0.7cm,y radius=1.6cm,rotate=-60];
		\end{tikzpicture}
		
	\end{tabular}
}
	\caption{
		\textbf{MAI and OCI.}
		MAI is sensitive to diffusion anisotropy but not orientation dispersion. OCI is sensitive to orientation heterogeneity.
		}
	\vspace{-1em}
	\label{fig:PAAOCI_cartoon}
\end{figure}

	\noindent\emph{Orientation Coherence ---}
	In case of full dispersion, orientation coherence is minimal and should correspond to a value of zero.
	We measure orientation coherence as the distance between the observed signal and the full dispersion signal:
	\begin{equation}\label{eq:CoherenceUnnormalized}
	\frac{1}{4\pi}\sum_{b}\int_{\mathbb{S}^{2}}[S_{b}(g) - \bar{S}_b]^2 dg.
	\end{equation}
	We normalize the coherence with the maximum coherence when there is no dispersion, giving the orientation coherence index (OCI):
	\begin{equation}
	\begin{split}
	\text{OCI} &= 
	\sqrt
	{
		\frac
		{ \left[ \frac{1}{4\pi}\sum_{b}\int_{\mathbb{S}^{2}}[S_{b}(g) - \bar{S}_b]^2 dg - \sigma^{2} \right]_{+} } 
		{ \frac{1}{4\pi}\sum_{b}\int_{\mathbb{S}^{2}}[S_{b}^{\uparrow}(g) - \bar{S}_{b}]^2 dg }
	}\\
	&\approx
	\sqrt
	{
		\frac
		{ \left[ \sum_{b}\int_{\mathcal{S}^{2}}[S_{b}(g) - \bar{S}_b]^2 dg - k\sigma^{2} \right]_{+} } 
		{ \sum_{b}\int_{\mathcal{S}^{2}}[S_{b}^{\uparrow}(g) - \bar{S}_{b}]^2 dg }
	},
	\end{split}
	\end{equation}
	where $\sigma$ is the noise standard deviation, which can be computed via maximum likelihood estimation using a set of b0 images \citep{kaden2016quantitative}, and $k$ is the total number of gradient directions across all shells. Operator $[z]_{+}$ returns $z$ if $z \geq 0$ and $0$ otherwise.
	OCI ranges from 0 for no coherence (full dispersion) to 1 for full coherence (no dispersion).
	Similar to MAI, the OCI definition is general and compatible among different models. 
	The relationship between MAI, OCI, and orientation heterogeneity is illustrated in Fig.~\ref{fig:PAAOCI_cartoon}.
	
	\smallskip
	
	\noindent\emph{Isotropic Diffusion Elimination ---}
	Isotropic diffusion signal can be removed to increase sensitivity to axonal changes \citep{pasternak2009free}. This is done for example via free-water elimination (FWE) indices \citep{pasternak2009free}. RSI models both free-water diffusivity, estimated from intra-ventricular space, and longitudinal diffusivity, estimated from white matter.  
	SMSI allows not only free water but the whole isotropic diffusion spectrum to be discarded, resulting in isotropic diffusion eliminated (IDE) indices. This is similar in spirit to DBSI \citep{wang2011quantification}.
	Table~\ref{tab:Indices} lists the SMSI indices. IDE indices are marked by symbol $\dagger$.

	\begin{table*}[t]
	\centering
\resizebox{\linewidth}{!}
{ 	
	
	\begin{scriptsize}
		\begin{threeparttable}
			\scriptsize
				\caption{SMSI indices.
				\label{tab:Indices}}
			\vspace{-1em}
		\begin{tabular}{cc|cc}
			\toprule
     \bfseries Description & \bfseries Indices & \bfseries Description & \bfseries Indices\\
      \midrule
			Anisotropic VF & 
			$\displaystyle
			v_{\text{a}} = \sum_{i\in \mathcal{A}} \nu[i]
			$
			&
			Intra-cellular AD & 
			$\displaystyle
			\mu\text{AD}_{\text{ic}}=\frac{\sum_{i\in\mathcal{R}}\nu[i]\lambda_{\parallel}[i]}{\sum_{i\in\mathcal{R}}\nu[i]}
			$
			\\
      		Intra-cellular VF & 
			$\displaystyle
			v_{\text{ic}} = \frac{\sum_{i\in \mathcal{R}} \nu[i]}{v_{\text{a}}}
			$
			&
			Intra-cellular RD & 
			$\displaystyle
			\mu\text{RD}_{\text{ic}}=\frac{\sum_{i\in\mathcal{R}}\nu[i]\lambda_{\bot}[i]}{\sum_{i\in\mathcal{R}}\nu[i]}
			$
			\\
			Extra-cellular VF & 
			$\displaystyle
			v_{\text{ec}} = \frac{\sum_{i\in \mathcal{H}} \nu[i]}{v_{\text{a}}}
			$
			&
			Extra-cellular AD & 
			$\displaystyle
			\mu\text{AD}_{\text{ec}}=\frac{\sum_{i\in\mathcal{H}}\nu[i]\lambda_{\parallel}[i]}{\sum_{i\in\mathcal{H}}\nu[i]}
			$
			\\
			Isotropic VF & 
			$\displaystyle
			v_{\text{iso}} = \sum_{i\in \mathcal{I}} \nu[i]
			$
			&
			Extra-cellular RD & 
			$\displaystyle
			\mu\text{RD}_{\text{ec}}=\frac{\sum_{i\in\mathcal{H}}\nu[i]\lambda_{\perp}[i]}{\sum_{i\in\mathcal{H}}\nu[i]}
			$
			\\
			Microscopic AD &
			$\displaystyle
			\mu\text{AD}=\frac{\sum_{i}\nu[i]\lambda_{\parallel}[i]}{\sum_{i}\nu[i]}
			$
			&
			Microscopic anisotropy index &
			$\displaystyle
			\text{MAI} = 
			\sqrt
			{
				\frac
				{ \sum_{b}\int_{\mathcal{S}^2}[S_{b}^{\uparrow}(g) - \bar{S}_{b}]^2 dg }
				{ \sum_{b}\int_{\mathcal{S}^2}[S_{b}^{\uparrow,*}(g) - \bar{S}_{b}^{*}]^2 dg }
			}.
			$
			\\
			Microscopic RD &
			$\displaystyle
			\mu\text{RD}=\frac{\sum_{i}\nu[i]\lambda_{\perp}[i]}{\sum_{i}\nu[i]}
			$
			&
			Orientation coherence index &
			$\displaystyle
			\text{OCI}\approx
			\sqrt
			{
				\frac
				{ \left[ \sum_{b}\int_{\mathcal{S}^2}[S_{b}(g) - \bar{S}_b]^2 dg - k\sigma^{2} \right]_{+} } 
				{ \sum_{b}\int_{\mathcal{S}^2}[S_{b}^{\uparrow}(g) - \bar{S}_{b}]^2 dg }
			} 
			$
			\\
			Microscopic MD &
			$\displaystyle
			\mu\text{MD} = \frac{\mu\text{AD} + 2\mu\text{RD}}{3} 
			$
			& 
			Microscopic sphericity &
			$\displaystyle
			\mu C_{\text{s}}=\frac{\mu\text{RD}}{\mu\text{MD}}
			$
			\\
			Microscopic FA &
			$\displaystyle
			\mu\text{FA} =
			\frac
			{\mu\text{AD} - \mu\text{RD}}
			{\sqrt{\mu\text{AD}^2+2\mu\text{RD}^2}}
			$
			&
			Microscopic linearity &
			$\displaystyle
			\mu C_{\text{l}}=\frac{\mu\text{AD}- \mu\text{RD}}{3\mu\text{MD}}
			$
			\\
			\bottomrule
		\end{tabular}
		  \begin{tablenotes}
		\scriptsize
    	\item VF: Volume fraction, AD/RD/MD: Axial/Radial/Mean diffusivity, FA: Fractional anisotropy
    	\item Trapped diffusion: $\mathcal{T} = \{i | \lambda_{\parallel}[i] = 0, \lambda_{\perp}[i] = 0 \}$
		\item Anisotropic diffusion: $\mathcal{A} = \{i | \lambda_{\parallel}[i] > \lambda_{\perp}[i] \}$, Isotropic diffusion: $\mathcal{I} = \{i | \lambda_{\parallel}[i] = \lambda_{\perp}[i] \}$
		\item Restricted diffusion: $\mathcal{R} = \{i | \lambda_{\perp}[i] \tau^2 \leq \lambda_{\parallel}[i], i \in \mathcal{A}, \tau>1 \}$, Hindered diffusion: $\mathcal{H} = \{i | \lambda_{\perp}[i] \tau^2 > \lambda_{\parallel}[i], i \in \mathcal{A}, \tau>1 \}$ 		
		\item $S_{b}$, $S_{b}^{\uparrow}(g)$ and $\bar{S}_{b}$ are the DW signal, the DW signal when all components are orientationally aligned, and the mean signal; $S_{b}^{\uparrow,*}(g)$ and $\bar{S}_{b}^{*}$ are the aligned signal and its mean when all $\lambda_{\bot}[i]=0$.
		\item $S_{b}^{\uparrow,\dagger}(g)$, $\bar{S}_{b}^{\dagger}$, $S_{b}^{\uparrow,*,\dagger}(g)$, and $\bar{S}_{b}^{*,\dagger}$ are $S_{b}^{\uparrow}(g)$, $\bar{S}_{b}$, $S_{b}^{\uparrow,*}(g)$, and $\bar{S}_{b}^{*}$, respectively, without isotropic compartments.
		\item $k$ is the total number of gradient directions, $\sigma$ is the noise standard deviation, and $\tau$ is the geometric tortuosity.
	\end{tablenotes}
	  \end{threeparttable}
  \end{scriptsize}
}
\end{table*}%

	\subsubsection{Implementation Details}\label{subsubsec:imple}
	The quantification of microstructure using 
	the spherical mean is affected by degeneracy
	in the sense that the spherical mean signal given by an anisotropic micro-environment may be indistinguishable from that of multiple isotropic micro-environments \citep{szczepankiewicz2015quantification}. 
	We describe in the following the implementation details of SMSI and how the degeneracy can be resolved using the full direction-sensitized signal from which the spherical mean signal is derived.

	\noindent\emph{Resolving Degeneracy via Full-Signal Spectrum (FSS) ---}
	The full signal, unlike the mean signal, is not ambiguous in distinguishing isotropic and anisotropic diffusion.
	
	Letting $\mathcal{H}(\lambda_\parallel [i],\lambda_\perp [i])$ be the matrix of rotational spherical harmonics (SHs) of  $h_{b}(g | \omega,\lambda_{\parallel},\lambda_{\perp})$, $\mathcal{Y}_{L}$ the spherical harmonics of even orders up to $L$, and $\varphi_{i}$ the SH coefficients of the fODF corresponding to $h_{b}(g | \omega,\lambda_{\parallel}[i],\lambda_{\perp}[i])$, \eqref{eq:SMSIRepresentation} can be discretized as \citep{tournier2004direct} 
	\begin{equation}
	\mathcal{S} \approx \sum_{i} \mathcal{H}(\lambda_\parallel [i],\lambda_\perp [i])\mathcal{Y}_{L}\varphi_{i} =  \mathcal{B}\Phi.
	\end{equation}
	Similar to \eqref{eq:augLasso},
	$\mathcal{B}$ can be seen as a dictionary matrix and $\Phi$ can be solved with Tikhonov regularization
	\begin{equation}\label{eq:RSIeq}
	\underset{\Phi}{
		\mathrm{min}}
	{\left\|
		\begin{pmatrix}
		\mathcal{B}\\
		\sqrt{\gamma_{3}}\text{diag}(w')
		\end{pmatrix}
		\Phi
		-
		\begin{pmatrix}
		\mathcal{S}\\
		0
		\end{pmatrix}
		\right\|_2^2
	}.
	\end{equation}
	From $\Phi$, the volume fraction of each compartment $i$ is the $0$-th order SH coefficient \citep{white2013probing}.

	Isotropic diffusion can be represented by isotropically-distributed anisotropic tensors, causing ambiguity.
	We prevent this by solving \eqref{eq:RSIeq} with weight vector $w'$ set to one for all atoms, identifying degenerate anisotropic atoms with generalized fractional anisotropy \citep{cohen2008detection} (GFA) smaller than $0.3$, and reapplying \eqref{eq:RSIeq} with higher penalization of the degenerate atoms. This is implemented by doubling the corresponding elements in $w'$. 
	The volume fractions obtained are denoted as $\nu_{\text{FSS}}$.

	\noindent\emph{Estimation of Isotropic Compartments ---}
	We use $b$-shells with $b \leq 1000\, \text{s}/\text{mm}^2$ for an initial estimation using \eqref{eq:augLasso} with $w$ set to one for all atoms. This improves the estimation of isotropic volume fractions. The volume fractions obtained are denoted as $\nu_{\text{SMS}}$.

	\noindent\emph{Iterative Reweighting ---}
	We then solve for the volume fractions using all $b$-shells via an iterative re-weighted elastic net, where at the $j$-th iteration we have
	\begin{equation}\label{eq:augLassoWeighted}
	\nu_j=\underset{\nu_j\succeq 0}{\mathrm{arg} \,\mathrm{min}}{\left\|
		\begin{pmatrix}
		A\\
		\sqrt{\gamma_{2}}I
		\end{pmatrix}
		\nu_j
		-
		\begin{pmatrix}
		\bar{S}\\
		0
		\end{pmatrix}
		\right\|_2^2
	}
	+\gamma _{1}\|\text{diag}(w_j) \nu_j\|_{1},
	\end{equation}
	where $w_j[i] =\frac{1}{\xi + \nu_{j-1}[i]}$ with $\xi$ being a constant and $\nu_0$ the geometric mean of $\nu_{\text{FSS}}$ and $\nu_{\text{SMS}}$. 
	The volume fractions estimated in this step can be utilized to set $w'$ in \eqref{eq:RSIeq} for alternating estimation of $\nu_{\text{FSS}}$ and $\nu_{\text{SMS}}$. However, we found that one round of estimation is sufficient to produce accurate results.

	The regularization parameters $\gamma_{1}$, $\gamma_{2}$, and $\gamma_{3}$ affect the estimation significantly. 
	We develop an adaptive framework to automatically select these parameters based on the data:
	\begin{enumerate}
		\item Select regions with ``simple'' microstructure (\eg, the body of the corpus callosum for anisotropic diffusion and the ventricles for isotropic diffusion). This can be done by selecting voxels with highest and lowest FA values.
		
		\item Perform SMSI estimation with initialization as described above in these regions using multiple combinations of $\gamma_{1}$'s, $\gamma_{2}$'s, and $\gamma_{3}$'s.
		\item For each combination of $\gamma$'s, substitute the obtained values for per-axon radial ($\mu$RD) and axial diffusivity ($\mu$AD) into \eqref{eq:SMTcost}.
		The optimal parameters are selected as those that minimize the difference between the predicted and the observed spherical mean signals.
	\end{enumerate}

	\subsubsection{Debiasing}\label{subsubsec:debias}
	Diffusion MRI signal is affected by Rician noise, especially at high $b$-value where the noise floor dominates the signal \citep{koay2009signal}. 
	To reduce potential effects of this noise-induced bias, we correct the measured signal using the following steps:
	\begin{enumerate}
		\item Estimate the noise level $\sigma$ voxel-wise via maximum-likelihood estimation (MLE) using a set of b0 images. This is based on the assumption that the SNR of the b0 images is high and therefore the noise distribution is approximately Gaussian.
		Only signal with $S < 5\sigma$ goes through the subsequent debiasing steps.
		\item Apply a 4-D smoothing filter to estimate $E[S^2]$. Using each measurement in each voxel in turn as a reference, the filter searches within a block of $3\times3\times3$ neighborhood and across all gradient directions for all measurements that differ from the reference measurement by less than $\sqrt2 \sigma$. The filtered value is the average of all the measurements that fall within the threshold.
		\item Estimate the true signal $\hat{S}_{R} = \sqrt{E[S^2] - 2\sigma^2}$. 
		\item Following \citep{koay2009signal}, obtain the debiased Gaussian-distributed signal $\hat{S}_G$ via
		$
		\hat{S}_G = P^{-1}_G\left(P_{R}(S| \hat{S}_R, \sigma) \big | \hat{S}_R, \sigma \right)
		$, where $P^{-1}_G$ is the inverse cumulative distribution function of a Gaussian distribution and $P_R$ is the cumulative probability function of a Rician distribution. 
	\end{enumerate} 
	These steps do not involve solving nonlinear problems and are therefore very fast.

	\section{Experiments}

	\subsection{SMSI Settings}
	To cover the whole diffusion spectrum, one can set the diffusivity from $0\,\text{mm}^{2}/\text{s}$ (no diffusion) to $3 \times 10^{-3}\,\text{mm}^{2}/\text{s}$ (free diffusion). 
	However, the portion of the spectrum that is not biologically meaningful can be removed to reduce computational complexity.
	For the anisotropic compartment, we determined using SMT the range of axial diffusivity based on the body of the corpus callosum. 
	For this purpose, we used adult data from the Human Connectome Project (HCP) \citep{van2013wu} and infant data from the Baby Connectome Project (BCP) \citep{howell2018unc} and found that the effective range for $\lambda_{\parallel}$ is from $1.5 \times 10^{-3}\,\text{mm}^{2}/\text{s}$ to $2.0 \times 10^{-3}\, \text{mm}^{2}/\text{s}$. 
	Radial diffusivity
	$\lambda_{\bot}$ was then set to satisfy ${\lambda_{\parallel}}/{\lambda_{\bot}} \geq 1.1$, as in \citep{white2013probing}. 
	For the isotropic compartment, we set the diffusivity $\lambda_{\parallel}=\lambda_{\perp}$ from $0\,\text{mm}^{2}/\text{s}$ to $3 \times 10^{-3}\,\text{mm}^{2}/\text{s}$ with step size $0.1 \times 10^{-3}\,\text{mm}^{2}/\text{s}$.
	Regularization parameters were automatically selected from the interval of $[10^{-5},1]$ as described in Section~\ref{subsubsec:imple}. %

	\subsection{Effects of Orientation Heterogeneity and Isotropic Diffusion}
	
	Simulated diffusion data were used to investigate the effects of orientation heterogeneity and free-water diffusion.
	We used a model consisting of intra-cellular (IC), extra-cellular (EC), and cerebrospinal fluid (CSF) compartments \citep{zhang2012noddi} with normalized signal defined as
	\begin{equation}
	E = v_{\text{iso}}E_{\text{iso}} + (1-v_{\text{iso}})(v_{\text{ic}}E_{\text{ic}}+v_{\text{ec}}E_{\text{ec}}),
	\end{equation}
	where $v_{\text{iso}}$, $v_{\text{ic}}$, and $v_{\text{ec}} = 1 - v_{\text{ic}}$ are the volume fractions of the isotropic, intra-cellular, and extra-cellular compartments, respectively.
	$E_{\text{iso}}$, $E_{\text{ic}}$, and $E_{\text{ec}}$ are the normalized signals of these compartments.
	Each compartment was represented by a tensor model: Intra-cellular compartment with $\lambda_{\parallel} = 1.7 \times 10^{-3}\,{\text{mm}}^{2}/\text{s}$, $\lambda_{\perp}=0\,{\text{mm}}^{2}/\text{s}$; extra-cellular compartment with $\lambda_{\parallel} = 1.7 \times 10^{-3}\,{\text{mm}}^{2}/\text{s}$, $\lambda_{\perp}=0.435 \times 10^{-3}\,{\text{mm}}^{2}/\text{s}$; and the isotropic compartment with $\lambda_{\parallel}=\lambda_{\perp}=3.0 \times 10^{-3}\,{\text{mm}}^{2}/\text{s}$.
	Unless mentioned otherwise, the signal for each shell ($b=1000,\,2000,\,3000\,\text{s}/\text{mm}^2$) was generated with $90$ non-collinear gradient directions, identical to the HCP protocol \citep{van2013wu}.%
	
	\subsubsection{Orientation Heterogeneity}
	
	To demonstrate that SMSI can correctly infer microscopic diffusivity in the presence of orientation heterogeneity, we simulated the signal from micro-environments %
	oriented in 1 to 10 directions distributed equally over a sphere. 
	Rician noise with signal-to-noise ratio (SNR) of 20, typical for HCP and BCP data, were added. We then compared the microscopic diffusion indices computed based on SMSI and DTI. Note that in this experiment, we included only the extra-cellular compartment because it can be sufficiently represented using DTI.
	Additionally, we also validated SMSI results with simulations including both intra- and extra-cellular compartments, each has volume fraction of 0.5.
	
	\subsubsection{Isotropic Diffusion}
	Free-water diffusion can confound estimation of microstructure \citep{bergamino2016applying}, especially in the infant brain typically with high water content \citep{mukherjee2001normal,lovblad2003isotropic,singhi1995body,forbes2002changes}.
	To demonstrate that SMSI can accurately estimate microstructural properties in the presence of isotropic diffusion, we simulated the signal with intra-cellular, extra-cellular, and isotropic compartments with $v_{\text{ic}} = v_{\text{ec}} =0.5$ and $v_{\text{iso}}$ ranging from 0 to 0.9 in steps of 0.1. 
	Rician noise with SNR of 20 was added. We validated the effectiveness of SMSI via microscopic FA and MD as well as extra-cellular, intra-cellular, and isotropic volume fractions.
	SMSI was compared with SMT \citep{kaden2016quantitative}, multi-compartment SMT (MC-SMT) \citep{kaden2016multi}, and NODDI \citep{daducci2015accelerated}.
	
	\subsection{Microscopic Anisotropy and Orientation Coherence}
	
	We compared the MAI and OCI values given by SMSI, SMT, MC-SMT, and NODDI. MAI$^\dagger$ was used for both SMSI and NODDI since both models account for the isotropic volume fraction. MAI was used for SMT and MC-SMT. MAI and MAI$^\dagger$ were validated with respect to different isotropic volume fractions. 
	OCI is intrinsically robust to isotropic diffusion and is computed for micro-environments with increasing number of directions.

	\subsection{Number of $b$-Shells}
	We evaluated the minimal number of $b$-shells needed for effective SMSI estimation.
	We used a 21-shell data of a healthy adult with
	$b$-values ranging from $500 \, \text{s}/\text{mm}^2$ to $3000 \, \text{s}/\text{mm}^2$ with step size $125 \, \text{s}/\text{mm}^2$, acquired with non-collinear gradients,
	\ie, 4 diffusion-weighted (DW) images for $b=500\, \text{s}/\text{mm}^2$, 5 for $b=625\, \text{s}/\text{mm}^2$, $\hdots$, 23 for $b=2875\, \text{s}/\text{mm}^2$, and 24 for $b=3000\, \text{s}/\text{mm}^2$, in addition to 13 non-DW images, resulting in a total of 307 volumes. The images were acquired with an \num{140 x 140} imaging matrix, \SI{1.5 x 1.5 x 1.5}{\milli\meter} resolution, TE=\SI{89}{\milli\second}, TR=\SI{2513}{\milli\second}, and multi-band factor 5. 
	We applied SMSI to 
	\begin{enumerate}
		\item The 21-shell dataset consisting of all images;
		\item The 11-shell dataset with $b$-values from $500 \, \text{s}/\text{mm}^2$ to $3000 \, \text{s}/\text{mm}^2$ with step size $250 \, \text{s}/\text{mm}^2$;
		\item The 6-shell dataset with $b$-values from $500 \, \text{s}/\text{mm}^2$ to $3000 \, \text{s}/\text{mm}^2$ with step size $500 \, \text{s}/\text{mm}^2$;
		\item The 3-shell-1000 with $b$-values from $1000 \, \text{s}/\text{mm}^2$ to $3000 \, \text{s}/\text{mm}^2$ with step size $1000 \, \text{s}/\text{mm}^2$; and 
		\item The 3-shell-500 dataset with $b$-values from $500 \, \text{s}/\text{mm}^2$ to $2500 \, \text{s}/\text{mm}^2$ with step size $1000 \, \text{s}/\text{mm}^2$.
	\end{enumerate}
	The different sampling schemes were compared with the 21-shell dataset as the reference.

	\subsection{Longitudinal Infant Data}
	To demonstrate the effectiveness of SMSI in probing microstructural changes in the early developing human brain, we used the longitudinal datasets of two infants from the Baby Connectome Project (BCP) \citep{howell2018unc}. The first subject was scanned at 54, 146, and 223 days after birth and the second subject were scanned at 318, 410, and 514 days after birth. 
	The diffusion data were
	acquired using a Siemens \num{3}T Magnetom Prisma MRI scanner with the following protocol: \num{140 x 140} imaging matrix, \SI{1.5 x 1.5 x 1.5}{\milli\meter} resolution, TE=\SI{88}{\milli\second}, TR=\SI{2365}{\milli\second}, \num{32}-channel receiver coil, and multi-band factor \num{5}. DW images for \num{9}, \num{12}, \num{17}, \num{24}, \num{34}, and \num{48} non-collinear gradient directions were collected respectively for $b=500, 1000, 1500, 2000, 2500, 3000\,\text{s}/\text{mm}^2$. A non-DW image $b=0\,\text{s}/\text{mm}^2$ was collected for every \num{24} images, giving a total of \num{6}.
	Image reconstruction was performed using SENSE1 \citep{sotiropoulos2013effects}, resulting in non-stationary Rician noise distribution. 
	The magnitude signal was debiased as described in Section~\ref{subsubsec:debias}.
	Diffusion indices were compared between SMSI, SMT, MC-SMT, and NODDI.

	\section{Results}

	\subsection{Orientation Heterogeneity}
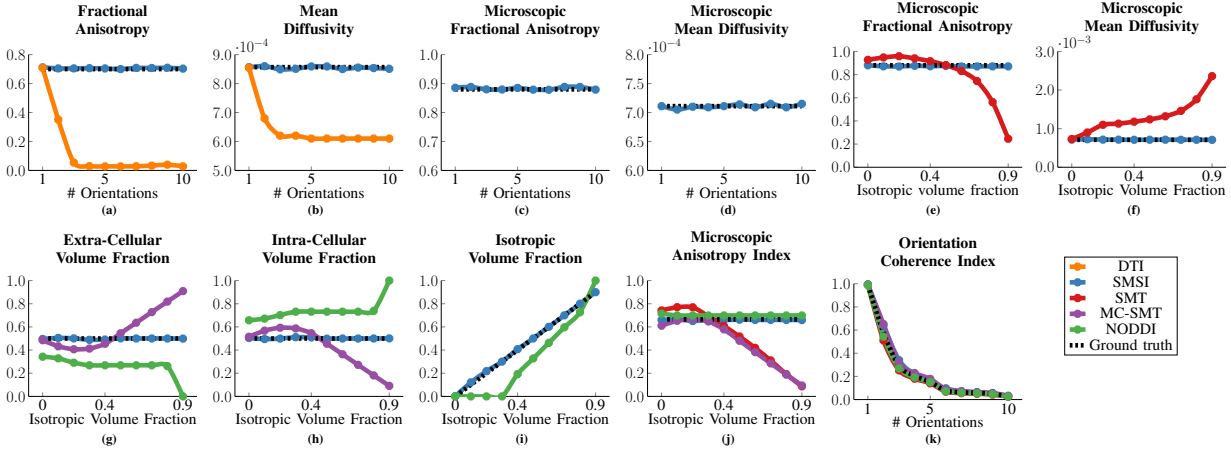
\begin{figure*}[t]
\definecolor{color1}{RGB}{55,126,184}
\definecolor{color2}{RGB}{215,26,28}
\definecolor{color3}{RGB}{77,175,74}
\definecolor{color4}{RGB}{152,78,163}
\definecolor{color5}{RGB}{255,127,0}
\pgfplotsset{every axis/.append style={
		align=center,
		label style={font=\LARGE},
		tick label style={font=\LARGE},
		title style={font=\bfseries\LARGE,yshift=2ex},
		legend style={font=\LARGE },
		width=8cm,height=6cm,
}}
	\newcommand{\myFontSize}[1]{
	\textcolor{black}{\textbf{\Large{#1}}}
}
\setlength{\tabcolsep}{0pt}
\resizebox{\linewidth}{!}
{ 
 	
\begin{tabular}{llllll}	
\begin{tikzpicture}[scale=1.0]
\begin{axis}[
  y tick label style={
	/pgf/number format/.cd,
	fixed,
	fixed zerofill,
	precision=1,
	/tikz/.cd
},
xlabel={\# Orientations},
ylabel=\empty,
ymin=0,ymax=0.8,
axis y line*=left, axis x line*=bottom,
xtick={1,5,10},
legend style={at={(1,0.6)},anchor=north east},
xticklabels={1,5,10},
title={Fractional \\ Anisotropy}
]

\addplot[smooth,mark=*,color=color1,line width=5pt] plot coordinates {
	(1,0.712)
(2,0.7046)
(3,0.70502)
(4,0.7063)
(5,0.70452)
(6,0.69998)
(7,0.70821)
(8,0.70743)
(9,0.70921)
(10,0.70314)

};

\addplot[smooth,color=color5,mark=*,line width=5pt]
plot coordinates {
	(1,0.708703231)
(2,0.3509303932)
(3,0.0541572)
(4,0.02936227)
(5,0.028)
(6,0.02782923)
(7,0.02868079)
(8,0.03289982)
(9,0.03972037)
(10,0.02831116)
};

\addplot[color=black,domain=1:10,dashed, line width =5pt] {0.7};

\end{axis}
\end{tikzpicture}
&
\begin{tikzpicture}[scale=1.0]
\begin{axis}[
  y tick label style={
	/pgf/number format/.cd,
	fixed,
	fixed zerofill,
	precision=1,
	/tikz/.cd
},
xlabel={\# Orientations},
ylabel=\empty,
ymin=0.0005,
ymax=0.0009,
axis y line*=left, axis x line*=bottom,
xtick={1,5,10},
legend style={at={(1,0.6)},anchor=north east},
xticklabels={1,5,10},
title={Mean \\ Diffusivity}
]
\addplot[smooth,mark=*,color=color1,line width=5pt] plot coordinates {
	(1,0.0008566)
(2,0.00086005)
(3,0.00084910)
(4,0.0008509)
(5,0.00085911)
(6,0.00085905)
(7,0.00085009)
(8,0.0008551)
(9,0.00085351)
(10,0.0008509)
};

\addplot[smooth,color=color5,mark=*,line width=5pt]
plot coordinates {
	(1,0.000855)
(2,0.00068)
(3,0.00062)
(4,0.00062)
(5,0.00061)
(6,0.00061)
(7,0.00061)
(8,0.00061)
(9,0.00061)
(10,0.00061)
};

\addplot[color=black,domain=1:10,dashed, line width =5pt] {0.00085667};

\end{axis}
\end{tikzpicture}
&
\begin{tikzpicture}[scale=1.0]
\begin{axis}[
  y tick label style={
	/pgf/number format/.cd,
	fixed,
	fixed zerofill,
	precision=1,
	/tikz/.cd
},
xlabel={\# Orientations},
ylabel=\empty,
ymin=0.6,ymax=1,
axis y line*=left, axis x line*=bottom,
xtick={1,5,10},
legend style={at={(1,0.5)},anchor=north east},
xticklabels={1,5,10},
title={Microscopic \\ Fractional Anisotropy}
]

\addplot[smooth,mark=*,color=color1,line width=5pt] plot coordinates {
	(1,0.8855)
	(2,0.8884)
	(3,0.8800)
	(4,0.8794)
	(5,0.8850)
	(6,0.8790)
	(7,0.8786)
	(8,0.8886)
	(9,0.8884)
	(10,0.8791)
	
};

\addplot[color=black,domain=1:10,dashed, line width =5pt] {0.88};

\end{axis}
\end{tikzpicture}
&
\begin{tikzpicture}[scale=1.0]
\begin{axis}[
  y tick label style={
	/pgf/number format/.cd,
	fixed,
	fixed zerofill,
	precision=1,
	/tikz/.cd
},
xlabel={\# Orientations},
ylabel=\empty,
ymin=0.0006,
ymax=0.0008,
axis y line*=left, axis x line*=bottom,
xtick={1,5,10},
legend style={at={(1,0.5)},anchor=north east},
xticklabels={1,5,10},
title={Microscopic \\ Mean Diffusivity}
]
\addplot[smooth,mark=*,color=color1,line width=5pt] plot coordinates {
	(1,0.000711)
	(2,0.000705)
	(3,0.000710)
	(4,0.000709)
	(5,0.000711)
	(6,0.000714)
	(7,0.000709)
	(8,0.000715)
	(9,0.000709)
	(10,0.000715)
};

\addplot[color=black,domain=1:10,dashed, line width =5pt] {0.000711};

\end{axis}
\end{tikzpicture}
&
\begin{tikzpicture}[scale=1.0]
\begin{axis}[
  y tick label style={
	/pgf/number format/.cd,
	fixed,
	fixed zerofill,
	precision=1,
	/tikz/.cd
},
xlabel={Isotropic volume fraction},
ylabel=\empty,
ymin=0,ymax=1,
axis y line*=left, axis x line*=bottom,
xtick={1,5,10},
legend style={at={(0,0)},anchor=south west},
xticklabels={0,0.4,0.9},
title={Microscopic \\ Fractional Anisotropy}
]

\addplot[smooth,mark=*,color=color1,line width=5pt] plot coordinates {
	(1,0.880322)
	(2,0.871331)
	(3,0.870047)
	(4,0.875910)
	(5,0.872211)
	(6,0.875113)
	(7,0.870011)
	(8,0.87076)
	(9,0.87083)
	(10,0.87091)
	
};

\addplot[smooth,color=color2,mark=*,line width=5pt]
plot coordinates {
	(1,0.92737)
	(2,0.94822)
	(3,0.95854)
	(4,0.94016)
	(5,0.91584)
	(6,0.88198)
	(7,0.83118)
	(8,0.74460)
	(9,0.56207)
	(10,0.24611)
};

\addplot[color=black,domain=1:10,dashed, line width=5pt] {0.88};
\end{axis}
\end{tikzpicture}
&
\begin{tikzpicture}[scale=1.0]
\begin{axis}[
  y tick label style={
	/pgf/number format/.cd,
	fixed,
	fixed zerofill,
	precision=1,
	/tikz/.cd
},
xlabel={Isotropic Volume Fraction},
ylabel=\empty,
ymin=0,ymax=0.003,
axis y line*=left, axis x line*=bottom,
xtick={1,5,10},
legend style={at={(0,1)},anchor=north west},
xticklabels={0,0.4,0.9},
title={Microscopic \\ Mean Diffusivity}
]
\addplot[smooth,mark=*,color=color1,line width=5pt] plot coordinates {
	(1,0.00071658)
	(2,0.0007252)
	(3,0.0007150)
	(4,0.00071451)
	(5,0.00071153)
	(6,0.00071252)
	(7,0.00071352)
	(8,0.00071153)
	(9,0.00071153)
	(10,0.00070954)
	
};

\addplot[smooth,color=color2,mark=*,line width=5pt]
plot coordinates {
	(1,0.00073)
	(2,0.00090)
	(3,0.00110)
	(4,0.00113)
	(5,0.00118)
	(6,0.00124)
	(7,0.00132)
	(8,0.00146)
	(9,0.00176)
	(10,0.00236)
};
\addplot[color=black,domain=1:10,dashed, line width =5pt] {0.000711};

\end{axis}
\end{tikzpicture}
\\
\multicolumn{1}{c}{\myFontSize{\quad\quad(a)}}&\multicolumn{1}{c}{\myFontSize{\quad\quad(b)}}&\multicolumn{1}{c}{\myFontSize{\quad\quad(c)}}&\multicolumn{1}{c}{\myFontSize{\quad\quad(d)}}&\multicolumn{1}{c}{\myFontSize{\quad\quad(e)}}&\multicolumn{1}{c}{\myFontSize{\quad\quad(f)}}
\quad
\\
\quad
\\

\begin{tikzpicture}[scale=1.0]
\begin{axis}[
  y tick label style={
	/pgf/number format/.cd,
	fixed,
	fixed zerofill,
	precision=1,
	/tikz/.cd
},
xlabel={Isotropic Volume Fraction},
ylabel=\empty,
ymin=0,ymax=1,
axis y line*=left, axis x line*=bottom,
xtick={1,5,10},
legend style={at={(0,1)},anchor=north west},
xticklabels={0,0.4,0.9},
title={Extra-Cellular \\ Volume Fraction}
]
\addplot[smooth,mark=*,color=color1,line width =5pt] plot coordinates {
	(1,0.49582)
	(2,0.50500)
	(3,0.50081)
	(4,0.48835)
	(5,0.49091)
	(6,0.50142)
	(7,0.50012)
	(8,0.50029)
	(9,0.49993)
	(10,0.49899)
};

\addplot[smooth,color=color4,mark=*,line width =5pt]
plot coordinates {
	(1,0.48518)
	(2,0.43245)
	(3,0.40713)
	(4,0.41301)
	(5,0.45453)
	(6,0.54483)
	(7,0.63615)
	(8,0.72759)
	(9,0.81866)
	(10,0.90911)
};

\addplot[smooth,color=color3,mark=*,line width =5pt]
plot coordinates {
	(1,0.34273)
	(2,0.32860)
	(3,0.29082)
	(4,0.26868)
	(5,0.26868)
	(6,0.26868)
	(7,0.26868)
	(8,0.26868)
	(9,0.26182)
	(10,0.0)
};

\addplot[color=black,domain=1:10,dashed, line width =5pt] {0.5};

\end{axis}
\end{tikzpicture}
&
\begin{tikzpicture}[scale=1.0]
\begin{axis}[
  y tick label style={
	/pgf/number format/.cd,
	fixed,
	fixed zerofill,
	precision=1,
	/tikz/.cd
},
xlabel={Isotropic Volume Fraction},
ylabel=\empty,
ymin=0,ymax=1,
axis y line*=left, axis x line*=bottom,
xtick={1,5,10},
legend style={at={(0,0)},anchor=south west},
xticklabels={0,0.4,0.9},
title={Intra-Cellular \\ Volume Fraction}
]
\addplot[smooth,mark=*,color=color1,line width =5pt] plot coordinates {
	(1,0.50418)
	(2,0.49500)
	(3,0.49919)
	(4,0.51165)
	(5,0.50909)
	(6,0.49858)
	(7,0.49988)
	(8,0.49971)
	(9,0.50007)
	(10,0.50101)
	
};

\addplot[smooth,color=color4,mark=*,line width =5pt]
plot coordinates {
	(1,0.51482)
	(2,0.56755)
	(3,0.59287)
	(4,0.58699)
	(5,0.54547)
	(6,0.45517)
	(7,0.36385)
	(8,0.27241)
	(9,0.18134)
	(10,0.09089)
};

\addplot[smooth,color=color3,mark=*,line width =5pt]
plot coordinates {
	(1,0.65727)
	(2,0.67140)
	(3,0.70198)
	(4,0.73132)
	(5,0.73132)
	(6,0.73132)
	(7,0.73132)
	(8,0.73132)
	(9,0.73818)
	(10,1)
};

\addplot[color=black,domain=1:10,dashed, line width =5pt] {0.5};
\end{axis}
\end{tikzpicture}
&
\begin{tikzpicture}[scale=1.0]
\begin{axis}[
  y tick label style={
	/pgf/number format/.cd,
	fixed,
	fixed zerofill,
	precision=1,
	/tikz/.cd
},
xlabel={Isotropic Volume Fraction},
ylabel=\empty,
ymin=0,ymax=1,
axis y line*=left, axis x line*=bottom,
xtick={1,5,10},
legend style={at={(0,1)},anchor=north west},
xticklabels={0,0.4,0.9},
title={Isotropic \\ Volume Fraction}
]
\addplot[smooth,mark=*,color=color1,line width =5pt] plot coordinates {
	(1,0.00135)
	(2,0.12092)
	(3,0.21781)
	(4,0.30051)
	(5,0.40812)
	(6,0.50009)
	(7,0.60052)
	(8,0.70000)
	(9,0.80000)
	(10,0.90000)
};

\addplot[smooth,color=color3,mark=*,line width =5pt]
plot coordinates {
	(1,0.0)
	(2,0.0)
	(3,0.0)
	(4,0.0)
	(5,0.19276)
	(6,0.32759)
	(7,0.46230)
	(8,0.59690)
	(9,0.72617)
	(10,1)
};

\addplot[color=black,dashed, line width =5pt]
plot coordinates {
	(1,0.0)
	(2,0.1)
	(3,0.2)
	(4,0.3)
	(5,0.4)
	(6,0.5)
	(7,0.6)
	(8,0.7)
	(9,0.8)
	(10,0.9)
};
\end{axis}
\end{tikzpicture}
&
\begin{tikzpicture}[scale=1.0]
\begin{axis}[
  y tick label style={
	/pgf/number format/.cd,
	fixed,
	fixed zerofill,
	precision=1,
	/tikz/.cd
},
xlabel={Isotropic Volume Fraction},
ylabel=\empty,
ymin=0,ymax=1,
axis y line*=left, axis x line*=bottom,
xtick={1,5,10},
legend style={at={(0,0)},anchor=south west},
xticklabels={0,0.4,0.9},
title={Microscopic \\ Anisotropy Index}
]
\addplot[smooth,mark=*,color=color1,line width =5pt] plot coordinates {
	(1,0.6601)
	(2,0.665500)
	(3,0.65081)
	(4,0.658835)
	(5,0.659091)
	(6,0.66242)
	(7,0.66012)
	(8,0.66029)
	(9,0.65993)
	(10,0.65899)
};

\addplot[smooth,mark=*,color=color2,line width =5pt] plot coordinates {
	(1,0.74202)
	(2,0.77032)
	(3,0.76933)
	(4,0.6913)
	(5,0.6076)
	(6,0.51706)
	(7,0.41875)
	(8,0.31041)
	(9,0.1915)
	(10,0.08703)
};

\addplot[smooth,color=color4,mark=*,line width =5pt]
plot coordinates {
	(1,0.61049)
	(2,0.65274)
	(3,0.66605)
	(4,0.64737)
	(5,0.57891)
	(6,0.4804)
	(7,0.38244)
	(8,0.28511)
	(9,0.1892)
	(10,0.09453)
};

\addplot[smooth,color=color3,mark=*,line width =5pt]
plot coordinates {
	(1,0.71474)
	(2,0.6987)
	(3,0.69882)
	(4,0.6988)
	(5,0.69879)
	(6,0.69878)
	(7,0.69877)
	(8,0.69876)
	(9,0.69880)
	(10,0.69879)
};

\addplot[color=black,domain=1:10,dashed, line width =5pt] {0.66501};

\end{axis}
\end{tikzpicture}
&
\begin{tikzpicture}[scale=1.0]
\begin{axis}[
y tick label style={
	/pgf/number format/.cd,
	fixed,
	fixed zerofill,
	precision=1,
	/tikz/.cd
},
xlabel={\# Orientations},
ylabel=\empty,
ymin=0,ymax=1,
axis y line*=left, axis x line*=bottom,
xtick={1,5,10},
legend style={at={(0,0)},anchor=south west},
xticklabels={1,5,10},
title={Orientation \\ Coherence Index}
]
\addplot[smooth,mark=*,color=color1,line width =5pt] plot coordinates {
	(1,0.9925)
	(2,0.63941)
	(3,0.34056)
	(4,0.21157)
	(5,0.1497)
	(6,0.0972)
	(7,0.069083)
	(8,0.06412)
	(9,0.05393)
	(10,0.03073)
	
};

\addplot[smooth,color=color2,mark=*,line width =5pt]
plot coordinates {
	(1,0.99)
	(2,0.5107)
	(3,0.25229)
	(4,0.18018)
	(5,0.14063)
	(6,0.06803)
	(7,0.05525)
	(8,0.05022)
	(9,0.04298)
	(10,0.02476)
};

\addplot[smooth,color=color4,mark=*,line width =5pt]
plot coordinates {
	(1,0.9951482)
	(2,0.65212)
	(3,0.31928)
	(4,0.22889)
	(5,0.1781)
	(6,0.08993)
	(7,0.07305)
	(8,0.06333)
	(9,0.05664)
	(10,0.03272)
};

\addplot[smooth,color=color3,mark=*,line width =5pt]
plot coordinates {
	(1,0.99)
	(2,0.54361)
	(3,0.26878)
	(4,0.19127)
	(5,0.14999)
	(6,0.072)
	(7,0.05849)
	(8,0.05371)
	(9,0.04539)
	(10,0.0262)
};

\addplot[smooth,color=black,dashed, line width =5pt]
plot coordinates {
	(1,0.99)
	(2,0.57845)
	(3,0.28832)
	(4,0.20501)
	(5,0.15067)
	(6,0.07723)
	(7,0.06274)
	(8,0.05761)
	(9,0.04864)
	(10,0.0281)
};
\end{axis}
\end{tikzpicture}
&
\begin{tikzpicture} 
\begin{axis}[%
hide axis,
xmin=10,
xmax=50,
ymin=0,
ymax=0.4,
legend style={at={(0.25,1.5)},anchor=north west},
]
\addlegendimage{color5,mark=*,line width =5pt}
\addlegendentry{DTI};
\addlegendimage{color1,mark=*,line width =5pt}
\addlegendentry{SMSI};
\addlegendimage{color2,mark=*,line width =5pt}
\addlegendentry{SMT};
\addlegendimage{color4,mark=*,line width =5pt}
\addlegendentry{MC-SMT};
\addlegendimage{color3,mark=*,line width =5pt}
\addlegendentry{NODDI};
\addlegendimage{color=black,dashed,line width =5pt}
\addlegendentry{Ground truth};
\end{axis}
\end{tikzpicture}
\\
\multicolumn{1}{c}{\myFontSize{\quad\quad(g)}}&\multicolumn{1}{c}{\myFontSize{\quad\quad(h)}}&\multicolumn{1}{c}{\myFontSize{\quad\quad(i)}}&\multicolumn{1}{c}{\myFontSize{\quad\quad(j)}}&\multicolumn{1}{c}{\myFontSize{\quad\quad(k)}}

\end{tabular}
}
\caption{
\textbf{Numerical Validations.}
Comparison of SMSI with DTI, SMT, MC-SMT, and NODDI.
(a) and (b): DTI FA and MD and SMSI $\mu\text{FA}$ and $\mu\text{MD}$ with respect to the number of crossing fibers.
(c) and (d): SMSI $\mu\text{FA}$ and $\mu\text{MD}$ with respect to orientation heterogeneity (with multiple compartments).
(e) and (f): SMT $\mu$FA and $\mu$MD and SMSI $\mu$FA$^{\dagger}$ and $\mu$MD$^{\dagger}$ with respect to isotropic volume fraction.
(g) and (h): Estimates of $v_{\text{ec}}$ and $v_{\text{ic}}$ given by SMSI, MC-SMT, and NODDI with respect to isotropic volume fraction.
(i): Estimates of $v_{\text{iso}}$ given by SMSI and NODDI with respect to isotropic volume fraction.
(j) and (k): Microscopic anisotropy index (MAI) with respect to isotropic volume fraction and orientation coherence index (OCI) with respect to the number of orientations given by SMSI, SMT, MC-SMT, and NODDI. MAI$^{\dagger}$ was calculated for SMSI and NODDI.
Values shown are the means of \num{1000} repetitions. Standard deviations are negligible and hence not displayed.
\label{fig:syn_combine}}
	\vspace{-1.0pt}
\end{figure*}

\begin{figure}[t]
\setlength{\iwidth}{0.3\textwidth}
\setlength{\iheight}{\iwidth}
\setlength{\tabcolsep}{0pt}
\setlength{\rowspacing}{-3pt}
\setlength{\colorbarlength}{0.5\iwidth}

\newcommand{\myFontSize}{\scriptsize}
\def \filePath {\figDir/uFAuMDvsCrossingFiber}
\renewcommand{\arraystretch}{1.5}
	
\newcommand\myPic[7]{
	\begin{tikzpicture}[baseline,trim left]
	\draw (0.0\iwidth, 0.0\iheight)[#5, ultra thick] rectangle(\iwidth, \iheight);
	#6 %
	\node[mylabel, right=0pt] at (0.0\iwidth, 1.0\iheight) {\textbf{\large{#7}}}; %
	
	\begin{pgfonlayer}{background}
	\fill (0.0\iwidth, 0.0\iheight)[black, ultra thick] rectangle(\iwidth, \iheight);
	\clip (0.0\iwidth, 0.0\iheight) rectangle (\iwidth, \iheight);
	\node at (#3\iwidth, #4\iheight) { \includegraphics[width=#2\iwidth, angle=0]{\filePath/#1} };
	\end{pgfonlayer}
	\end{tikzpicture}
}

\newcommand\columnText[1]{
	\begin{tikzpicture}[baseline,trim left]
		\node[rotate=90] at (0.05\iwidth, 0.5\iheight) {\myFontSize{#1}};
	\end{tikzpicture}
}

\newcommand\colorBar[5]{		
	\begin{tikzpicture}
	\node (bar) at (#3\iwidth, #4\iheight) { \includegraphics[width=0.16\colorbarlength,height=1.5\colorbarlength]{\filePath/#5} };
	\node [below=-1mm of bar] {\myFontSize{#1}};
	\node [above=-1mm of bar] {\myFontSize{#2}};
	\end{tikzpicture}
}

\newcommand\voidAnn{}

\centering
\resizebox{\linewidth}{!}
{
\begin{tabular}{c@{~~~~~~~~~~~~~~~}cc}
{\Large \textbf{DTI FA}} & {\Large \textbf{SMSI $\boldsymbol\mu$FA}} & {\Large \textbf{SMSI OCI}}\\
\myPic{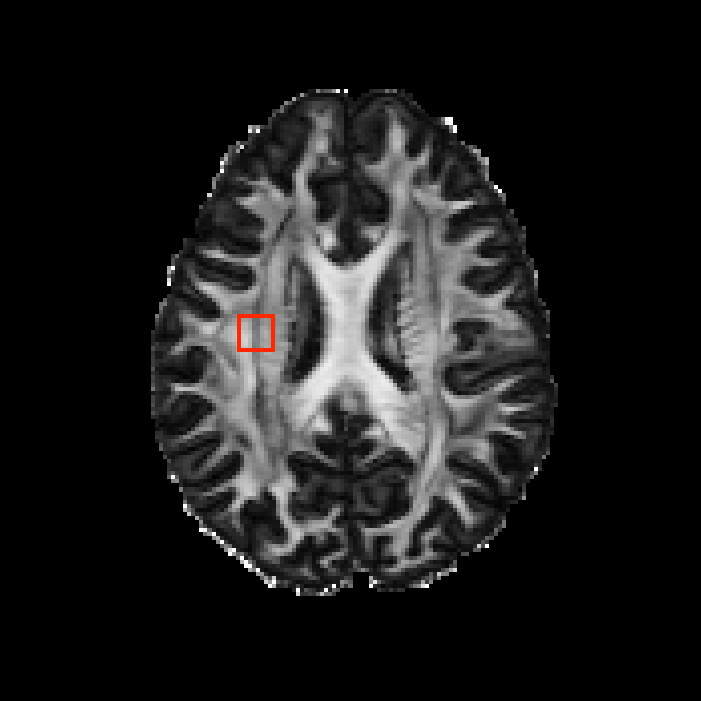}{1.2}{0.5}{0.5}{black}{\voidAnn}{} & \myPic{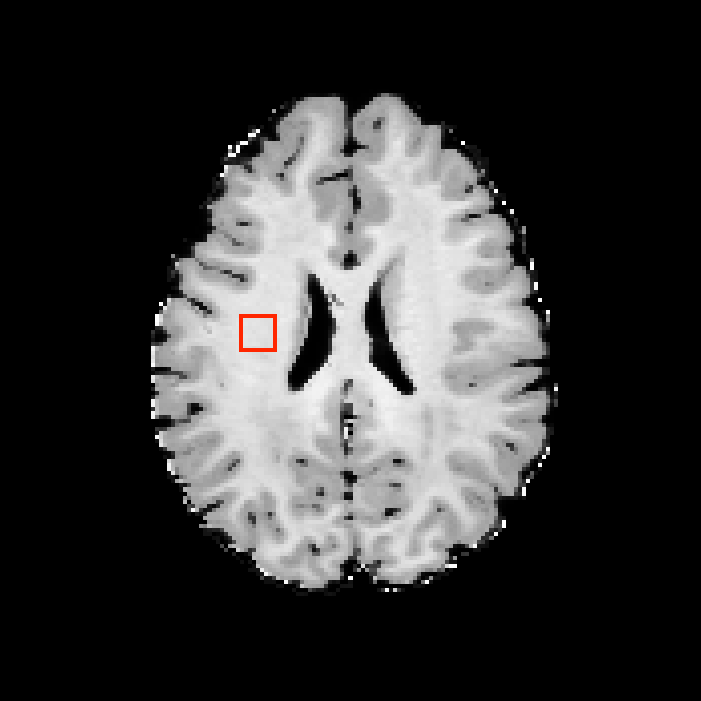}{1.2}{0.5}{0.5}{black}{\voidAnn}{} & \myPic{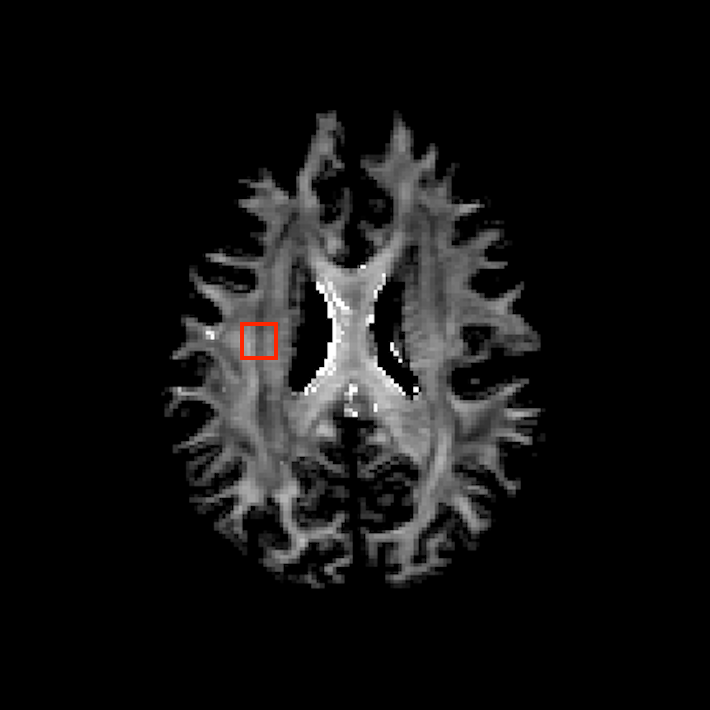}{1.2}{0.5}{0.5}{black}{\voidAnn}{}\\
\myPic{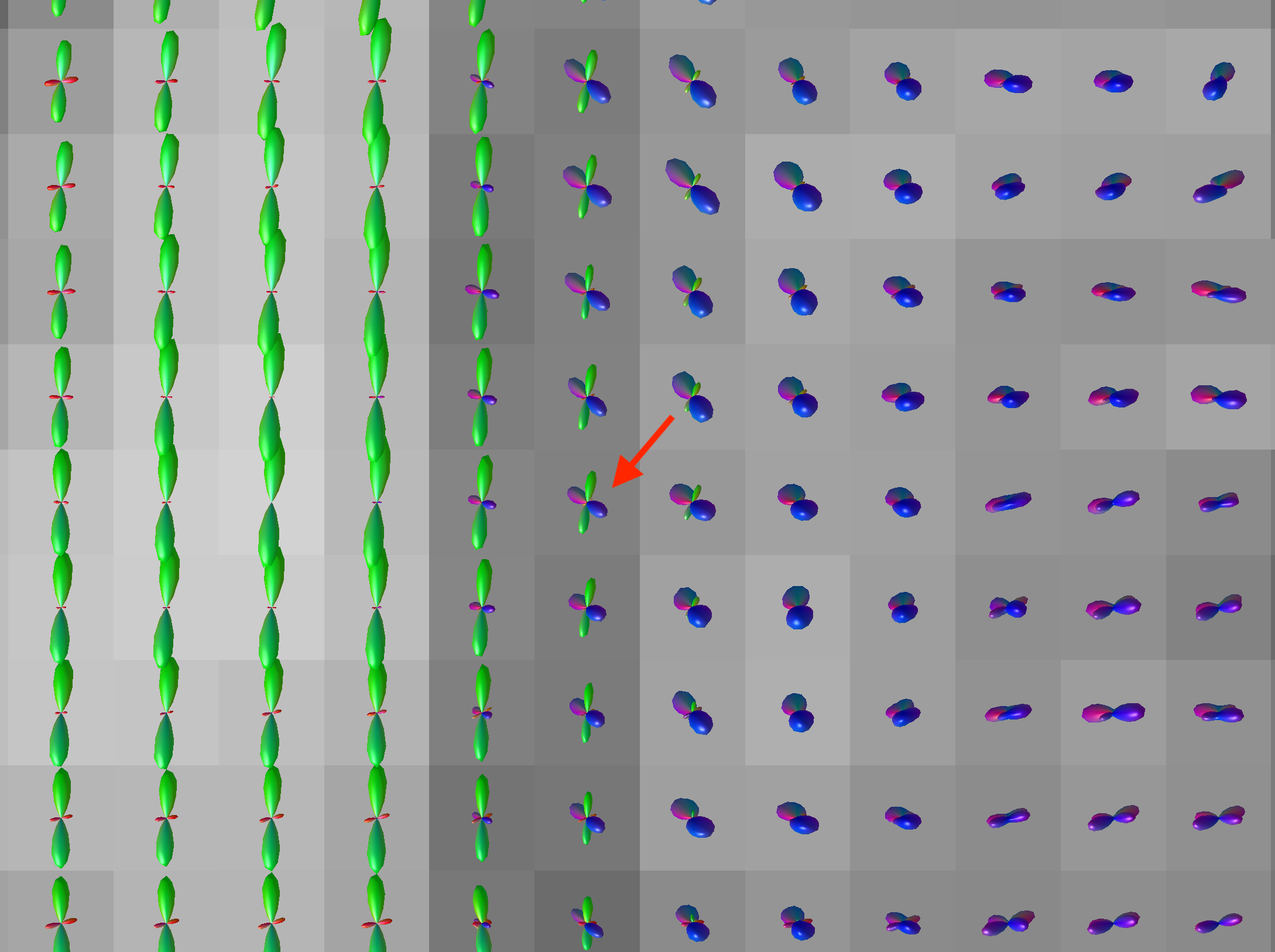}{1.8}{0.6}{0.5}{black}{\voidAnn}{} & \myPic{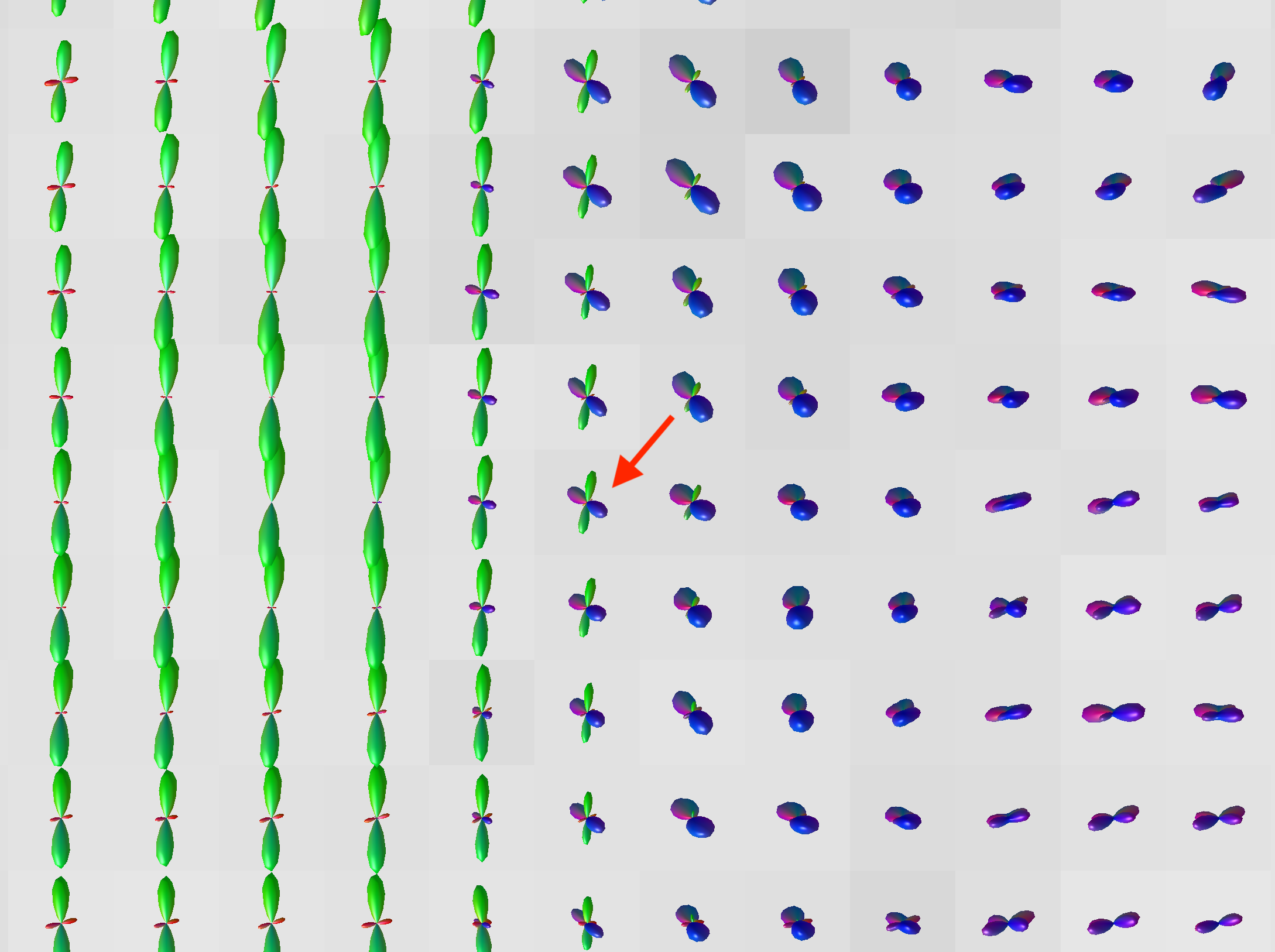}{1.8}{0.6}{0.5}{black}{\voidAnn}{} & \myPic{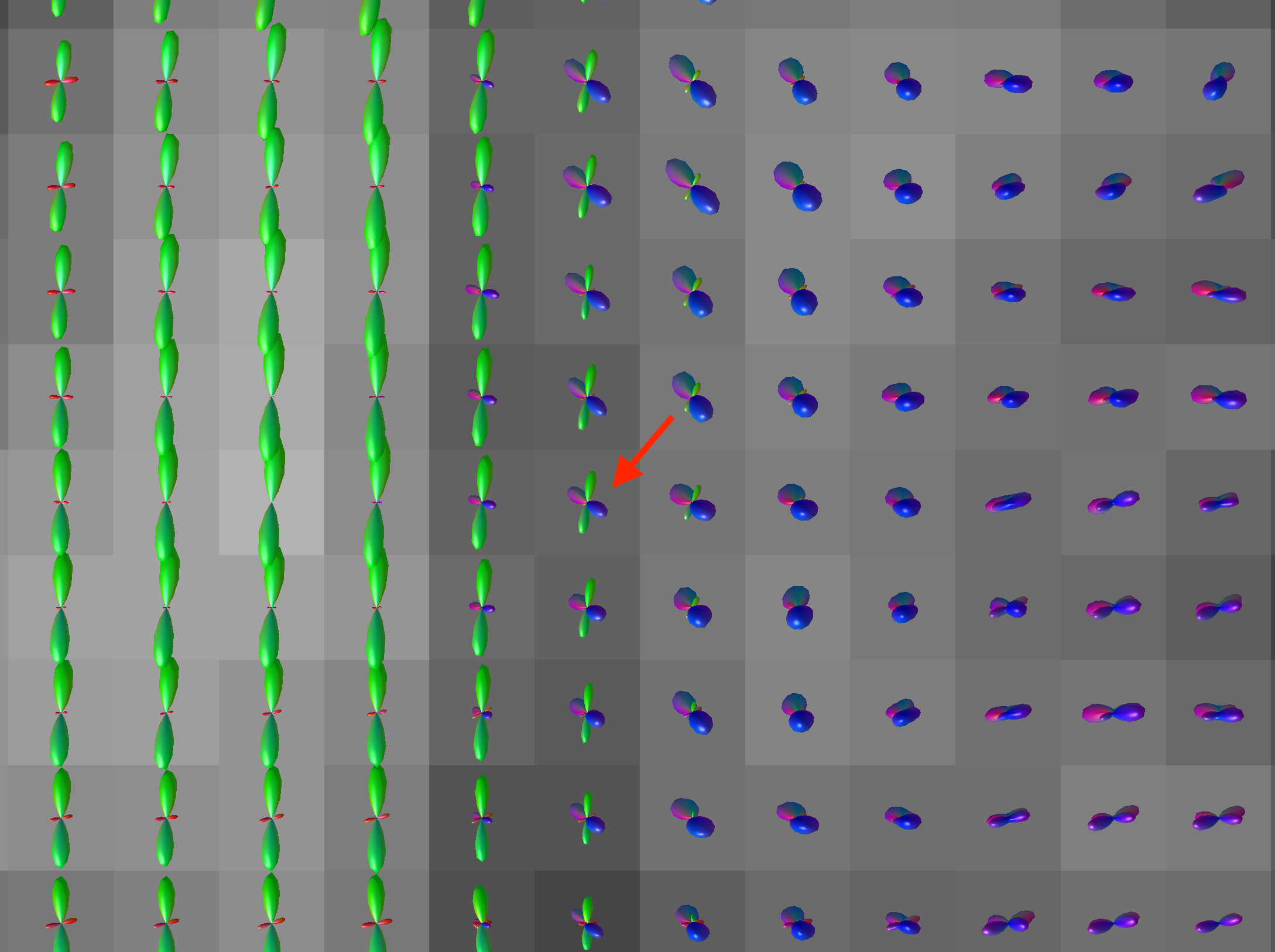}{1.8}{0.6}{0.5}{black}{\voidAnn}{}\\

\end{tabular}
}
\caption{\textbf{Voxel and Microscopic FA.} 
Top: DTI FA, SMSI $\mu$FA, and SMSI OCI. Bottom: Close-up view with fiber ODF overlaid. Red arrows mark the region with crossing fibers.}
\label{fig:odf}
\end{figure}

	From Fig.~\ref{fig:syn_combine} (a) and (b), one can appreciate that DTI FA and MD decrease with the number of orientations whereas SMSI $\mu$FA and $\mu$MD remain consistent.
	Similarly, Fig.~\ref{fig:syn_combine} (c) and (d) confirm the robustness of SMSI to orientation heterogeneity in case of multiple compartments.
	Fig.~\ref{fig:odf} shows FA (top left) and $\mu$FA (top right) of a representative HCP subject.
	DTI FA results in a dark band due to lower anisotropy caused by fiber crossings.
	SMSI $\mu$FA reveals the true anisotropy unconfounded by fiber dispersion. SMSI OCI quantifies orientation dispersion.
	A close-up view of a region with three-way crossings as shown by the fiber orientation distribution functions (ODFs) confirms this observation.
	

	\subsection{Isotropic Diffusion}
	
	\subsubsection{Microscopic FA and MD}
	Fig.~\ref{fig:syn_combine} (e) and (f) show the microstructural properties estimated using SMT and SMSI.
	The SMT model is a single-compartment model and does not account for isotropic diffusion. 
	Hence, SMT $\mu$FA and $\mu$MD are significantly affected by the isotropic volume fraction. 
	Note that even when the isotropic volume fraction is low, the results given by SMT, unlike SMSI, deviate from  the ground truth. 
	SMSI $\mu$FA$^{\dagger}$ and $\mu$MD$^{\dagger}$ are robust to isotropic diffusion.
	
	\subsubsection{Extra- and Intra-Cellular Volume Fractions}
	Fig.~\ref{fig:syn_combine} (g) and (h) show that NODDI underestimates the extra-cellular volume fraction and overestimates the intra-cellular volume fraction for all isotropic volume fractions. The bias is due to the fixed intrinsic parallel diffusivity assumption in the NODDI implementation \citep{guerrero2019optimizing}.
	MC-SMT produces correct estimates when the isotropic volume fraction is 0. 
	However, when isotropic volume fraction increases, MC-SMT fails to yield accurate results as it does not account for isotropic diffusion and its tortuosity assumption on the extra-cellular radial diffusivity \citep{kaden2016multi}. 
	SMSI gives correct and consistent results. 
	Notice that estimation bias occurs even when the isotropic volume fraction is small. 
	We will show that for in vivo data MC-SMT and NODDI exhibit similar bias in underestimating the extra-cellular volume fraction and overestimating the intra-cellular volume fraction.

	\subsubsection{Isotropic Diffusion Estimation}
	Fig.~\ref{fig:syn_combine} (i) shows that SMSI yields accurate estimates of the isotropic volume fraction, which 
	NODDI however tends to underestimate, especially when the actual value is less then 0.3. 
	
	\subsection{Microscopic Anisotropy and Orientation Coherence}
	Fig.~\ref{fig:syn_combine} (j) shows the MAI values given by SMSI, SMT, MC-SMT, and NODDI. 
	MAI$^{\dagger}$ was computed for SMSI and NODDI since they explicitly considers isotropic diffusion.
	Similar to the trend of $\mu$FA, SMT overestimates/underestimates MAI when the isotropic volume fraction is low/high. 
	MC-SMT exhibits a similar trend but the bias is smaller thanks to the two-compartment model. 
	NODDI is more stable but introduces a systematic bias across isotropic volume fractions. 
	SMSI yields results that are close to the ground truth.
	Fig.~\ref{fig:syn_combine} (k) shows that all methods produce OCI values that are close to the ground truth and decrease with increasing number of orientations.

\begin{figure}[t]
	\newcommand\myPic[7]{
		\begin{tikzpicture}[baseline,trim left]
		\draw (0.0\iwidth, 0.0\iheight)[#5, ultra thick] rectangle(\iwidth, \iheight);
		#6 %
		\node[mylabel, right=0pt] at (0.0\iwidth, 1.0\iheight) {\textbf{\large{#7}}}; %
		
		\begin{pgfonlayer}{background}
		\fill (0.0\iwidth, 0.0\iheight)[black, ultra thick] rectangle(\iwidth, \iheight);
		\clip (0.0\iwidth, 0.0\iheight) rectangle (\iwidth, \iheight);
		\node at (#3\iwidth, #4\iheight) { \includegraphics[width=#2\iwidth, angle=0]{\filePath/#1} };
		\end{pgfonlayer}
		\end{tikzpicture}
	}
	
	\newcommand\columnText[1]{
		\begin{tikzpicture}[baseline,trim left]
		\node[rotate=90] at (0.1\iwidth, 0.5\iheight) {\myFontSize{#1}};
		\end{tikzpicture}
	}
	
\newcommand\colorBar[5]{		
	\begin{tikzpicture}
	\node (bar) at (#3\iwidth, #4\iheight) { \includegraphics[width=0.16\colorbarlength,height=1.5\colorbarlength]{\filePath/#5} };
	\node [below=-1mm of bar] {\myFontSize{#1}};
	\node [above=-1mm of bar] {\myFontSize{#2}};
	\end{tikzpicture}
}
\mdfdefinestyle{mdfexample1}{leftmargin=1.5cm,rightmargin=1.5cm}
	
	\begin{mdframed}[backgroundcolor=black,style=mdfexample1]
	\newcommand\voidAnn{}

\setlength{\iwidth}{0.2\textwidth}
\setlength{\iheight}{\iwidth}
\setlength{\tabcolsep}{-2pt}
\setlength{\rowspacing}{0pt}
\setlength{\colorbarlength}{0.2\iwidth}
\newcommand{\myFontSize}[1]{
	\textcolor{white}{\textbf{\scriptsize{#1}}}
}
\def \filePath {./Figures/PAAOCI}

\renewcommand{\arraystretch}{1.25}
\renewcommand{\arrayrulewidth}{0pt}
\centering

\resizebox{\linewidth}{!}
{

\begin{tabular}{lccccc}

& \myFontSize{SMSI} & \myFontSize{SMT} & \myFontSize{MC-SMT} & \myFontSize{NODDI} \\

\columnText{OCI} & \myPic{SMSI_AOCI_clean.png}{1.35}{0.5}{0.48}{black}{\voidAnn}{} & \myPic{SMT_OCI_clean.png}{1.35}{0.5}{0.48}{black}{\voidAnn}{} & \myPic{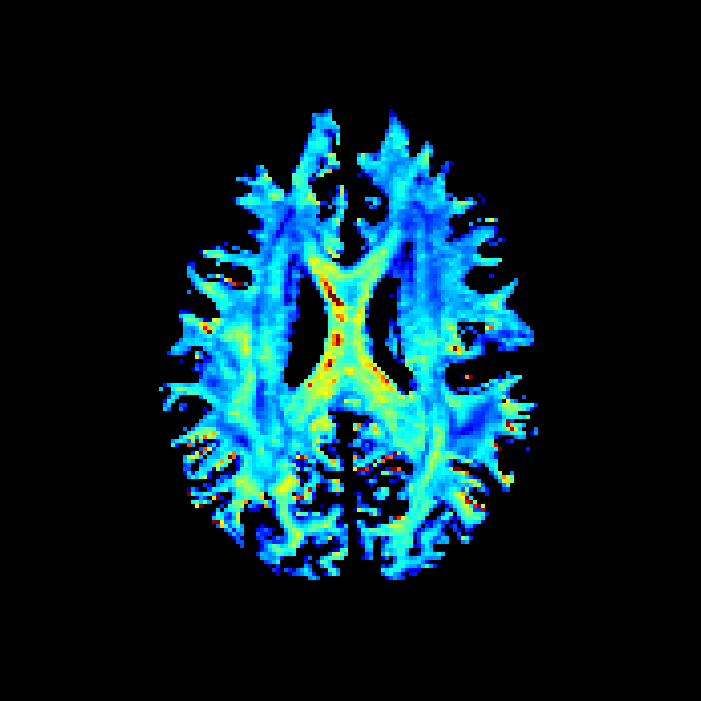}{1.35}{0.5}{0.48}{black}{\voidAnn}{} & \myPic{NODDI_OCI_clean.png}{1.35}{0.5}{0.48}{black}{\voidAnn}{} & \colorBar{\tiny0}{\tiny1}{0.5}{0.5}{colorbar}\\
[-0.275em]
\columnText{MAI} & \myPic{SMSI_MAI_Norm.png}{1.35}{0.5}{0.48}{black}{\voidAnn}{} & \myPic{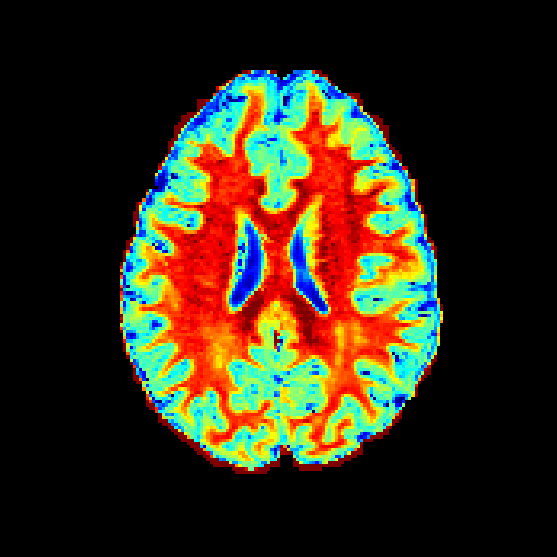}{1.35}{0.5}{0.48}{black}{\voidAnn}{} & \myPic{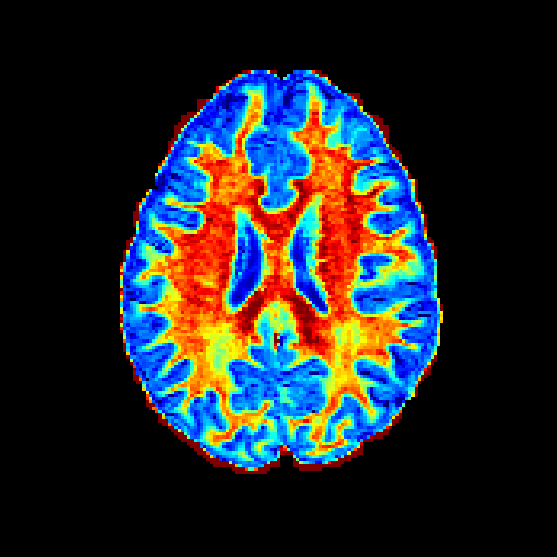}{1.35}{0.5}{0.48}{black}{\voidAnn}{} & \myPic{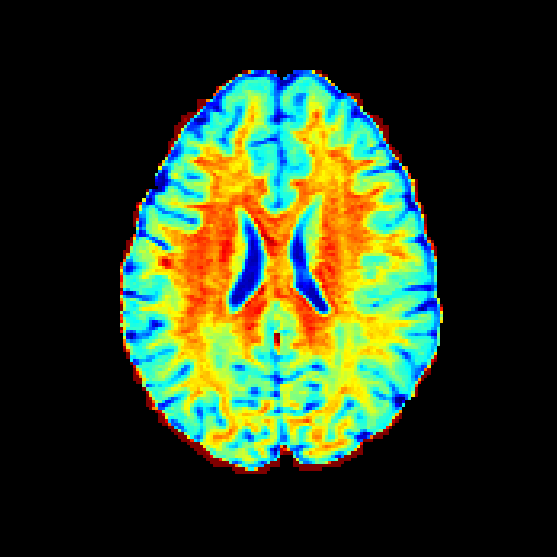}{1.35}{0.5}{0.48}{black}{\voidAnn}{} & \colorBar{\tiny0}{\tiny1}{0.5}{0.5}{colorbar}\\
[-0.3em]
\columnText{MAI$^{\dagger}$} & \myPic{SMSI_AMAI_Norm.png}{1.35}{0.5}{0.48}{black}{\voidAnn}{} & \myPic{black.png}{1.35}{0.5}{0.48}{black}{\voidAnn}{} & \myPic{black.png}{1.35}{0.5}{0.48}{black}{\voidAnn}{} & \myPic{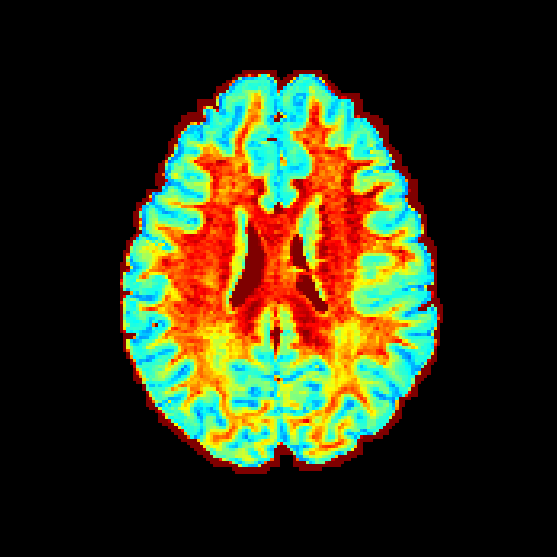}{1.35}{0.5}{0.48}{black}{\voidAnn}{} & \colorBar{\tiny0}{\tiny1}{0.5}{0.5}{colorbar}\\
\end{tabular}
	
}
\end{mdframed}
\caption{
	\textbf{Microscopic Anisotropy and Orientation Coherence.}
	Microscopic anisotropy (MAI) and orientation coherence index (OCI) maps given by SMSI, SMT, MC-SMT, and NODDI. MAI$^{\dagger}$ is computed only for SMSI and NODDI. A subject from the HCP was used.
}
\label{fig:PAAOCI}
\end{figure}

	Fig.~\ref{fig:PAAOCI} shows similar trends for in vivo data. 
	In white matter where isotropic volume fraction is low, SMT and MC-SMT yield significantly higher MAI values than SMSI and NODDI. The MAI$^{\dagger}$ values given by SMSI and NODDI in superficial white matter are higher as both methods eliminate the isotropic diffusion contamination. NODDI returns slightly higher MAI$^{\dagger}$ than SMSI. OCI values, on the other hand, are almost similar for all methods, with SMT giving slightly lower values. All these observations are consistent with Fig.~\ref{fig:syn_combine} (j) and (k).
	
	\subsection{Diffusion Indices}
\begin{figure}[t]
	\newcommand\myPic[7]{
		\begin{tikzpicture}[baseline,trim left]
		\draw (0.0\iwidth, 0.0\iheight)[#5, ultra thick] rectangle(\iwidth, \iheight);
		#6 %
		\node[mylabel, right=0pt] at (0.0\iwidth, 1.0\iheight) {\textbf{\large{#7}}}; %
		
		\begin{pgfonlayer}{background}
		\fill (0.0\iwidth, 0.0\iheight)[black, ultra thick] rectangle(\iwidth, \iheight);
		\clip (0.0\iwidth, 0.0\iheight) rectangle (\iwidth, \iheight);
		\node at (#3\iwidth, #4\iheight) { \includegraphics[width=#2\iwidth, angle=0]{\filePath/#1} };
		\end{pgfonlayer}
		\end{tikzpicture}
	}
	
	\newcommand\columnText[1]{
		\begin{tikzpicture}[baseline,trim left]
		\node[rotate=90] at (0.5\iwidth, 0.5\iheight) {\myFontSize{#1}};
		\end{tikzpicture}
	}
	
	\newcommand\colorBar[5]{		
		\begin{tikzpicture}
		\node (bar) at (#3\iwidth, #4\iheight) { \includegraphics[width=\colorbarlength,height=0.16\colorbarlength]{\filePath/#5} };
		\node [left=-1mm of bar] {\myFontSize{#1}};
		\node [right=-1mm of bar] {\myFontSize{#2}};
		\end{tikzpicture}
	}

	\newcommand\voidAnn{}
\begin{mdframed}[backgroundcolor=black]
\setlength{\iwidth}{0.3\textwidth}
\setlength{\iheight}{\iwidth}
\setlength{\tabcolsep}{-4pt}
\setlength{\rowspacing}{0pt}
\setlength{\colorbarlength}{0.5\iwidth}
\newcommand{\myFontSize}[1]{
	\textcolor{white}{\LARGE{#1}}
}
\def \filePath {./Figures/showcase}
\renewcommand{\arraystretch}{1.5}
\renewcommand{\arrayrulewidth}{0pt}
\centering

\resizebox{\linewidth}{!}
{

\begin{tabular}{c@{~~~~~~~~}c@{~~~~~~~~}c@{~~~~~~~~}cccccc}
	\myFontSize{\textbf{\Large{SMT}}} & \myFontSize{\textbf{\Large{MC-SMT}}} & \myFontSize{\textbf{\Large{NODDI}}} & \multicolumn{6}{c}{\myFontSize{\textbf{\Large{SMSI}}}}\\
	
	\myPic{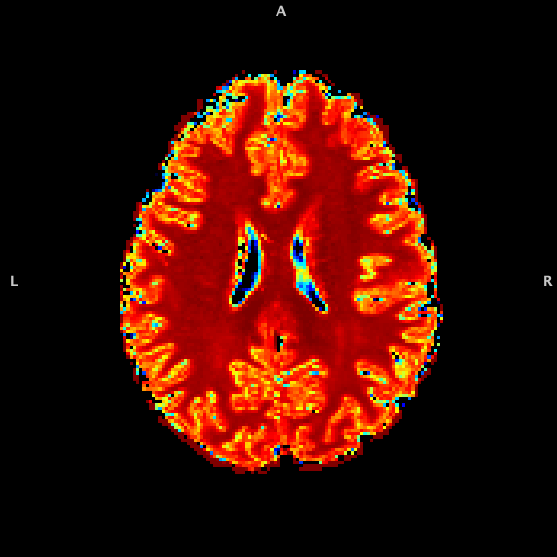}{1.35}{0.5}{0.48}{black}{\voidAnn}{} & \myPic{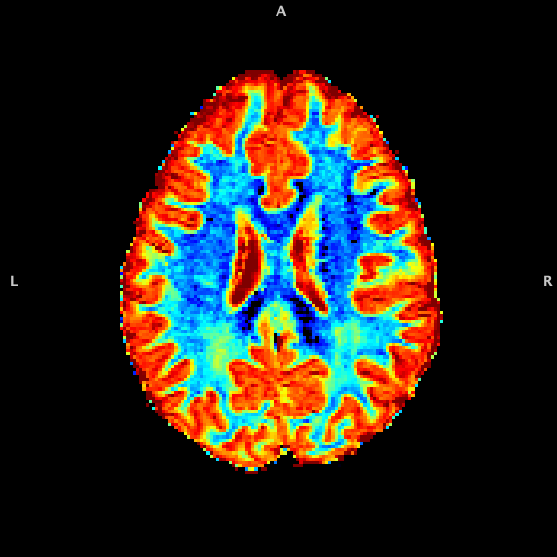}{1.35}{0.5}{0.48}{black}{\voidAnn}{} & \myPic{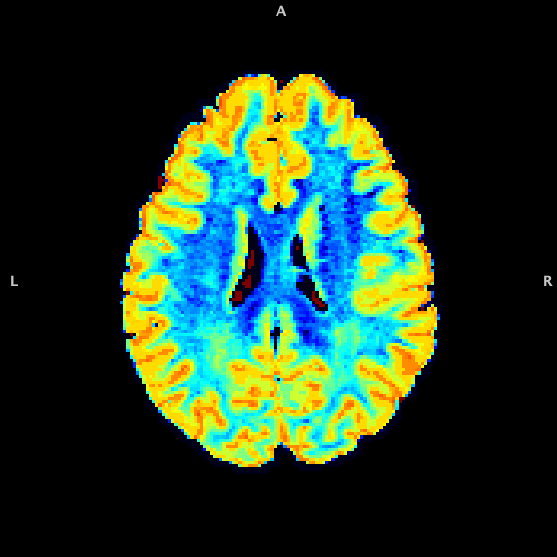}{1.35}{0.5}{0.48}{black}{\voidAnn}{} & \myPic{SMSI_mFA.png}{1.35}{0.5}{0.48}{black}{\voidAnn}{} & \myPic{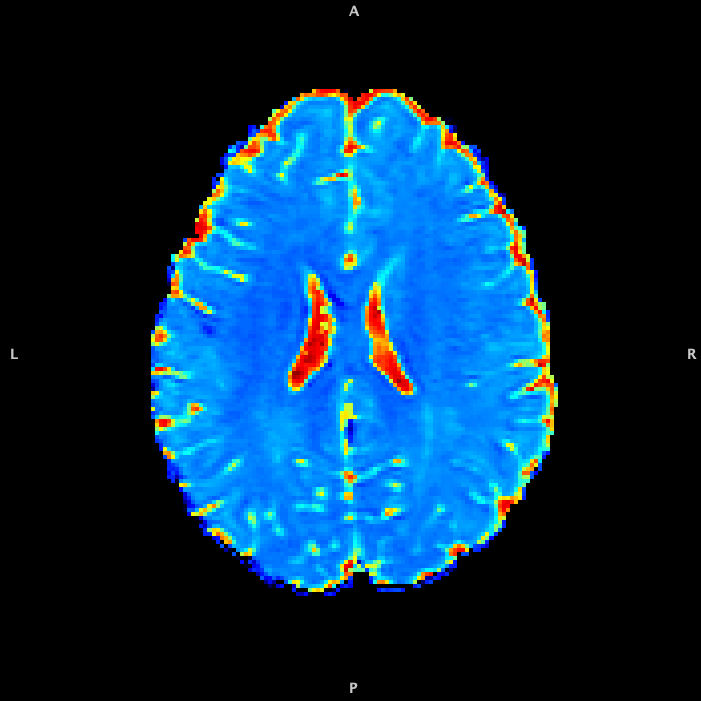}{1.35}{0.5}{0.48}{black}{\voidAnn}{} & \myPic{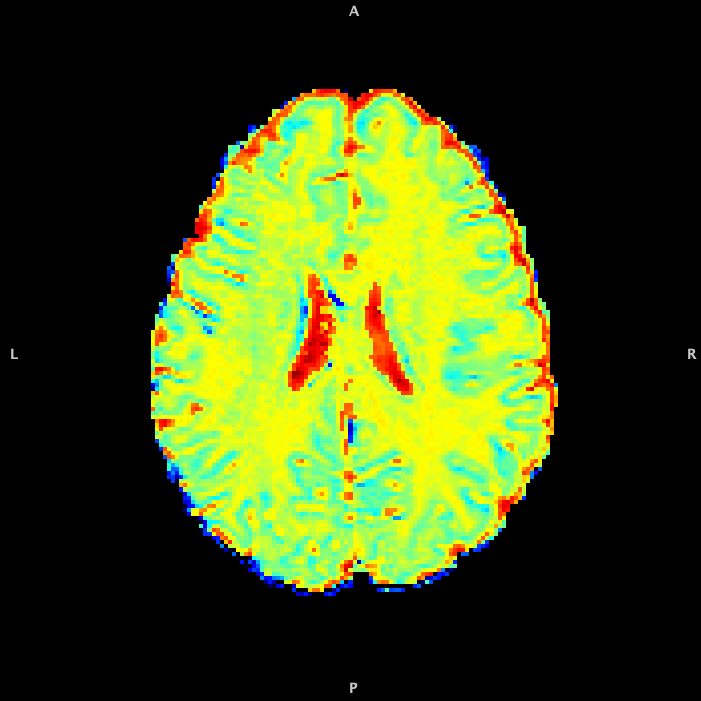}{1.35}{0.5}{0.48}{black}{\voidAnn}{} & \myPic{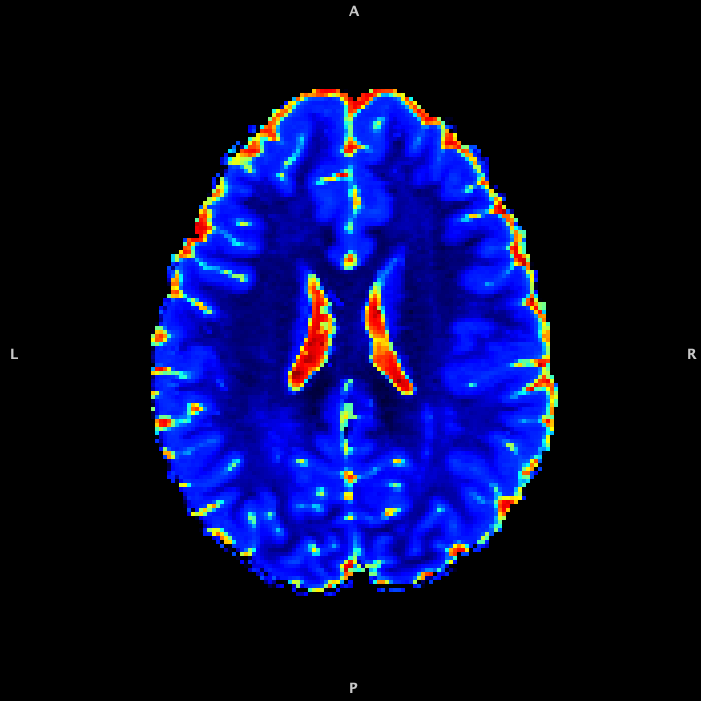}{1.35}{0.5}{0.48}{black}{\voidAnn}{} & \myPic{SMSI_MAI_Norm.png}{1.35}{0.5}{0.48}{black}{\voidAnn}{} & \myPic{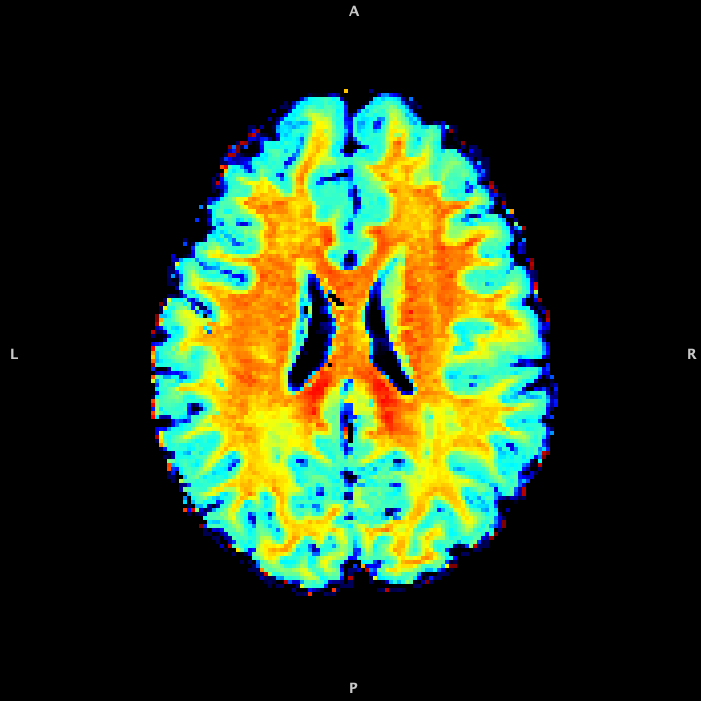}{1.35}{0.5}{0.48}{black}{\voidAnn}{}\\
	
	\myFontSize{$\mu$FA} & \myFontSize{$v_{\text{ec}}$} & \myFontSize{$v_{\text{ec}}$} & \myFontSize{$\mu$FA} & \myFontSize{$\mu$MD} & \myFontSize{$\mu$AD} & \myFontSize{$\mu$RD}  & \myFontSize{MAI} & \myFontSize{$\mu C_{\text{l}}$}\\

	\textcolor{black}{aaaa}\\

	\myPic{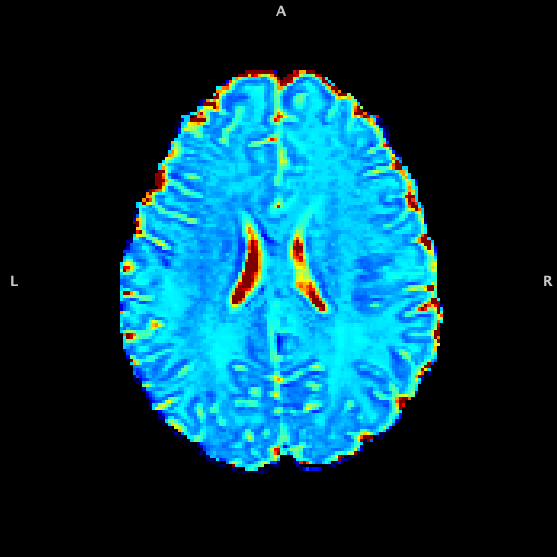}{1.35}{0.5}{0.48}{black}{\voidAnn}{} & \myPic{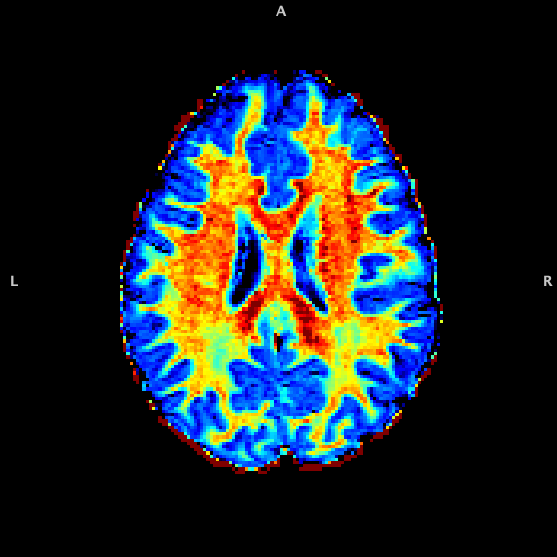}{1.35}{0.5}{0.48}{black}{\voidAnn}{} & \myPic{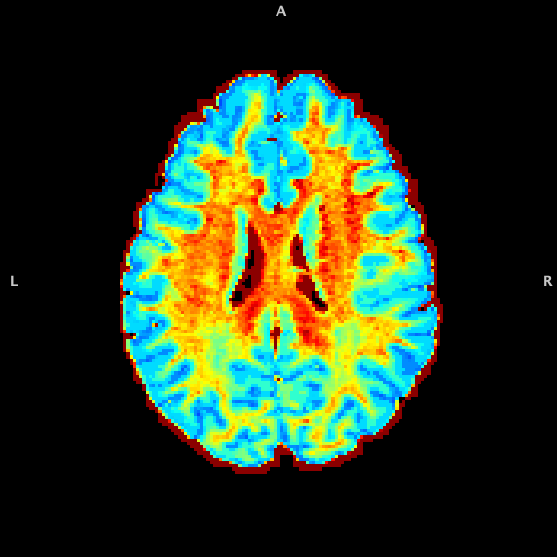}{1.35}{0.5}{0.48}{black}{\voidAnn}{}  &\myPic{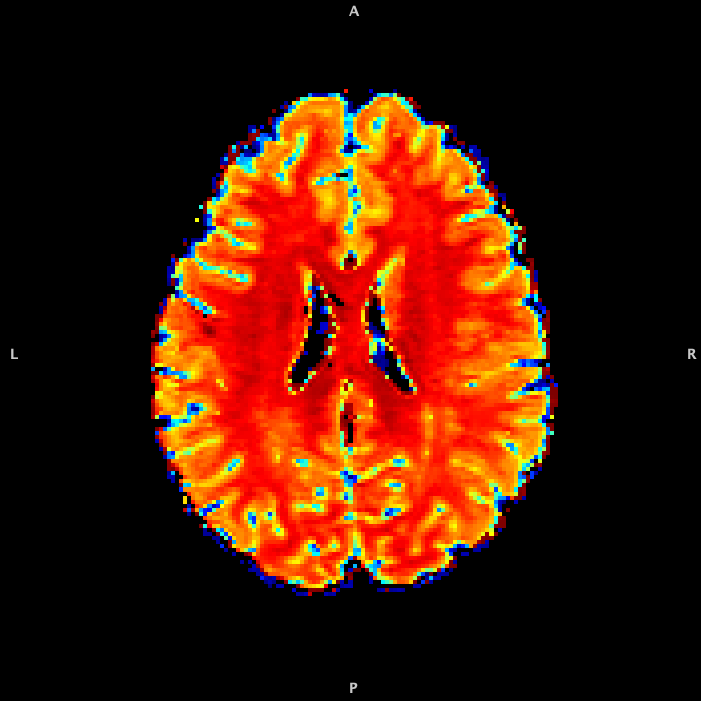}{1.35}{0.5}{0.48}{black}{\voidAnn}{} & \myPic{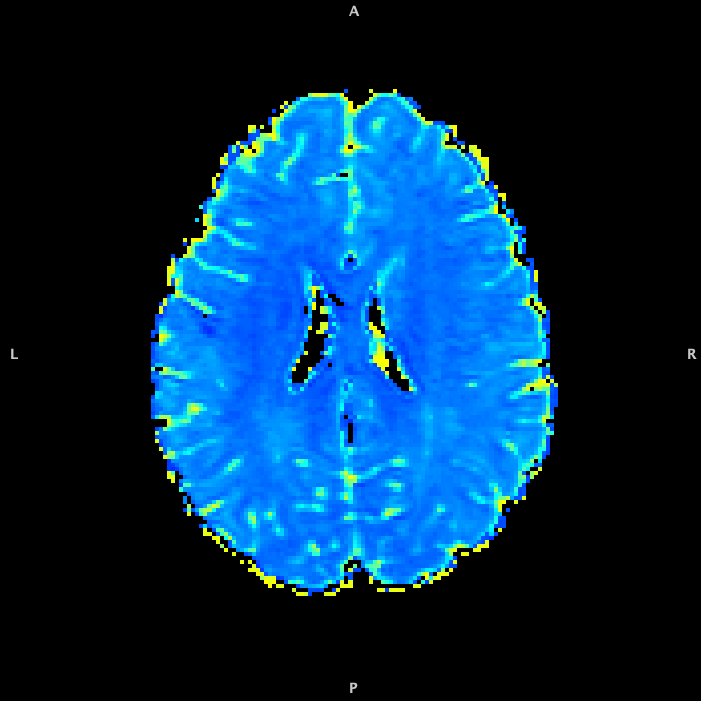}{1.35}{0.5}{0.48}{black}{\voidAnn}{} & \myPic{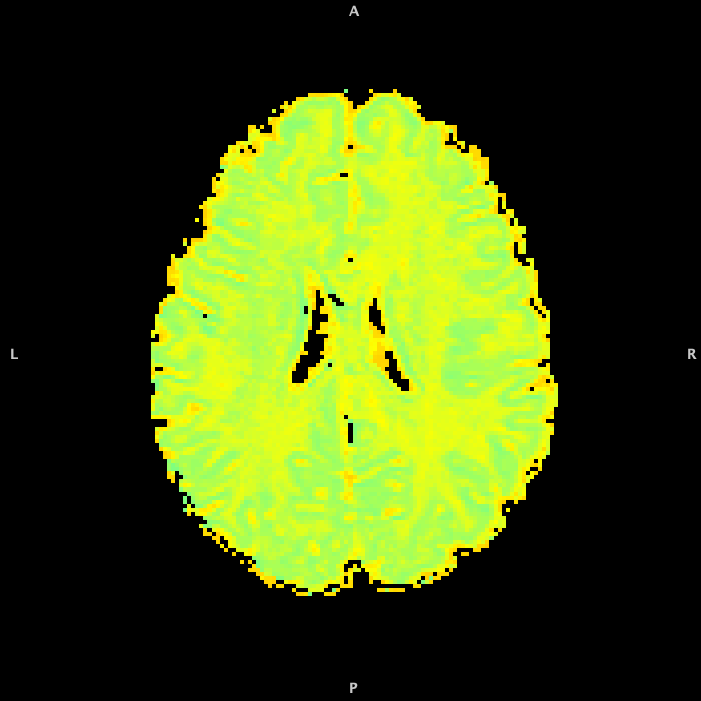}{1.35}{0.5}{0.48}{black}{\voidAnn}{} & \myPic{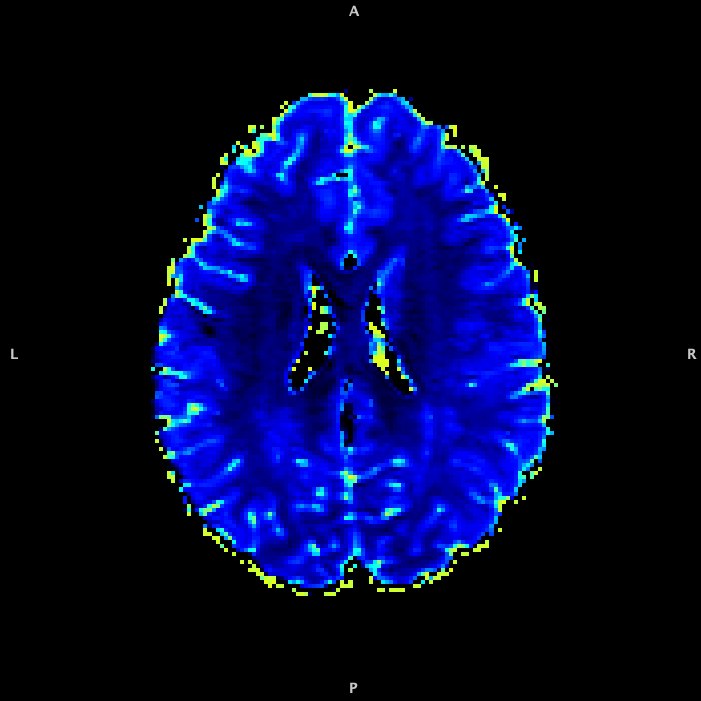}{1.35}{0.5}{0.48}{black}{\voidAnn}{} & \myPic{SMSI_AMAI_Norm.png}{1.35}{0.5}{0.48}{black}{\voidAnn}{} & \myPic{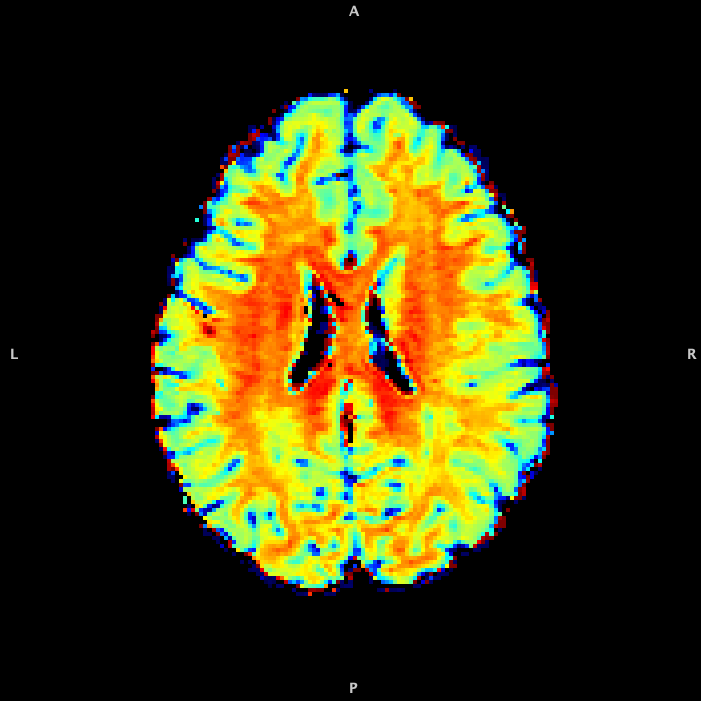}{1.35}{0.5}{0.48}{black}{\voidAnn}{}\\
	
	\myFontSize{$\mu$MD} & \myFontSize{$v_{\text{ic}}$} & \myFontSize{$v_{\text{ic}}$} &  \myFontSize{$\mu$FA$^{\dagger}$} & \myFontSize{$\mu$MD$^{\dagger}$} & \myFontSize{$\mu$AD$^{\dagger}$} & \myFontSize{$\mu$RD$^{\dagger}$} & \myFontSize{MAI$^{\dagger}$} & \myFontSize{$\mu C_{\text{l}} ^\dagger$}\\
	
	\textcolor{black}{aaaa}\\
	
	\myPic{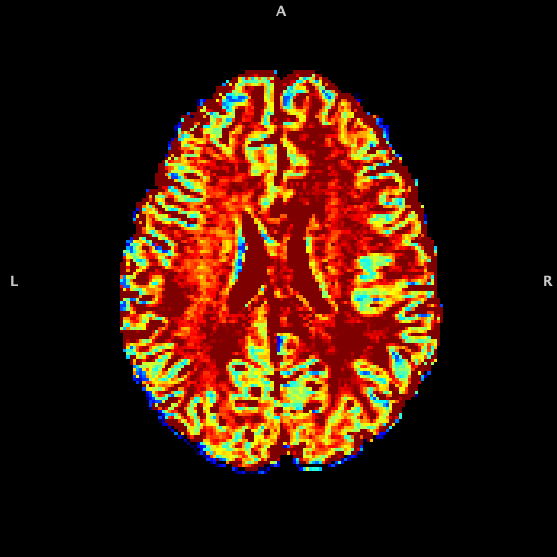}{1.35}{0.5}{0.48}{black}{\voidAnn}{} & \myPic{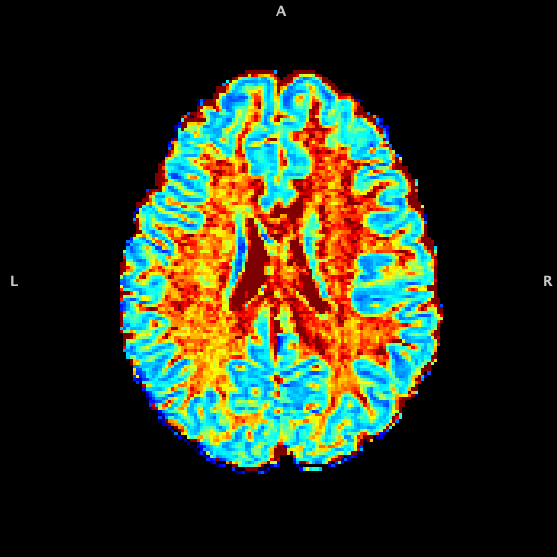}{1.35}{0.5}{0.48}{black}{\voidAnn}{} & \myPic{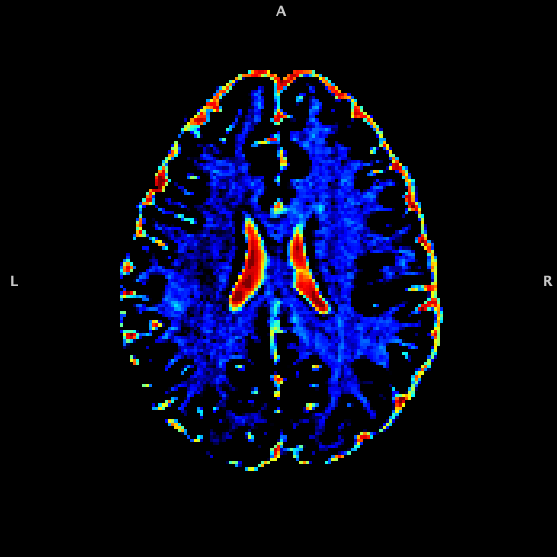}{1.35}{0.5}{0.48}{black}{\voidAnn}{} & \myPic{SMSI_ECVF.png}{1.35}{0.5}{0.48}{black}{\voidAnn}{} & \myPic{SMSI_ICVF.png}{1.35}{0.5}{0.48}{black}{\voidAnn}{} & \myPic{SMSI_IVF.png}{1.35}{0.5}{0.48}{black}{\voidAnn}{} & \myPic{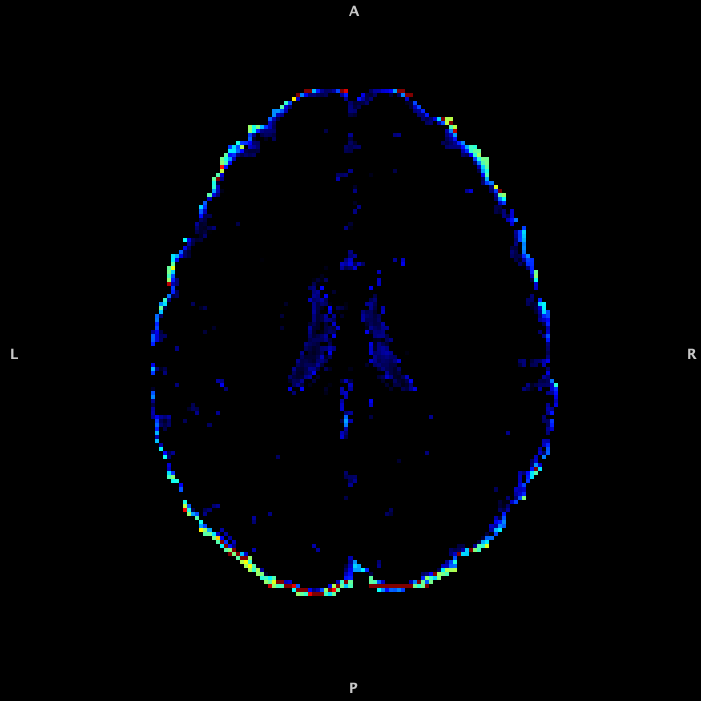}{1.35}{0.5}{0.48}{black}{\voidAnn}{} & \myPic{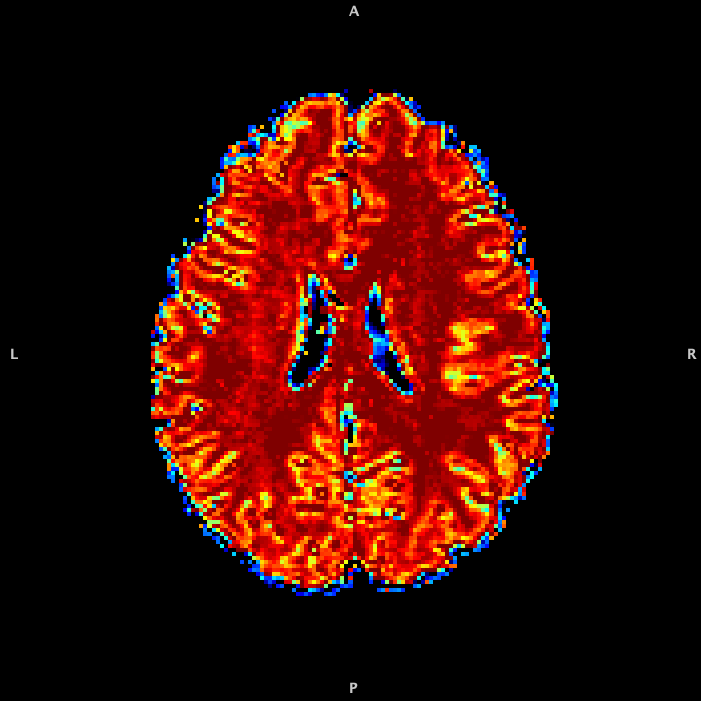}{1.35}{0.5}{0.48}{black}{\voidAnn}{} & \myPic{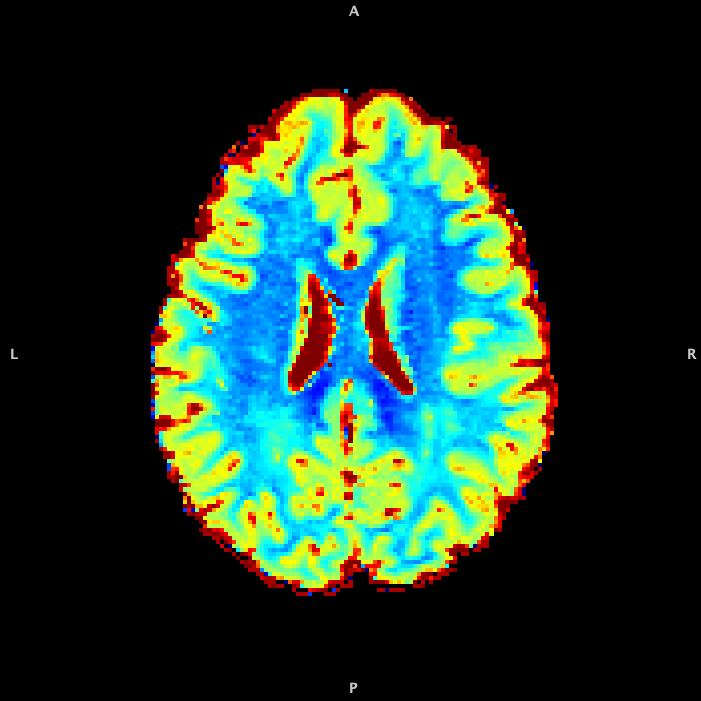}{1.35}{0.5}{0.48}{black}{\voidAnn}{}\\
	
	\myFontSize{$\mu$AD} & \myFontSize{Ins. Diff.} & \myFontSize{$v_{\text{iso}}$} & \myFontSize{$v_{\text{ec}}$} & \myFontSize{$v_{\text{ic}}$} & \myFontSize{$v_{\text{iso}}$} & \myFontSize{$v_{\text{tw}}$} & \myFontSize{$v_{\text{a}}$} & \myFontSize{$\mu C_{\text{s}}$}\\
	
	\textcolor{black}{aaaa}\\
	
	\myPic{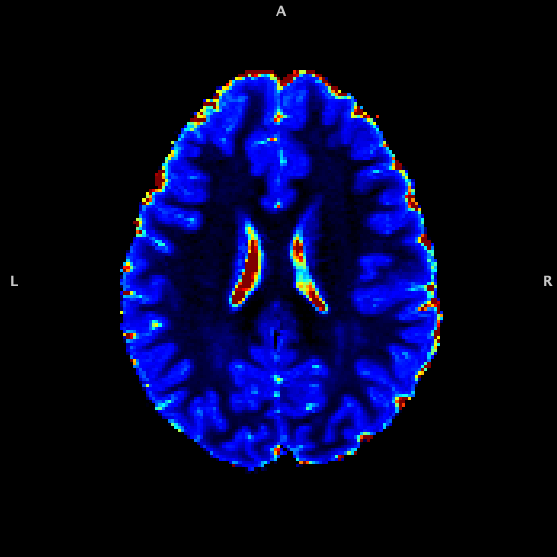}{1.35}{0.5}{0.48}{black}{\voidAnn}{} & \myPic{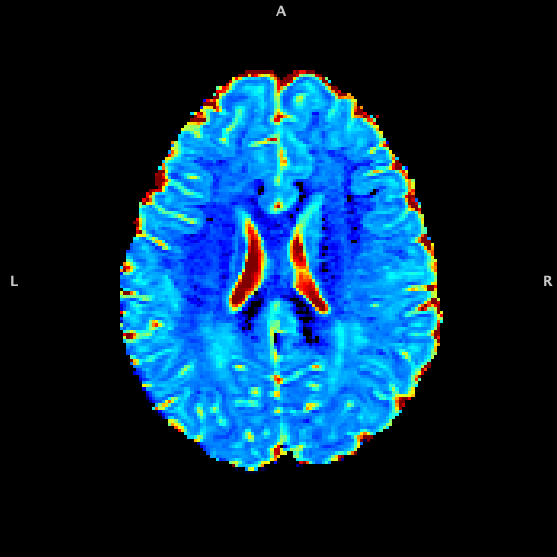}{1.35}{0.5}{0.48}{black}{\voidAnn}{} & \myPic{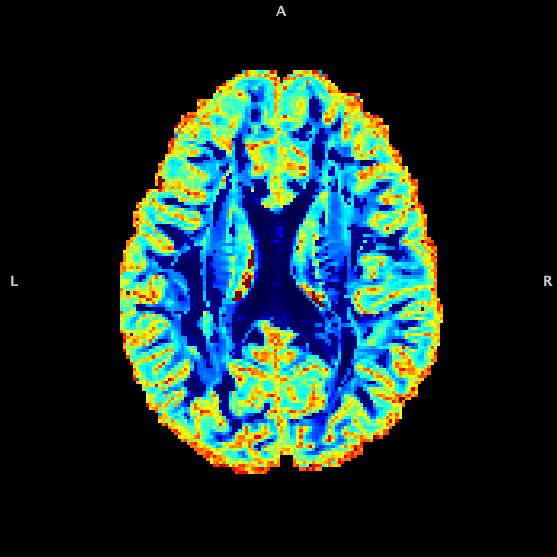}{1.35}{0.5}{0.48}{black}{\voidAnn}{} & \myPic{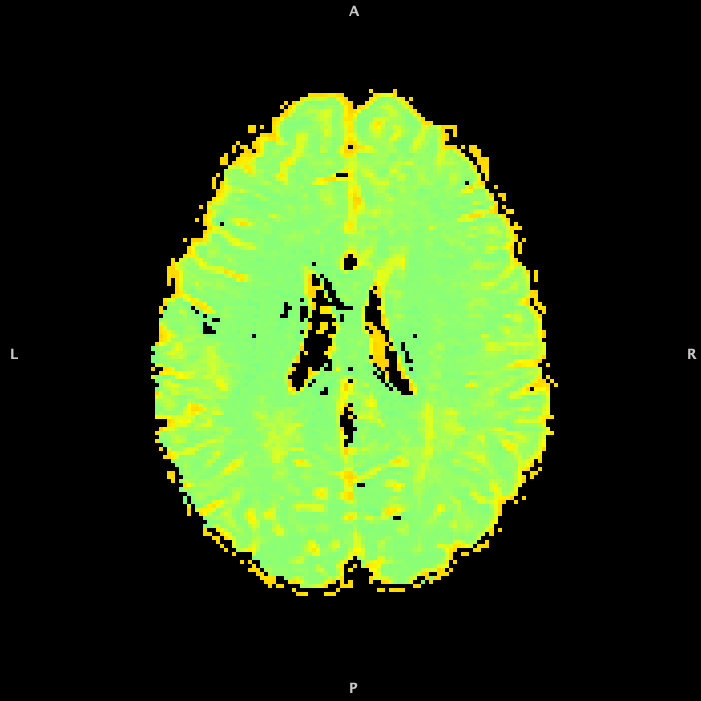}{1.35}{0.5}{0.48}{black}{\voidAnn}{} & \myPic{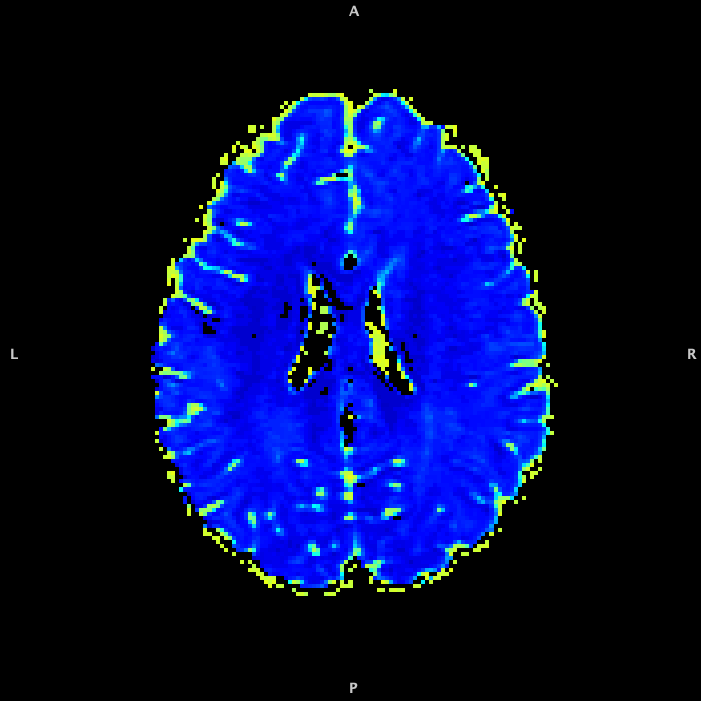}{1.35}{0.5}{0.48}{black}{\voidAnn}{} & \myPic{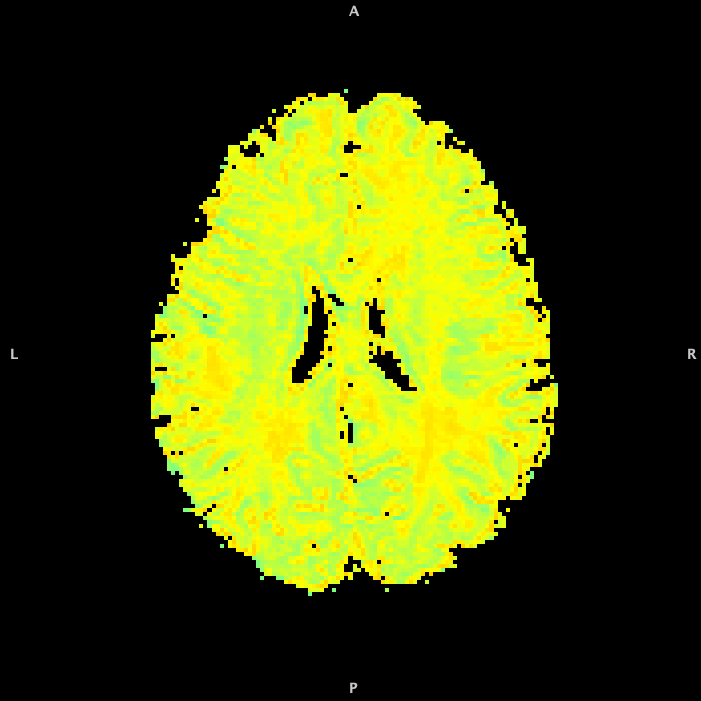}{1.35}{0.5}{0.48}{black}{\voidAnn}{} & \myPic{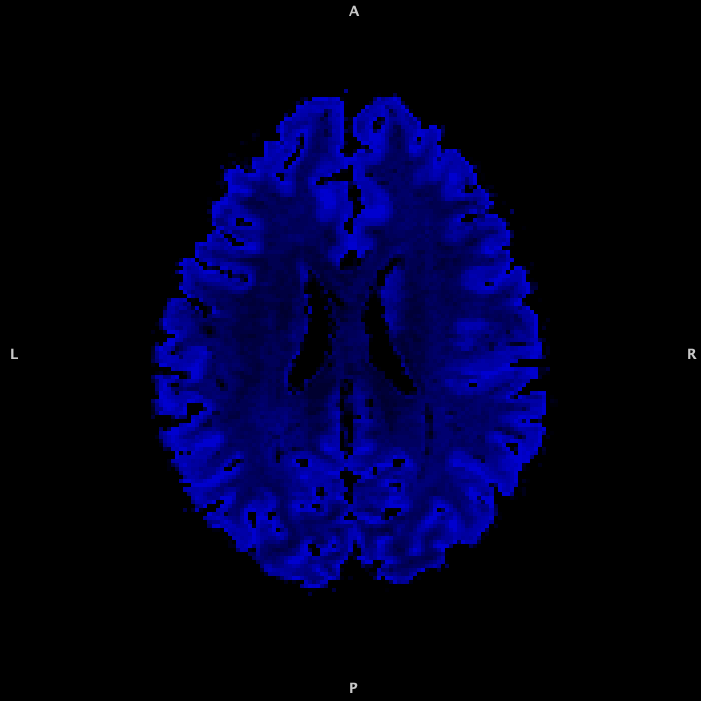}{1.35}{0.5}{0.48}{black}{\voidAnn}{} & \myPic{SMSI_AOCI_clean.png}{1.35}{0.5}{0.48}{black}{\voidAnn}{} & \myPic{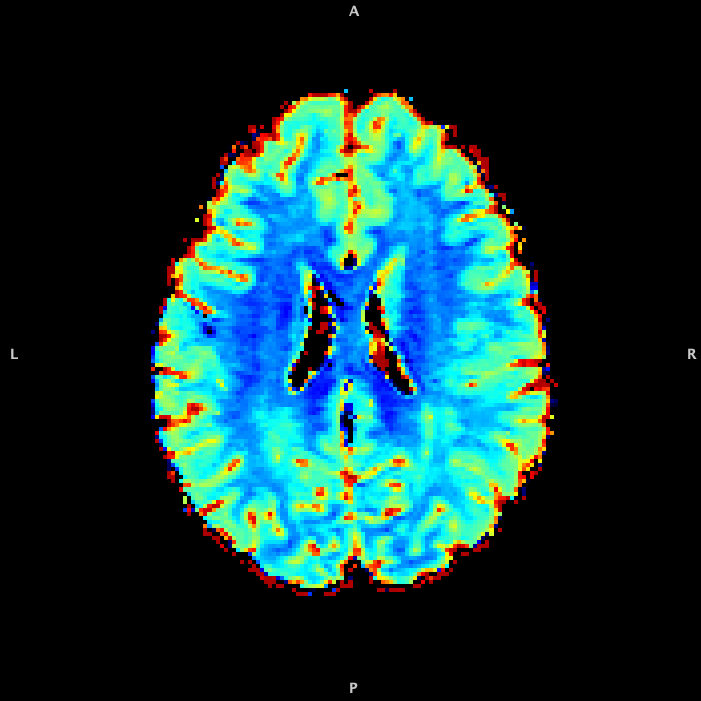}{1.35}{0.5}{0.48}{black}{\voidAnn}{}\\
	
	\myFontSize{$\mu$RD} & \myFontSize{$\mu\text{RD}_{\text{ec}}$} & \myFontSize{ODI} & \myFontSize{$\mu\text{AD}_{\text{ec}}$} & \myFontSize{$\mu\text{RD}_{\text{ec}}$} & \myFontSize{$\mu\text{AD}_{\text{ic}}$} & \myFontSize{$\mu\text{RD}_{\text{ic}}$}  & \myFontSize{OCI} & \myFontSize{$\mu C_{\text{s}} ^\dagger$}\\
	&&&&&&&&  \colorBar{low}{high}{0.5}{0.5}{colorbar}\\	
	
\end{tabular}
}
\end{mdframed}
\caption{
	\textbf{Diffusion Indices.}
	Diffusion indices of SMSI, SMT, MC-SMT, and NODDI. 
  The intrinsic diffusivity (Ins. Diff.) of MC-SMT is the longitudinal diffusivity common for both extra- and intra-cellular compartments. Jet color mapping, with cool colors for low values and warm colors for high values, is used. The values range from 0 to 0.003 for diffusivity-based indices and 0 to 1 for other indices. Please refer to Table~\ref{tab:Indices} for the definitions of the indices.
}
	\label{fig:showcase}
\end{figure}

	Fig.~\ref{fig:showcase} shows that SMSI provides a wider range of diffusion indices than SMT, MC-SMT, and NODDI, allowing greater specificity in characterizing tissue microstructure. 
	The discrepancies between SMSI and the other methods can be explained based on our previous observations from the synthetic data experiments. 
	For instance, one can observe that $\mu$FA given by SMT is higher than SMSI in gray matter. 
	This is consistent with our previous observation that SMT overestimates $\mu$FA when the actual value is low (Fig.~\ref{fig:syn_combine} (e) and (f)).
	MC-SMT overestimates and NODDI underestimates the extra-cellular volume fraction when its actual value is high (Fig.~\ref{fig:syn_combine} (g) and (h)), such as in gray matter. 
	Additionally, 
	NODDI yields higher isotropic volume fraction in deep white matter than gray matter (Fig.~\ref{fig:syn_combine} (i)), which does not reflect the fact that isotropic diffusion should be less prominent in deep white matter in view of the tightly packed micro-architecture,
	particularly in the adult brain \citep{barr1974human,blumenfeld2010neuroanatomy}.
	On the other hand, SMSI gives more biologically feasible results with lower isotropic volume fraction in white matter than gray matter. Note that isotropic diffusion in gray matter is in part intra-soma diffusion \citep{palombo2019sandi,huynh2020somamiccai}. 
	
	\subsection{Number of $b$-Shells}
\begin{figure}[!htb]

\setlength{\iwidth}{0.3\textwidth}
\setlength{\iheight}{\iwidth}
\setlength{\tabcolsep}{0pt}
\setlength{\rowspacing}{0pt}
\setlength{\colorbarlength}{0.5\iwidth}
\newcommand{\myFontSize}{\normalsize}
\def \filePath {./Figures/24shells}
\renewcommand{\arraystretch}{0}
\renewcommand{\arrayrulewidth}{0pt}
\centering

\subfigure[Isotropic volume fraction ($v_{\text{iso}}$)]{\label{fig:24shellsIVF}\includegraphics[width=0.9\linewidth]{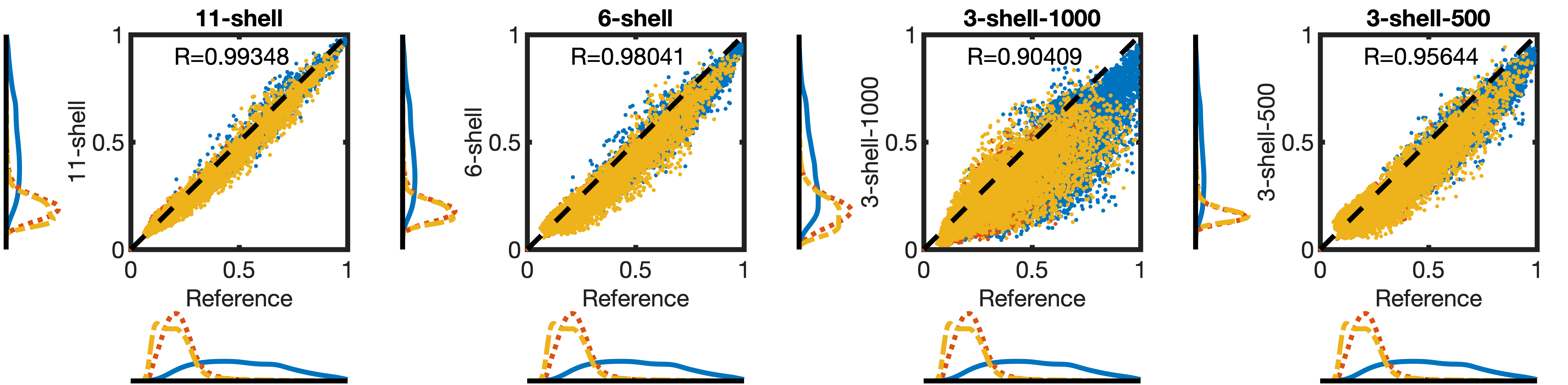}}
\\
\subfigure[Microscopic FA ($\mu$FA)]{\label{fig:24shellsmFA}\includegraphics[width=0.9\linewidth]{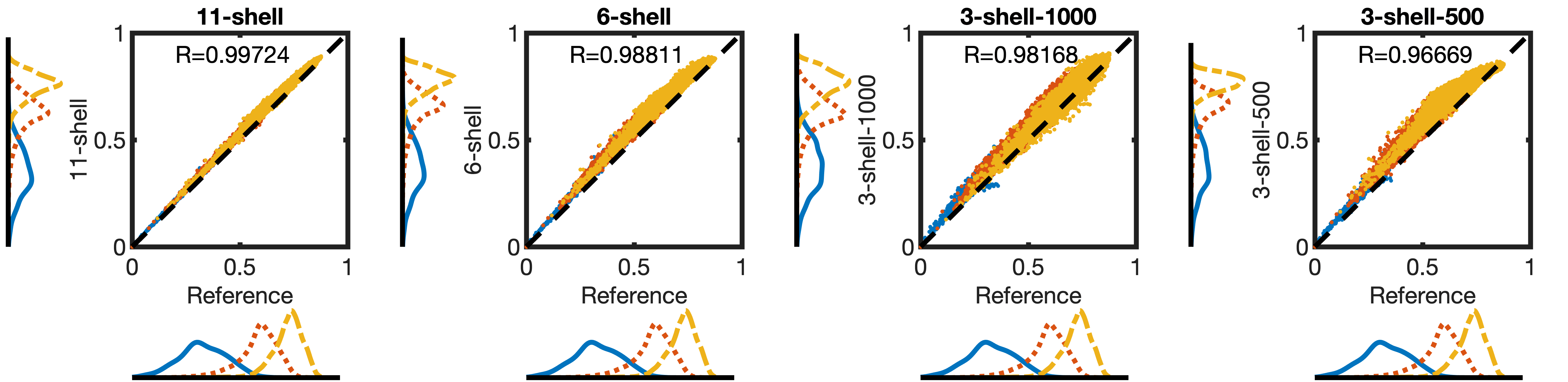}}
\\
\subfigure[Microscopic anisotropy index (MAI)]{\label{fig:24shellsPAA}\includegraphics[width=0.9\linewidth]{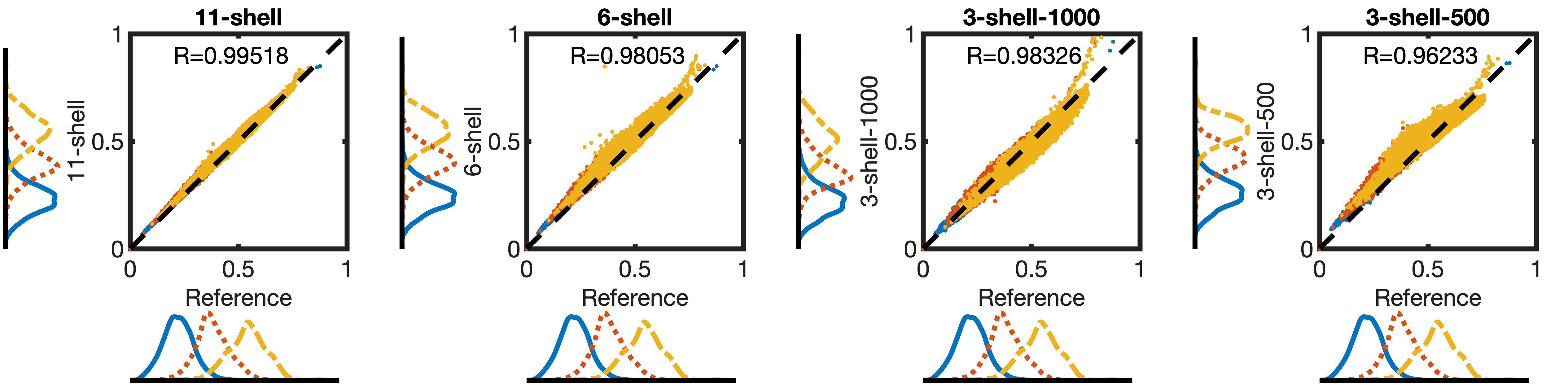}}
\\
\subfigure[Orientation coherence index (OCI)]{\label{fig:24shellsOCI}\includegraphics[width=0.9\linewidth]{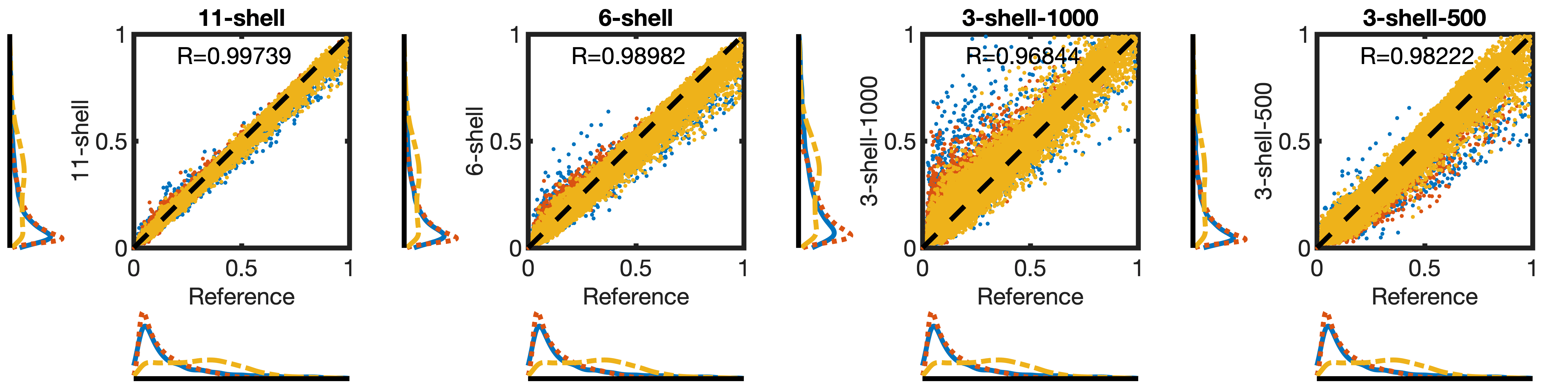}}

\caption{
	\textbf{Number of b-Shells.}
	Scatter plots and histograms of representative SMSI scalars indices of sampling schemes 11-shell, 6-shell, 3-shell-1000, and 3-shell-500 with 21-shell as the reference. Voxels are classified as CSF (blue), gray matter (red), or white matter (yellow). For better visibility, only one in every six voxels is shown. 
	\label{fig:24shells-full}}
\end{figure}

	Fig.~\ref{fig:24shells-full} shows the scatter plots and histograms of representative SMSI indices of different sampling schemes with `21-shell' as the reference. 
	The 11-shell sampling scheme produces results closest to the reference as shown by the high histogram similarity and  the high correlation coefficient. 
	Fewer number of shells still yield reasonable results with correlation coefficient $R>$ 0.9.
	The 3-shell-500 scheme is better than 3-shell-1000 in estimating the isotropic volume fraction thanks to the $b = 500 \, \text{s}/\text{mm}^2$ shell as the signal of free water decays significantly at $b = 1000 \, \text{s}/\text{mm}^2$.
	SMSI is hence applicable to many public datasets, such as the HCP (3 shells) and the BCP (6 shells) datasets.

	\subsection{Longitudinal Infant Data}
\begin{figure}[tb]
	
	\newcommand\myPic[7]{
		\begin{tikzpicture}[baseline,trim left]
		\draw (0.0\iwidth, 0.0\iheight)[#5, ultra thick] rectangle(\iwidth, \iheight);
		#6 %
		\node[mylabel, right=0pt] at (0.0\iwidth, 1.0\iheight) {\textbf{\large{#7}}}; %
		
		\begin{pgfonlayer}{background}
		\fill (0.0\iwidth, 0.0\iheight)[black, ultra thick] rectangle(\iwidth, \iheight);
		\clip (0.0\iwidth, 0.0\iheight) rectangle (\iwidth, \iheight);
		\node at (#3\iwidth, #4\iheight) { \includegraphics[width=#2\iwidth, angle=0]{\filePath/#1} };
		\end{pgfonlayer}
		\end{tikzpicture}
	}
	
	\newcommand\columnText[1]{
	\begin{tikzpicture}[baseline,trim left]
	\node[rotate=90] at (0.1\iwidth, 0.5\iheight) {\myFontSize{#1}};
	\end{tikzpicture}
}
	\newcommand\columnTextRight[1]{
	\begin{tikzpicture}[baseline,trim left]
	\node[rotate=270] at (0.5\iwidth, 0.5\iheight) {\myFontSize{#1}};
	\end{tikzpicture}
}
	
	\newcommand\colorBar[5]{		
		\begin{tikzpicture}
		\node (bar) at (#3\iwidth, #4\iheight) { \includegraphics[width=\colorbarlength,height=0.16\colorbarlength]{\filePath/#5} };
		\node [left=-1mm of bar] {\myFontSize{#1}};
		\node [right=-1mm of bar] {\myFontSize{#2}};
		\end{tikzpicture}
	}

	\newcolumntype{L}[1]{>{\raggedright\let\newline\\\arraybackslash\hspace{0pt}}m{#1}}
	\newcolumntype{C}[1]{>{\centering\let\newline\\\arraybackslash\hspace{0pt}}m{#1}}
	\newcolumntype{R}[1]{>{\raggedleft\let\newline\\\arraybackslash\hspace{0pt}}m{#1}}
	
	\begin{mdframed}[backgroundcolor=black]		
	\newcommand\voidAnn{}

\setlength{\iwidth}{0.3\textwidth}
\setlength{\iheight}{\iwidth}
\setlength{\tabcolsep}{0pt}
\setlength{\rowspacing}{0pt}
\setlength{\colorbarlength}{0.5\iwidth}
\newcommand{\myFontSize}[1]{
	\textcolor{white}{\textbf{\Large{#1}}}
}
\def \filePath {./Figures/BCP}
\renewcommand{\arraystretch}{1.25}
\renewcommand{\arrayrulewidth}{0pt}
\centering

\resizebox{\linewidth}{!}
{

\begin{tabular}{ccccccccc}
& \myFontSize{MAI} & \myFontSize{MAI$^{\dagger}$} & \myFontSize{$\mu$FA} & \myFontSize{$\mu$FA$^{\dagger}$} & \myFontSize{$v_{\text{iso}}$} & \myFontSize{$v_{\text{ic}}$} & \myFontSize{$v_{\text{ec}}$} & \myFontSize{OCI}  \\
\columnText{54}&\myPic{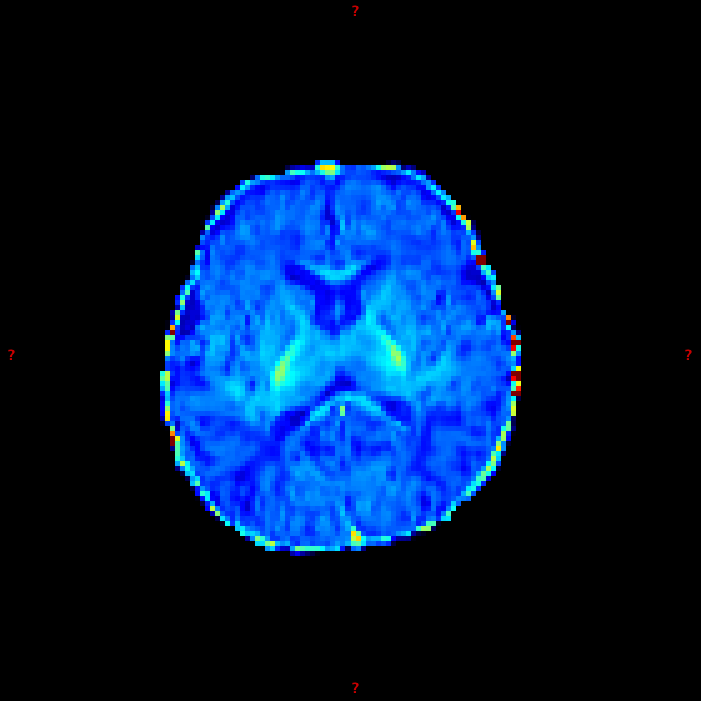}{0.9}{0.5}{0.52}{black}{\voidAnn}{}&\myPic{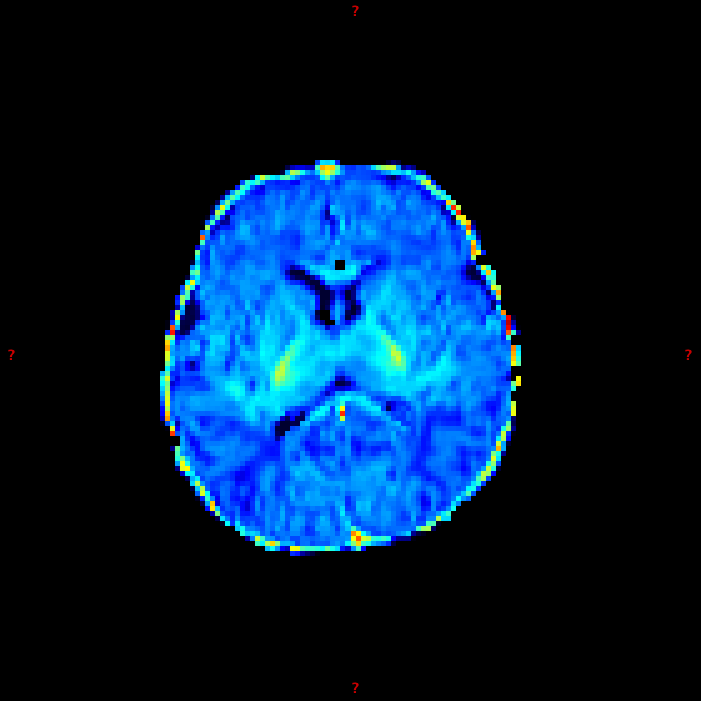}{0.9}{0.5}{0.52}{black}{\voidAnn}{}&\myPic{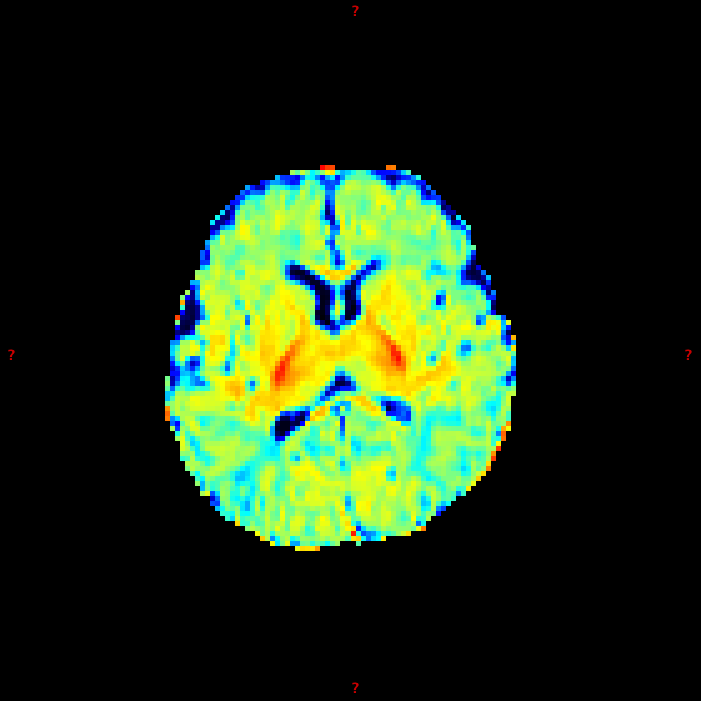}{0.9}{0.5}{0.52}{black}{\voidAnn}{}&\myPic{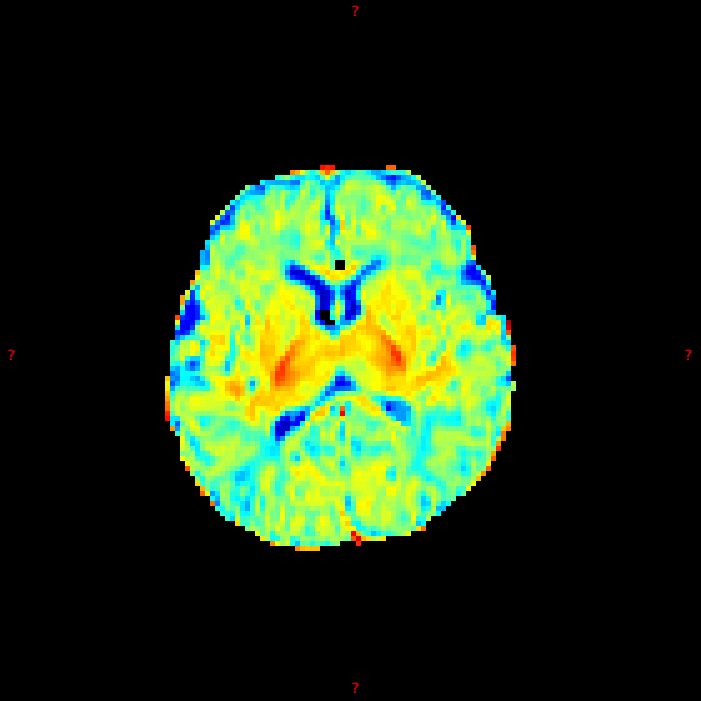}{0.9}{0.5}{0.52}{black}{\voidAnn}{}&\myPic{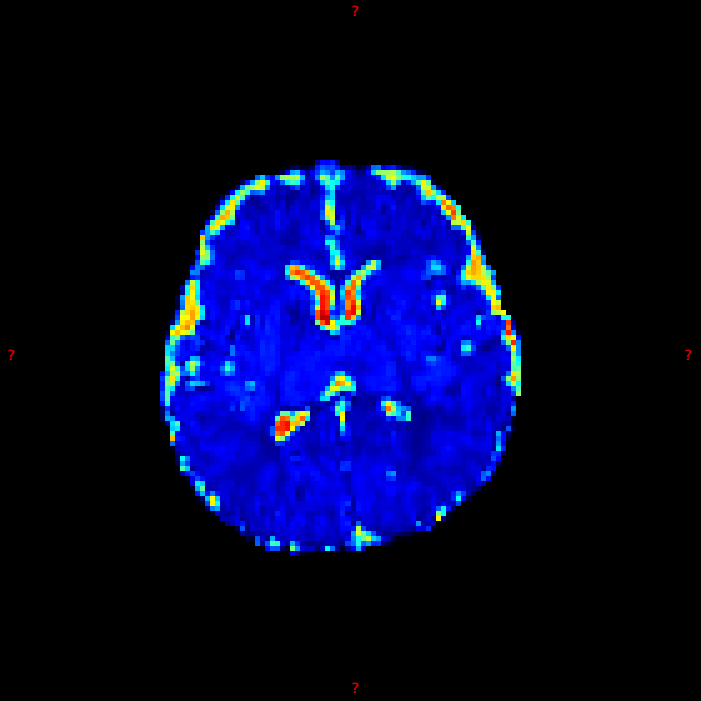}{0.9}{0.5}{0.52}{black}{\voidAnn}{}&\myPic{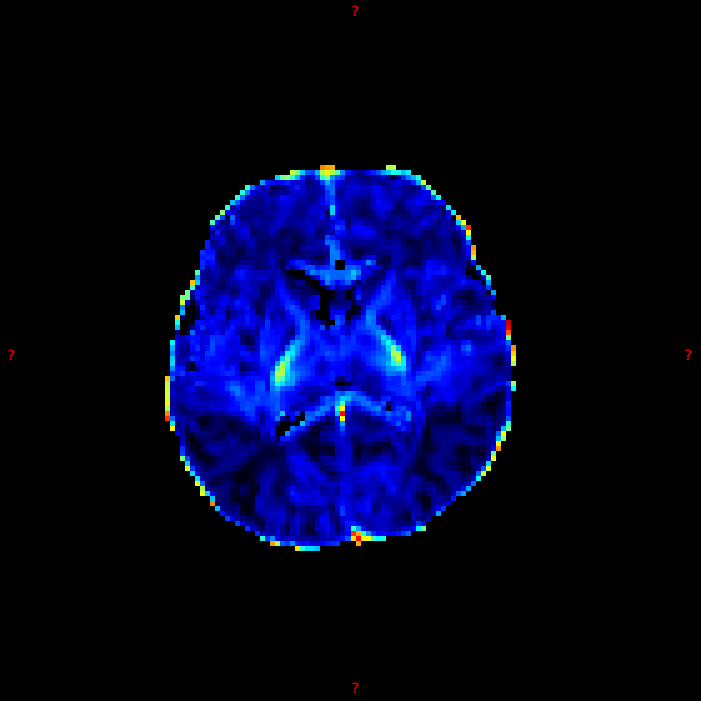}{0.9}{0.5}{0.52}{black}{\voidAnn}{}&\myPic{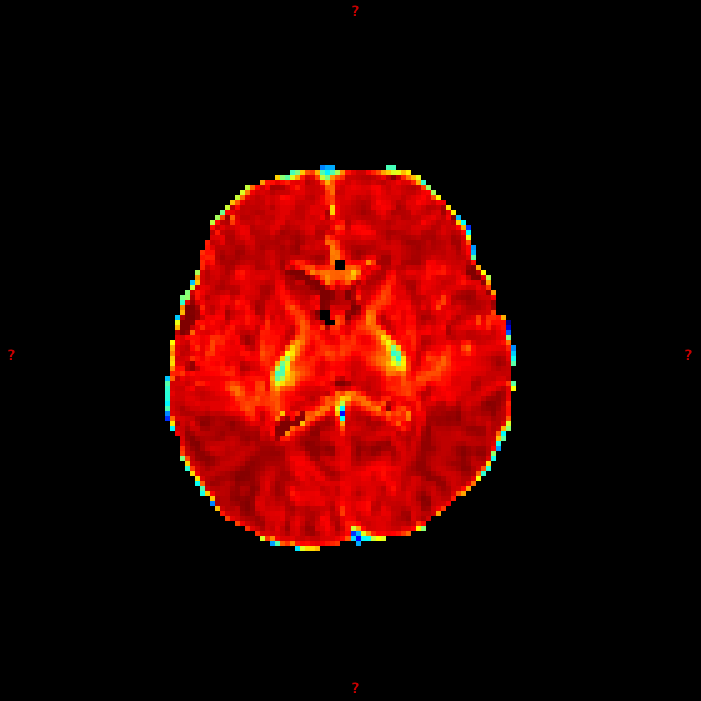}{0.9}{0.5}{0.52}{black}{\voidAnn}{}&\myPic{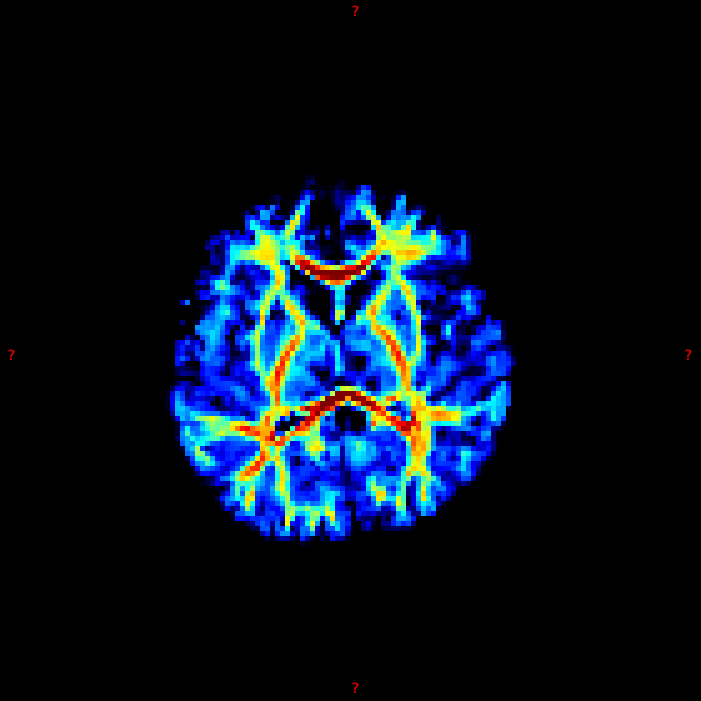}{0.9}{0.5}{0.52}{black}{\voidAnn}{}\\

\columnText{146}&\myPic{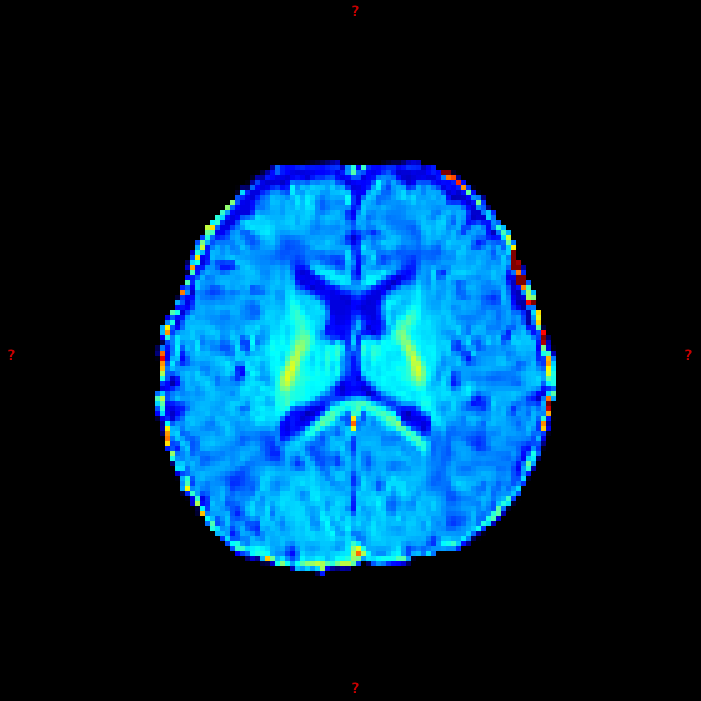}{1.0}{0.5}{0.56}{black}{\voidAnn}{}&\myPic{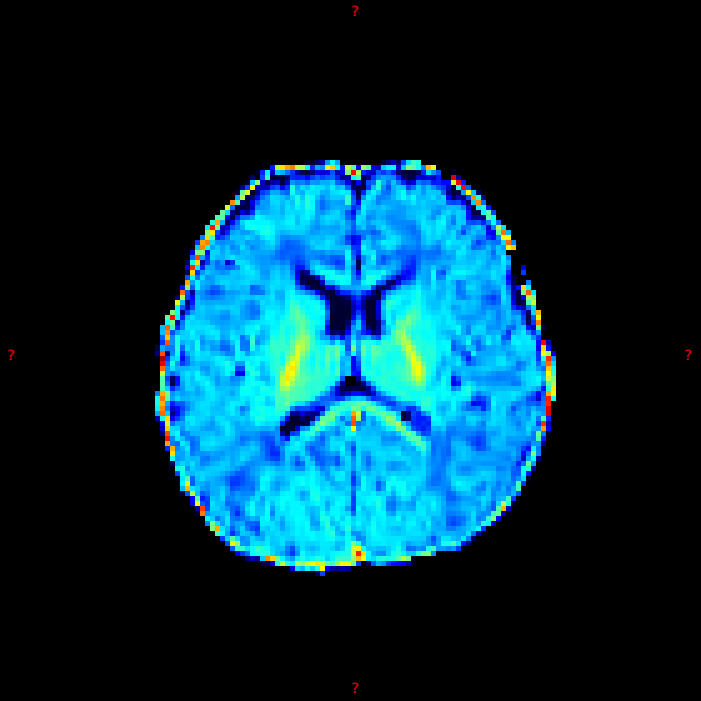}{1.0}{0.5}{0.56}{black}{\voidAnn}{}&\myPic{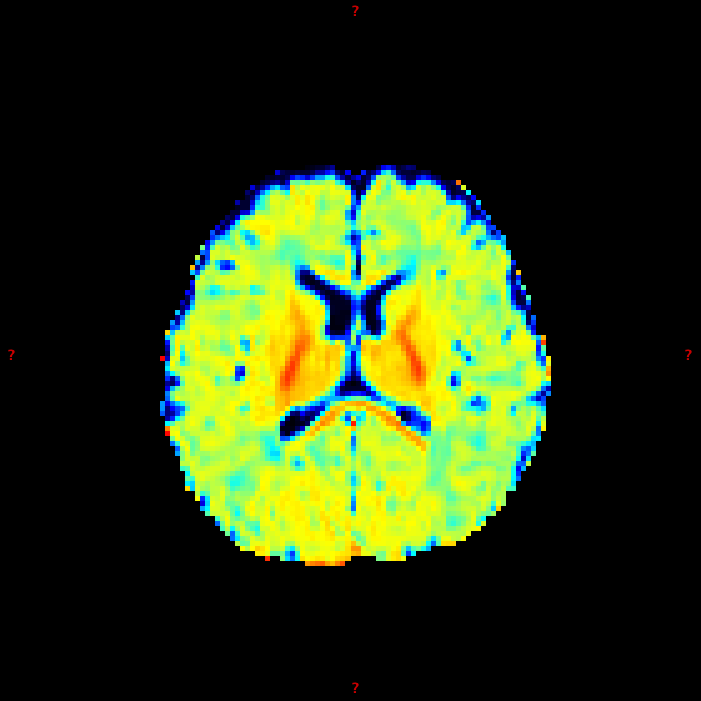}{1.0}{0.5}{0.56}{black}{\voidAnn}{}&\myPic{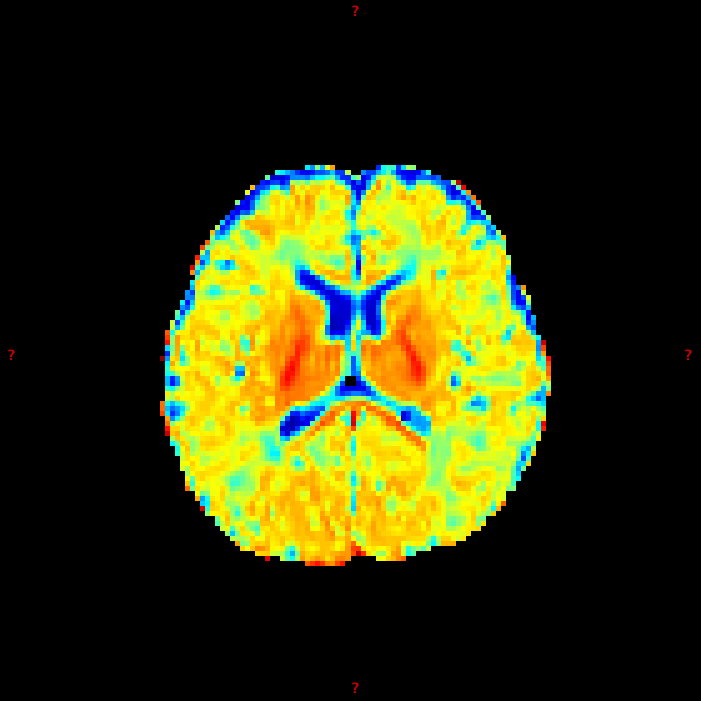}{1.0}{0.5}{0.56}{black}{\voidAnn}{}&\myPic{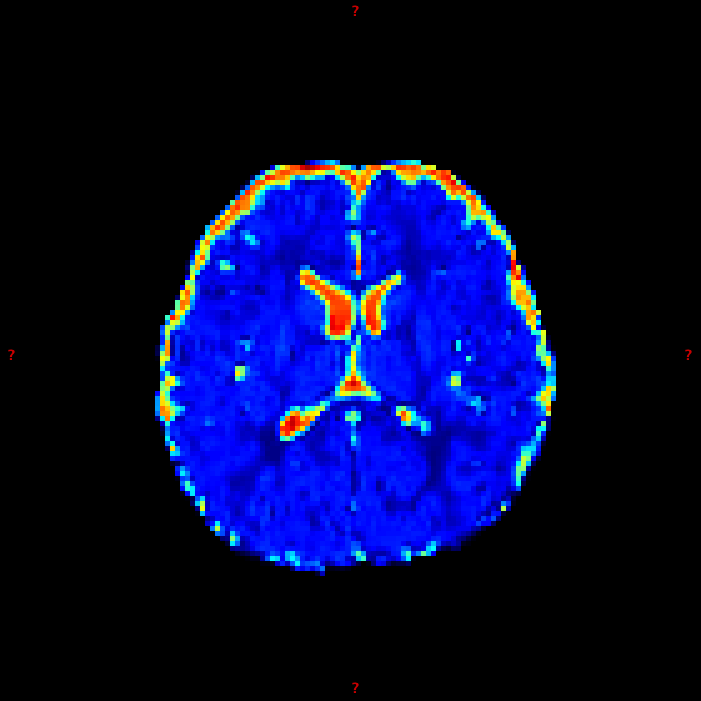}{1.0}{0.5}{0.56}{black}{\voidAnn}{}&\myPic{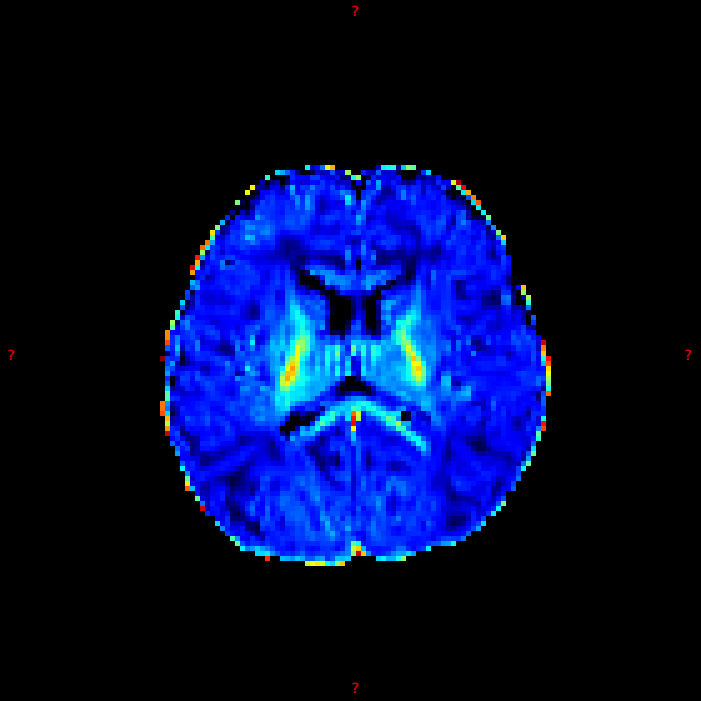}{1.0}{0.5}{0.56}{black}{\voidAnn}{}&\myPic{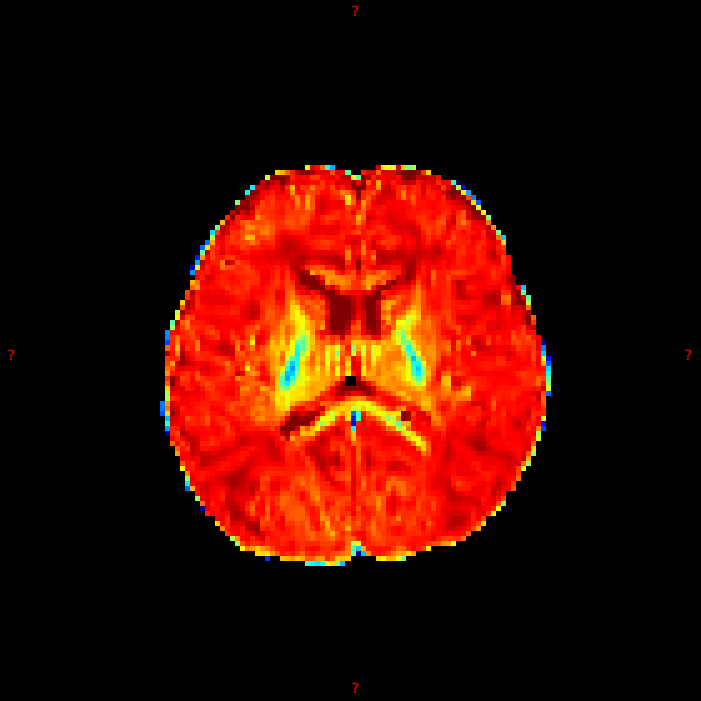}{1.0}{0.5}{0.56}{black}{\voidAnn}{}&\myPic{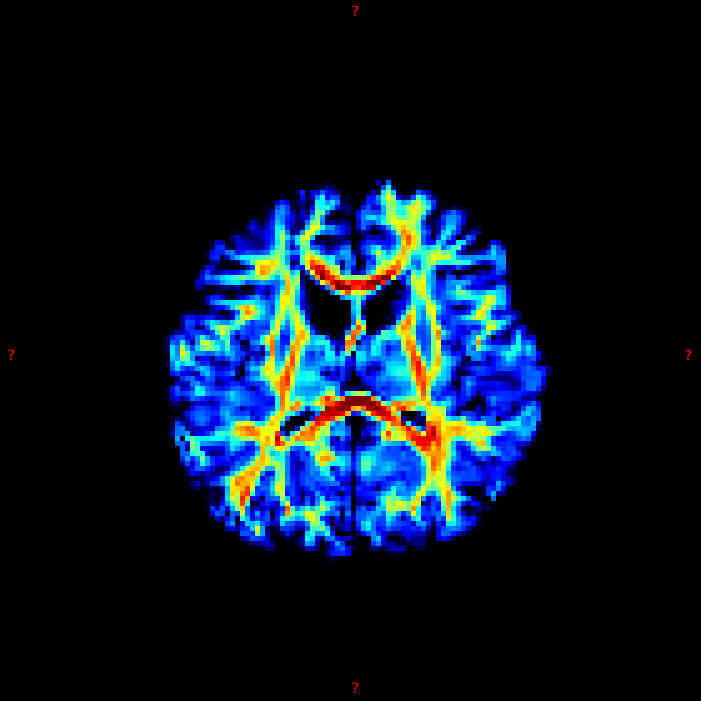}{1.0}{0.5}{0.56}{black}{\voidAnn}{}\\

\columnText{223}&\myPic{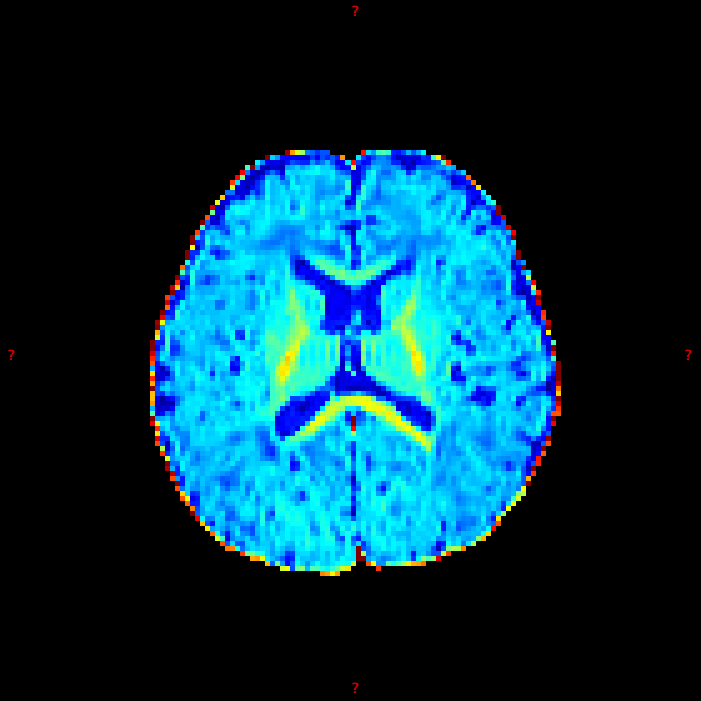}{1.15}{0.5}{0.52}{black}{\voidAnn}{}&\myPic{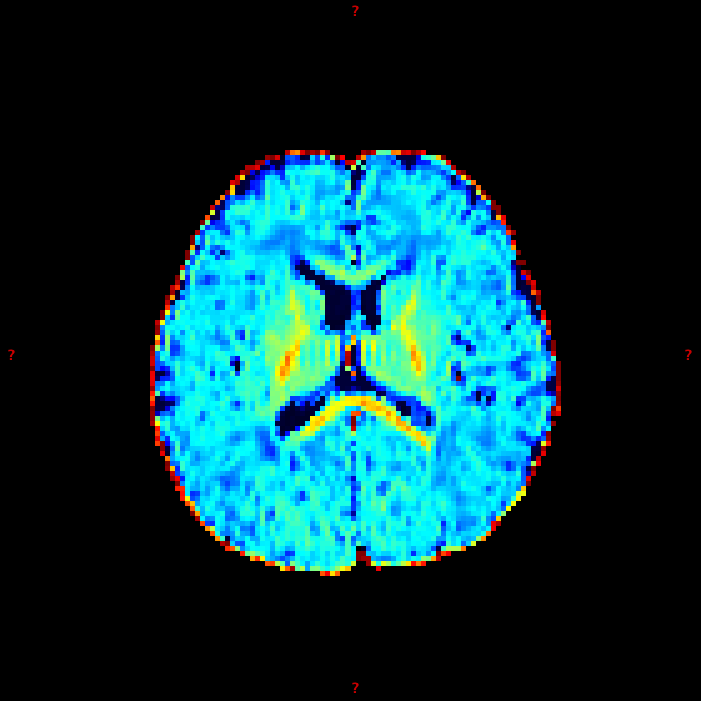}{1.15}{0.5}{0.52}{black}{\voidAnn}{}&\myPic{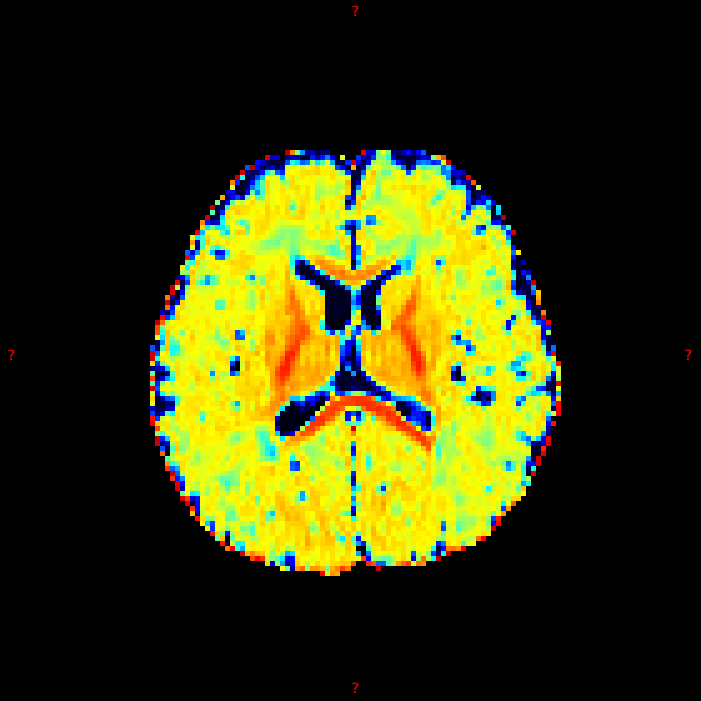}{1.15}{0.5}{0.52}{black}{\voidAnn}{}&\myPic{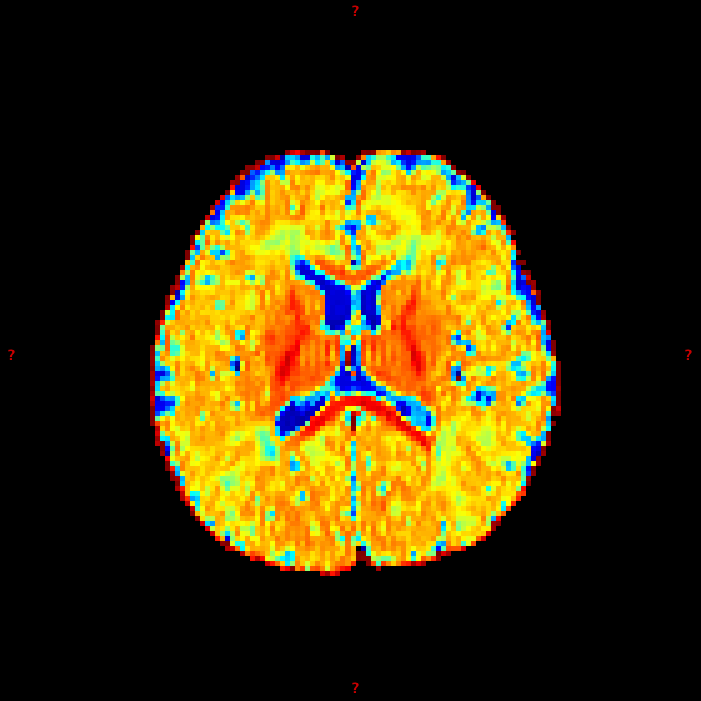}{1.15}{0.5}{0.52}{black}{\voidAnn}{}&\myPic{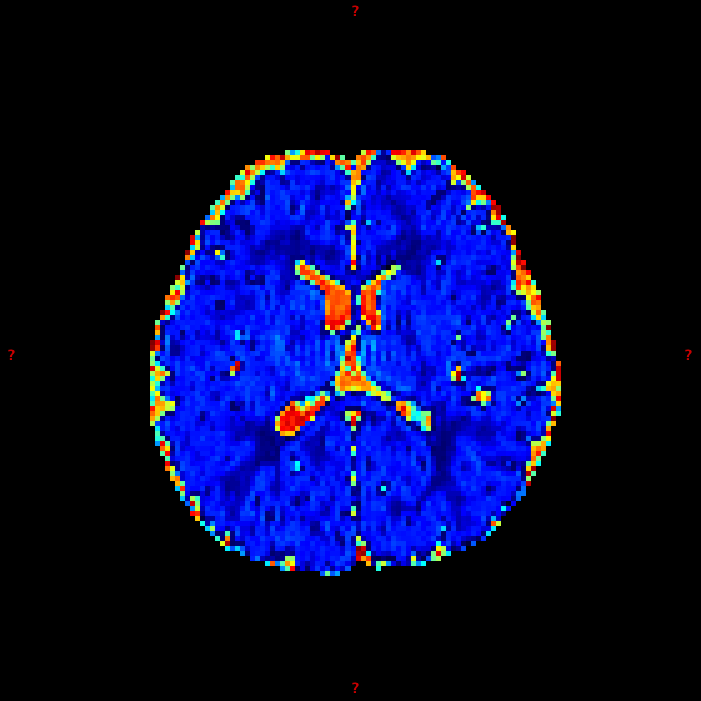}{1.15}{0.5}{0.52}{black}{\voidAnn}{}&\myPic{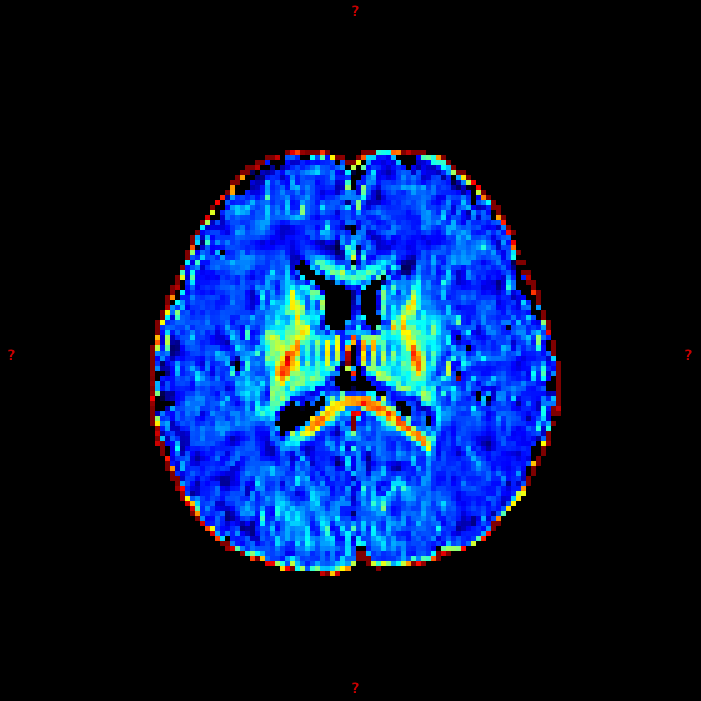}{1.15}{0.5}{0.52}{black}{\voidAnn}{}&\myPic{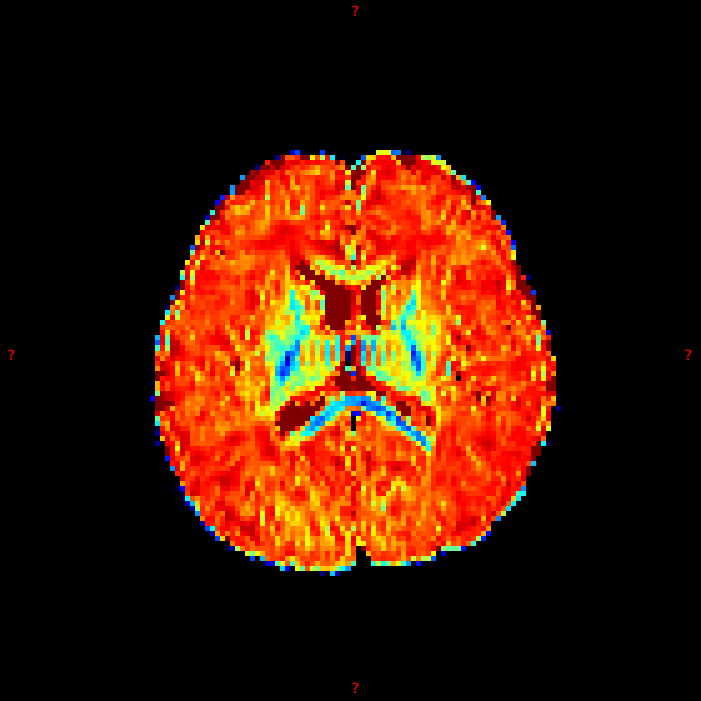}{1.15}{0.5}{0.52}{black}{\voidAnn}{}&\myPic{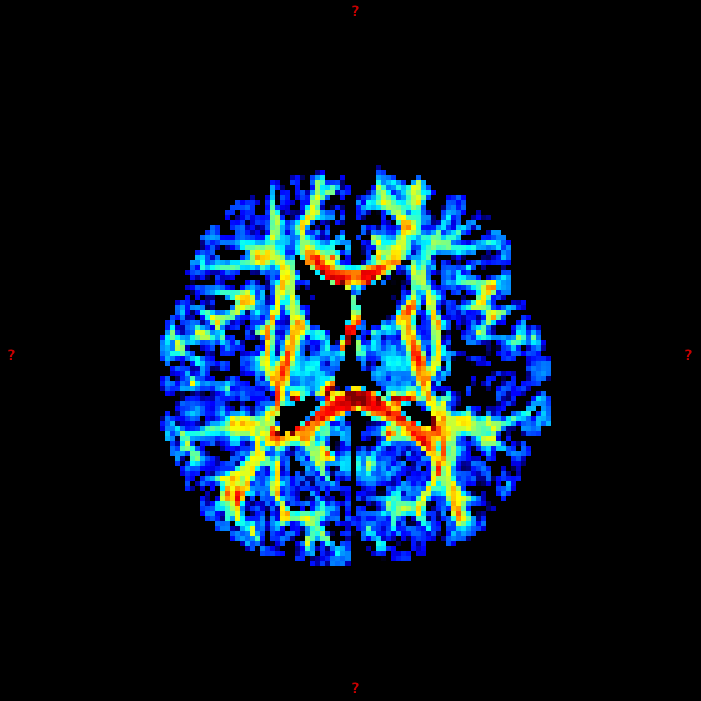}{1.15}{0.5}{0.52}{black}{\voidAnn}{}\\

&\textcolor{black}{a}
\\
&\textcolor{black}{a}
\\

\columnText{318}&\myPic{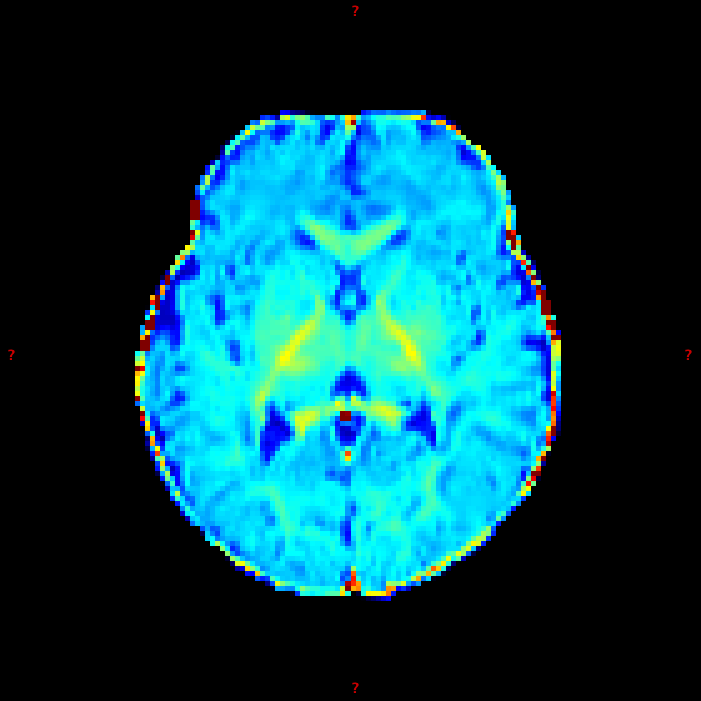}{1.1}{0.5}{0.56}{black}{\voidAnn}{}&\myPic{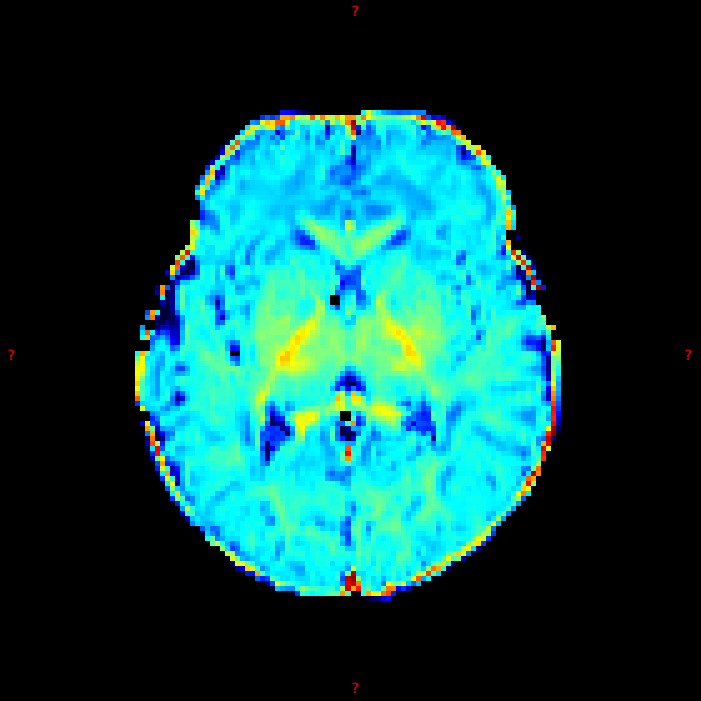}{1.1}{0.5}{0.56}{black}{\voidAnn}{}&\myPic{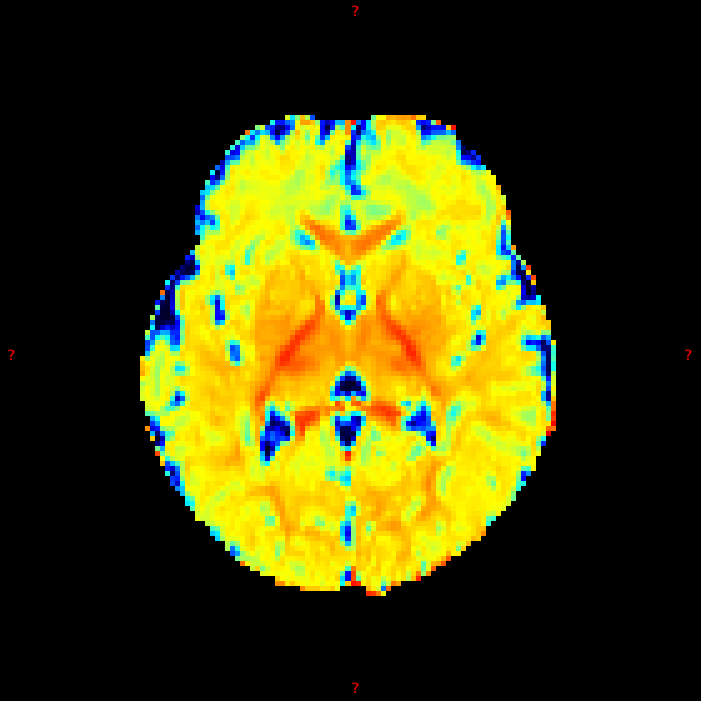}{1.1}{0.5}{0.56}{black}{\voidAnn}{}&\myPic{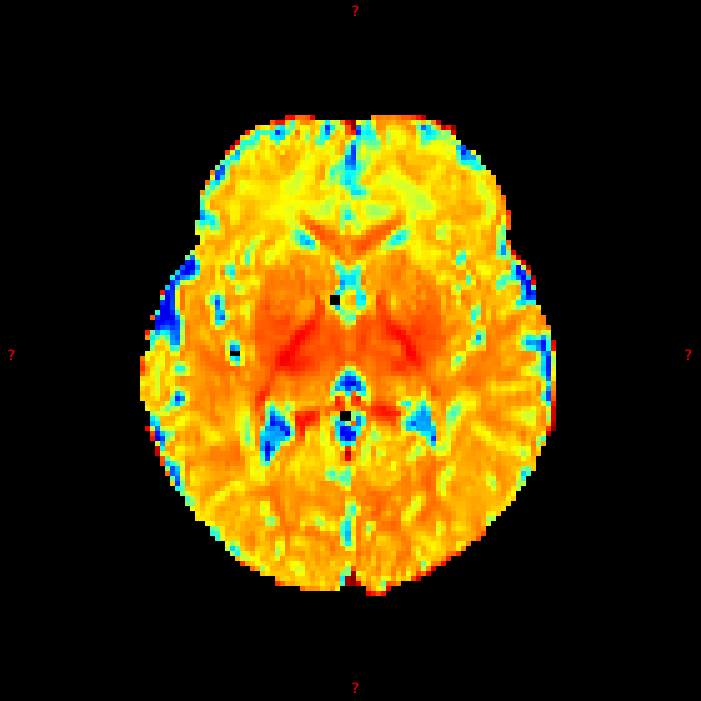}{1.1}{0.5}{0.56}{black}{\voidAnn}{}&\myPic{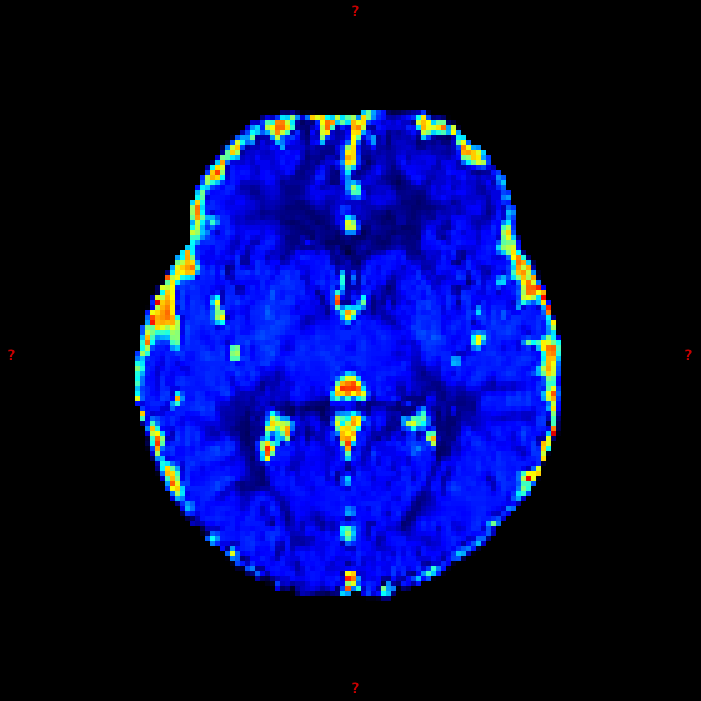}{1.1}{0.5}{0.56}{black}{\voidAnn}{}&\myPic{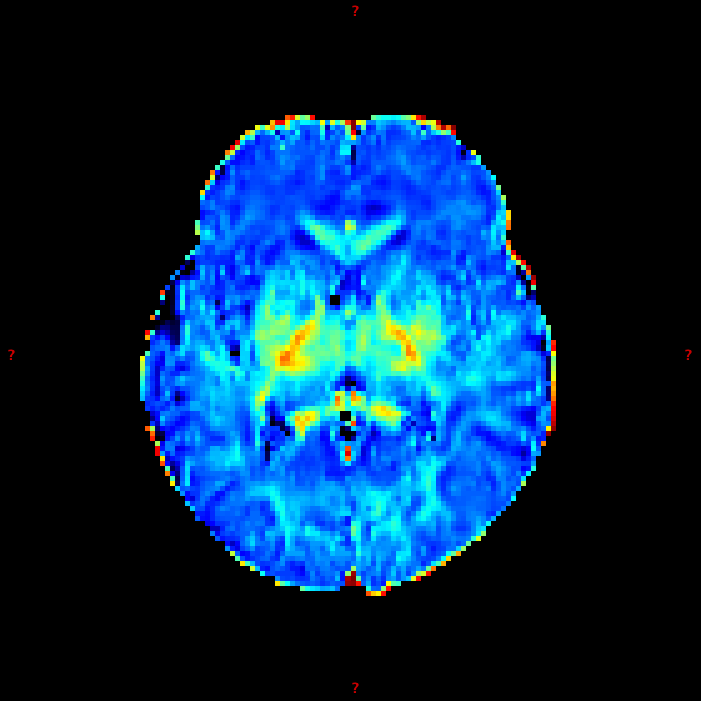}{1.1}{0.5}{0.56}{black}{\voidAnn}{}&\myPic{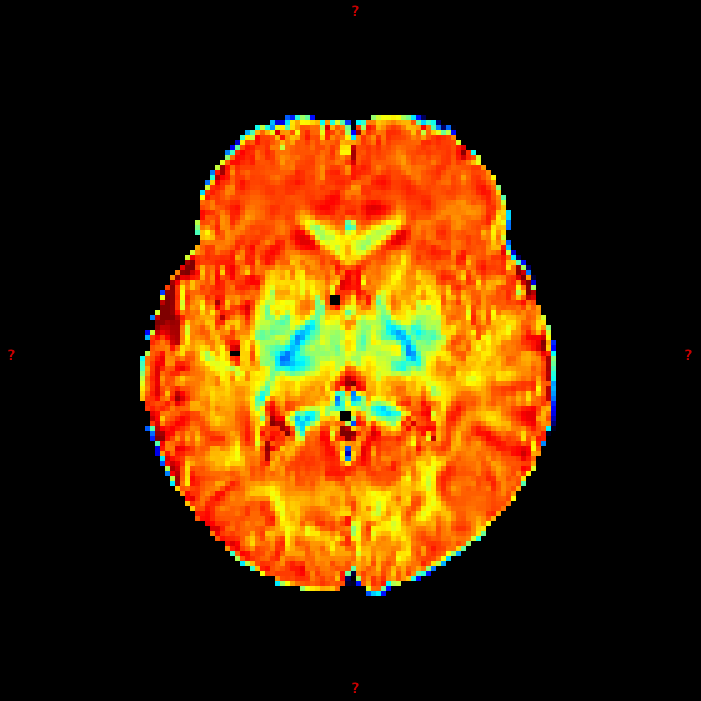}{1.1}{0.5}{0.56}{black}{\voidAnn}{}&\myPic{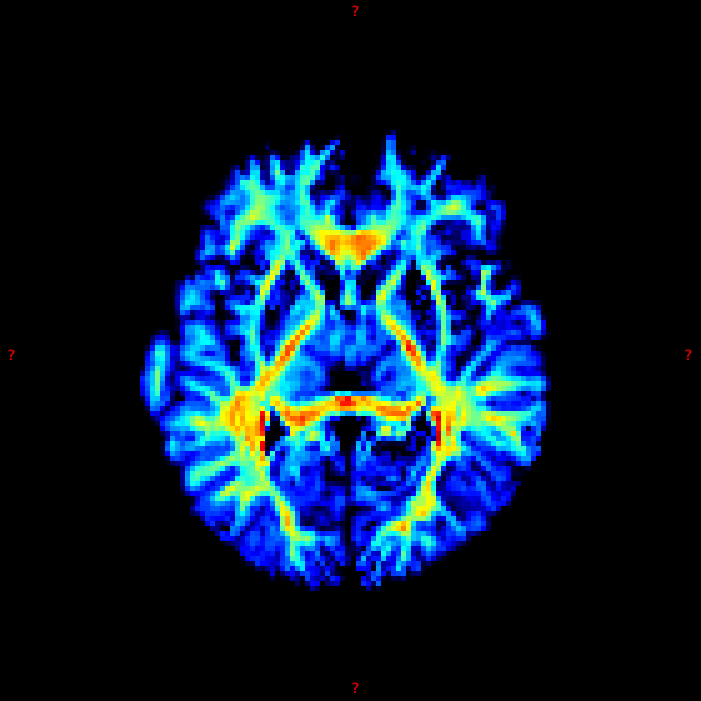}{1.1}{0.5}{0.56}{black}{\voidAnn}{}\\

\columnText{410}&\myPic{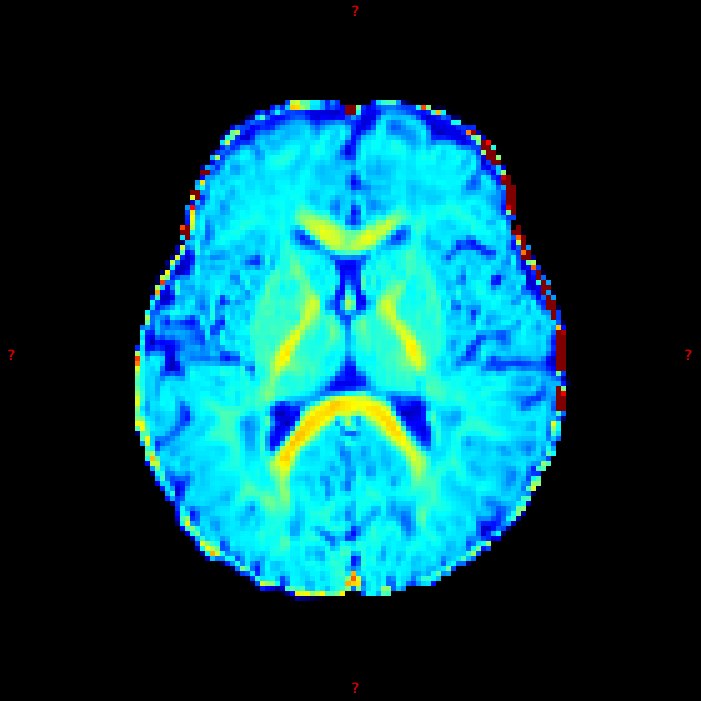}{1.2}{0.5}{0.56}{black}{\voidAnn}{}&\myPic{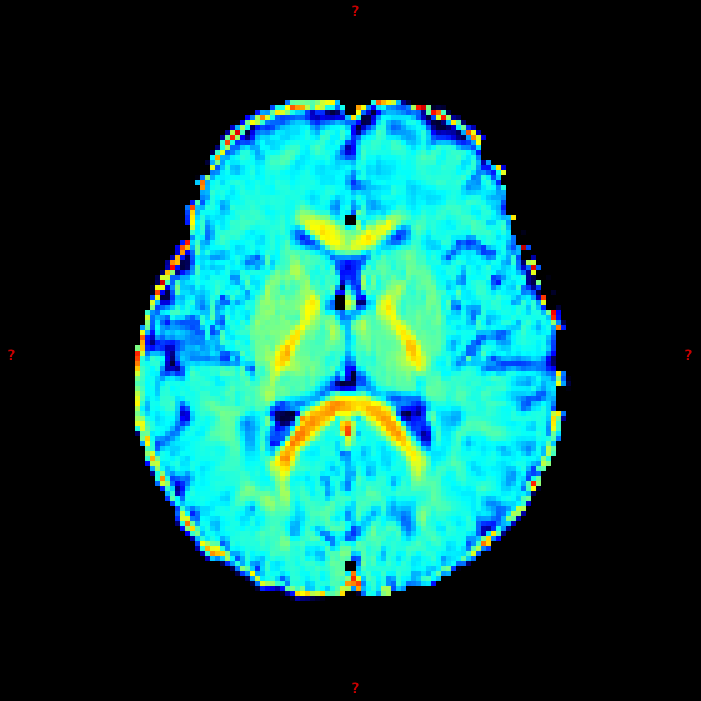}{1.2}{0.5}{0.56}{black}{\voidAnn}{}&\myPic{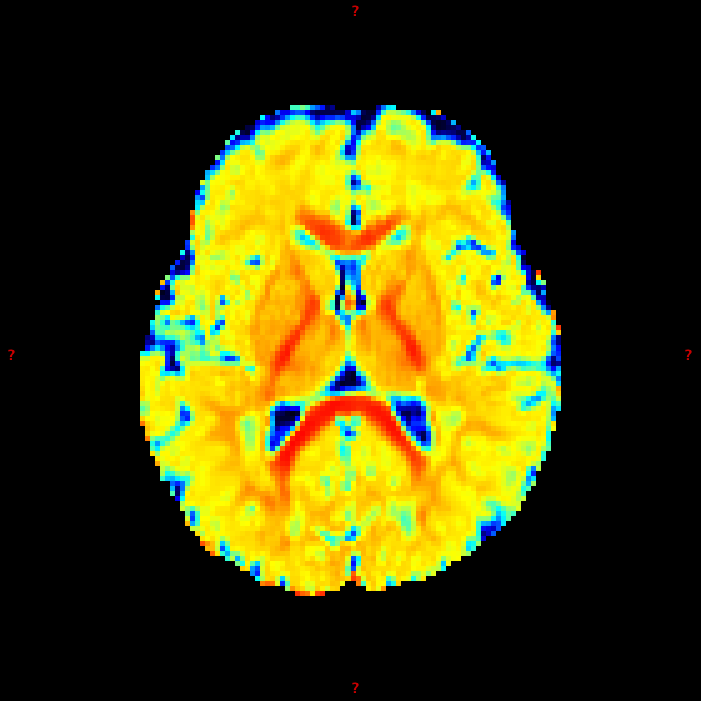}{1.2}{0.5}{0.56}{black}{\voidAnn}{}&\myPic{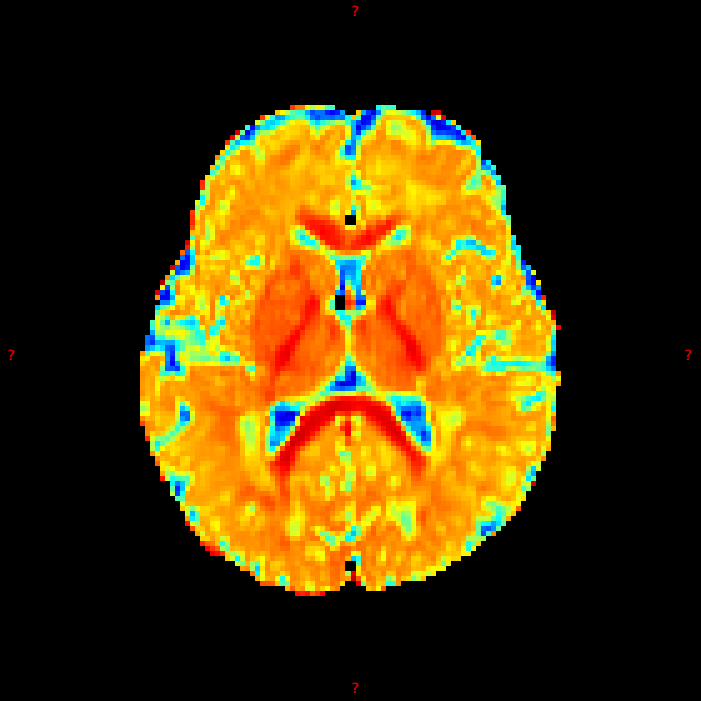}{1.2}{0.5}{0.56}{black}{\voidAnn}{}&\myPic{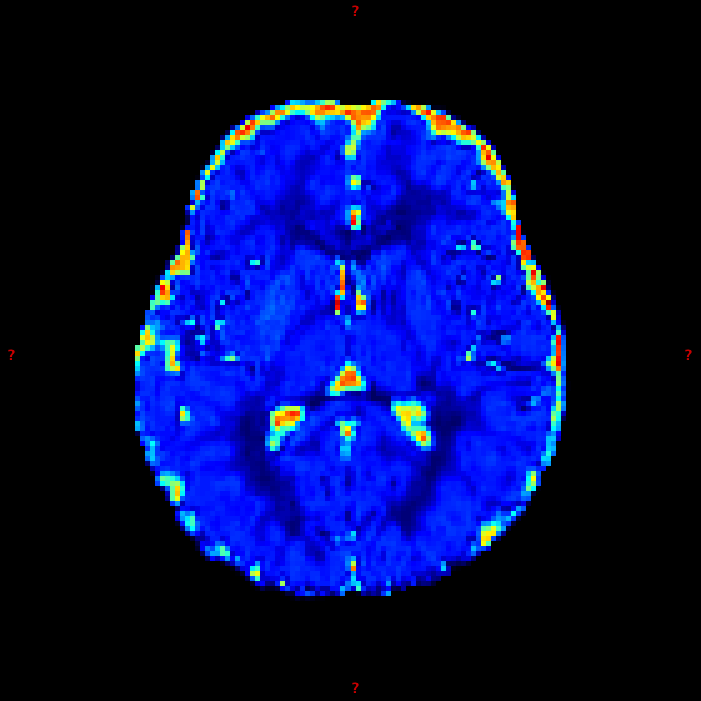}{1.2}{0.5}{0.56}{black}{\voidAnn}{}&\myPic{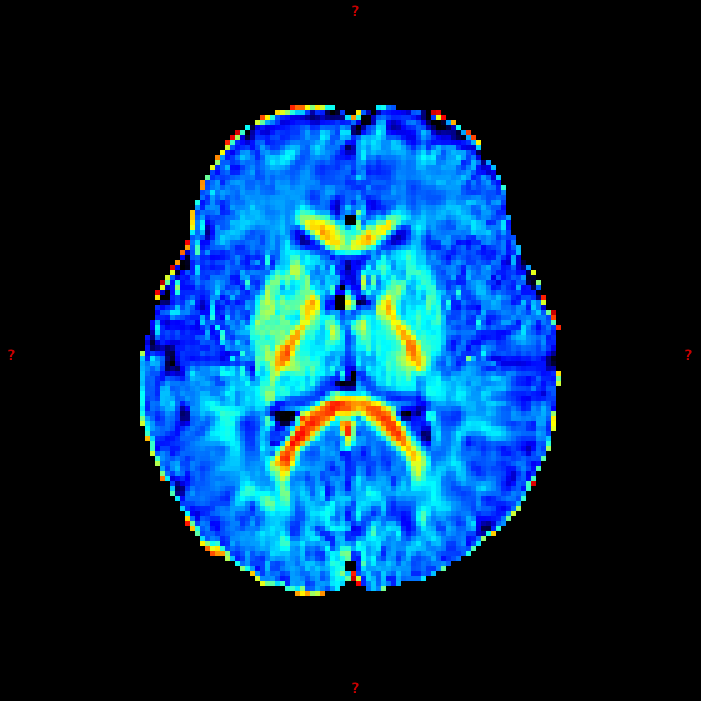}{1.2}{0.5}{0.56}{black}{\voidAnn}{}&\myPic{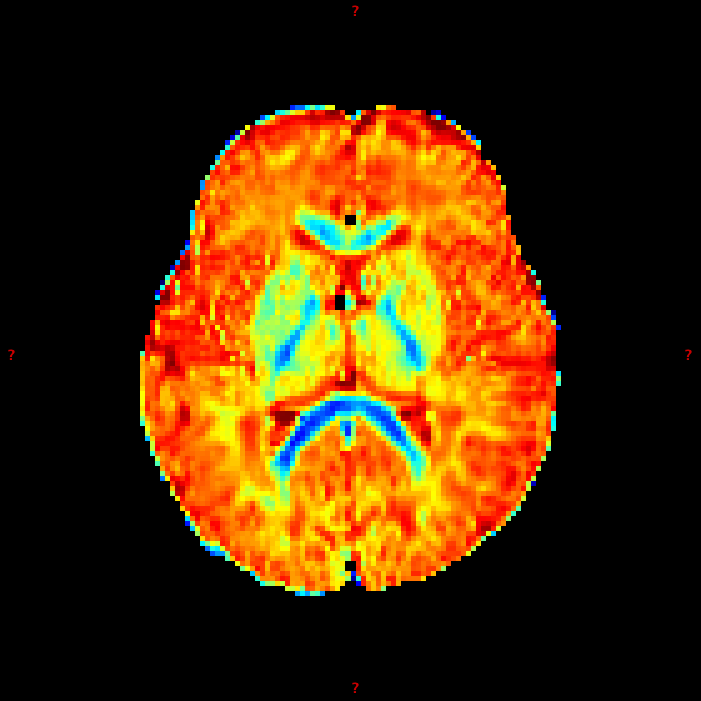}{1.2}{0.5}{0.56}{black}{\voidAnn}{}&\myPic{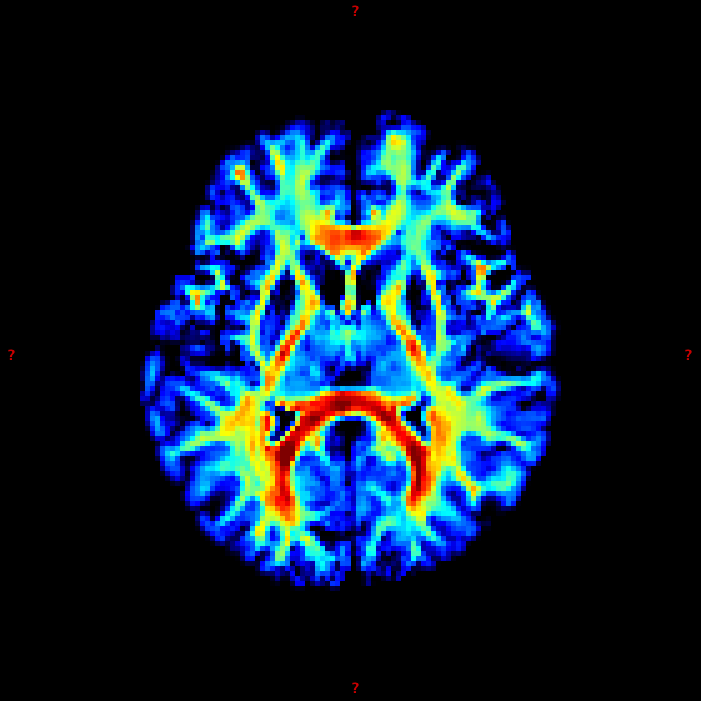}{1.2}{0.5}{0.56}{black}{\voidAnn}{}\\

\columnText{514}&\myPic{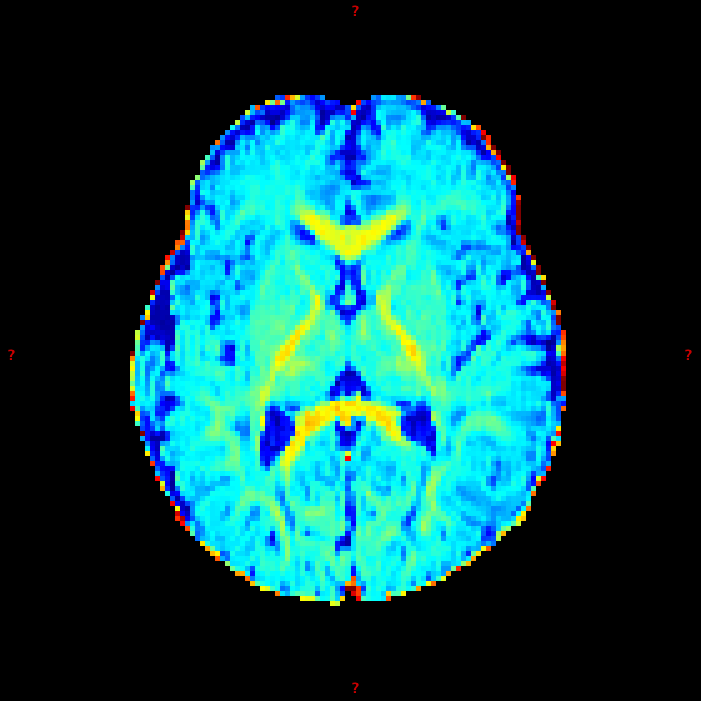}{1.21}{0.5}{0.56}{black}{\voidAnn}{}&\myPic{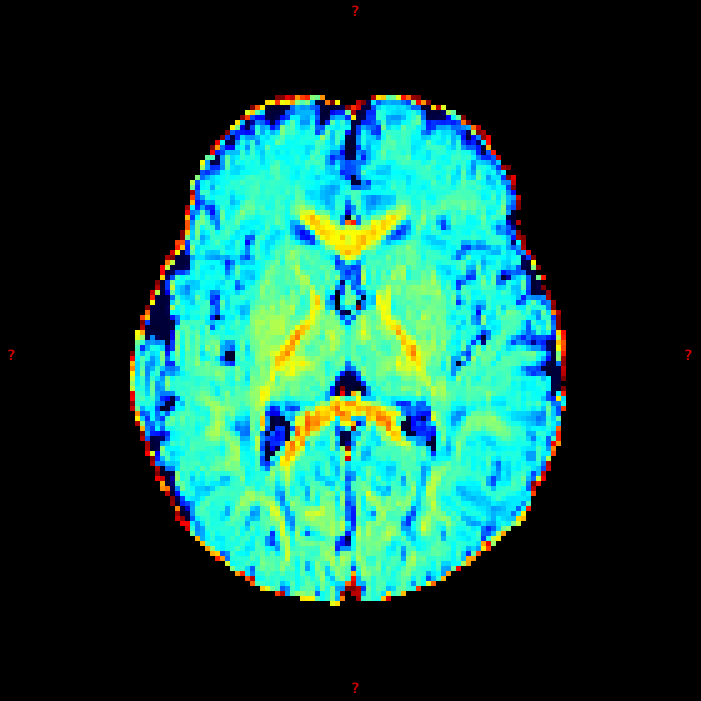}{1.21}{0.5}{0.56}{black}{\voidAnn}{}&\myPic{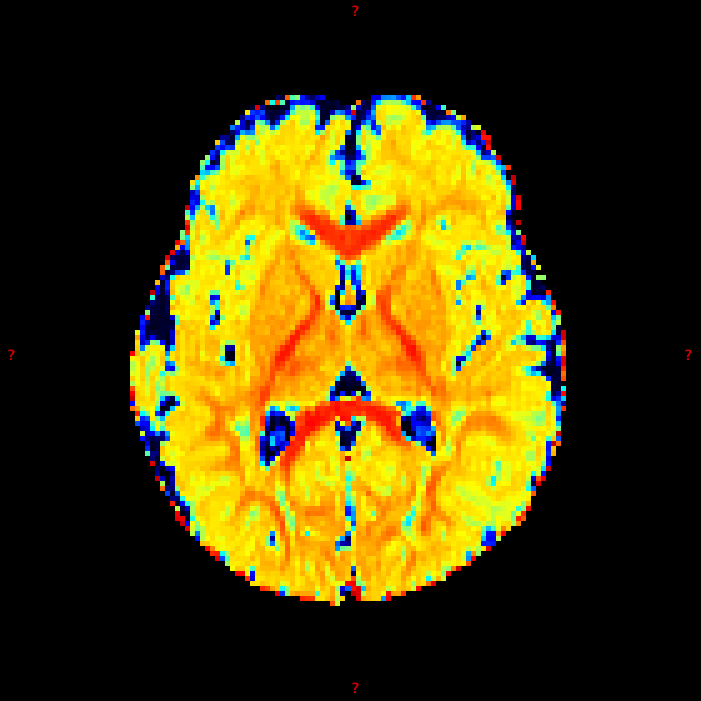}{1.21}{0.5}{0.56}{black}{\voidAnn}{}&\myPic{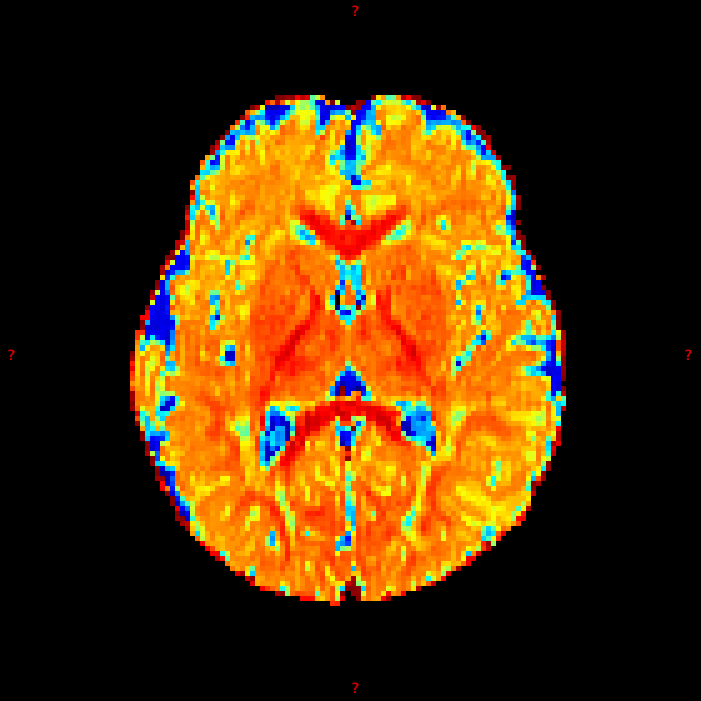}{1.21}{0.5}{0.56}{black}{\voidAnn}{}&\myPic{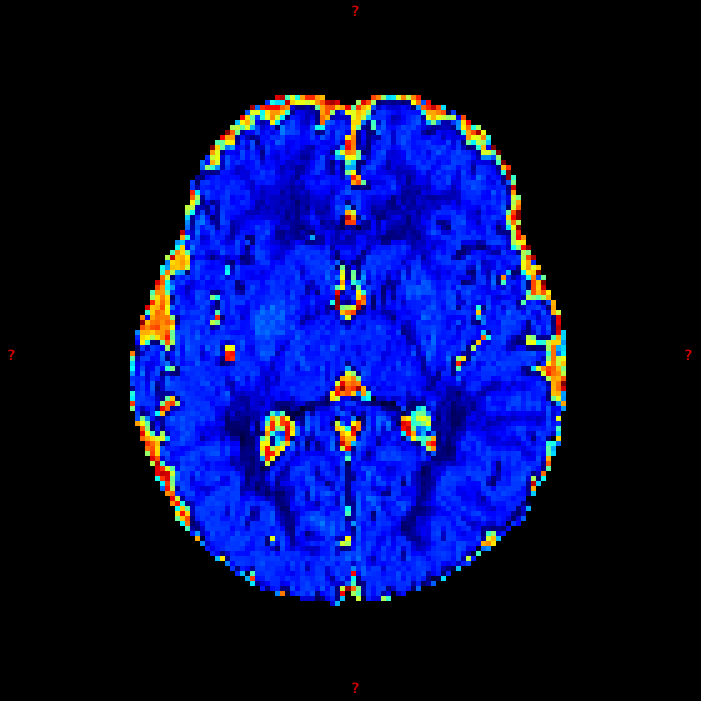}{1.21}{0.5}{0.56}{black}{\voidAnn}{}&\myPic{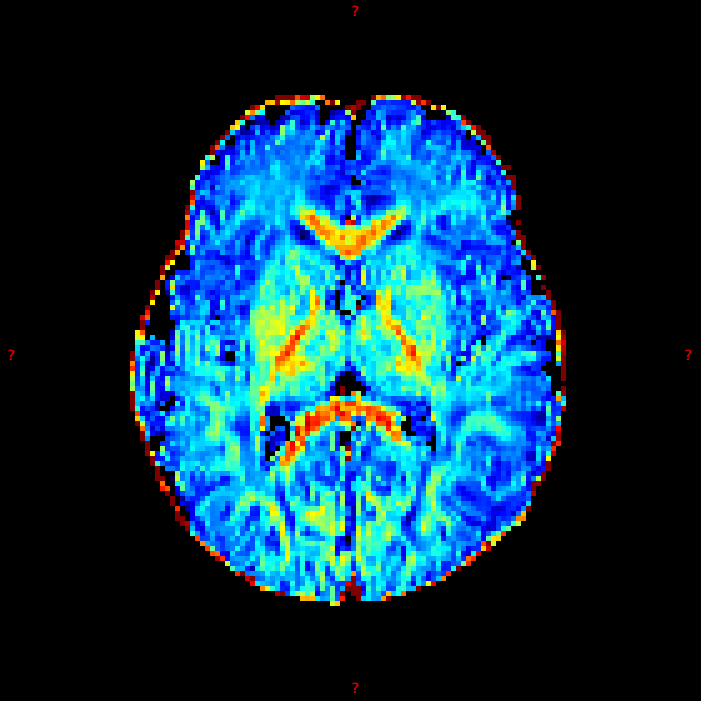}{1.21}{0.5}{0.56}{black}{\voidAnn}{}&\myPic{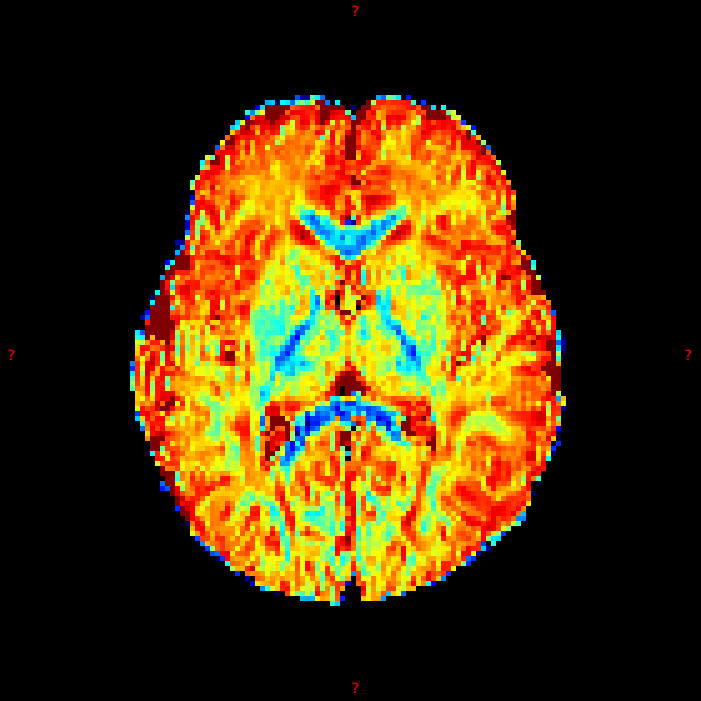}{1.21}{0.5}{0.56}{black}{\voidAnn}{}&\myPic{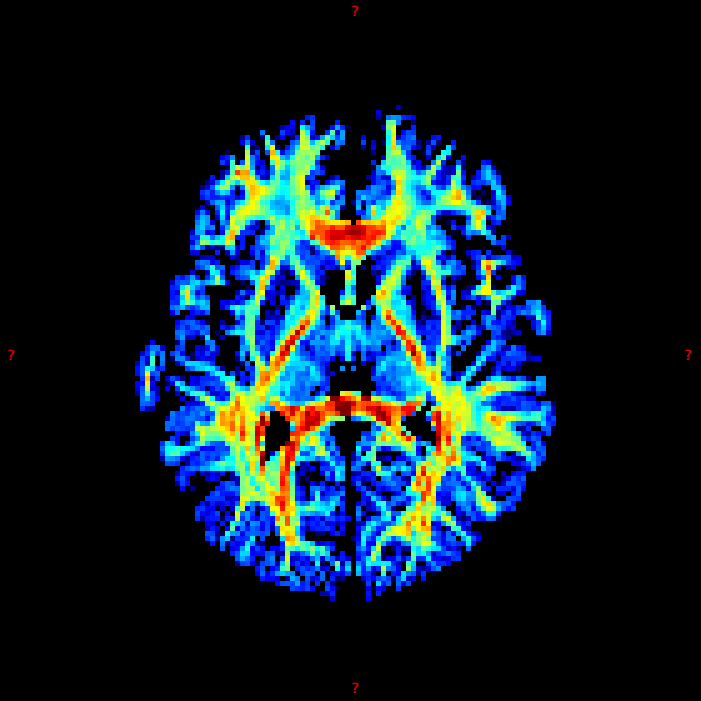}{1.3}{0.5}{0.52}{black}{\voidAnn}{}\\

&  \colorBar{0}{1}{0.5}{0.5}{colorbar} & \colorBar{0}{1}{0.5}{0.5}{colorbar} & \colorBar{0}{1}{0.5}{0.5}{colorbar} & \colorBar{0}{1}{0.5}{0.5}{colorbar} & \colorBar{0}{1}{0.5}{0.5}{colorbar} & \colorBar{0}{1}{0.5}{0.5}{colorbar}  & \colorBar{0}{1}{0.5}{0.5}{colorbar} & \colorBar{0}{1}{0.5}{0.5}{colorbar}\\

\end{tabular}

}
\end{mdframed}
\caption{
	\textbf{Longitudinal Development of Microstructure.}
	Microstructural development of two BCP subjects: one scanned at 54, 146, and 223 days after birth (top panel) and the other at 318, 410, and 514 days after birth (bottom panel). 
	\label{fig:BCP}}
\end{figure}

	Fig.~\ref{fig:BCP} shows longitudinal microstructural changes quantified via SMSI indices. 
	Note that IDE anisotropy indices (third and forth columns) give higher values than non-IDE indices (first and second columns) since isotropic diffusion lowers anisotropy.
	
	With brain development, anisotropy, coherence, and intra-cellular volume fraction increase and isotropic and extra-cellular volume fraction decrease. Spatially, development progresses from center to peripheral, and from posterior to anterior.
	This is line with prior knowledge about myelination and axon maturation \citep{dubois2014early}.
	
\begin{figure}[!htb]

	\newcommand\myPic[7]{
		\begin{tikzpicture}[baseline,trim left]
		\draw (0.0\iwidth, 0.0\iheight)[#5, ultra thick] rectangle(\iwidth, \iheight);
		#6 %
		\node[mylabel, right=0pt] at (0.0\iwidth, 1.0\iheight) {\textbf{\large{#7}}}; %
		
		\begin{pgfonlayer}{background}
		\fill (0.0\iwidth, 0.0\iheight)[black, ultra thick] rectangle(\iwidth, \iheight);
		\clip (0.0\iwidth, 0.0\iheight) rectangle (\iwidth, \iheight);
		\node at (#3\iwidth, #4\iheight) { \includegraphics[width=#2\iwidth, angle=0]{\filePath/#1} };
		\end{pgfonlayer}
		\end{tikzpicture}
	}
	
	\newcommand\columnText[1]{
	\begin{tikzpicture}[baseline,trim left]
	\node[rotate=90] at (0.1\iwidth, 0.5\iheight) {\myFontSize{#1}};
	\end{tikzpicture}
}
	\newcommand\columnTextRight[1]{
	\begin{tikzpicture}[baseline,trim left]
	\node[rotate=270] at (0.5\iwidth, 0.5\iheight) {\myFontSize{#1}};
	\end{tikzpicture}
}
	
	\newcommand\colorBar[5]{		
		\begin{tikzpicture}
		\node (bar) at (#3\iwidth, #4\iheight) { \includegraphics[width=\colorbarlength,height=0.16\colorbarlength]{\filePath/#5} };
		\node [left=-1mm of bar] {\myFontSize{#1}};
		\node [right=-1mm of bar] {\myFontSize{#2}};
		\end{tikzpicture}
	}

	\newcolumntype{L}[1]{>{\raggedright\let\newline\\\arraybackslash\hspace{0pt}}m{#1}}
\newcolumntype{C}[1]{>{\centering\let\newline\\\arraybackslash\hspace{0pt}}m{#1}}
\newcolumntype{R}[1]{>{\raggedleft\let\newline\\\arraybackslash\hspace{0pt}}m{#1}}

\begin{mdframed}[backgroundcolor=black]		
	\newcommand\voidAnn{}

\setlength{\iwidth}{0.3\textwidth}
\setlength{\iheight}{\iwidth}
\setlength{\tabcolsep}{0pt}
\setlength{\rowspacing}{0pt}
\setlength{\colorbarlength}{0.5\iwidth}
\newcommand{\myFontSize}[1]{
	\textcolor{white}{\textbf{\Large{#1}}}
}
\def \filePath {./Figures/BCP}
\renewcommand{\arraystretch}{1.25}
\renewcommand{\arrayrulewidth}{0pt}
\centering

\resizebox{0.85\linewidth}{!}
{

\begin{tabular}{lc@{~~~}c@{~~~}c@{~~~}c@{~~~}c@{~~~}c}
& \myFontSize{SMT $\mu$FA} & \myFontSize{MC-SMT  $v_{\text{ic}}$} & \myFontSize{MC-SMT  $v_{\text{ec}}$} & \myFontSize{NODDI $v_{\text{ic}}$} & \myFontSize{NODDI $v_{\text{ec}}$} & \myFontSize{NODDI $v_{\text{iso}}$}\\

\columnText{54}&\myPic{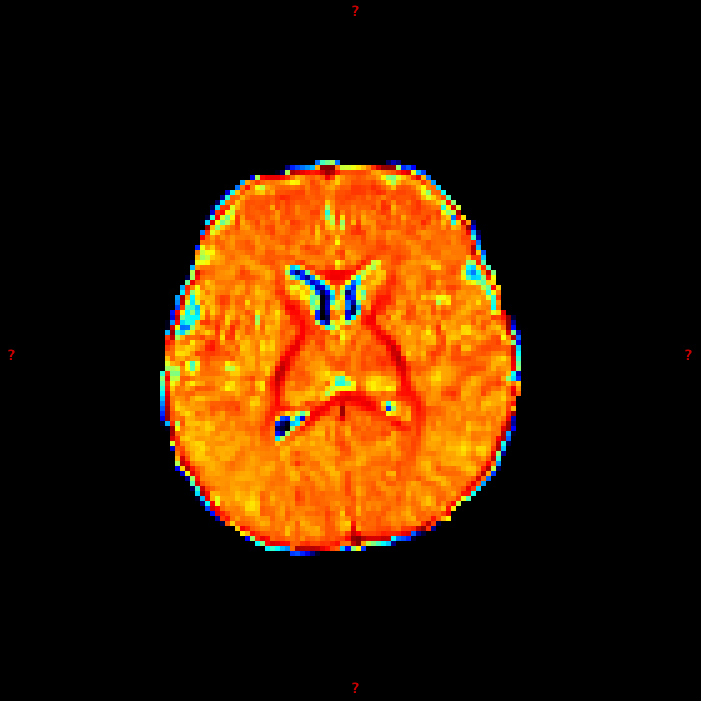}{0.9}{0.5}{0.52}{black}{\voidAnn}{}&\myPic{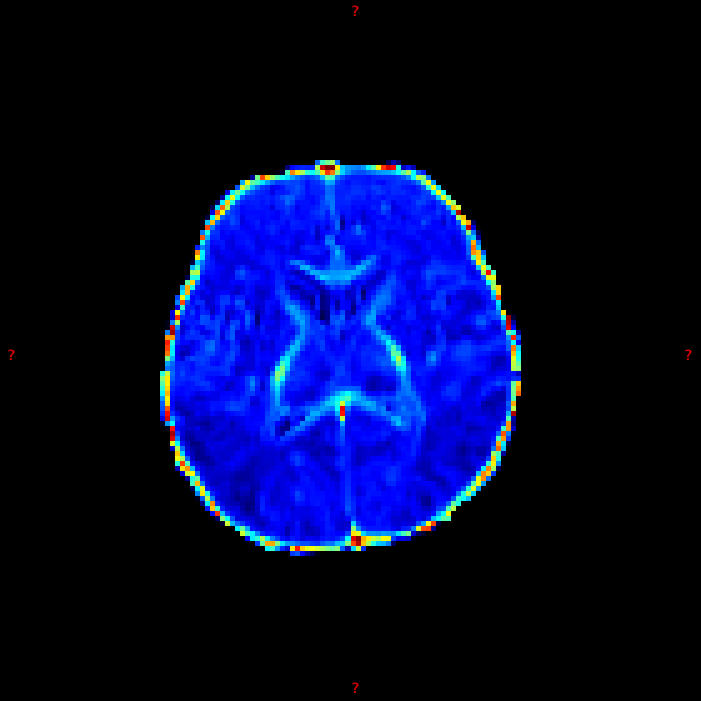}{0.9}{0.5}{0.52}{black}{\voidAnn}{}&\myPic{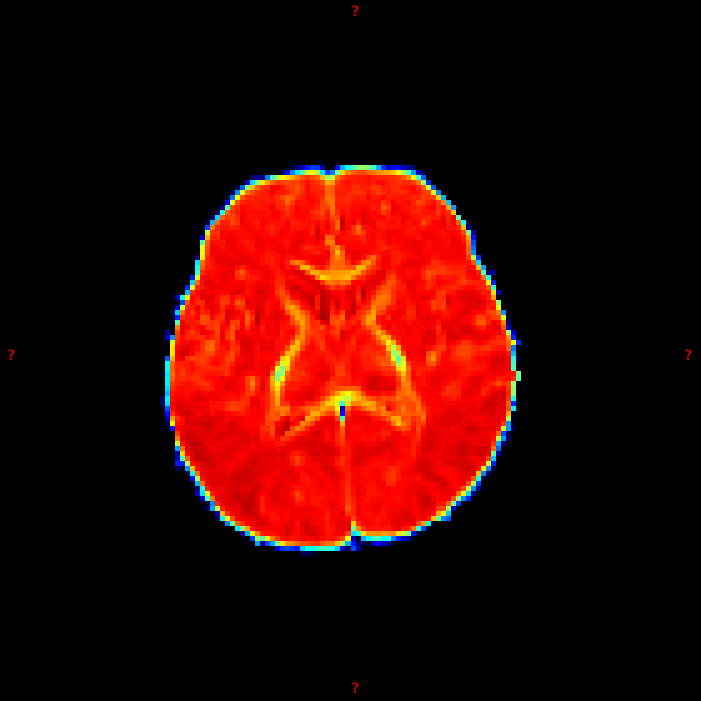}{0.9}{0.5}{0.52}{black}{\voidAnn}{}&\myPic{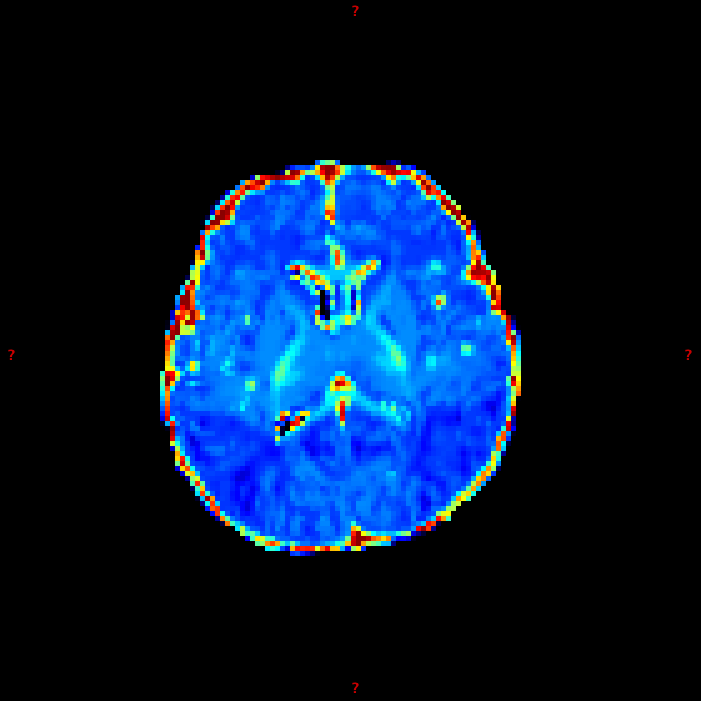}{0.9}{0.5}{0.52}{black}{\voidAnn}{}&\myPic{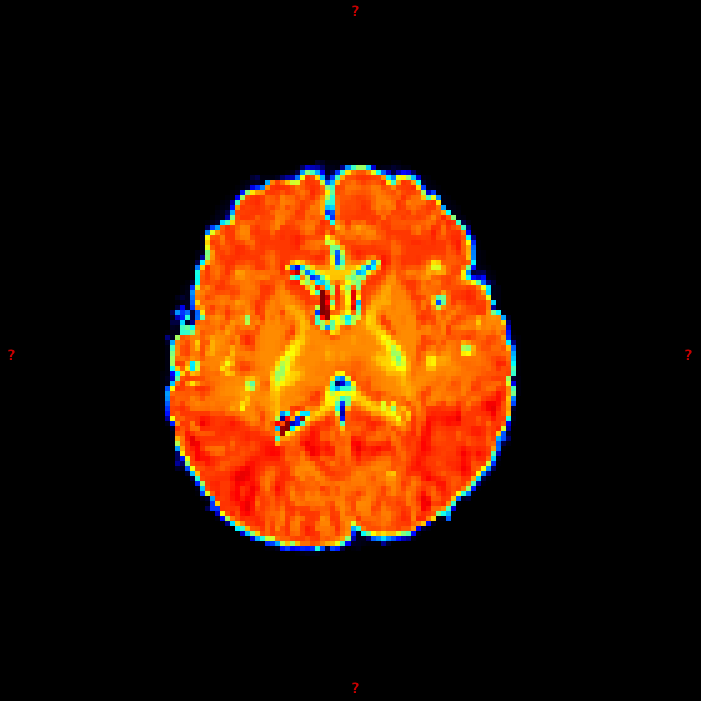}{0.9}{0.5}{0.52}{black}{\voidAnn}{}&\myPic{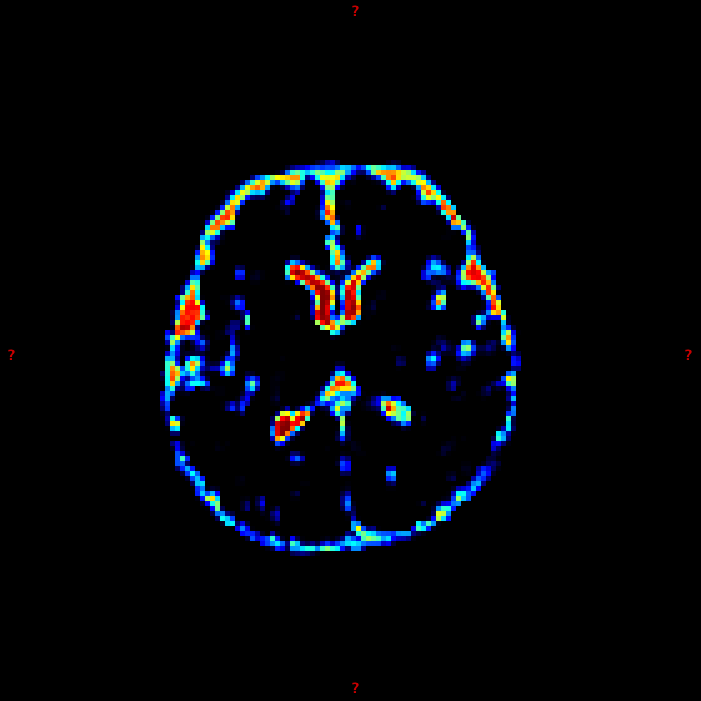}{0.9}{0.5}{0.52}{black}{\voidAnn}{}\\

\columnText{146}&\myPic{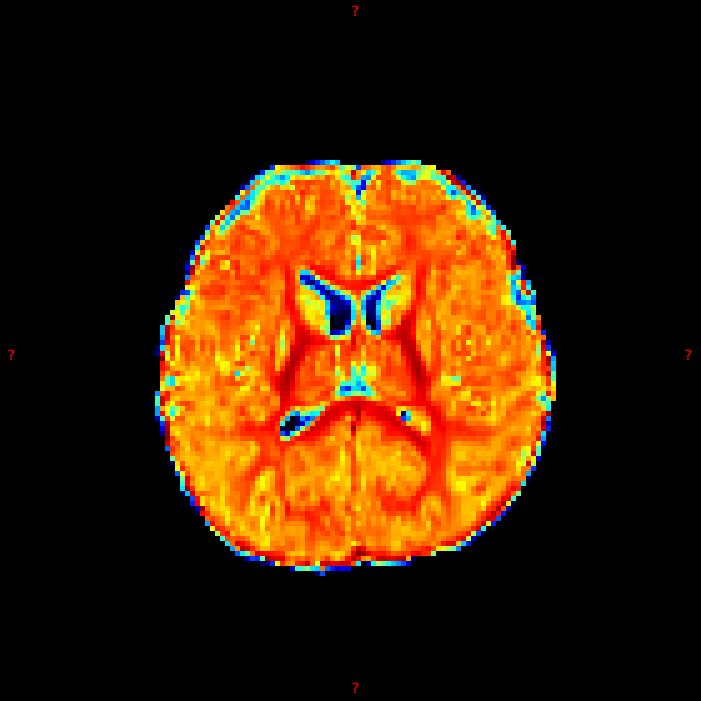}{1.0}{0.5}{0.56}{black}{\voidAnn}{}&\myPic{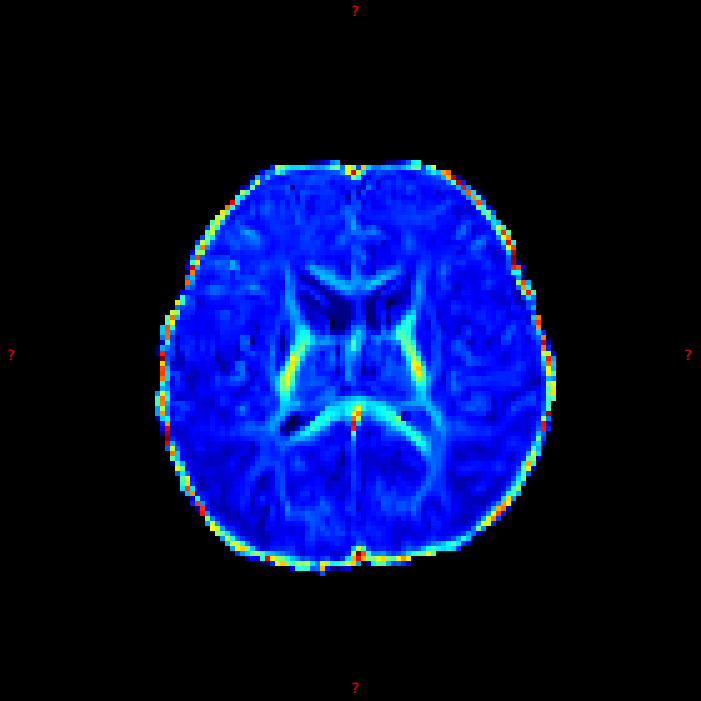}{1.0}{0.5}{0.56}{black}{\voidAnn}{}&\myPic{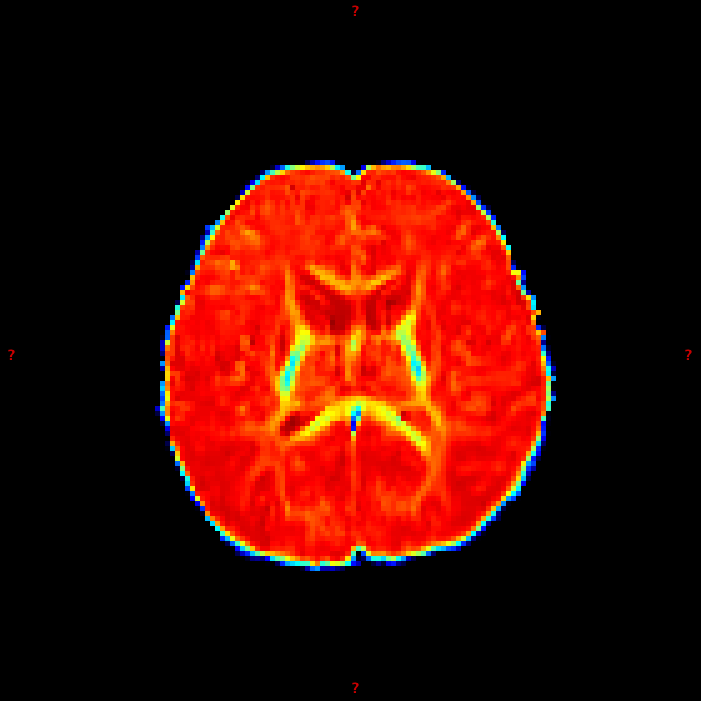}{1.0}{0.5}{0.56}{black}{\voidAnn}{}&\myPic{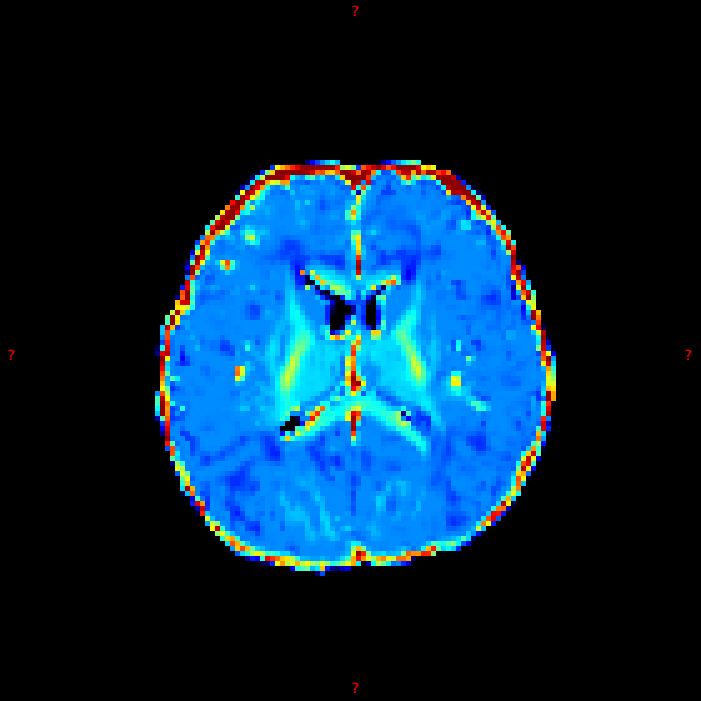}{1.0}{0.5}{0.56}{black}{\voidAnn}{}&\myPic{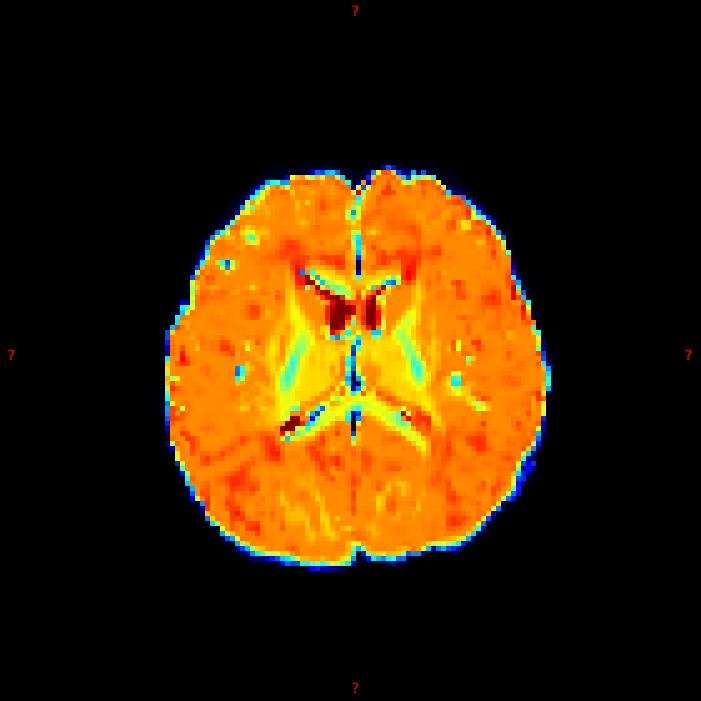}{1.0}{0.5}{0.56}{black}{\voidAnn}{}&\myPic{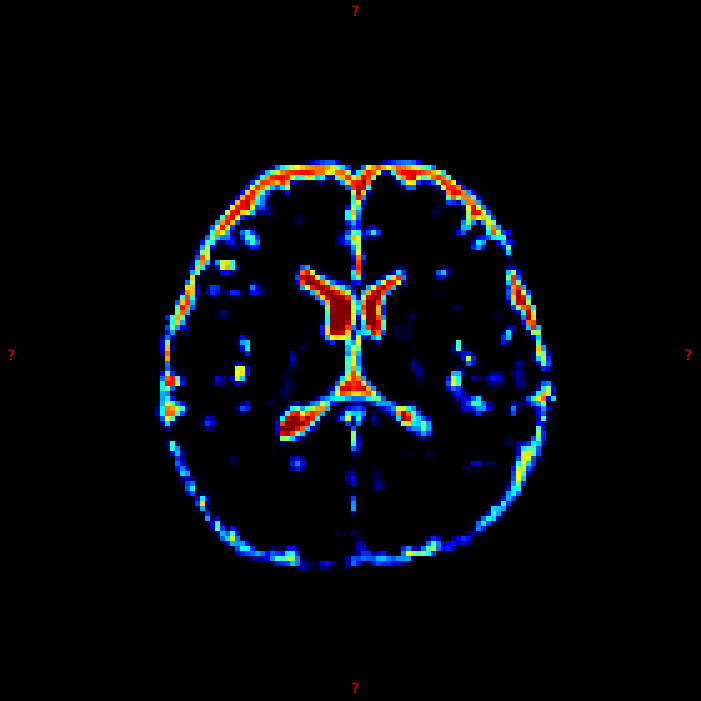}{1.0}{0.5}{0.56}{black}{\voidAnn}{}\\

\columnText{223}&\myPic{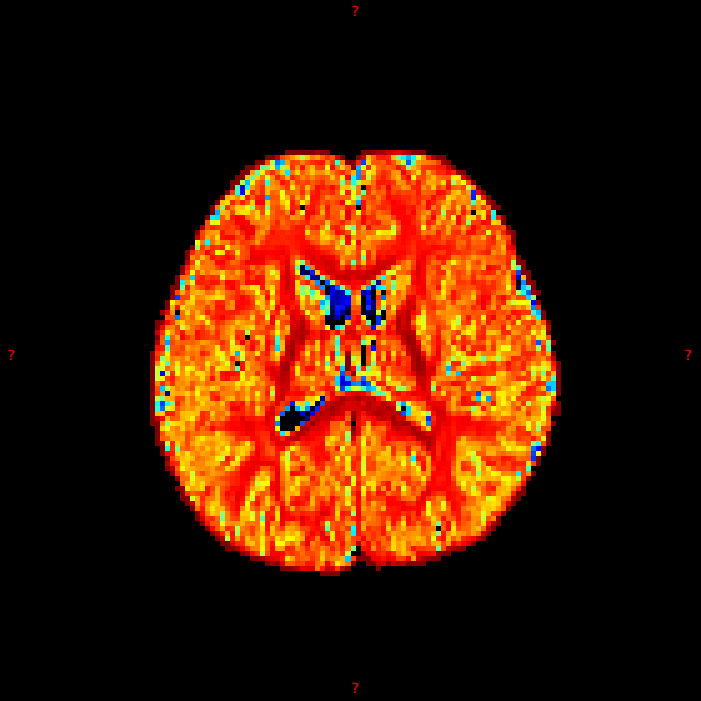}{1.15}{0.5}{0.52}{black}{\voidAnn}{}&\myPic{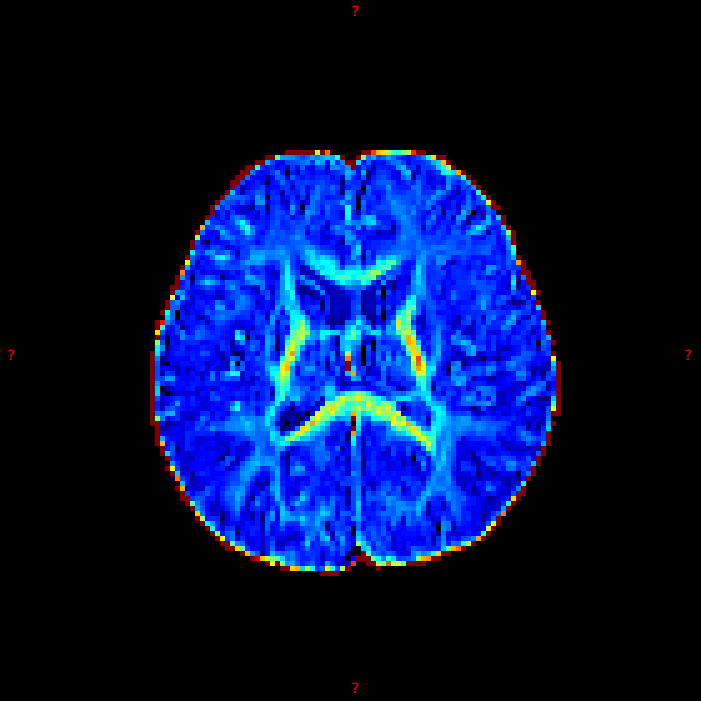}{1.15}{0.5}{0.52}{black}{\voidAnn}{}&\myPic{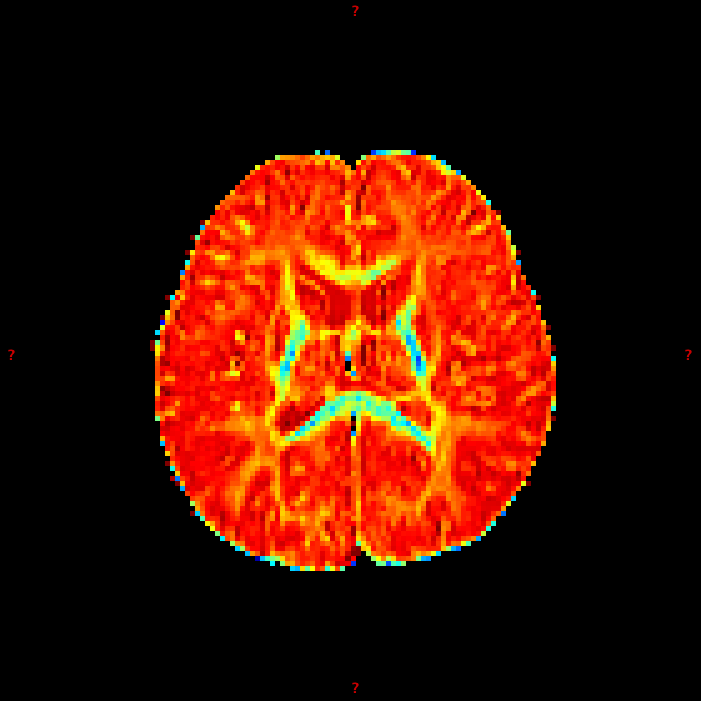}{1.15}{0.5}{0.52}{black}{\voidAnn}{}&\myPic{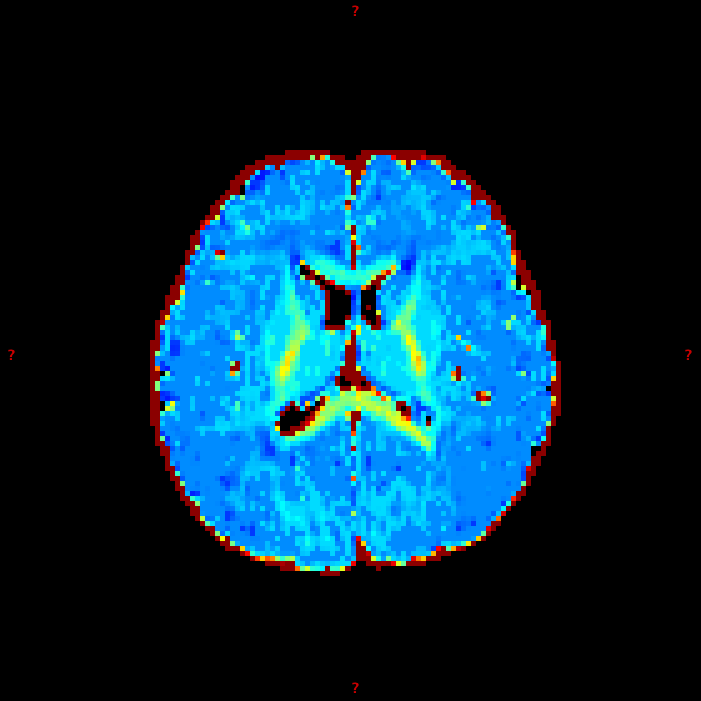}{1.15}{0.5}{0.52}{black}{\voidAnn}{}&\myPic{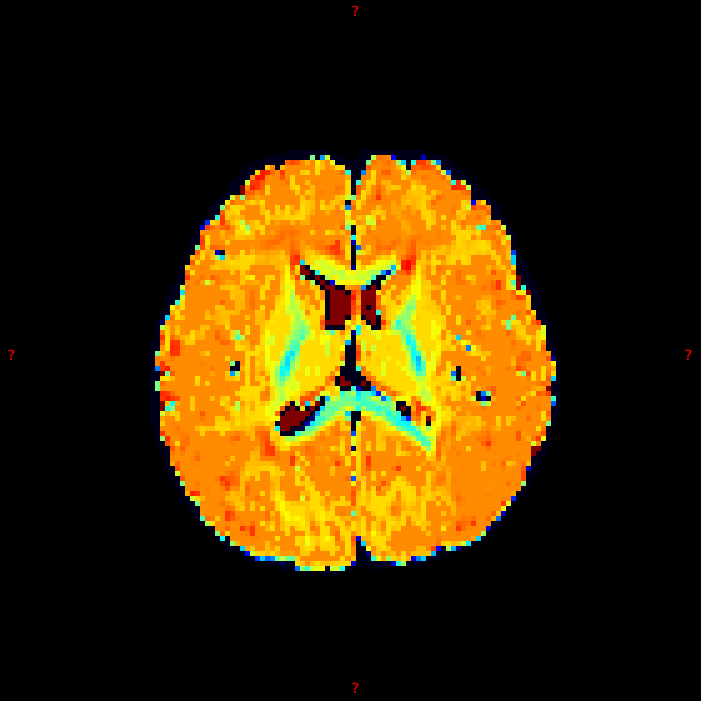}{1.15}{0.5}{0.52}{black}{\voidAnn}{}&\myPic{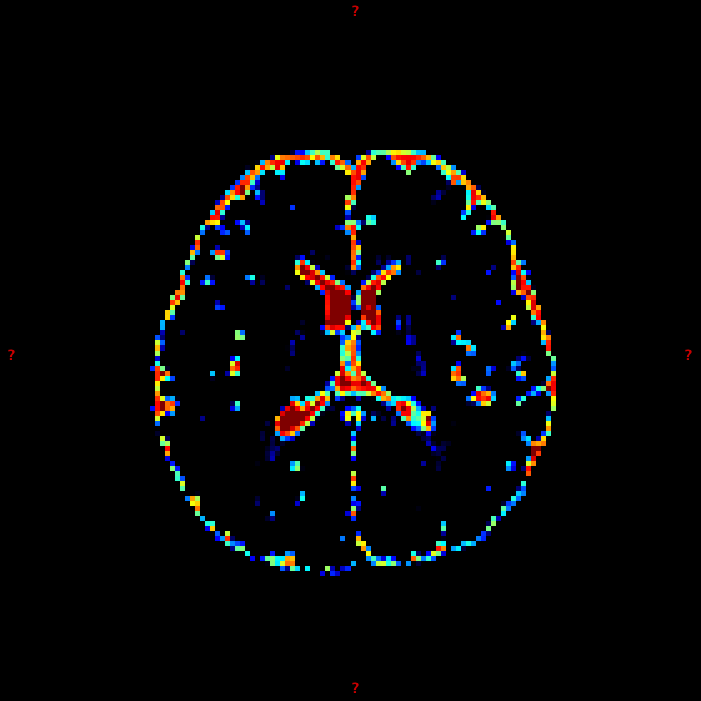}{1.15}{0.5}{0.52}{black}{\voidAnn}{}\\

&\textcolor{black}{a}
\\
&\textcolor{black}{a}
\\

\columnText{318}&\myPic{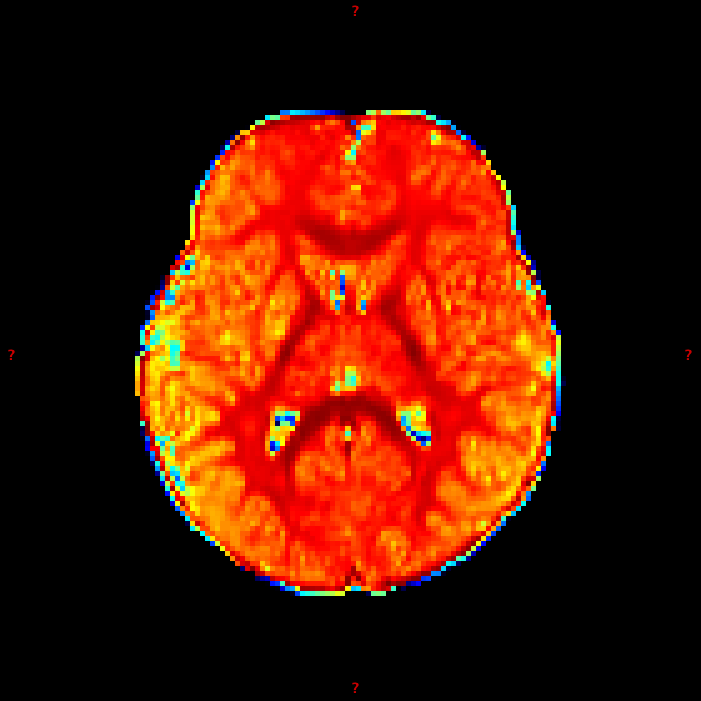}{1.1}{0.5}{0.56}{black}{\voidAnn}{}&\myPic{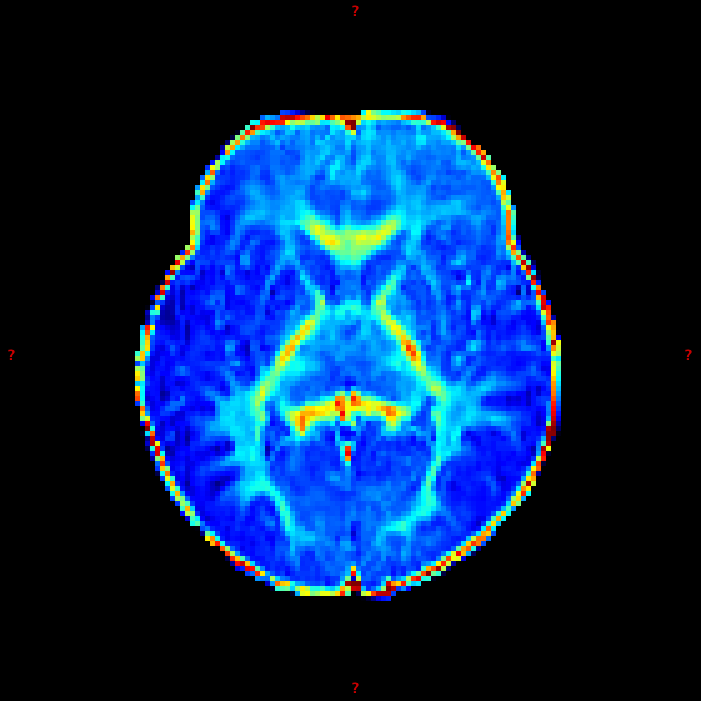}{1.1}{0.5}{0.56}{black}{\voidAnn}{}&\myPic{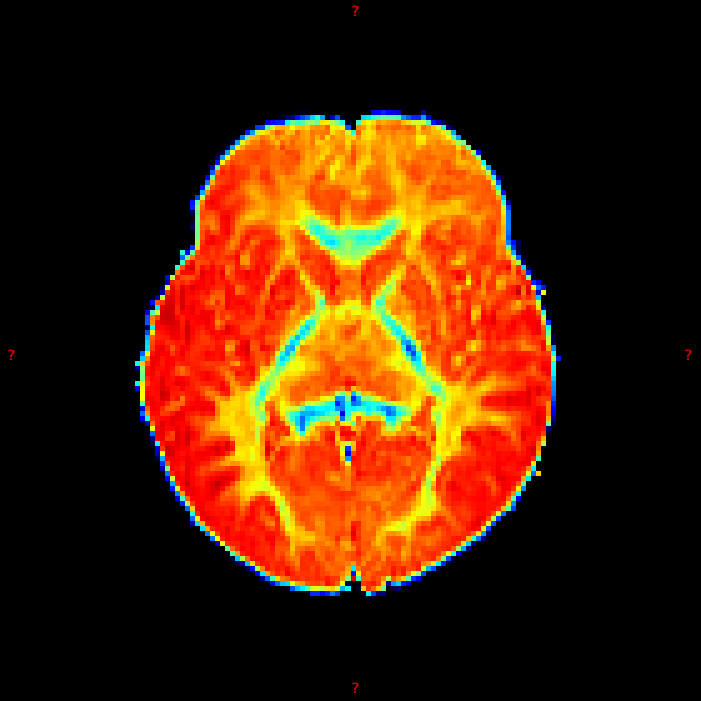}{1.1}{0.5}{0.56}{black}{\voidAnn}{}&\myPic{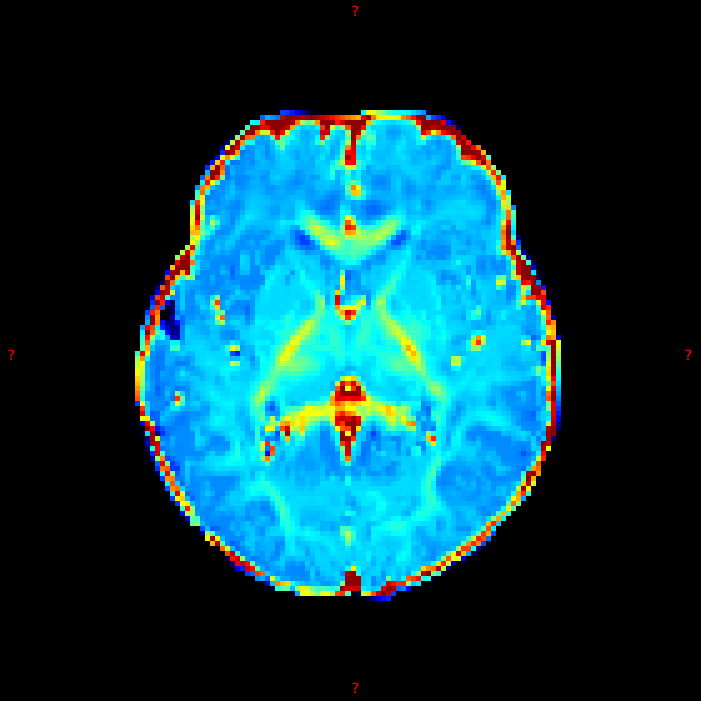}{1.1}{0.5}{0.56}{black}{\voidAnn}{}&\myPic{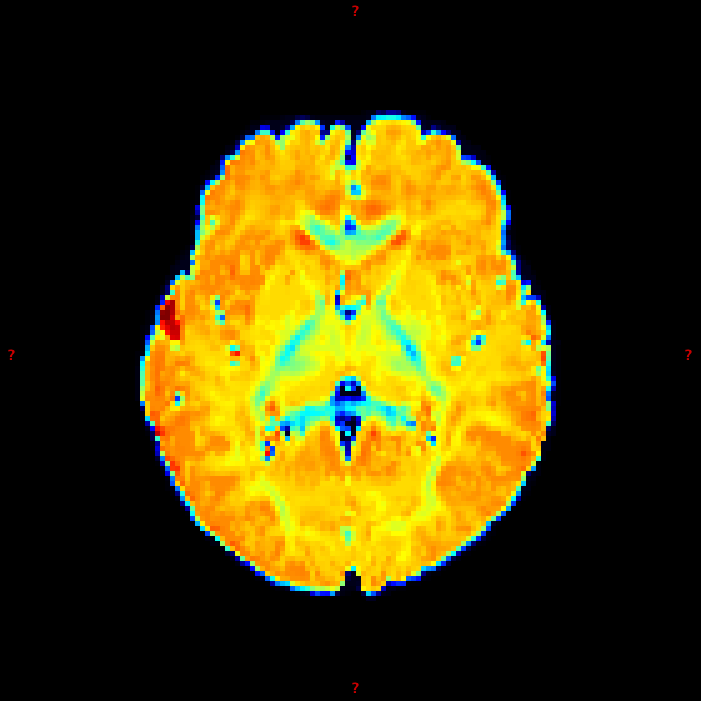}{1.1}{0.5}{0.56}{black}{\voidAnn}{}&\myPic{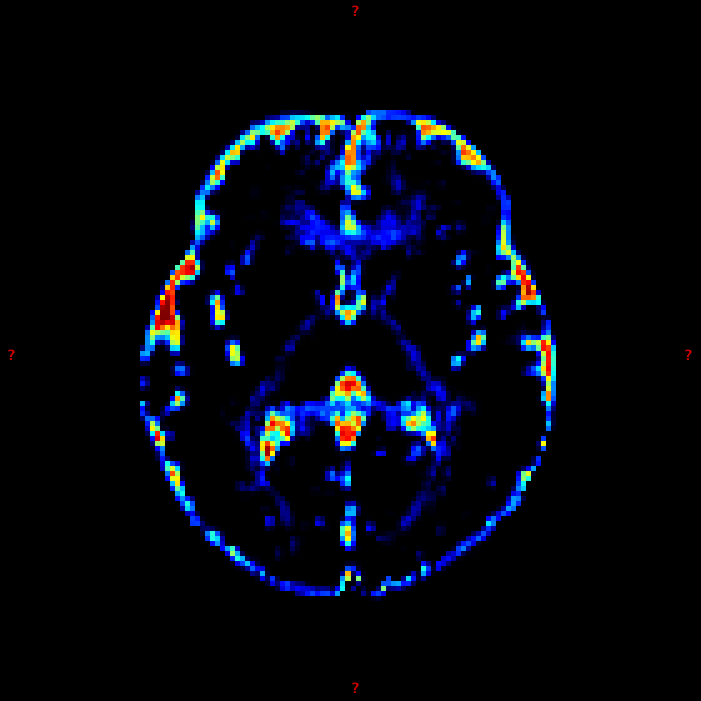}{1.1}{0.5}{0.56}{black}{\voidAnn}{}\\

\columnText{410}&\myPic{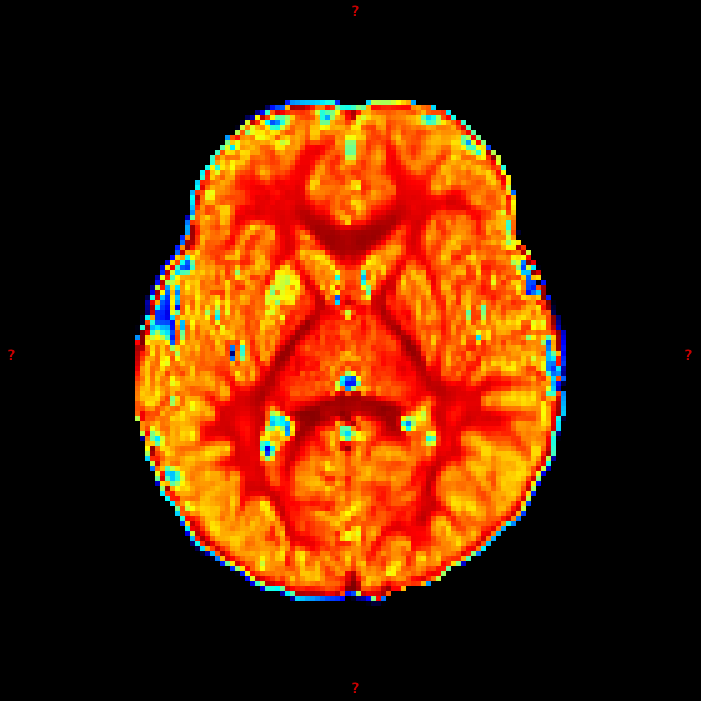}{1.2}{0.5}{0.56}{black}{\voidAnn}{}&\myPic{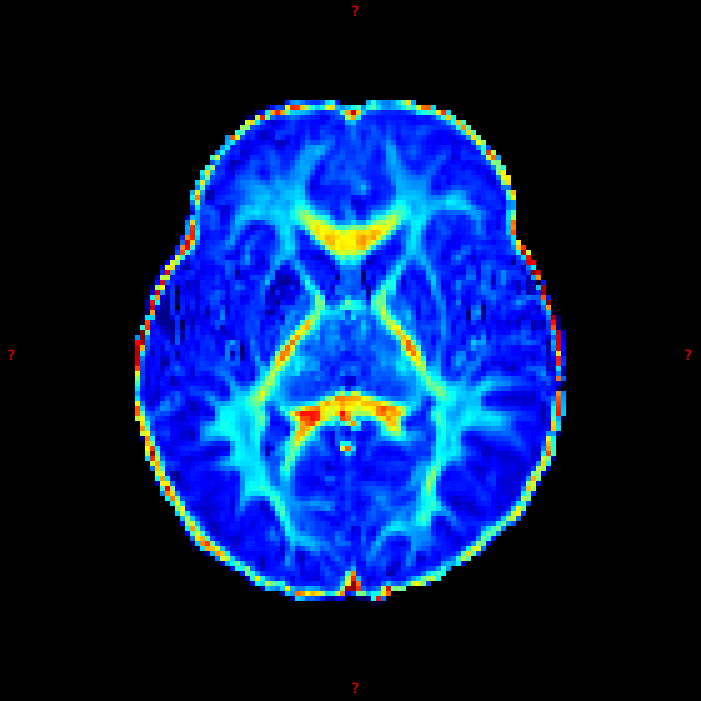}{1.2}{0.5}{0.56}{black}{\voidAnn}{}&\myPic{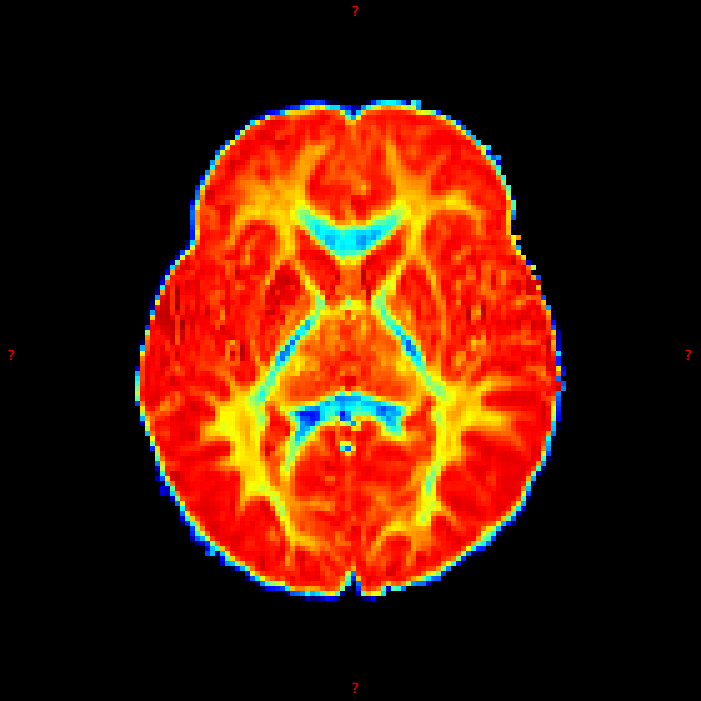}{1.2}{0.5}{0.56}{black}{\voidAnn}{}&\myPic{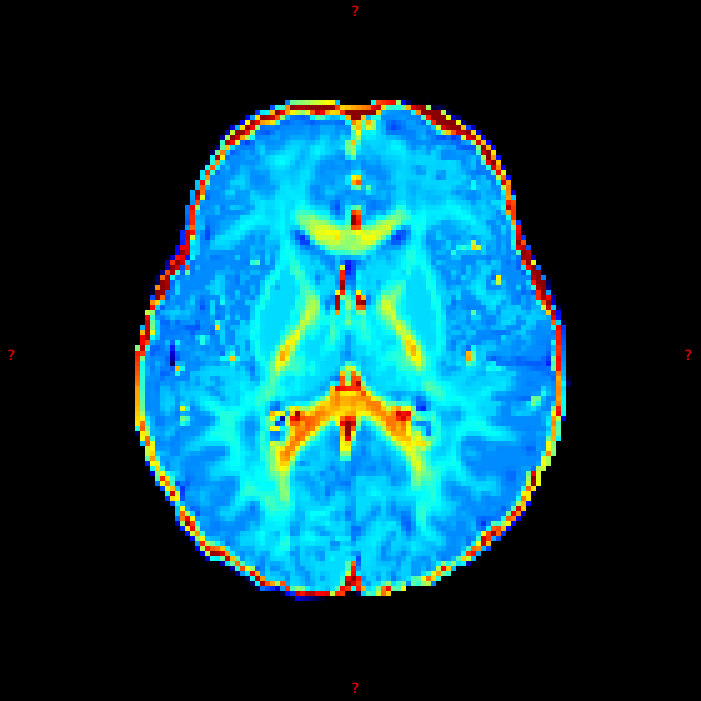}{1.2}{0.5}{0.56}{black}{\voidAnn}{}&\myPic{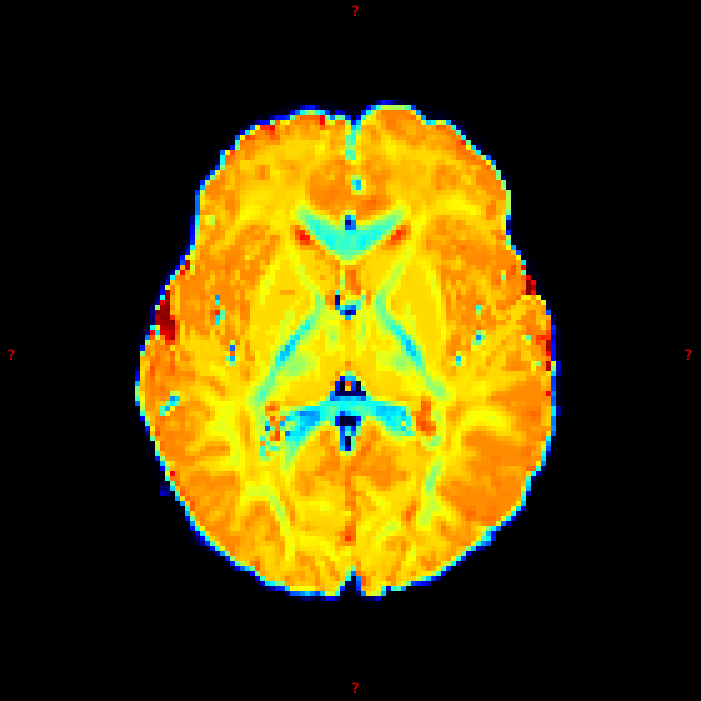}{1.2}{0.5}{0.56}{black}{\voidAnn}{}&\myPic{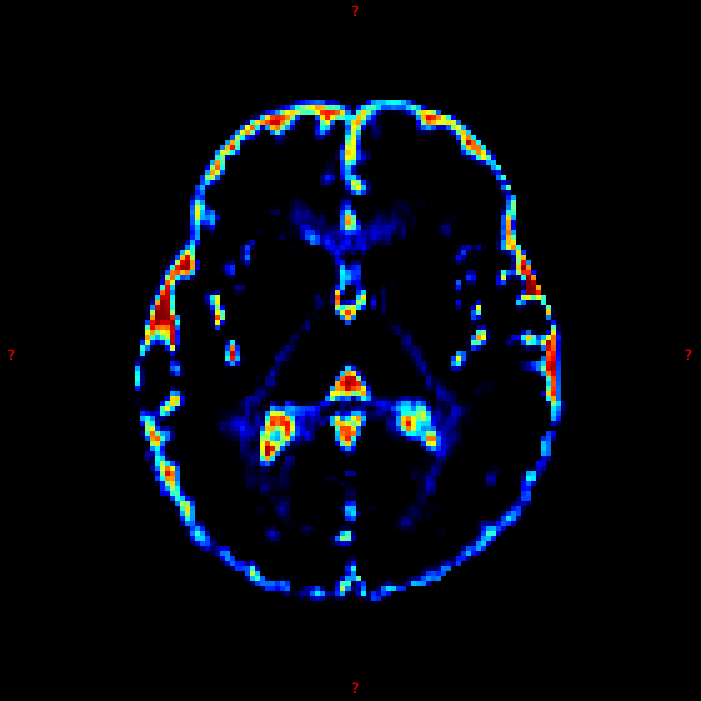}{1.2}{0.5}{0.56}{black}{\voidAnn}{}\\

\columnText{514}&\myPic{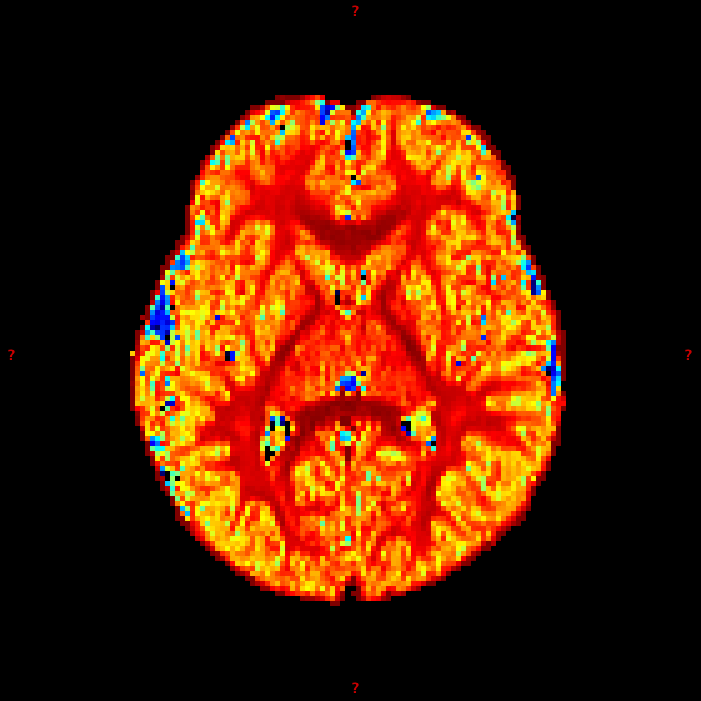}{1.21}{0.5}{0.56}{black}{\voidAnn}{}&\myPic{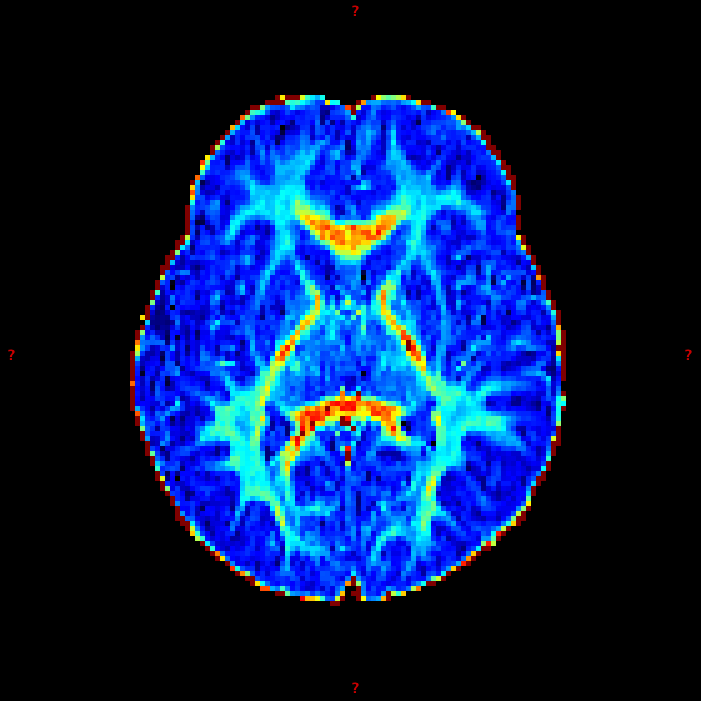}{1.21}{0.5}{0.56}{black}{\voidAnn}{}&\myPic{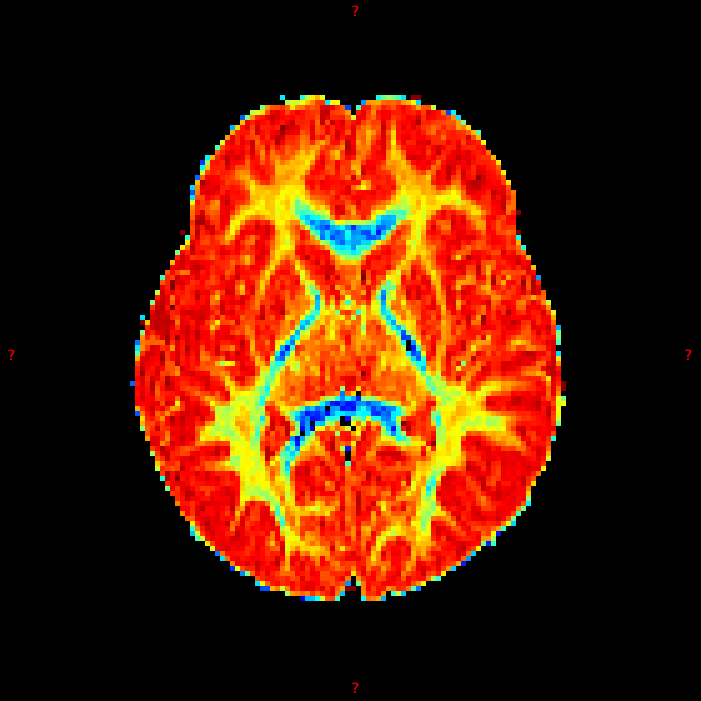}{1.21}{0.5}{0.56}{black}{\voidAnn}{}&\myPic{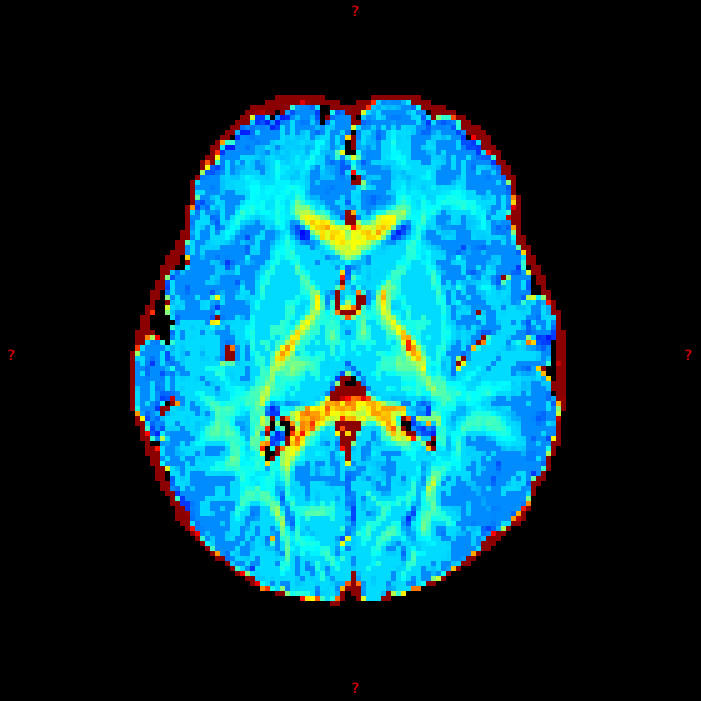}{1.21}{0.5}{0.56}{black}{\voidAnn}{}&\myPic{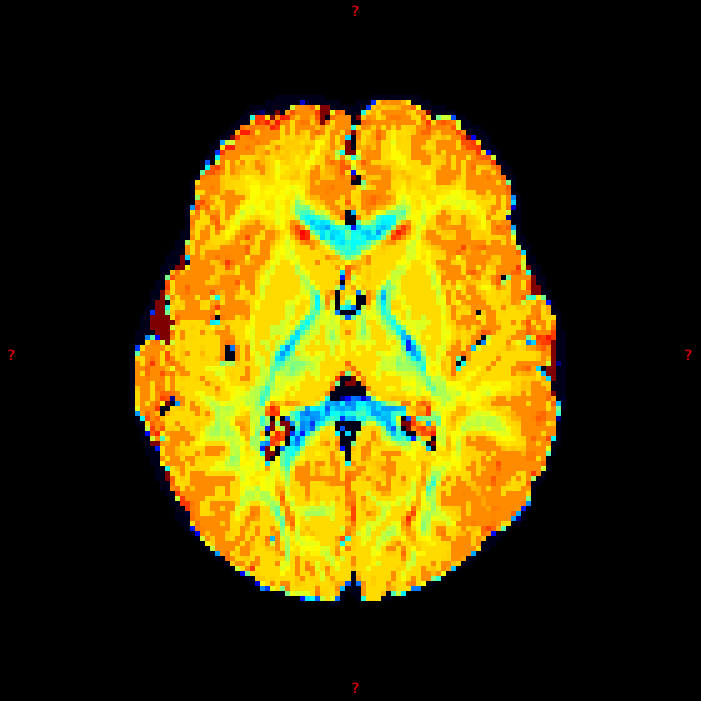}{1.21}{0.5}{0.56}{black}{\voidAnn}{}&\myPic{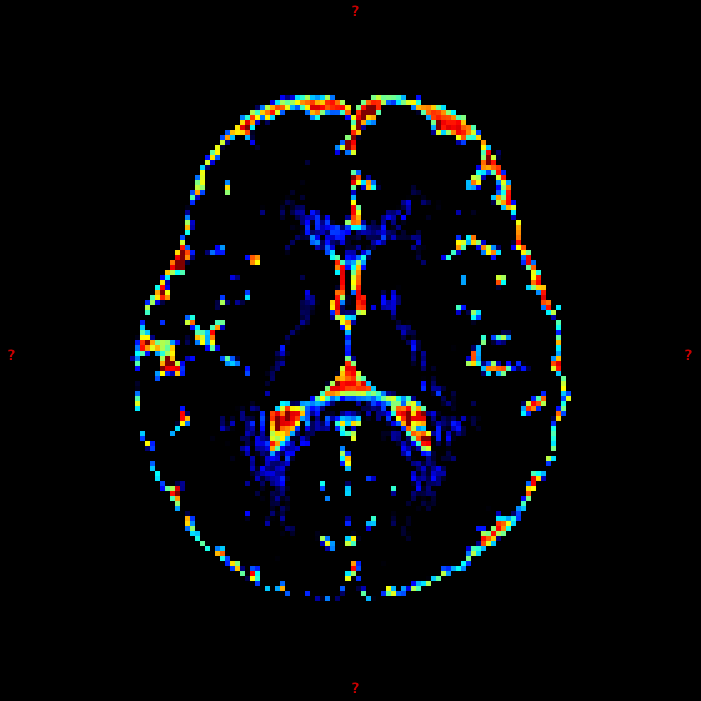}{1.21}{0.5}{0.56}{black}{\voidAnn}{}\\

&  \colorBar{0}{1}{0.5}{0.5}{colorbar} & \colorBar{0}{1}{0.5}{0.5}{colorbar} & \colorBar{0}{1}{0.5}{0.5}{colorbar} & \colorBar{0}{1}{0.5}{0.5}{colorbar} & \colorBar{0}{1}{0.5}{0.5}{colorbar} & \colorBar{0}{1}{0.5}{0.5}{colorbar} \\

\end{tabular}

}
\end{mdframed}
\caption{
	\textbf{Longitudinal SMT, MC-SMT, and NODDI Indices.}
	Similar to Fig.~\ref{fig:BCP} but based on SMT, MC-SMT, and NODDI.
	\label{fig:BCP_othermodel}}
\end{figure}

	Fig.~\ref{fig:BCP_othermodel} presents results given by SMT, MC-SMT, and NODDI. 
	Comparing with Fig.~\ref{fig:BCP}, a noteworthy difference is 
	NODDI significantly underestimates the isotropic volume fraction (Fig.~\ref{fig:syn_combine} (i)), giving zero values in most of gray matter across all time points. 
	This is contradictory to the observation that infant brains typically have higher water content during early development, which decreases later during brain maturation \citep{mukherjee2001normal,lovblad2003isotropic} due to a combination of multiple factors such as natural water  reduction in the body \citep{singhi1995body}, the growth of neuronal and glial cells \citep{forbes2002changes}%
	, and myelination \citep{sakuma1991adult,wimberger1995identification}.
	Note also that MC-SMT and NODDI give higher intra-cellular fraction and lower extra-cellular volume fraction (Fig.~\ref{fig:syn_combine} (g) and (h)).
	SMT overestimates $\mu$FA (Fig.~\ref{fig:syn_combine} (e) and (f)) especially in deep white matter regions. For example, at the splenium of the corpus callosum, the values are almost always maximum, i.e., 1, across all time points. 
	This observation contradicts with previous findings that these regions are immature at birth and undergo a progressive development during infancy
	\citep{lebel2017review,huang2006white}.%
	
	\section{Discussion}\label{sec:Discussion}
	
	Heterogeneously oriented micro-environments are ubiquitous in brain tissues. We have introduced SMSI as a flexible tool for quantification of microarchitecture, unconfounded by orientation heterogeneity.
	Unlike SMT, MC-SMT, and NODDI, SMSI does not rely on a model that is based on a predefined number of compartments. SMSI allows the data to speak for themselves via diffusion characterization using the spherical mean spectrum. We have shown that proper modeling of isotropic diffusion is of paramount importance for accurate characterization of microstructural properties. Failure to do so significantly biases microstructural estimates.

	In addition to the infant brain, SMSI, owing to its ability in characterizing the whole diffusion spectrum, can be employed to quantify adult brain changes and pathologies, such as increased cellularity and vasogenic oedema associated with inflammatory demyelination and axonal injury common in multiple sclerosis \citep{wang2011quantification}. SMSI can also be applied to organs beyond the brain.
	
	Microstructural estimation has been reported to be affected by degeneracy caused by 
	\begin{inparaenum}[(i)]
		\item the interplay of orientation dispersion and diffusion anisotropy \citep{cottaar2019accurate,howard2019joint}, and
		\item the inability of the spherical mean in distinguishing some cases of anisotropic diffusion from isotropic diffusion \citep{szczepankiewicz2015quantification,szczepankiewicz2016imaging}.%
	\end{inparaenum}
	In the first type of degeneracy, a set of orientation coherent low-anisotropy tensors could result in the same diffusion signal as a set of dispersed high-anisotropy tensors.
	This degeneracy can be resolved by using the spherical mean due to its invariance to the orientation distribution \cite{kaden2016quantitative}.
	In the second type of degeneracy, the spherical mean signal of an anisotropic tensor can be indistinguishable from that of a combination of multiple isotropic tensors with different diffusivity \cite{szczepankiewicz2015quantification}.
	This degeneracy can be mitigated by additional data acquired for example via spherical tensor encoding (STE) \citep{westin2014measurement,szczepankiewicz2016imaging}.
	STE can be used in combination with linear tensor encoding (LTE) to provide a means to measure microscopic anisotropy independent of the orientation distribution \citep{cottaar2019accurate}. 
	While effective, STE data are not commonly available.
	We discuss next how SMSI, which utilizes the full direction-sensitize signal in addition to the spherical mean signal (see Section~\ref{subsubsec:imple}), can be used to resolve this degeneracy. The full signal, unlike the mean signal, is not ambiguous in distinguishing isotropic and anisotropic diffusion.

	Ideally, different combinations of diffusion compartments are expected to give different mean signals. However, in practice, 
	the difference between the mean signals of an anisotropic diffusion compartment and a combination of multiple isotropic diffusion compartments at different scales might be negligibly small \cite{szczepankiewicz2015quantification}. 
	
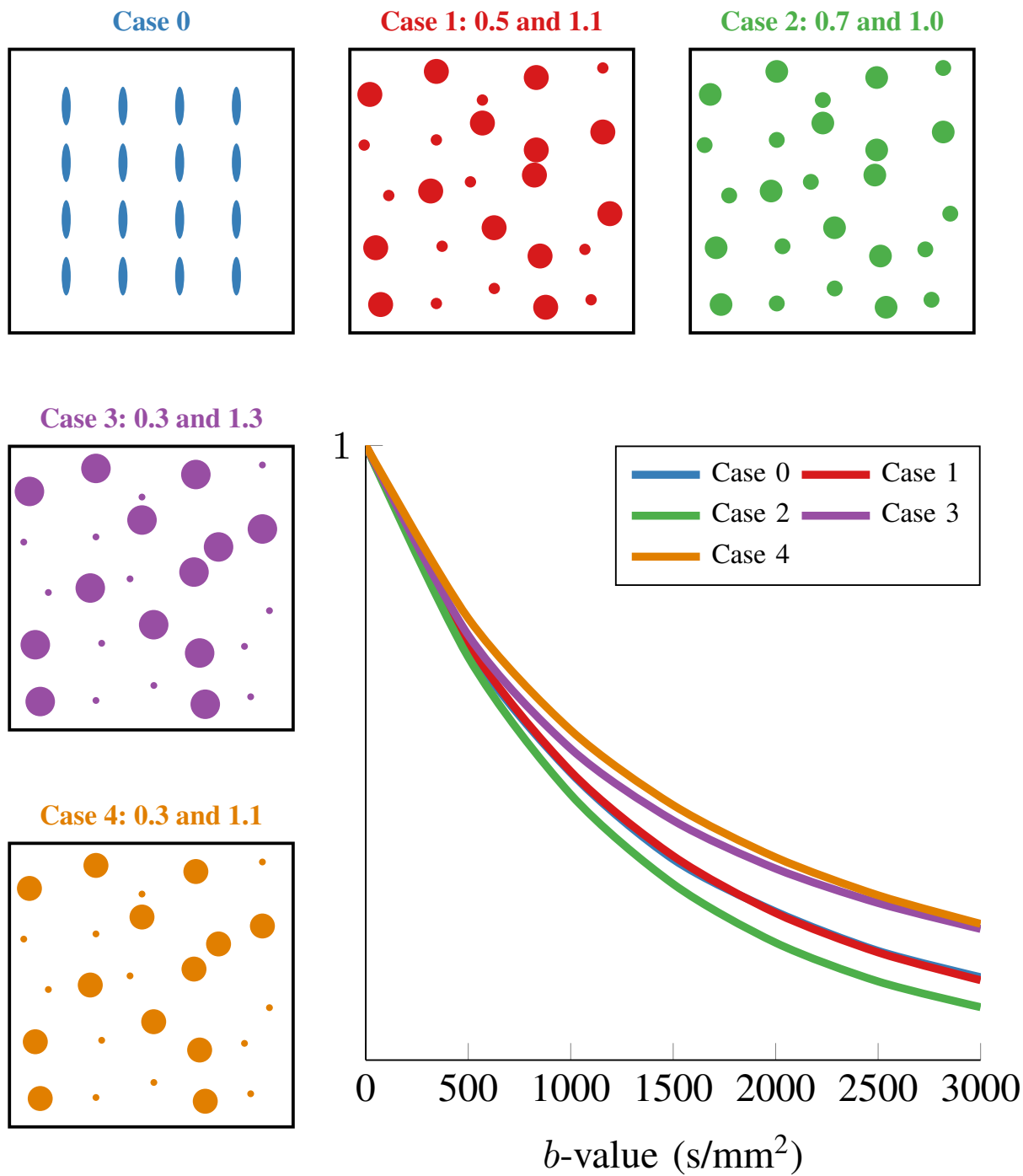
\begin{figure}[t]

	\centering
	\resizebox{\linewidth}{!}
	{
		\begin{tikzpicture}
		
		\def\ColorA{red!100!blue}
		\def\ColorB{red!75!blue}
		\def\ColorC{red!50!blue}
		\def\DirectionColor{green!60!blue}
		\definecolor{color1}{RGB}{55,126,184}
		\definecolor{color2}{RGB}{215,26,28}
		\definecolor{color3}{RGB}{77,175,74}
		\definecolor{color4}{RGB}{152,78,163}
		\definecolor{color5}{RGB}{225,127,0}
		\definecolor{colorecic}{RGB}{146,100,51}
		\definecolor{colorall}{RGB}{51,109,95}
		
		\definecolor{p1}{RGB}{215,26,28}
		\definecolor{p2}{RGB}{251,154,153}
		\definecolor{p3}{RGB}{77,175,74}
		\definecolor{p4}{RGB}{178,223,138}
		\definecolor{p5}{RGB}{255,127,0}
		\definecolor{p6}{RGB}{106,61,154}
		\definecolor{p7}{RGB}{166,206,227}
		\definecolor{p8}{RGB}{55,126,184}
		\definecolor{p9}{RGB}{253,191,111}
		\definecolor{p10}{RGB}{202,178,214}
		\definecolor{p11}{RGB}{180,180,100}
		\definecolor{p12}{RGB}{177,89,40}

		\let\iwidth\relax
		\let\iheight\relax
		\newlength{\iwidth}
		\newlength{\iheight}
		\setlength{\iwidth}{0.095\textwidth}
		\setlength{\iheight}{1.33\iwidth}
		\setlength{\iwidth}{0.1\textwidth}
		
		\tikzstyle{EC} = [x radius=.08cm,y radius=0.34cm]
		\tikzstyle{ellipsecolorec} = [color=color1]

		\tikzstyle{iso3} = [x radius=0.06cm,y radius=0.06cm]
		\tikzstyle{iso5} = [x radius=0.10cm,y radius=0.10cm]
		\tikzstyle{iso7} = [x radius=0.14cm,y radius=0.14cm]
		\tikzstyle{iso10} = [x radius=0.20cm,y radius=0.20cm]
		\tikzstyle{iso11} = [x radius=0.22cm,y radius=0.22cm]
		\tikzstyle{iso13} = [x radius=0.26cm,y radius=0.26cm]
		
		\newcommand{\drawEC}[0]
		{
			\fill[ellipsecolorec] (1,1) ellipse [EC,rotate=0];
			\fill[ellipsecolorec] (1,2) ellipse [EC,rotate=0];
			\fill[ellipsecolorec] (1,3) ellipse [EC,rotate=0];
			\fill[ellipsecolorec] (1,4) ellipse [EC,rotate=0];
			
			\fill[ellipsecolorec] (2,1) ellipse [EC,rotate=0];
			\fill[ellipsecolorec] (2,2) ellipse [EC,rotate=0];
			\fill[ellipsecolorec] (2,3) ellipse [EC,rotate=0];
			\fill[ellipsecolorec] (2,4) ellipse [EC,rotate=0];
			
			\fill[ellipsecolorec] (3,1) ellipse [EC,rotate=0];
			\fill[ellipsecolorec] (3,2) ellipse [EC,rotate=0];
			\fill[ellipsecolorec] (3,3) ellipse [EC,rotate=0];
			\fill[ellipsecolorec] (3,4) ellipse [EC,rotate=0];
			
			\fill[ellipsecolorec] (4,1) ellipse [EC,rotate=0];
			\fill[ellipsecolorec] (4,2) ellipse [EC,rotate=0];
			\fill[ellipsecolorec] (4,3) ellipse [EC,rotate=0];
			\fill[ellipsecolorec] (4,4) ellipse [EC,rotate=0];
		}
	
					\newcommand{\drawconfusion}[0]
	{
		\fill[color=color5] (0.542,0.5) ellipse [EC,rotate=0];
		\fill[color=color5] (1.5245,0.5184) ellipse [iso5];
		\fill[color=color5] (2.545,0.7832) ellipse [EC,rotate=0];
		\fill[color=color5] (3.451,0.452) ellipse [iso11];
		\fill[color=color5] (4.2514,0.584) ellipse [iso5];
		
		\fill[color=color5] (0.4554,1.5011) ellipse [iso11];
		\fill[color=color5] (1.6254,1.5271) ellipse [EC,rotate=0];
		\fill[color=color5] (2.542,1.8542) ellipse [iso11];
		\fill[color=color5] (3.352,1.3552) ellipse [EC,rotate=0];
		\fill[color=color5] (4.1425,1.4724) ellipse [iso5];
		
		\fill[color=color5] (0.6855,2.4222) ellipse [EC,rotate=0];
		\fill[color=color5] (1.42427,2.5012) ellipse [iso11];
		\fill[color=color5] (2.125,2.6627) ellipse [iso5];
		\fill[color=color5] (3.2525,2.7825) ellipse [EC,rotate=0];
		\fill[color=color5] (4.5824,2.102) ellipse [iso11];
		
		\fill[color=color5] (0.252,3.3112) ellipse [EC,rotate=0];
		\fill[color=color5] (1.5242,3.4024) ellipse [iso5];
		\fill[color=color5] (2.33587,3.7001) ellipse [EC,rotate=0];
		\fill[color=color5] (3.2852,3.5252) ellipse [iso11];
		\fill[color=color5] (4.4582,3.5421) ellipse [iso11];
		
		\fill[color=color5] (0.352,4.2045) ellipse [EC,rotate=0];
		\fill[color=color5] (1.5242,4.6102) ellipse [iso11];
		\fill[color=color5] (2.33587,4.5054) ellipse [iso5];
		\fill[color=color5] (3.2852,4.5025) ellipse [iso11];
		\fill[color=color5] (4.4582,4.4708) ellipse [EC,rotate=0];
	}

				\newcommand{\drawisofiveeleven}[0]
		{
			\fill[color=color2] (0.542,0.5) ellipse [iso11];
			\fill[color=color2] (1.5245,0.5184) ellipse [iso5];
			\fill[color=color2] (2.545,0.7832) ellipse [iso5];
			\fill[color=color2] (3.451,0.452) ellipse [iso11];
			\fill[color=color2] (4.2514,0.584) ellipse [iso5];
			
			\fill[color=color2] (0.4554,1.5011) ellipse [iso11];
			\fill[color=color2] (1.6254,1.5271) ellipse [iso5];
			\fill[color=color2] (2.542,1.8542) ellipse [iso11];
			\fill[color=color2] (3.352,1.3552) ellipse [iso11];
			\fill[color=color2] (4.1425,1.4724) ellipse [iso5];
			
			\fill[color=color2] (0.6855,2.4222) ellipse [iso5];
			\fill[color=color2] (1.42427,2.5012) ellipse [iso11];
			\fill[color=color2] (2.125,2.6627) ellipse [iso5];
			\fill[color=color2] (3.2525,2.7825) ellipse [iso11];
			\fill[color=color2] (4.5824,2.102) ellipse [iso11];
			
			\fill[color=color2] (0.252,3.3112) ellipse [iso5];
			\fill[color=color2] (1.5242,3.4024) ellipse [iso5];
			\fill[color=color2] (2.33587,3.7001) ellipse [iso11];
			\fill[color=color2] (3.2852,3.2252) ellipse [iso11];
			\fill[color=color2] (4.4582,3.5421) ellipse [iso11];
			
			\fill[color=color2] (0.352,4.2045) ellipse [iso11];
			\fill[color=color2] (1.5242,4.6102) ellipse [iso11];
			\fill[color=color2] (2.33587,4.1054) ellipse [iso5];
			\fill[color=color2] (3.2852,4.5025) ellipse [iso11];
			\fill[color=color2] (4.4582,4.6708) ellipse [iso5];
		}
	
				\newcommand{\drawisoseventen}[0]
	{
		\fill[color=color3] (0.542,0.5) ellipse [iso10];
		\fill[color=color3] (1.5245,0.5184) ellipse [iso7];
		\fill[color=color3] (2.545,0.7832) ellipse [iso7];
		\fill[color=color3] (3.451,0.452) ellipse [iso10];
		\fill[color=color3] (4.2514,0.584) ellipse [iso7];
		
		\fill[color=color3] (0.4554,1.5011) ellipse [iso10];
		\fill[color=color3] (1.6254,1.5271) ellipse [iso7];
		\fill[color=color3] (2.542,1.8542) ellipse [iso10];
		\fill[color=color3] (3.352,1.3552) ellipse [iso10];
		\fill[color=color3] (4.1425,1.4724) ellipse [iso7];
		
		\fill[color=color3] (0.6855,2.4222) ellipse [iso7];
		\fill[color=color3] (1.42427,2.5012) ellipse [iso10];
		\fill[color=color3] (2.125,2.6627) ellipse [iso7];
		\fill[color=color3] (3.2525,2.7825) ellipse [iso10];
		\fill[color=color3] (4.5824,2.102) ellipse [iso7];
		
		\fill[color=color3] (0.252,3.3112) ellipse [iso7];
		\fill[color=color3] (1.5242,3.4024) ellipse [iso7];
		\fill[color=color3] (2.33587,3.7001) ellipse [iso10];
		\fill[color=color3] (3.2852,3.2252) ellipse [iso10];
		\fill[color=color3] (4.4582,3.5421) ellipse [iso10];
		
		\fill[color=color3] (0.352,4.2045) ellipse [iso10];
		\fill[color=color3] (1.5242,4.6102) ellipse [iso10];
		\fill[color=color3] (2.33587,4.1054) ellipse [iso7];
		\fill[color=color3] (3.2852,4.5025) ellipse [iso10];
		\fill[color=color3] (4.4582,4.6708) ellipse [iso7];
	}
		\newcommand{\drawisothreethirdteen}[0]
		{
			\fill[color=color4] (0.542,0.5) ellipse [iso13];
			\fill[color=color4] (1.5245,0.5184) ellipse [iso3];
			\fill[color=color4] (2.545,0.7832) ellipse [iso3];
			\fill[color=color4] (3.451,0.452) ellipse [iso13];
			\fill[color=color4] (4.2514,0.584) ellipse [iso3];
			
			\fill[color=color4] (0.4554,1.5011) ellipse [iso13];
			\fill[color=color4] (1.6254,1.5271) ellipse [iso3];
			\fill[color=color4] (2.542,1.8542) ellipse [iso13];
			\fill[color=color4] (3.352,1.3552) ellipse [iso13];
			\fill[color=color4] (4.1425,1.4724) ellipse [iso3];
			
			\fill[color=color4] (0.6855,2.4222) ellipse [iso3];
			\fill[color=color4] (1.42427,2.5012) ellipse [iso13];
			\fill[color=color4] (2.125,2.6627) ellipse [iso3];
			\fill[color=color4] (3.2525,2.7825) ellipse [iso13];
			\fill[color=color4] (4.5824,2.102) ellipse [iso3];
			
			\fill[color=color4] (0.252,3.3112) ellipse [iso3];
			\fill[color=color4] (1.5242,3.4024) ellipse [iso3];
			\fill[color=color4] (2.33587,3.7001) ellipse [iso13];
			\fill[color=color4] (3.6852,3.2252) ellipse [iso13];
			\fill[color=color4] (4.4582,3.5421) ellipse [iso13];
			
			\fill[color=color4] (0.352,4.2045) ellipse [iso13];
			\fill[color=color4] (1.5242,4.6102) ellipse [iso13];
			\fill[color=color4] (2.33587,4.1054) ellipse [iso3];
			\fill[color=color4] (3.2852,4.5025) ellipse [iso13];
			\fill[color=color4] (4.4582,4.6708) ellipse [iso3];
		}
		
		\newcommand{\drawisothreeeleven}[0]
		{
			\fill[color=color5] (0.542,0.5) ellipse [iso11];
			\fill[color=color5] (1.5245,0.5184) ellipse [iso3];
			\fill[color=color5] (2.545,0.7832) ellipse [iso3];
			\fill[color=color5] (3.451,0.452) ellipse [iso11];
			\fill[color=color5] (4.2514,0.584) ellipse [iso3];
			
			\fill[color=color5] (0.4554,1.5011) ellipse [iso11];
			\fill[color=color5] (1.6254,1.5271) ellipse [iso3];
			\fill[color=color5] (2.542,1.8542) ellipse [iso11];
			\fill[color=color5] (3.352,1.3552) ellipse [iso11];
			\fill[color=color5] (4.1425,1.4724) ellipse [iso3];
			
			\fill[color=color5] (0.6855,2.4222) ellipse [iso3];
			\fill[color=color5] (1.42427,2.5012) ellipse [iso11];
			\fill[color=color5] (2.125,2.6627) ellipse [iso3];
			\fill[color=color5] (3.2525,2.7825) ellipse [iso11];
			\fill[color=color5] (4.5824,2.102) ellipse [iso3];
			
			\fill[color=color5] (0.252,3.3112) ellipse [iso3];
			\fill[color=color5] (1.5242,3.4024) ellipse [iso3];
			\fill[color=color5] (2.33587,3.7001) ellipse [iso11];
			\fill[color=color5] (3.6852,3.2252) ellipse [iso11];
			\fill[color=color5] (4.4582,3.5421) ellipse [iso11];
			
			\fill[color=color5] (0.352,4.2045) ellipse [iso11];
			\fill[color=color5] (1.5242,4.6102) ellipse [iso11];
			\fill[color=color5] (2.33587,4.1054) ellipse [iso3];
			\fill[color=color5] (3.2852,4.5025) ellipse [iso11];
			\fill[color=color5] (4.4582,4.6708) ellipse [iso3];
		}
		
		\begin{scope}
		\draw[ultra thick] (0,0) rectangle (5,5);
		\drawEC
		\end{scope}

		\begin{scope} [yshift=0cm, xshift=6cm]
		\draw[ultra thick] (0,0) rectangle (5,5);
		\drawisofiveeleven
		\end{scope}
		
		\begin{scope} [yshift=0cm, xshift=12cm]
		\draw[ultra thick] (0,0) rectangle (5,5);
		\drawisoseventen
		\end{scope}
		
		\begin{scope} [yshift=-7cm, xshift= 0cm]
		\draw[ultra thick] (0,0) rectangle (5,5);
		\drawisothreethirdteen
		\end{scope}

		\begin{scope} [yshift=-14cm, xshift= 0cm]
		\draw[ultra thick] (0,0) rectangle (5,5);
		\drawisothreeeleven
		\end{scope}

		\begin{scope} [yshift=-11cm, xshift= 13cm]
		\node at (-1.25,2.7)
		{
			\begin{tikzpicture}[scale=2,smooth]
			\begin{axis}[
			legend style={font=\scriptsize},
			legend columns=2,
			xlabel={$b$-value (s/{mm}\textsuperscript{2})},
			xmin=0,xmax=6,
			ymin=0,ymax=1,
			axis y line*=left, 
			axis x line*=bottom,
			xtick={0,1,2,3,4,5,6},
			legend style={at={(1,1)},anchor=north east},
			xticklabels={0,500,1000,1500,2000,2500,3000},
			ytick={1},
			height=7cm,width=7cm,
			]
			\addplot[smooth,line width =2pt,mark=none,color=color1] plot coordinates {
				(0,1)
				(1,0.67235)
				(2,0.4674)
				(3,0.32673)
				(4,0.24157)
				(5,0.1777)
				(6,0.13488)
			};
			\addlegendentry{Case 0}
			
			\addplot[smooth,line width =2pt,mark=none,color=color2] plot coordinates {
				(0,1)
				(1,0.67788)
				(2,0.4697)
				(3,0.33221)
				(4,0.23934)
				(5,0.17522)
				(6,0.13001)
			};
			\addlegendentry{Case 1}
			
			\addplot[smooth,line width =2pt,mark=none,color=color3] plot coordinates {
				(0,1)
				(1,0.65561)
				(2,0.43223)
				(3,0.28653)
				(4,0.19097)
				(5,0.12793)
				(6,0.08612)
			};
			\addlegendentry{Case 2}
			
			\addplot[smooth,line width =2pt,mark=none,color=color4] plot coordinates {
				(0,1)
				(1,0.69138)
				(2,0.50668)
				(3,0.38995)
				(4,0.31154)
				(5,0.25557)
				(6,0.21341)
			};
			\addlegendentry{Case 3}
			
			\addplot[smooth,line width =2pt,mark=none,color=color5] plot coordinates {
				(0,1)
				(1,0.71883)
				(2,0.53684)
				(3,0.41484)
				(4,0.32981)
				(5,0.26815)
				(6,0.22173)
			};
			\addlegendentry{Case 4}

			\end{axis}
			\end{tikzpicture}
		};
	\end{scope}

	\node at (2.5,5.5) {\textbf{\textcolor{color1}{\Large{Case 0}}}};
	\node at (8.5,5.5) {\textbf{\textcolor{color2}{\Large{Case 1: 0.5 and 1.1}}}};
	\node at (14.5,5.5) {\textbf{\textcolor{color3}{\Large{Case 2: 0.7 and 1.0}}}};
	\node at (2.5,-1.5){\textbf{\textcolor{color4}{\Large{Case 3: 0.3 and 1.3}}}};
	\node at (2.5,-8.5){\textbf{\textcolor{color5}{\Large{Case 4: 0.3 and 1.1}}}};

\end{tikzpicture}
}
\caption{\textbf{Degeneracy.} Spherical mean signals of anisotropic (Case 0) and isotropic (Cases 1--4) configurations. Case 1 has spherical mean signal almost identical to Case 0.}\label{fig:syntheticdegeneracy}
\end{figure}

	To illustrate the degeneracy problem, we simulated DW signals from different configurations:
	\begin{enumerate}
		\item Case~0: Zeppelins with $\text{AD}=1.7\,\mu \text{m}^2\text{/}\text{ms}$ and $\text{RD}=0.4\,\mu \text{m}^2\text{/}\text{ms}$.
		\item Case~1: Isotropic diffusion at two different scales with $\text{AD}=\text{RD}=0.5\,\mu \text{m}^2\text{/}\text{ms}$ and $\text{AD}=\text{RD}=1.1\,\mu \text{m}^2\text{/}\text{ms}$ with equal volume fractions.
		\item Case~2: Same as Case 1 but with $\text{AD}=\text{RD}=0.7\,\mu \text{m}^2\text{/}\text{ms}$ and $\text{AD}=\text{RD}=1.0\,\mu \text{m}^2\text{/}\text{ms}$.
		\item Case~3: Same as Case 1 but with $\text{AD}=\text{RD}=0.3\,\mu \text{m}^2\text{/}\text{ms}$ and $\text{AD}=\text{RD}=1.3\,\mu \text{m}^2\text{/}\text{ms}$.
		\item Case~4: 
		Same as Case 1 but with $\text{AD}=\text{RD}=0.3\,\mu \text{m}^2\text{/}\text{ms}$ and $\text{AD}=\text{RD}=1.1\,\mu \text{m}^2\text{/}\text{ms}$.
	\end{enumerate}
	Fig.~\ref{fig:syntheticdegeneracy} illustrates the cases and their respective spherical mean signals. The signal of Case 1 is almost identical to Case 0, causing degeneracy. The signals of the other cases differ from Case 0, but the small differences can still cause degeneracy, especially when the SNR is low.
	We also tested our method on mixtures of Case 0 and Case 1 with varying volume fractions.

	As described in Section~\ref{subsubsec:imple}, 
	multi-scale isotropic diffusion may degenerate  anisotropic fODFs to isotropic fODFs. Hence, a good way to identify this FSS degeneracy is to gauge the isotropy of each anisotropic fODF, which can be measured with the help of generalized fractional anisotropy (GFA) \citep{cohen2008detection}:
	\begin{equation}
	\text{ISO} = \sqrt{1-\text{GFA}^2}.
	\end{equation}
	Anisotropic atoms with ISO$\geq$0.95 (GFA$<$0.3) are considered degenerate. This is captured by an indicator function for the $i$-th atom:
	\begin{equation}
	\Upsilon[i]=
	\begin{cases}
	1, & \text{ISO}[i] \geq 0.95,\\
	0, & \text{otherwise}.\\
	\end{cases}
	\end{equation} 
	Note that even if an atom is affected by degeneracy, the effect on microstructural estimation might be minimal if the volume fraction associated with the atom is small.
	We assess the severity of degeneracy via a degeneracy index (DI):
	\begin{equation}
	\text{DI} = \sum_{i}\nu[i]\Upsilon[i],
	\end{equation}
	which is a linear combination of the elements in $\Upsilon$ weighted by the corresponding volume fractions $\nu$ of the anisotropic atoms.
	DI ranges from 0 to 1, with higher values indicating greater degeneracy. 
	If either isotropy or volume fraction is low, the DI is low and the effect of degeneracy is negligible.
	
	Our implementation of SMSI breaks the degeneracy by utilizing complementary information from the full diffusion signal and the spherical mean signal.
	This full signal captures directional information and can hence distinguish between isotropic and anisotropic diffusion even if the spherical mean signals of the two cases are identical.
	Fig.~\ref{fig:DI}
	indicates that the full signal alone is unable to fully resolve the degeneracy, yielding high DI (Fig.~\ref{fig:DI}(a)) and inaccurate volume fraction estimates (Fig.~\ref{fig:DI}(c)). 
	Case 1 has the highest DI since, instead of isotropic tensors, it can be represented equally with degenerate anisotropic tensors. 
	SMSI suppresses the degenerate atoms and lowers the DI (Fig.~\ref{fig:DI}(b)), resulting in accurate volume fraction estimates (Fig.~\ref{fig:DI}(d)).
	Fig.~\ref{fig:DI}(e) shows the results for different mixtures of Case 0 and Case 1.
	
	Figure~\ref{fig:DImap} shows the DI maps given by the FSS on a HCP dMRI dataset \cite{van2013wu}. Less than 1\% of the total brain voxels, mostly beyond the brain parenchyma, are affected by degeneracy, implying that the  problem is not severe in practice, at least for the healthy adult brain.
	Degeneracy suppression with SMSI results in less than 0.1\% voxels with non-zero DI values.
	The DI statistics for 20 HCP subjects are summarized in Table~\ref{tab:DI_tab}.

	\pgfplotsset{every axis/.append style={
		label style={font=\large},
		tick label style={font=\large},
		title style={font=\bfseries\large},
		compat=1.14,
}}
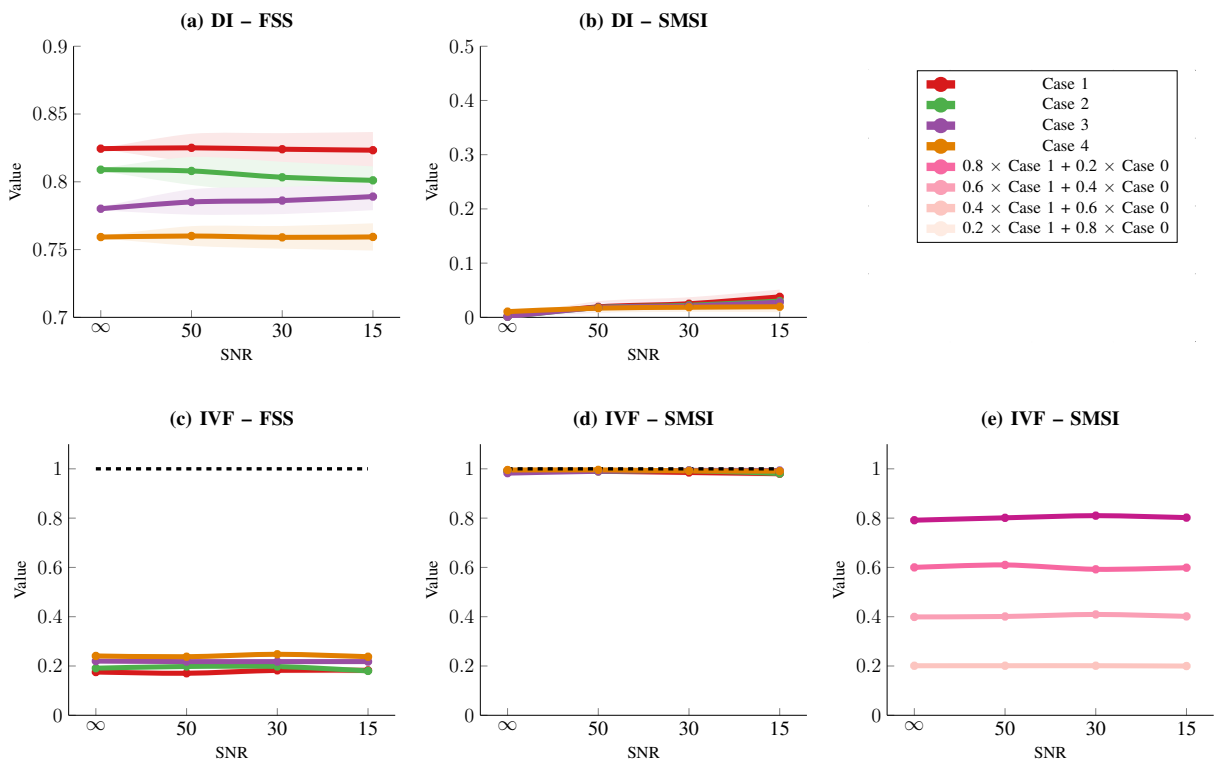
\begin{figure}[t]
	\centering
	\definecolor{color1}{RGB}{55,126,184}
\definecolor{color2}{RGB}{215,26,28}
\definecolor{color3}{RGB}{77,175,74}
\definecolor{color4}{RGB}{152,78,163}
\definecolor{color5}{RGB}{225,127,0}

\definecolor{color6}{RGB}{254,235,226}
\definecolor{color7}{RGB}{252,197,192}	
\definecolor{color8}{RGB}{250,159,181}	
\definecolor{color9}{RGB}{247,104,161}	
\definecolor{color10}{RGB}{197,27,138}
\definecolor{color11}{RGB}{122,1,119}	
	
	\usepgfplotslibrary{fillbetween}
	
	\pgfplotstableread
	{
	x case1 err1 case2 err2 case3 err3 case4 err4
1 0.82444 0.00012 0.80893 0.00081 0.78013 0.0012 0.75923 0.000812
2 0.82506 0.01033 0.80801 0.01033 0.78512 0.00933 0.76001 0.007233
3 0.82403 0.01184 0.80332 0.01144 0.78615 0.009944 0.75901 0.008344
4 0.82331 0.0134 0.80108 0.01031 0.78910 0.010023 0.75933 0.0100331
	}\DI

	\pgfplotstableread
{
	x case1 err1 case2 err2 case3 err3 case4 err4
	1 0.001524 0.00012 0.001042 0.00081 0.00124 0.0012 0.01055 0.000812
	2 0.019344 0.01033 0.018072 0.001033 0.0178512 0.00933 0.01701 0.007233
	3 0.025100 0.01184 0.022182 0.010144 0.02078615 0.009944 0.0185901 0.008344
	4 0.0378231 0.0134 0.0300108 0.001031 0.0278910 0.010023 0.0195933 0.0100331
}\DISMSI

	\pgfplotstableread
{
	x case1 err1 case2 err2 case3 err3 case4 err4
	1 0.1755 0.00012 0.19107 0.00081 0.21987 0.0012 0.24077 0.000812
	2 0.1705 0.0103 0.198080 0.01033 0.2178512 0.00933 0.2376001 0.007233
	3 0.182403 0.01184 0.1980332 0.01144 0.2178615 0.009944 0.2475901 0.008344
	4 0.182331 0.0134 0.180108 0.01031 0.2178910 0.010023 0.2375933 0.0100331
}\IVF

\pgfplotstableread
{
	x case1 err1 case2 err2 case3 err3 case4 err4
1 0.9925 0.00012 0.9913 0.00081 0.9825 0.0012 0.9952 0.000812
2 0.9901 0.01033 0.9901 0.001033 0.9900 0.00933 0.9957 0.007233
3 0.9852 0.01184 0.9927 0.010144 0.9937 0.009944 0.9917 0.008344
4 0.9802 0.0134 0.9809 0.001031 0.9932 0.010023 0.9906 0.0100331
}\IVFSMSI

	\pgfplotstableread
{
	x case1 err1 case2 err2 case3 err3 case4 err4 case5 err5 case6 err6
	1 0.82444 0.00012 0.0152 0.00081 0.0013 0.00052 0.0023 0.000812 0.0012 0.00032 0.00011 0.00017
	2 0.82506 0.01033 0.0174 0.005033 0.0212 0.000533 0.026001 0.001233 0.0011 0.00013 0.0010 0.00018
	3 0.82403 0.01184 0.0112 0.005144 0.0215 0.008944 0.025901 0.002344 0.0018 0.0027 0.0009 0.00023
	4 0.82331 0.0134 0.0221 0.005031 0.02910 0.00023 0.03033 0.000331 0.0019 0.0039 0.0011 0.0038
}\DIConfusion
	
		\pgfplotstableread
	{
		x case1 err1 case2 err2 case3 err3 case4 err4 case5 err5 case6 err6
		1 0.001524 0.00012 0.0152 0.00081 0.0013 0.00052 0.0023 0.000812 0.0012 0.00032 0.00011 0.00017
		2 0.001952 0.01033 0.0174 0.005033 0.0212 0.000533 0.026001 0.001233 0.0011 0.00013 0.0010 0.00018
		3 0.002532 0.01184 0.0112 0.005144 0.0215 0.008944 0.025901 0.002344 0.0018 0.0027 0.0009 0.00023
		4 0.03702 0.0134 0.0221 0.005031 0.02910 0.00023 0.03033 0.000331 0.0019 0.0039 0.0011 0.0038
	}\DISMSIConfusion

	\pgfplotstableread
{
	x case1 err1 case2 err2 case3 err3 case4 err4 case5 err5 case6 err6
	1 0.1755 0.00012 0.7928 0.00103 0.6010 0.00052 0.4001 0.000812 0.2002 0.00032 0.00928 0.00017
	2 0.1705 0.01033 0.7998 0.005033 0.6122 0.000533 0.4028 0.001233 0.2103 0.00013 0.008934 0.00018
	3 0.182403 0.01184 0.8101 0.005 0.5991 0.008944 0.4125 0.002344 0.1908 0.0027 0.005908 0.00023
	4 0.182331 0.0134 0.8003 0.0050 0.6001 0.00023 0.3812 0.000331 0.2402 0.0039 0.010827 0.0038
}\IVFConfusion

	\pgfplotstableread
{
	x case1 err1 case2 err2 case3 err3 case4 err4 case5 err5 case6 err6
	1 0.9925 0.00012 0.79152 0.00081 0.6001 0.00052 0.3991 0.000812 0.2010 0.00032 0.00011 0.00017
	2 0.9901 0.01033 0.80121 0.005033 0.6101 0.000533 0.4009 0.001233 0.2011 0.00013 0.0010 0.00018
	3 0.9852 0.01184 0.81001 0.005144 0.5923 0.008944 0.4092 0.002344 0.2011 0.0027 0.0009 0.00023
	4 0.9802 0.0134 0.80201 0.005031 0.5988 0.00023 0.4013 0.000331 0.1998 0.0039 0.0011 0.0038
}\IVFSMSIConfusion
	
	\resizebox{1\linewidth}{!}
	{
		
 		\begin{tabular}{ccc}
			\begin{tikzpicture}[scale=1]
			\begin{axis}[
			xlabel={SNR},
			ylabel= {Value},
			ymin=0.7,ymax=0.9,
			axis y line*=left, axis x line*=bottom,
			xtick={1,2,3,4},
			legend style={at={(0.05,1)},anchor=north west},
			xticklabels={$\infty$,50,30,15},
			title={(a) DI -- FSS}
			]
			
			\addplot [color=color2, mark=*,mark options={scale=0.5},line width=3pt, smooth] table [y index=1] {\DI};
			\addplot [name path=upper1,draw=none, smooth, forget plot] table[y expr=\thisrow{case1}+\thisrow{err1}] {\DI};
			\addplot [name path=lower1,draw=none, smooth, forget plot] table[y expr=\thisrow{case1}-\thisrow{err1}] {\DI};
			\addplot [fill=color2!10, forget plot] fill between[of=upper1 and lower1];
			
			\addplot [color=color3, mark=*,mark options={scale=0.5},line width=3pt, smooth] table [y index=3] {\DI};
			\addplot [name path=upper1,draw=none, smooth, forget plot] table[y expr=\thisrow{case2}+\thisrow{err2}] {\DI};
			\addplot [name path=lower1,draw=none, smooth, forget plot] table[y expr=\thisrow{case2}-\thisrow{err2}] {\DI};
			\addplot [fill=color3!10, forget plot] fill between[of=upper1 and lower1];

			\addplot [color=color4, mark=*,mark options={scale=0.5},line width=3pt, smooth] table [y index=5] {\DI};
			\addplot [name path=upper1,draw=none, smooth, forget plot] table[y expr=\thisrow{case3}+\thisrow{err3}] {\DI};
			\addplot [name path=lower1,draw=none, smooth, forget plot] table[y expr=\thisrow{case3}-\thisrow{err3}] {\DI};
			\addplot [fill=color4!10, forget plot] fill between[of=upper1 and lower1];

			\addplot [color=color5, mark=*,mark options={scale=0.5},line width=3pt, smooth] table [y index=7] {\DI};
			\addplot [name path=upper1,draw=none, smooth, forget plot] table[y expr=\thisrow{case4}+\thisrow{err4}] {\DI};
			\addplot [name path=lower1,draw=none, smooth, forget plot] table[y expr=\thisrow{case4}-\thisrow{err4}] {\DI};
			\addplot [fill=color5!10, forget plot] fill between[of=upper1 and lower1];

			\end{axis}
			\end{tikzpicture}
			
			&
			
				\begin{tikzpicture}[scale=1]
			\begin{axis}[
			xlabel={SNR},
			ylabel= {Value},
			ymin=0,ymax=0.5,
			axis y line*=left, axis x line*=bottom,
			xtick={1,2,3,4},
			legend style={at={(0.05,1)},anchor=north west},
			xticklabels={$\infty$,50,30,15},
			title={(b) DI -- SMSI}
			]
			
			\addplot [color=color2, mark=*,mark options={scale=0.5},line width=3pt, smooth] table [y index=1] {\DISMSI};
			\addplot [name path=upper1,draw=none, smooth, forget plot] table[y expr=\thisrow{case1}+\thisrow{err1}] {\DISMSI};
			\addplot [name path=lower1,draw=none, smooth, forget plot] table[y expr=\thisrow{case1}-\thisrow{err1}] {\DISMSI};
			\addplot [fill=color2!10, forget plot] fill between[of=upper1 and lower1];
			
			\addplot [color=color3, mark=*,mark options={scale=0.5},line width=3pt, smooth] table [y index=3] {\DISMSI};
			\addplot [name path=upper1,draw=none, smooth, forget plot] table[y expr=\thisrow{case2}+\thisrow{err2}] {\DISMSI};
			\addplot [name path=lower1,draw=none, smooth, forget plot] table[y expr=\thisrow{case2}-\thisrow{err2}] {\DISMSI};
			\addplot [fill=color3!10, forget plot] fill between[of=upper1 and lower1];

			\addplot [color=color4, mark=*,mark options={scale=0.5},line width=3pt, smooth] table [y index=5] {\DISMSI};
			\addplot [name path=upper1,draw=none, smooth, forget plot] table[y expr=\thisrow{case3}+\thisrow{err3}] {\DISMSI};
			\addplot [name path=lower1,draw=none, smooth, forget plot] table[y expr=\thisrow{case3}-\thisrow{err3}] {\DISMSI};
			\addplot [fill=color4!10, forget plot] fill between[of=upper1 and lower1];

			\addplot [color=color5, mark=*,mark options={scale=0.5},line width=3pt, smooth] table [y index=7] {\DISMSI};
			\addplot [name path=upper1,draw=none, smooth, forget plot] table[y expr=\thisrow{case4}+\thisrow{err4}] {\DISMSI};
			\addplot [name path=lower1,draw=none, smooth, forget plot] table[y expr=\thisrow{case4}-\thisrow{err4}] {\DISMSI};
			\addplot [fill=color5!10, forget plot] fill between[of=upper1 and lower1];

			\end{axis}
			\end{tikzpicture}
			
			& 
			\begin{tikzpicture} 
			\begin{axis}[%
			xmin=10,
			xmax=50,
			ymin=0,
			ymax=0.4,
			legend style={at={(0.15,1.00)},anchor=north west},
			axis line style={white},
			every axis label/.append style ={white},
			every tick label/.append style={white},
			x tick label style={major tick length=0pt},
			]
			\addlegendimage{color2,mark=*,line width =5pt}
			\addlegendentry{Case 1};
			\addlegendimage{color3,mark=*,line width =5pt}
			\addlegendentry{Case 2};
			\addlegendimage{color4,mark=*,line width =5pt}
			\addlegendentry{Case 3};
			\addlegendimage{color5,mark=*,line width =5pt}
			\addlegendentry{Case 4};
			\addlegendimage{color9,mark=*,line width =5pt}
			\addlegendentry{0.8 $\times$ Case 1 + 0.2 $\times$ Case 0}
			\addlegendimage{color8,mark=*,line width =5pt}
			\addlegendentry{0.6 $\times$ Case 1 + 0.4 $\times$ Case 0}
			\addlegendimage{color7,mark=*,line width =5pt}
			\addlegendentry{0.4 $\times$ Case 1 + 0.6 $\times$ Case 0}
			\addlegendimage{color6,mark=*,line width =5pt}
			\addlegendentry{0.2 $\times$ Case 1 + 0.8 $\times$ Case 0}
			\end{axis}
			\end{tikzpicture}

			\\[20pt]
			
				\begin{tikzpicture}[scale=1]
			\begin{axis}[
			xlabel={SNR},
			ylabel= {Value},
			ymin=0,ymax=1.1,
			axis y line*=left, axis x line*=bottom,
			xtick={1,2,3,4},
			legend style={at={(0.05,0.6)},anchor=north west},
			xticklabels={$\infty$,50,30,15},
			title={(c) IVF -- FSS}
			]
			
			\addplot [color=color2, mark=*,mark options={scale=0.5},line width=3pt, smooth] table [y index=1] {\IVF};
			\addplot [name path=upper1,draw=none, smooth, forget plot] table[y expr=\thisrow{case1}+\thisrow{err1}] {\IVF};
			\addplot [name path=lower1,draw=none, smooth, forget plot] table[y expr=\thisrow{case1}-\thisrow{err1}] {\IVF};
			\addplot [fill=color2!10, forget plot] fill between[of=upper1 and lower1];
			
			\addplot [color=color3, mark=*,mark options={scale=0.5},line width=3pt, smooth] table [y index=3] {\IVF};
			\addplot [name path=upper1,draw=none, smooth, forget plot] table[y expr=\thisrow{case2}+\thisrow{err2}] {\IVF};
			\addplot [name path=lower1,draw=none, smooth, forget plot] table[y expr=\thisrow{case2}-\thisrow{err2}] {\IVF};
			\addplot [fill=color3!10, forget plot] fill between[of=upper1 and lower1];

			\addplot [color=color4, mark=*,mark options={scale=0.5},line width=3pt, smooth] table [y index=5] {\IVF};
			\addplot [name path=upper1,draw=none, smooth, forget plot] table[y expr=\thisrow{case3}+\thisrow{err3}] {\IVF};
			\addplot [name path=lower1,draw=none, smooth, forget plot] table[y expr=\thisrow{case3}-\thisrow{err3}] {\IVF};
			\addplot [fill=color4!10, forget plot] fill between[of=upper1 and lower1];

			\addplot [color=color5, mark=*,mark options={scale=0.5},line width=3pt, smooth] table [y index=7] {\IVF};
			\addplot [name path=upper1,draw=none, smooth, forget plot] table[y expr=\thisrow{case4}+\thisrow{err4}] {\IVF};
			\addplot [name path=lower1,draw=none, smooth, forget plot] table[y expr=\thisrow{case4}-\thisrow{err4}] {\IVF};
			\addplot [fill=color5!10, forget plot] fill between[of=upper1 and lower1];
			
			\addplot[color=black,domain=1:4,dashed, line width =2pt] {1};
			
			\end{axis}
			\end{tikzpicture}
			
			&
			
			\begin{tikzpicture}[scale=1]
			\begin{axis}[
			xlabel={SNR},
			ylabel= {Value},
			ymin=0,ymax=1.1,
			axis y line*=left, axis x line*=bottom,
			xtick={1,2,3,4},
			legend style={at={(0.05,0.025)},anchor=south west},
			xticklabels={$\infty$,50,30,15},
			title={(d) IVF -- SMSI}
			]
			
			\addplot [color=color2, mark=*,mark options={scale=0.5},line width=3pt, smooth] table [y index=1] {\IVFSMSI};
			\addplot [name path=upper1,draw=none, smooth, forget plot] table[y expr=\thisrow{case1}+\thisrow{err1}] {\IVFSMSI};
			\addplot [name path=lower1,draw=none, smooth, forget plot] table[y expr=\thisrow{case1}-\thisrow{err1}] {\IVFSMSI};
			\addplot [fill=color2!10, forget plot] fill between[of=upper1 and lower1];
			
			\addplot [color=color3, mark=*,mark options={scale=0.5},line width=3pt, smooth] table [y index=3] {\IVFSMSI};
			\addplot [name path=upper1,draw=none, smooth, forget plot] table[y expr=\thisrow{case2}+\thisrow{err2}] {\IVFSMSI};
			\addplot [name path=lower1,draw=none, smooth, forget plot] table[y expr=\thisrow{case2}-\thisrow{err2}] {\IVFSMSI};
			\addplot [fill=color3!10, forget plot] fill between[of=upper1 and lower1];

			\addplot [color=color4, mark=*,mark options={scale=0.5},line width=3pt, smooth] table [y index=5] {\IVFSMSI};
			\addplot [name path=upper1,draw=none, smooth, forget plot] table[y expr=\thisrow{case3}+\thisrow{err3}] {\IVFSMSI};
			\addplot [name path=lower1,draw=none, smooth, forget plot] table[y expr=\thisrow{case3}-\thisrow{err3}] {\IVFSMSI};
			\addplot [fill=color4!10, forget plot] fill between[of=upper1 and lower1];

			\addplot [color=color5, mark=*,mark options={scale=0.5},line width=3pt, smooth] table [y index=7] {\IVFSMSI};
			\addplot [name path=upper1,draw=none, smooth, forget plot] table[y expr=\thisrow{case4}+\thisrow{err4}] {\IVFSMSI};
			\addplot [name path=lower1,draw=none, smooth, forget plot] table[y expr=\thisrow{case4}-\thisrow{err4}] {\IVFSMSI};
			\addplot [fill=color5!10, forget plot] fill between[of=upper1 and lower1];
			
			\addplot[color=black,domain=1:4,dashed, line width =2pt] {1};
			
			\end{axis}
			\end{tikzpicture}

			&
			
			\begin{tikzpicture}[scale=1]
			\begin{axis}[
			xlabel={SNR},
			ylabel= {Value},
			ymin=0.0,ymax=1.1,
			axis y line*=left, axis x line*=bottom,
			xtick={1,2,3,4},
			xticklabels={$\infty$,50,30,15},
			title={(e) IVF -- SMSI}
			]

			\addplot [color=color10, mark=*,mark options={scale=0.5},line width=3pt, smooth] table [y index=3] {\IVFSMSIConfusion};
			\addplot [name path=upper1,draw=none, smooth, forget plot] table[y expr=\thisrow{case2}+\thisrow{err2}] {\IVFSMSIConfusion};
			\addplot [name path=lower1,draw=none, smooth, forget plot] table[y expr=\thisrow{case2}-\thisrow{err2}] {\IVFSMSIConfusion};
			\addplot [fill=color10!10, forget plot] fill between[of=upper1 and lower1];

			\addplot [color=color9, mark=*,mark options={scale=0.5},line width=3pt, smooth] table [y index=5] {\IVFSMSIConfusion};
			\addplot [name path=upper1,draw=none, smooth, forget plot] table[y expr=\thisrow{case3}+\thisrow{err3}] {\IVFSMSIConfusion};
			\addplot [name path=lower1,draw=none, smooth, forget plot] table[y expr=\thisrow{case3}-\thisrow{err3}] {\IVFSMSIConfusion};
			\addplot [fill=color9!10, forget plot] fill between[of=upper1 and lower1];

			\addplot [color=color8, mark=*,mark options={scale=0.5},line width=3pt, smooth] table [y index=7] {\IVFSMSIConfusion};
			\addplot [name path=upper1,draw=none, smooth, forget plot] table[y expr=\thisrow{case4}+\thisrow{err4}] {\IVFSMSIConfusion};
			\addplot [name path=lower1,draw=none, smooth, forget plot] table[y expr=\thisrow{case4}-\thisrow{err4}] {\IVFSMSIConfusion};
			\addplot [fill=color8!10, forget plot] fill between[of=upper1 and lower1];
			
			\addplot [color=color7, mark=*,mark options={scale=0.5},line width=3pt, smooth] table [y index=9] {\IVFSMSIConfusion};
			\addplot [name path=upper1,draw=none, smooth, forget plot] table[y expr=\thisrow{case5}+\thisrow{err5}] {\IVFSMSIConfusion};
			\addplot [name path=lower1,draw=none, smooth, forget plot] table[y expr=\thisrow{case5}-\thisrow{err5}] {\IVFSMSIConfusion};
			\addplot [fill=color7!10, forget plot] fill between[of=upper1 and lower1];
			
			\end{axis}
			\end{tikzpicture}
		
		\end{tabular}
	}
	\caption{\textbf{Degeneracy.}
    DI values and IVF estimates given by FSS and SMSI for the different configurations in  Fig.~\ref{fig:syntheticdegeneracy} and different combinations of Case 0 and Case 1.
The dashed lines represent the ground truth. Shaded regions represent standard deviations computed based on 1000 instances for each SNR. The DI of Case 0 is 0 and is hence not shown. The standard deviations of the IVFs are negligible.}\label{fig:DI}
\end{figure}

	\begin{figure}[htb]
	
	\newcommand\myPic[7]{
		\begin{tikzpicture}[baseline,trim left]
		\draw (0.0\iwidth, 0.0\iheight)[#5, ultra thick] rectangle(\iwidth, \iheight);
		#6 %
		\node[mylabel, right=0pt] at (0.0\iwidth, 1.0\iheight) {\textbf{\large{#7}}}; %
		
		\begin{pgfonlayer}{background}
		\fill (0.0\iwidth, 0.0\iheight)[black, ultra thick] rectangle(\iwidth, \iheight);
		\clip (0.0\iwidth, 0.0\iheight) rectangle (\iwidth, \iheight);
		\node at (#3\iwidth, #4\iheight) { \includegraphics[width=#2\iwidth, angle=0]{\filePath/#1} };
		\end{pgfonlayer}
		\end{tikzpicture}
	}

\newcommand\columnText[1]{
	\begin{tikzpicture}[baseline,trim left]
	\node[rotate=90] at (0.1\iwidth, 0.5\iheight) {\myFontSize{#1}};
	\end{tikzpicture}
}
	
\newcommand\colorBar[5]{		
	\begin{tikzpicture}
	\node (bar) at (#3\iwidth, #4\iheight) { \includegraphics[width=0.16\colorbarlength,height=1.5\colorbarlength]{\filePath/#5} };
	\node [below=-1mm of bar] {\myFontSize{#1}};
	\node [above=-1mm of bar] {\myFontSize{#2}};
	\end{tikzpicture}
}

\newcommand\myFontSize[1]{
\textbf{\large{#1}}
}

	\newcommand\voidAnn{}

\setlength{\iwidth}{0.3\textwidth}
\setlength{\iheight}{\iwidth}
\setlength{\tabcolsep}{0pt}
\setlength{\rowspacing}{0pt}
\setlength{\colorbarlength}{0.4\iwidth}
\def \filePath {./Figures/syntheticdegeneracy}
\renewcommand{\arraystretch}{0}
\renewcommand{\arrayrulewidth}{0pt}
\centering

\resizebox{\linewidth}{!}
{
\begin{tabular}{ccccc}
	
\columnText{FSS} & \myPic{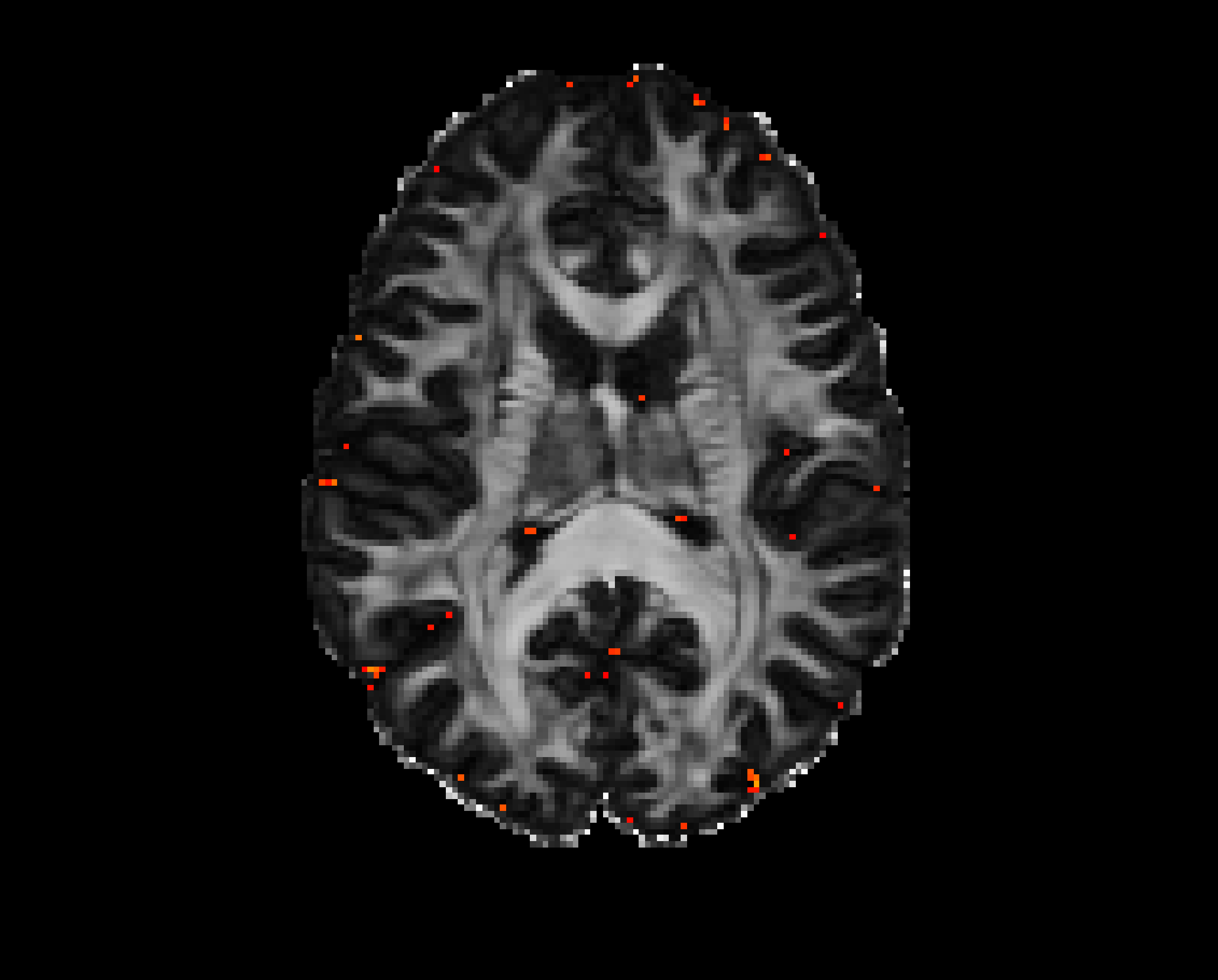}{1.4}{0.5}{0.48}{black}{\voidAnn}{}&\myPic{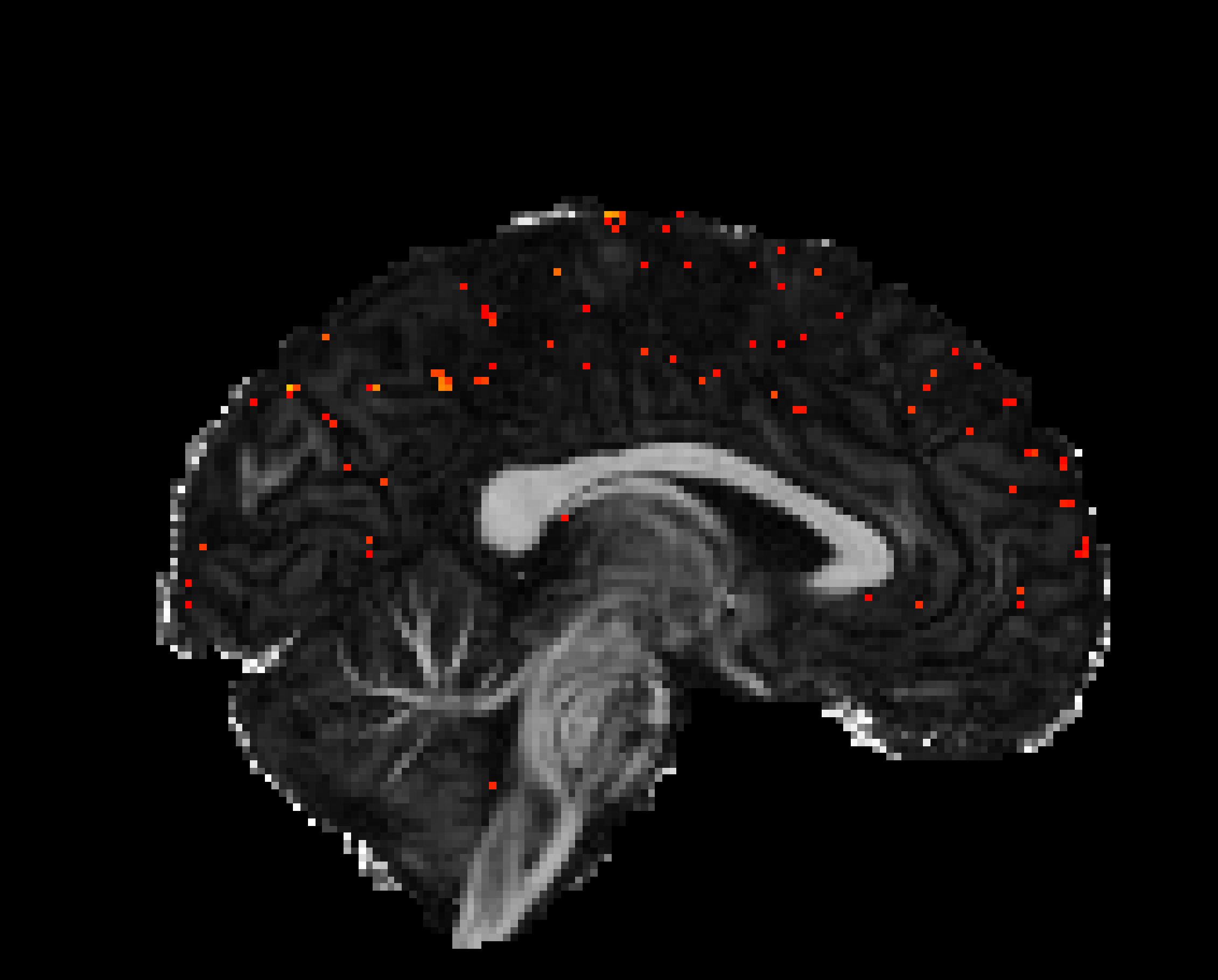}{1.2}{0.48}{0.6}{black}{\voidAnn}{}&\myPic{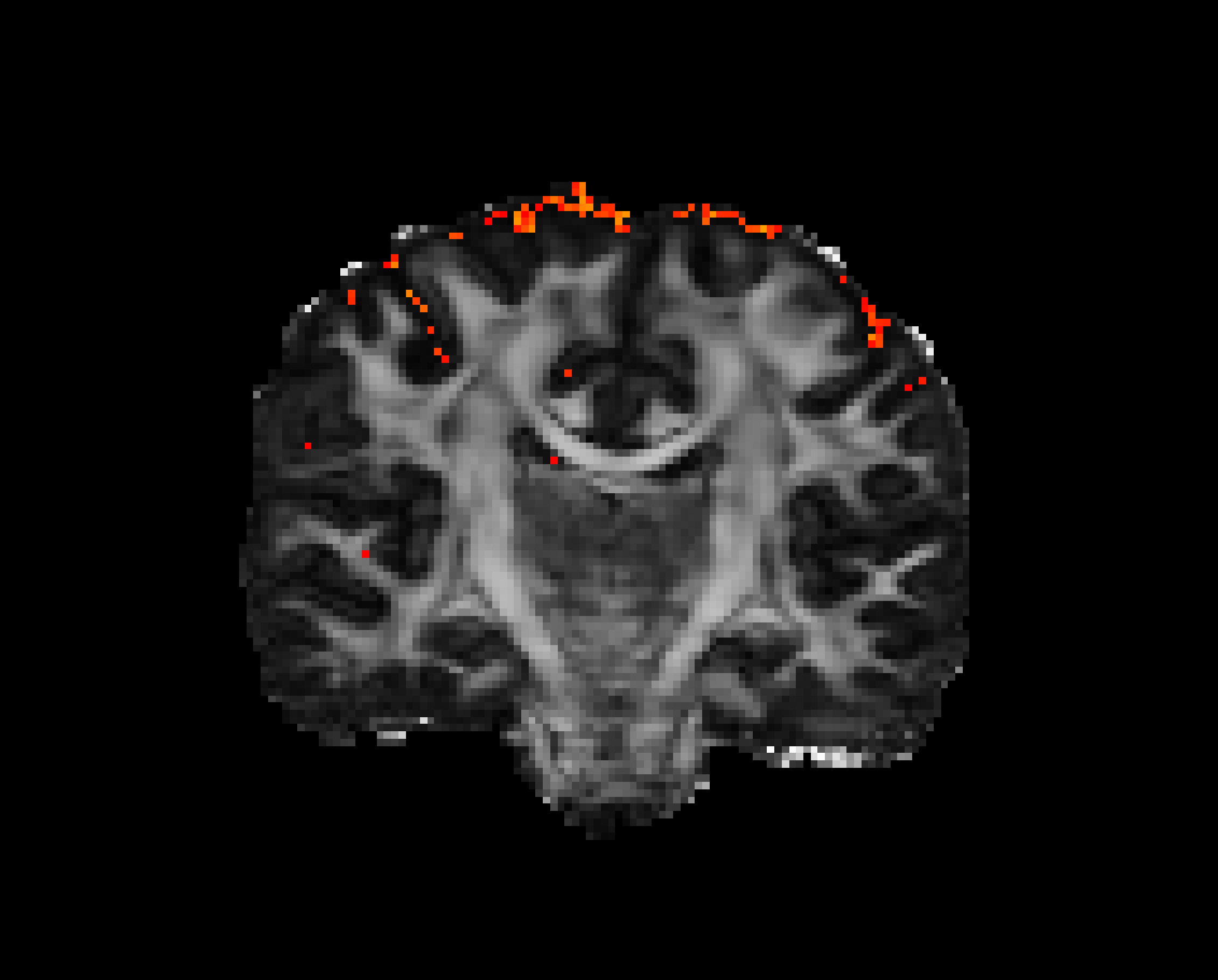}{1.4}{0.5}{0.6}{black}{\voidAnn}{} & \colorBar{\small0}{\small0.6}{0.5}{0.5}{colorbar}\\
\columnText{SMSI} & \myPic{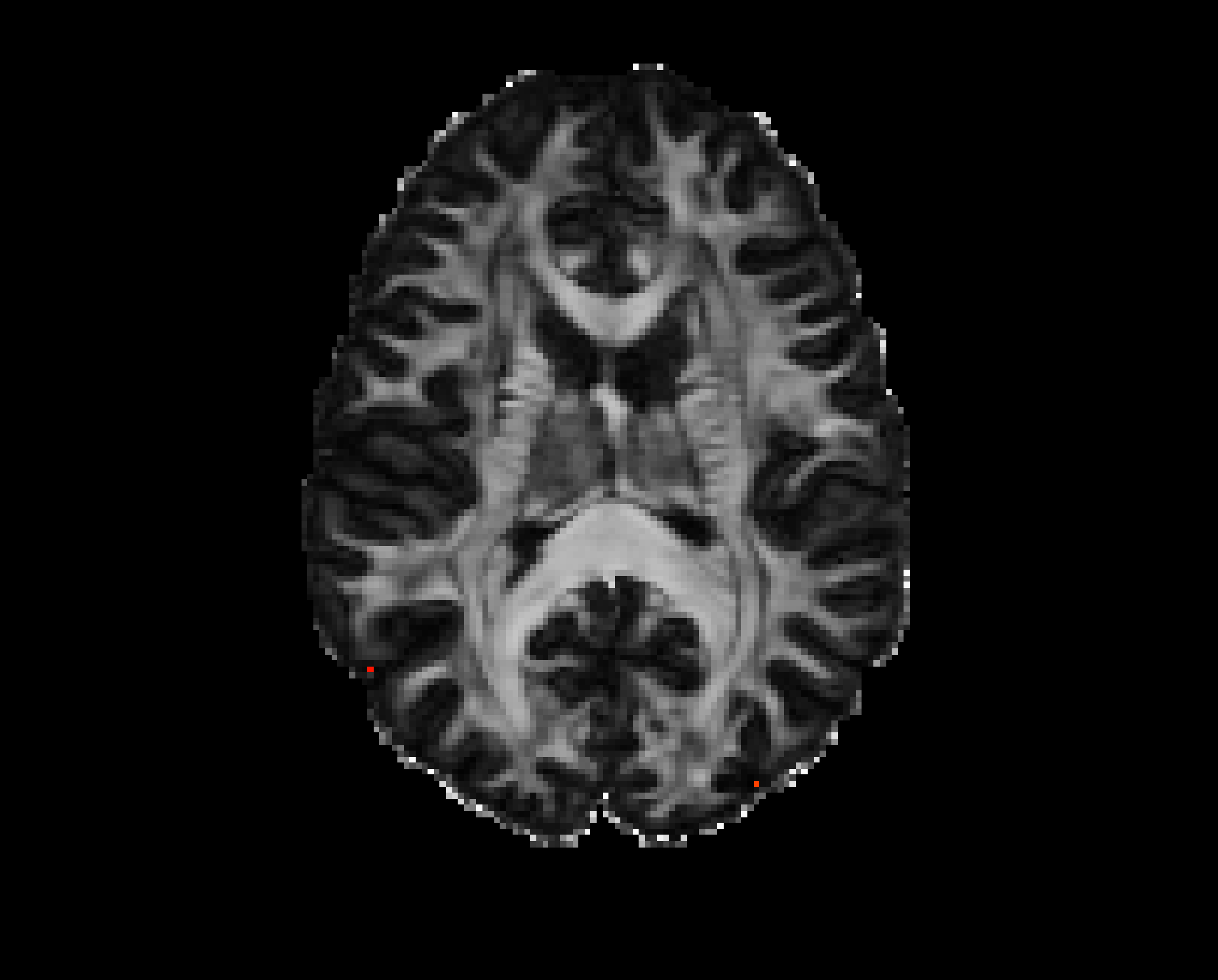}{1.4}{0.5}{0.48}{black}{\voidAnn}{}&\myPic{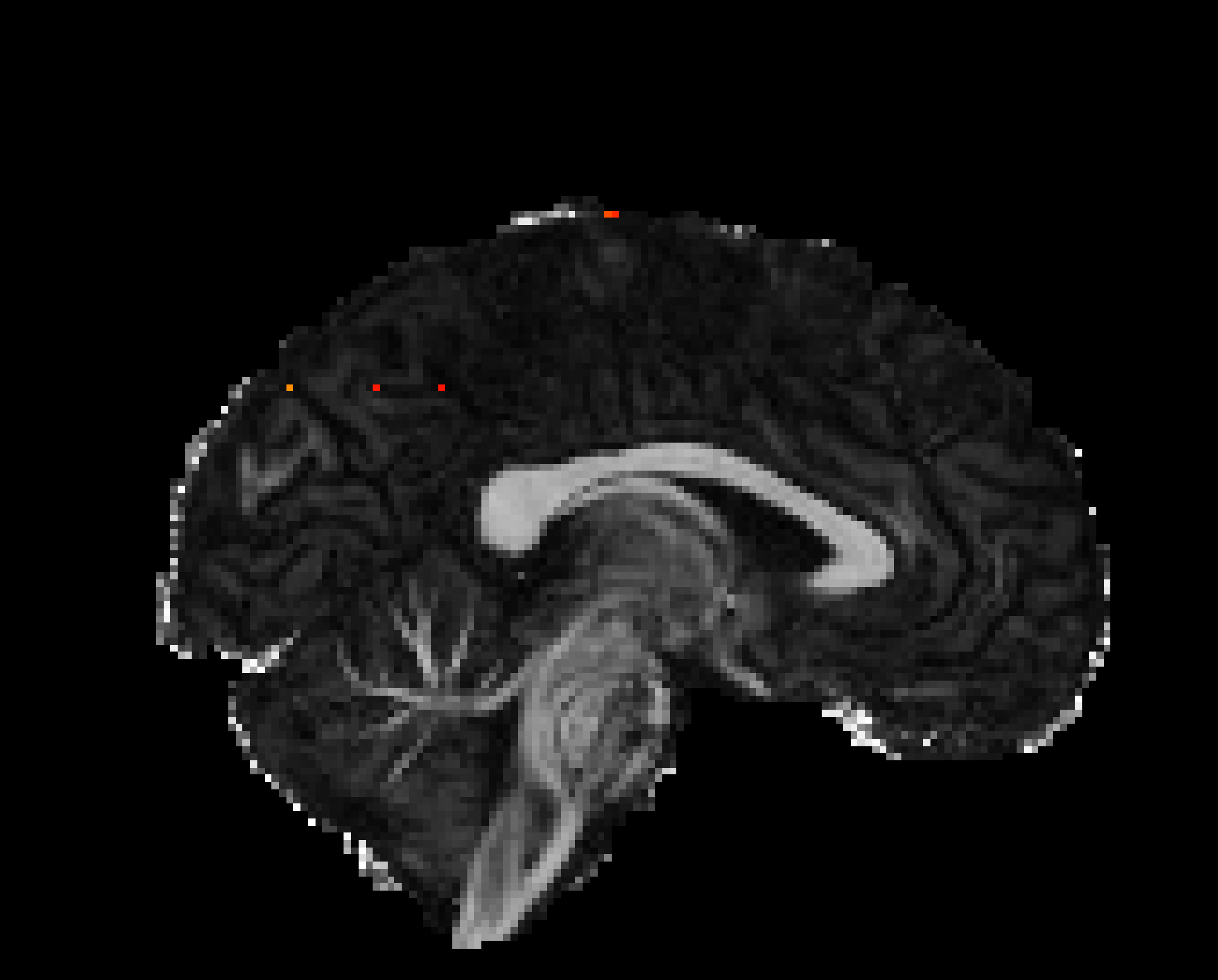}{1.2}{0.48}{0.6}{black}{\voidAnn}{}&\myPic{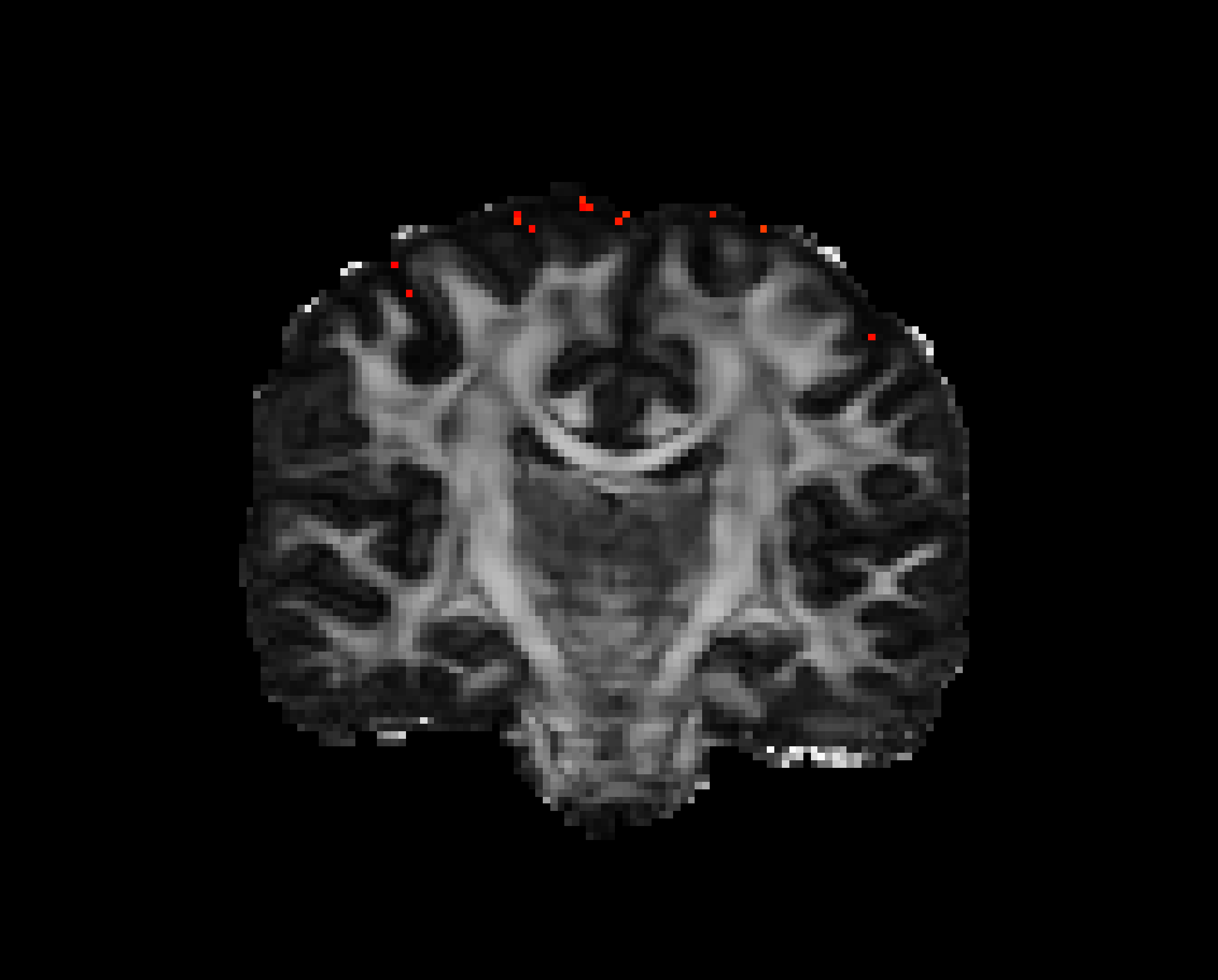}{1.4}{0.5}{0.6}{black}{\voidAnn}{} & \colorBar{\small0}{\small0.6}{0.5}{0.5}{colorbar}\\
	
\end{tabular}
}

\caption{\textbf{Degeneracy Index.} DI maps, overlaid on FA images, given by FSS only and SMSI for a representative HCP subject.}\label{fig:DImap}
\end{figure}

	\begin{table}[t]
	
	\setlength\tabcolsep{1.2pt}
	\renewcommand{\arraystretch}{0.8}
	
	\newcolumntype{L}[1]{>{\raggedright\let\newline\\\arraybackslash\hspace{0pt}}m{#1}}
	\newcolumntype{C}[1]{>{\centering\let\newline\\\arraybackslash\hspace{0pt}}m{#1}}
	\newcolumntype{R}[1]{>{\raggedleft\let\newline\\\arraybackslash\hspace{0pt}}m{#1}}

  \caption{Degeneracy index statistics for 20 HCP subjects.}
\makebox[\linewidth][c]
{
   \begin{tabular}{C{3.5cm}|C{2.5cm}|C{2.5cm}}
   	\toprule
   	
   	& FSS Only & SMSI\\
   	\cmidrule{1-3}
   	
   	\% non-zero DI voxels & 0.813 & 0.056\\
   	
   	DI range & 0.001 - 0.712 & 0.002 - 0.558\\
   	
   	DI mean & 0.073 & 0.017 \\
   	
   	DI standard deviation & 0.101 & 0.072 \\
   	
   	\bottomrule
   \end{tabular}
}

 \label{tab:DI_tab}

\end{table}

	Similar to AMICO \citep{daducci2015accelerated}, SMSI estimation can be potentially improved via deep learning, as demonstrated by Microstructure Estimation using a Deep Network (MEDN) \citep{ye2017tissue}. This will allow the estimation of tissue microstructure properties using of clinical dMRI acquired with a limited number of diffusion gradients.
	
	SMSI involves convex and fast numerical optimization. Based on our preliminary MATLAB implementation, running SMSI on an \SI{1.5}{\milli\meter} isotropic resolution infant dataset for the whole brain on a 4.2GHz Core i7 machine typically takes \num{15} minutes. 
	Implementation with C++ will likely further significantly improve the speed. SMSI is therefore well suited for large-scale studies.

	\section{Conclusion}
	We have presented in this paper a flexible method for quantification of microarchitecture, called spherical mean spectrum imaging (SMSI). The SMS encodes the volume fractions associated with a spectrum of diffusion scales. This allows a wide variety of features to be computed for comprehensive microstructural analysis. We have demonstrated the utility of SMSI in probing the tissue microarchitecture of the developing brain. 
	We have in fact demonstrated that SMSI shows greater sensitivity to brain development \citep{huynh2020mahalohbm} and reveals distinct developmental patterns of cortical microstructure \citep{huynh2020corticalohbm}.
	Future work entails applying SMSI to investigating brain pathologies and potentially identifying sensitive and specific biomarkers for disease diagnosis.

	\clearpage
	\bibliographystyle{IEEEtran}
	\scriptsize
	{\bibliography{References}}

\begin{thebibliography}{10}
\providecommand{\url}[1]{#1}
\csname url@samestyle\endcsname
\providecommand{\newblock}{\relax}
\providecommand{\bibinfo}[2]{#2}
\providecommand{\BIBentrySTDinterwordspacing}{\spaceskip=0pt\relax}
\providecommand{\BIBentryALTinterwordstretchfactor}{4}
\providecommand{\BIBentryALTinterwordspacing}{\spaceskip=\fontdimen2\font plus
\BIBentryALTinterwordstretchfactor\fontdimen3\font minus
  \fontdimen4\font\relax}
\providecommand{\BIBforeignlanguage}[2]{{%
\expandafter\ifx\csname l@#1\endcsname\relax
\typeout{** WARNING: IEEEtran.bst: No hyphenation pattern has been}%
\typeout{** loaded for the language `#1'. Using the pattern for}%
\typeout{** the default language instead.}%
\else
\language=\csname l@#1\endcsname
\fi
#2}}
\providecommand{\BIBdecl}{\relax}
\BIBdecl

\bibitem{kunz2014assessing}
N.~Kunz, H.~Zhang, L.~Vasung, K.~R. O'Brien, Y.~Assaf, F.~Lazeyras, D.~C.
  Alexander, and P.~S. H{\"u}ppi, ``Assessing white matter microstructure of
  the newborn with multi-shell diffusion {MRI} and biophysical compartment
  models.'' \emph{NeuroImage}, vol.~96, no.~8, pp. 288--299, 2014.

\bibitem{lampinen2017neurite}
B.~Lampinen, F.~Szczepankiewicz, J.~M\r{a}rtensson, D.~van Westen, P.~C.
  Sundgren, and M.~Nilsson, ``Neurite density imaging versus imaging of
  microscopic anisotropy in diffusion {MRI}: A model comparison using spherical
  tensor encoding,'' \emph{NeuroImage}, vol. 147, pp. 517--531, 2017.

\bibitem{assaf2005composite}
Y.~Assaf and P.~J. Basser, ``Composite hindered and restricted model of
  diffusion ({CHARMED}) {MR} imaging of the human brain,'' \emph{NeuroImage},
  vol.~27, no.~1, pp. 48--58, 2005.

\bibitem{assaf2008axcaliber}
Y.~Assaf, T.~Blumenfeld-Katzir, Y.~Yovel, and P.~J. Basser, ``{AxCaliber}: {A}
  method for measuring axon diameter distribution from diffusion {MRI},''
  \emph{Magnetic Resonance in Medicine}, vol.~59, no.~6, pp. 1347--1354, 2008.

\bibitem{alexander2010orientationally}
D.~C. Alexander, P.~L. Hubbard, M.~G. Hall, E.~A. Moore, M.~Ptito, G.~J.
  Parker, and T.~B. Dyrby, ``Orientationally invariant indices of axon diameter
  and density from diffusion {MRI},'' \emph{NeuroImage}, vol.~52, no.~4, pp.
  1374--1389, 2010.

\bibitem{fieremans2011white}
E.~Fieremans, J.~H. Jensen, and J.~A. Helpern, ``White matter characterization
  with diffusional kurtosis imaging,'' \emph{NeuroImage}, vol.~58, no.~1, pp.
  177--188, 2011.

\bibitem{zhang2012noddi}
H.~Zhang, T.~Schneider, C.~A. Wheeler-Kingshott, and D.~C. Alexander,
  ``{NODDI}: {Practical} in vivo neurite orientation dispersion and density
  imaging of the human brain,'' \emph{NeuroImage}, vol.~61, no.~4, pp.
  1000--1016, 2012.

\bibitem{daducci2015accelerated}
A.~Daducci, E.~J. Canales-Rodr{\'\i}guez, H.~Zhang, T.~B. Dyrby, D.~C.
  Alexander, and J.-P. Thiran, ``Accelerated microstructure imaging via convex
  optimization ({AMICO}) from diffusion {MRI} data,'' \emph{NeuroImage}, vol.
  105, pp. 32--44, 2015.

\bibitem{white2013probing}
N.~S. White, T.~B. Leergaard, H.~D'arceuil, J.~G. Bjaalie, and A.~M. Dale,
  ``Probing tissue microstructure with restriction spectrum imaging:
  {Histological} and theoretical validation,'' \emph{Human Brain Mapping},
  vol.~34, no.~2, pp. 327--346, 2013.

\bibitem{tournier2004direct}
J.-D. Tournier, F.~Calamante, D.~G. Gadian, and A.~Connelly, ``Direct
  estimation of the fiber orientation density function from diffusion-weighted
  {MRI} data using spherical deconvolution,'' \emph{NeuroImage}, vol.~23,
  no.~3, pp. 1176--1185, 2004.

\bibitem{tournier2007robust}
J.-D. Tournier, F.~Calamante, and A.~Connelly, ``Robust determination of the
  fibre orientation distribution in diffusion {MRI}: Non-negativity constrained
  super-resolved spherical deconvolution,'' \emph{NeuroImage}, vol.~35, pp.
  1459--1472, 2007.

\bibitem{kaden2016quantitative}
E.~Kaden, F.~Kruggel, and D.~C. Alexander, ``Quantitative mapping of the
  per-axon diffusion coefficients in brain white matter,'' \emph{Magnetic
  Resonance in Medicine}, vol.~75, no.~4, pp. 1752--1763, 2016.

\bibitem{kaden2016multi}
E.~Kaden, N.~D. Kelm, R.~P. Carson, M.~D. Does, and D.~C. Alexander,
  ``Multi-compartment microscopic diffusion imaging,'' \emph{NeuroImage}, vol.
  139, pp. 346--359, 2016.

\bibitem{scherrer2015characterizing}
B.~Scherrer, A.~Schwartzman, M.~Taquet, M.~Sahin, S.~P. Prabhu, and S.~K.
  Warfield, ``Characterizing brain tissue by assessment of the distribution of
  anisotropic microstructural environments in diffusion-compartment imaging
  ({DIAMOND}),'' \emph{Magnetic Resonance in Medicine}, vol.~76, no.~3, pp.
  963--977, 2016.

\bibitem{yablonskiy03statistical}
D.~Yablonskiy, G.~Bretthorst, and J.~Ackerman, ``Statistical model for
  diffusion attenuated {MR} signal,'' \emph{Magnetic Resonance in Medicine},
  vol.~50, pp. 664--669, 2003.

\bibitem{wang2011quantification}
Y.~Wang, Q.~Wang, J.~P. Haldar, F.-C. Yeh, M.~Xie, P.~Sun, T.-W. Tu,
  K.~Trinkaus, R.~S. Klein, A.~H. Cross, and S.-K. Song, ``Quantification of
  increased cellularity during inflammatory demyelination,'' \emph{Brain}, vol.
  134, pp. 3590--3601, 2011.

\bibitem{paus2001maturation}
T.~Paus, D.~Collins, A.~Evans, G.~Leonard, B.~Pike, and A.~Zijdenbos,
  ``Maturation of white matter in the human brain: a review of magnetic
  resonance studies,'' \emph{Brain Research Bulletin}, vol.~54, no.~3, pp.
  255--266, 2001.

\bibitem{partridge2004diffusion}
S.~C. Partridge, P.~Mukherjee, R.~G. Henry, S.~P. Miller, J.~I. Berman, H.~Jin,
  Y.~Lu, O.~A. Glenn, D.~M. Ferriero, A.~J. Barkovich, and B.~V. Daniel,
  ``Diffusion tensor imaging: serial quantitation of white matter tract
  maturity in premature newborns,'' \emph{NeuroImage}, vol.~22, no.~3, pp.
  1302--1314, 2004.

\bibitem{dubois2006assessment}
J.~Dubois, L.~Hertz-Pannier, G.~Dehaene-Lambertz, Y.~Cointepas, and
  D.~Le~Bihan, ``Assessment of the early organization and maturation of
  infants' cerebral white matter fiber bundles: a feasibility study using
  quantitative diffusion tensor imaging and tractography,'' \emph{NeuroImage},
  vol.~30, no.~4, pp. 1121--1132, 2006.

\bibitem{anjari2007diffusion}
M.~Anjari, L.~Srinivasan, J.~M. Allsop, J.~V. Hajnal, M.~A. Rutherford, A.~D.
  Edwards, and S.~J. Counsell, ``Diffusion tensor imaging with tract-based
  spatial statistics reveals local white matter abnormalities in preterm
  infants,'' \emph{NeuroImage}, vol.~35, no.~3, pp. 1021--1027, 2007.

\bibitem{kersbergen2014microstructural}
K.~J. Kersbergen, A.~Leemans, F.~Groenendaal, N.~E. van~der Aa, M.~A.
  Viergever, L.~S. de~Vries, and M.~J. Benders, ``Microstructural brain
  development between 30 and 40 weeks corrected age in a longitudinal cohort of
  extremely preterm infants,'' \emph{NeuroImage}, vol. 103, pp. 214--224, 2014.

\bibitem{rose2014brain}
J.~Rose, R.~Vassar, K.~Cahill-Rowley, X.~S. Guzman, D.~K. Stevenson, and
  N.~Barnea-Goraly, ``Brain microstructural development at near-term age in
  very-low-birth-weight preterm infants: {An} atlas-based diffusion imaging
  study,'' \emph{NeuroImage}, vol.~86, pp. 244--256, 2014.

\bibitem{dubois2008asynchrony}
J.~Dubois, G.~Dehaene-Lambertz, M.~Perrin, J.-F. Mangin, Y.~Cointepas,
  E.~Duchesnay, D.~Le~Bihan, and L.~Hertz-Pannier, ``Asynchrony of the early
  maturation of white matter bundles in healthy infants: {Q}uantitative
  landmarks revealed noninvasively by diffusion tensor imaging,'' \emph{Human
  Brain Mapping}, vol.~29, no.~1, pp. 14--27, 2008.

\bibitem{sullivan2006diffusion}
E.~V. Sullivan and A.~Pfefferbaum, ``Diffusion tensor imaging and aging,''
  \emph{Neuroscience \& Biobehavioral Reviews}, vol.~30, no.~6, pp. 749--761,
  2006.

\bibitem{jelescu2015one}
I.~O. Jelescu, J.~Veraart, V.~Adisetiyo, S.~S. Milla, D.~S. Novikov, and
  E.~Fieremans, ``One diffusion acquisition and different white matter models:
  {H}ow does microstructure change in human early development based on {WMTI}
  and {NODDI}?'' \emph{NeuroImage}, vol. 107, pp. 242--256, 2015.

\bibitem{dean2016mapping}
D.~C. Dean, J.~O'muircheartaigh, H.~Dirks, B.~G. Travers, N.~Adluru, A.~L.
  Alexander, and S.~C. Deoni, ``Mapping an index of the myelin g-ratio in
  infants using magnetic resonance imaging,'' \emph{NeuroImage}, vol. 132, pp.
  225--237, 2016.

\bibitem{jones2010diffusion}
D.~K. Jones, \emph{Diffusion {MRI}}.\hskip 1em plus 0.5em minus 0.4em\relax
  Oxford University Press, 2010.

\bibitem{huppi1998microstructural}
P.~S. H{\"u}ppi, S.~E. Maier, S.~Peled, G.~P. Zientara, P.~D. Barnes, F.~A.
  Jolesz, and J.~J. Volpe, ``Microstructural development of human newborn
  cerebral white matter assessed in vivo by diffusion tensor magnetic resonance
  imaging,'' \emph{Pediatric research}, vol.~44, no.~4, p. 584, 1998.

\bibitem{dubois2014early}
J.~Dubois, G.~Dehaene-Lambertz, S.~Kulikova, C.~Poupon, P.~S. H{\"u}ppi, and
  L.~Hertz-Pannier, ``The early development of brain white matter: a review of
  imaging studies in fetuses, newborns and infants,'' \emph{Neuroscience}, vol.
  276, pp. 48--71, 2014.

\bibitem{huynh2019probing}
K.~M. Huynh, T.~Xu, Y.~Wu, G.~Chen, K.-H. Thung, H.~Wu, W.~Lin, D.~Shen, P.-T.
  Yap, and {the UNC/UMN Baby Connectome Project Consortium}, ``Probing brain
  micro-architecture by orientation distribution invariant identification of
  diffusion compartments,'' in \emph{Medical Image Computing and
  Computer-Assisted Intervention (MICCAI)}, 2019, pp. 547--555.

\bibitem{anderson2005measurement}
A.~W. Anderson, ``Measurement of fiber orientation distributions using high
  angular resolution diffusion imaging,'' \emph{Magnetic Resonance in
  Medicine}, vol.~54, no.~5, pp. 1194--1206, 2005.

\bibitem{wu2018multi}
Y.~Wu, Y.~Feng, D.~Shen, and P.-T. Yap, ``A multi-tissue global estimation
  framework for asymmetric fiber orientation distributions,'' in
  \emph{International Conference on Medical Image Computing and
  Computer-Assisted Intervention}.\hskip 1em plus 0.5em minus 0.4em\relax
  Springer, 2018, pp. 45--52.

\bibitem{zou2005regularization}
H.~Zou and T.~Hastie, ``Regularization and variable selection via the elastic
  net,'' \emph{Journal of the Royal Statistical Society: Series B (Statistical
  Methodology)}, vol.~67, no.~2, pp. 301--320, 2005.

\bibitem{bihan1995molecular}
D.~L. Bihan, ``Molecular diffusion, tissue microdynamics and microstructure,''
  \emph{NMR in Biomedicine}, vol.~8, no.~7, pp. 375--386, 1995.

\bibitem{jespersen2007modeling}
S.~N. Jespersen, C.~D. Kroenke, L.~{\O}stergaard, J.~J. Ackerman, and D.~A.
  Yablonskiy, ``Modeling dendrite density from magnetic resonance diffusion
  measurements,'' \emph{NeuroImage}, vol.~34, pp. 1473--1486, 2007.

\bibitem{assaf2004new}
Y.~Assaf, R.~Z. Freidlin, G.~K. Rohde, and P.~J. Basser, ``New modeling and
  experimental framework to characterize hindered and restricted water
  diffusion in brain white matter,'' \emph{Magnetic Resonance in Medicine},
  vol.~52, no.~5, pp. 965--978, 2004.

\bibitem{pasternak2009free}
O.~Pasternak, N.~Sochen, Y.~Gur, N.~Intrator, and Y.~Assaf, ``Free water
  elimination and mapping from diffusion {MRI},'' \emph{Magnetic Resonance in
  Medicine: An Official Journal of the International Society for Magnetic
  Resonance in Medicine}, vol.~62, no.~3, pp. 717--730, 2009.

\bibitem{szczepankiewicz2015quantification}
F.~Szczepankiewicz, S.~Lasi{\v{c}}, D.~van Westen, P.~C. Sundgren, E.~Englund,
  C.-F. Westin, F.~St{\aa}hlberg, J.~L{\"a}tt, D.~Topgaard, and M.~Nilsson,
  ``Quantification of microscopic diffusion anisotropy disentangles effects of
  orientation dispersion from microstructure: applications in healthy
  volunteers and in brain tumors,'' \emph{NeuroImage}, vol. 104, pp. 241--252,
  2015.

\bibitem{cohen2008detection}
J.~Cohen-Adad, M.~Descoteaux, S.~Rossignol, R.~D. Hoge, R.~Deriche, and
  H.~Benali, ``Detection of multiple pathways in the spinal cord using q-ball
  imaging,'' \emph{Neuroimage}, vol.~42, no.~2, pp. 739--749, 2008.

\bibitem{koay2009signal}
C.~G. Koay, E.~{\"O}zarslan, and P.~J. Basser, ``A signal transformational
  framework for breaking the noise floor and its applications in {MRI},''
  \emph{Journal of Magnetic Resonance}, vol. 197, no.~2, pp. 108--119, 2009.

\bibitem{van2013wu}
D.~C. Van~Essen, S.~M. Smith, D.~M. Barch, T.~E. Behrens, E.~Yacoub, and
  K.~Ugurbil, ``The {WU}-{M}inn human connectome project: {An} overview,''
  \emph{NeuroImage}, vol.~80, pp. 62--79, 2013.

\bibitem{howell2018unc}
B.~R. Howell, M.~A. Styner, W.~Gao, P.-T. Yap, L.~Wang, K.~Baluyot, E.~Yacoub,
  G.~Chen, T.~Potts, A.~Salzwedel, G.~Li, J.~H. Gilmore, J.~Piven, J.~K. Smith,
  D.~Shen, K.~Ugurbil, H.~Zhu, W.~Lin, and J.~T. Elison, ``The {UNC/UMN} {Baby
  Connectome Project} ({BCP}): An overview of the study design and protocol
  development,'' \emph{NeuroImage}, 2018.

\bibitem{bergamino2016applying}
M.~Bergamino, O.~Pasternak, M.~Farmer, M.~E. Shenton, and J.~P. Hamilton,
  ``Applying a free-water correction to diffusion imaging data uncovers
  stress-related neural pathology in depression,'' \emph{NeuroImage: Clinical},
  vol.~10, pp. 336--342, 2016.

\bibitem{mukherjee2001normal}
P.~Mukherjee, J.~H. Miller, J.~S. Shimony, T.~E. Conturo, B.~C. Lee, C.~R.
  Almli, and R.~C. McKinstry, ``Normal brain maturation during childhood:
  developmental trends characterized with diffusion-tensor mr imaging,''
  \emph{Radiology}, vol. 221, no.~2, pp. 349--358, 2001.

\bibitem{lovblad2003isotropic}
K.~L{\"o}vblad, J.~Schneider, K.~Ruoss, M.~Steinlin, C.~Fusch, and G.~Schroth,
  ``Isotropic apparent diffusion coefficient mapping of postnatal cerebral
  development,'' \emph{Neuroradiology}, vol.~45, no.~6, pp. 400--403, 2003.

\bibitem{singhi1995body}
S.~Singhi, N.~Ganguli, O.~Bhakoo, K.~Dhall, V.~Sood, and A.~Kaur, ``Body water
  distribution in newborn infants appropriate for gestational age.'' \emph{The
  Indian journal of medical research}, vol. 101, pp. 193--200, 1995.

\bibitem{forbes2002changes}
K.~P. Forbes, J.~G. Pipe, and C.~R. Bird, ``Changes in brain water diffusion
  during the 1st year of life,'' \emph{Radiology}, vol. 222, no.~2, pp.
  405--409, 2002.

\bibitem{sotiropoulos2013effects}
S.~N. Sotiropoulos, S.~Moeller, S.~Jbabdi, J.~Xu, J.~L. Andersson, E.~J.
  Auerbach, E.~Yacoub, D.~Feinberg, K.~Setsompop, L.~L. Wald, T.~E.~J. Behrens,
  K.~Ugurbil, and C.~Lenglet, ``Effects of image reconstruction on fiber
  orientation mapping from multichannel diffusion {MRI}: Reducing the noise
  floor using {SENSE},'' \emph{Magnetic Resonance in Medicine}, vol.~70, pp.
  1682--1689, 2013.

\bibitem{guerrero2019optimizing}
J.~M. Guerrero, N.~Adluru, B.~B. Bendlin, H.~H. Goldsmith, S.~M. Schaefer,
  R.~J. Davidson, S.~R. Kecskemeti, H.~Zhang, and A.~L. Alexander, ``Optimizing
  the intrinsic parallel diffusivity in {NODDI}: An extensive empirical
  evaluation,'' \emph{PloS one}, vol.~14, no.~9, 2019.

\bibitem{barr1974human}
M.~L. Barr, \emph{The human nervous system: An anatomical viewpoint}.\hskip 1em
  plus 0.5em minus 0.4em\relax Harper \& Row, 1974.

\bibitem{blumenfeld2010neuroanatomy}
H.~Blumenfeld, \emph{Neuroanatomy through clinical cases}, 2nd~ed.\hskip 1em
  plus 0.5em minus 0.4em\relax Oxford University Press, 2010.

\bibitem{palombo2019sandi}
M.~Palombo, A.~Ianus, D.~Nunes, M.~Guerreri, D.~C. Alexander, N.~Shemesh, and
  H.~Zhang, ``{SANDI}: a compartment-based model for non-invasive apparent soma
  and neurite imaging by diffusion {MRI},'' \emph{arXiv preprint
  arXiv:1907.02832}, 2019.

\bibitem{huynh2020somamiccai}
K.~M. Huynh, Y.~Wu, K.-H. Thung, S.~Ahmad, H.~P. {Taylor IV}, D.~Shen, and
  P.-T. Yap, ``Characterizing intra-soma diffusion with spherical mean spectrum
  imaging,'' in \emph{Medical Image Computing and Computer-Assisted
  Intervention (MICCAI)}, 2020.

\bibitem{sakuma1991adult}
H.~Sakuma, Y.~Nomura, K.~Takeda, T.~Tagami, T.~Nakagawa, Y.~Tamagawa, Y.~Ishii,
  and T.~Tsukamoto, ``Adult and neonatal human brain: diffusional anisotropy
  and myelination with diffusion-weighted {MR} imaging.'' \emph{Radiology},
  vol. 180, no.~1, pp. 229--233, 1991.

\bibitem{wimberger1995identification}
D.~M. Wimberger, T.~P. Roberts, A.~J. Barkovich, L.~M. Prayer, M.~E. Moseley,
  and J.~Kucharczyk, ``Identification of" premyelination" by diffusion-weighted
  {MRI}.'' \emph{Journal of computer assisted tomography}, vol.~19, no.~1, pp.
  28--33, 1995.

\bibitem{lebel2017review}
C.~Lebel, S.~Treit, and C.~Beaulieu, ``A review of diffusion {MRI} of typical
  white matter development from early childhood to young adulthood,'' \emph{NMR
  in Biomedicine}, p. e3778, 2017.

\bibitem{huang2006white}
H.~Huang, J.~Zhang, S.~Wakana, W.~Zhang, T.~Ren, L.~J. Richards, P.~Yarowsky,
  P.~Donohue, E.~Graham, P.~C. van Zijl, and S.~Mori, ``White and gray matter
  development in human fetal, newborn and pediatric brains,''
  \emph{Neuroimage}, vol.~33, no.~1, pp. 27--38, 2006.

\bibitem{cottaar2019accurate}
M.~Cottaar, F.~Szczepankiewicz, M.~Bastiani, M.~Hernandez-Fernandez, S.~N.
  Sotiropoulos, M.~Nilsson, and S.~Jbabdi, ``Accurate fibre dispersion from
  multiple diffusion encoding,'' \emph{arXiv preprint arXiv:1901.05820}, 2019.

\bibitem{howard2019joint}
A.~F. Howard, J.~Mollink, M.~Kleinnijenhuis, M.~Pallebage-Gamarallage,
  M.~Bastiani, M.~Cottaar, K.~L. Miller, and S.~Jbabdi, ``Joint modelling of
  diffusion {MRI} and microscopy,'' \emph{bioRxiv}, p. 563809, 2019.

\bibitem{szczepankiewicz2016imaging}
F.~Szczepankiewicz, ``Imaging diffusional variance by {MRI}: The role of
  tensor-valued diffusion encoding and tissue heterogeneity,'' Ph.D.
  dissertation, Lund University, 2016.

\bibitem{westin2014measurement}
C.-F. Westin, F.~Szczepankiewicz, O.~Pasternak, E.~{\"O}zarslan, D.~Topgaard,
  H.~Knutsson, and M.~Nilsson, ``Measurement tensors in diffusion {MRI}:
  generalizing the concept of diffusion encoding,'' in \emph{International
  conference on medical image computing and computer-assisted
  intervention}.\hskip 1em plus 0.5em minus 0.4em\relax Springer, 2014, pp.
  209--216.

\bibitem{ye2017tissue}
C.~Ye, ``Tissue microstructure estimation using a deep network inspired by a
  dictionary-based framework,'' \emph{Medical image analysis}, vol.~42, pp.
  288--299, 2017.

\bibitem{huynh2020mahalohbm}
K.~M. Huynh, Y.~Wu, K.-H. Thung, S.~Ahmad, H.~P. {Taylor IV}, W.~Lin, P.-T.
  Yap, and {the UNC/UMN Baby Connectome Project Consortium}, ``Multivariate
  quantification of brain development during the first two years of life with
  mahalanobis distance based on microstructure features,'' in \emph{Proceedings
  of the Annual Meeting of the Organization for Human Brain Mapping (OHBM)},
  2020.

\bibitem{huynh2020corticalohbm}
K.~M. Huynh, Y.~Wu, K.-H. Thung, S.~Ahmad, Z.~Wu, G.~Li, W.~Lin, P.-T. Yap, and
  {the UNC/UMN Baby Connectome Project Consortium}, ``Densely mapping cortical
  micro-architecture properties of the developing brain,'' in \emph{Proceedings
  of the Annual Meeting of the Organization for Human Brain Mapping (OHBM)},
  2020.

\end{thebibliography}

	\pagebreak
	\setcounter{equation}{0}
	\setcounter{figure}{0}
	\setcounter{table}{0}
	\setcounter{page}{1}
	\makeatletter
	\renewcommand{\theequation}{S\arabic{equation}}
	\renewcommand{\thefigure}{S\arabic{figure}}
	\renewcommand{\bibnumfmt}[1]{[S#1]}
	\renewcommand{\citenumfont}[1]{S#1}
	
	\newcounter{cases}
	\newcounter{subcases}[cases]
	\newenvironment{mycases}
	{%
		\setcounter{cases}{0}%
		\setcounter{subcases}{0}%
		\def\case
		{%
			\par\noindent
			\refstepcounter{cases}%
			\textbf{Case \thecases.}
		}%
		\def\subcase
		{%
			\par\noindent
			\refstepcounter{subcases}%
			\underline{\textit{Subcase (\thesubcases):}}
		}%
	}
	{%
		\par
	}
	\renewcommand*\thecases{\arabic{cases}}
	\renewcommand*\thesubcases{\roman{subcases}}

	\setlistdepth{9}
	
	\setlist[itemize,1]{label=\textbullet}
	\setlist[itemize,2]{label=$\circ$ }
	\setlist[itemize,3]{label=\normalfont \bfseries \textendash }
	\setlist[itemize,4]{label=\textdaggerdbl}
	\setlist[itemize,5]{label=\textasteriskcentered}
	\setlist[itemize,6]{label=\textperiodcentered}
	
	\renewlist{itemize}{itemize}{9}

	\section*{Appendix}
	\noindent 
	\subsection*{Linear Independence}\label{sec:linear}
	\noindent 
	For $b\geq 0$ and $\lambda_{\parallel} \geq \lambda_{\perp} \geq 0$, given
	\begin{equation}
		h_{b}(x, \lambda_{\parallel}, \lambda_{\perp}) = 
		\exp\left(-b \lambda_{\perp}\right) \exp\left( -b( \lambda_{\parallel} - \lambda_{\perp}) x^{2} \right)
	\end{equation}
	and
	\begin{equation}\label{eq:hbar}
		\bar{h}_{b}(\lambda_{\parallel}, \lambda_{\perp}) = \int_{0}^{1}h_{b}(x, \lambda_{\parallel}, \lambda_{\perp})dx,
	\end{equation}
	prove that $\bar{h}_{b}(\lambda_{\parallel_i}, \lambda_{\perp_i})$ for $i=1,\cdots n$ are linearly independent with each other with ($\lambda_{\parallel_{i}}\neq \lambda_{\parallel_{j}} \lor \lambda_{\perp_{i}} \neq \lambda_{\perp_{j}}$) for all $i \neq j$.

	\begin{proof}
		\begin{mycases}
			
			Given
			
			\begin{equation}\label{eq:LinearIndependence}
				\sum\limits_{i=1}^{n} \mu_i \bar{h}_{b}(\lambda_{\parallel_{i}}, \lambda_{\perp_{i}})=0.
			\end{equation}

			We move the term with $\mu_i\geq0$ to the left and $\mu_i\leq0$ to the right and rewrite \eqref{eq:LinearIndependence} as 
			
			\begin{equation}\label{eq:twosides}
				\sum\limits_{i=1}^{n_1} \mu_i \bar{h}_{b}(\lambda_{\parallel_{i}}, \lambda_{\perp_{i}})
				=
				\sum\limits_{j=1}^{n_2} \gamma_j \bar{h}_{b}(\widetilde{\lambda}_{\parallel_{j}}, \widetilde{\lambda}_{\perp_{j}}).
			\end{equation}
			
			where $\mu_i$, $\gamma_j$ are all non-negative.
			
			As long as we can show any of the $\mu_i$ or  $\gamma_j$ is zero then we can finish the proof by induction. Throughout the proof, we add an indicator $(\star\star\star)$ when reaching this terminal condition.
			
			\medskip
			
			Let $\lambda_1=\min(\{\lambda_{\perp_{i}}:i=1,\hdots, n_1\})$ and $\lambda_2=\min(\{\widetilde{\lambda}_{\perp_{j}}:j=1,\hdots, n_2\})$.
			\case
			$\lambda_1\neq\lambda_2$, W.L.O.G, suppose $\lambda_1<\lambda_2$
			
			Dividing \eqref{eq:twosides} by $\frac{\exp(-b\lambda_1)}{\sqrt{b}}$ yield
			
			\begin{equation}
				\begin{aligned}
					&\sum\limits_{\{i: \lambda_{\perp_{i}} = \lambda_{1}   \}}^{} \mu_i \int_{0}^{1} \sqrt{b} \mathrm{e}^{-b(\lambda_{\parallel_i}-\lambda_{\perp_i})x^2}dx
					+
					\sum\limits_{\{i: \lambda_{\perp_{i}} > \lambda_{1}   \}}^{} \mu_i \mathrm{e}^{-b(\lambda_{\perp_i}-\lambda_{1})}\int_{0}^{1} \sqrt{b} \mathrm{e}^{-b(\lambda_{\parallel_i}-\lambda_{\perp_i})x^2}dx\\
					=
					&\sum\limits_{j=1}^{n_2} \gamma_j \mathrm{e}^{-b(\widetilde{\lambda}_{\perp_j}-\lambda_{1})}\int_{0}^{1} \sqrt{b} \mathrm{e}^{-b(\lambda_{\parallel_j}-\lambda_{\perp_j})x^2}dx
				\end{aligned}
			\end{equation}
			
			Note that
			
			\begin{equation}
				\lim_{b \to \infty} \int_{0}^{1} \sqrt{b} \mathrm{e}^{-b(\lambda_{\parallel}-\lambda_{\perp})x^2}dx 
				= \lim_{b \to \infty}
				\frac{\pi\erf(\sqrt{b(\lambda_{\parallel}-\lambda_{\perp})})} {2\sqrt{(\lambda_{\parallel}-\lambda_{\perp})}}
				= \frac{\pi}{2\sqrt{(\lambda_{\parallel}-\lambda_{\perp})}} \neq 0
			\end{equation}
			where $\erf(\cdot)$ is the error function.
			
			Thus, take $b \to\infty$, the left hand side goes to a non-zero value while the right hand side goes to zero.
			This implies that $\mu_i=0$ for $\{i:\lambda_{\perp_{i}}=\lambda_1\}$ $(\star\star\star)$.
			
			\case
			$\lambda_1=\lambda_2$, multiply \eqref{eq:twosides} by $\sqrt{b}\exp(b\lambda_1)$ and take the derivative:
			\begin{equation}\label{eq:twosidesderi}
				\begin{aligned}
					&\sum\limits_{i=1}^{n_1} \mu_i(\frac{1}{2\sqrt{b}}\mathrm{e}^{-b(\lambda_{\parallel_i} - \lambda_{1})}-\sqrt{b}
					(\lambda_{\perp_{i}} - \lambda_{1})\mathrm{e}^{-b(\lambda_{\perp_{i}} - \lambda_{1})}
					\int_{0}^{1} \mathrm{e}^{-b( \lambda_{\parallel_i} - \lambda_{\perp_i}) x^{2} }dx)\\
					=
					&\sum\limits_{j=1}^{n_2} \gamma_j(\frac{1}{2\sqrt{b}}\mathrm{e}^{-b(\widetilde{\lambda}_{\parallel_j} - \lambda_{1})}-\sqrt{b}
					(\widetilde{\lambda}_{\perp_{j}} - \lambda_{1})
					\mathrm{e}^{-b(\widetilde{\lambda}_{\perp_{i}} - \lambda_{1})}
					\int_{0}^{1} \mathrm{e}^{-b( \widetilde{\lambda}_{\parallel_j} - \widetilde{\lambda}_{\perp_j}) x^{2} }dx)
				\end{aligned}
			\end{equation}
			rewrite as 
			\begin{equation}\label{eq:derirearrange}
				\begin{aligned}
					&\sum\limits_{\{i:\lambda_{\perp_{i}}=\lambda_1\}} \mu_i\frac{1}{2\sqrt{b}}\mathrm{e}^{-b\lambda_{\parallel_i}}+\sum\limits_{\{i:\lambda_{\perp_{i}}>\lambda_1\}}\mu_i\frac{1}{2\sqrt{b}}\mathrm{e}^{-b\lambda_{\parallel_i}}
					+\sum\limits_{\{j:\widetilde{\lambda}_{\perp_{j}}>\lambda_1\}}\gamma_j\sqrt{b}
					(\widetilde{\lambda}_{\perp_{j}} - \lambda_{1})
					\mathrm{e}^{-b\widetilde{\lambda}_{\perp_{j}}}
					\int_{0}^{1} \mathrm{e}^{-b( \widetilde{\lambda}_{\parallel_j} - \widetilde{\lambda}_{\perp_j}) x^{2}}dx\\
					=&\sum\limits_{\{j:\widetilde{\lambda}_{\perp_{j}}=\lambda_1\}}\gamma_j\frac{1}{2\sqrt{b}}\mathrm{e}^{-b\widetilde{\lambda}_{\parallel_j}}+\sum\limits_{\{j:\widetilde{\lambda}_{\perp_{j}}>\lambda_1\}}\gamma_j\frac{1}{2\sqrt{b}}\mathrm{e}^{-b\widetilde{\lambda}_{\parallel_j}}
					+\sum\limits_{\{i:\lambda_{\perp_{i}}>\lambda_1\}} \mu_i\sqrt{b}
					(\lambda_{\perp_{i}} - \lambda_{1})
					\mathrm{e}^{-b\lambda_{\perp_{i}}}
					\int_{0}^{1} \mathrm{e}^{-b( \lambda_{\parallel_i} - \lambda_{\perp_i}) x^{2} }dx.
				\end{aligned}
			\end{equation}
			
			Denote $\lambda_{11}=\min(\{\lambda_{\parallel_{i}}:i=1,\hdots, n_1\}\cup\{\widetilde{\lambda}_{\perp_{j}}:\widetilde{\lambda}_{\perp_{j}}>\lambda_1\})$, the minimum of exponent on the left and $\lambda_{21}=\min(\{\widetilde{\lambda}_{\parallel_{j}}:j=1,\hdots, n_2\})\cup\{\lambda_{\perp_{i}}:\lambda_{\perp_{i}}>\lambda_1\})$, the minimum of exponent on the right. We also denote $A_1=\{\lambda_{\parallel_i}:\lambda_{\perp_{i}}=\lambda_1\}$, $A_2=\{\lambda_{\parallel_i}:\lambda_{\perp_{i}}>\lambda_1\}$, $A_3=\{\widetilde{\lambda}_{\perp_{j}}:\widetilde{\lambda}_{\perp_{j}}>\lambda_1\}$, $B_1=\{\widetilde{\lambda}_{\parallel_{j}}:\widetilde{\lambda}_{\perp_{j}}=\lambda_1\}$, $B_2=\{\widetilde{\lambda}_{\parallel_{j}}:\widetilde{\lambda}_{\perp_{j}}>\lambda_1\}$, $B_3=\{\lambda_{\perp_i}:\lambda_{\perp_{i}}>\lambda_1\}$, $M_1=\{i:\lambda_{\perp_{i}}>\lambda_1\}, N_1=\{j:\widetilde{\lambda}_{\perp_{j}}>\lambda_1\}$.
			
			\medskip
			
			\subcase $M_1 \equiv N_1 \equiv \emptyset$,  \eqref{eq:derirearrange} becomes
			
			\begin{equation}\label{eq:MNempty}
				\sum\limits_{\{i:\lambda_{\perp_{i}}=\lambda_1\}} \mu_i\frac{1}{2\sqrt{b}}\mathrm{e}^{-b\lambda_{\parallel_i}}=\sum\limits_{\{j:\widetilde{\lambda}_{\perp_{j}}=\lambda_1\}}\gamma_j\frac{1}{2\sqrt{b}}\mathrm{e}^{-b\widetilde{\lambda}_{\parallel_j}}.
			\end{equation}
			Since $(A_1\cap B_1)\equiv\emptyset$, $\lambda_{11}\neq\lambda_{21}$. W.L.O.G, suppose $\lambda_{11}<\lambda_{21}$. 
			Dividing \eqref{eq:MNempty} by $\frac{\exp(-b\lambda_{11})}{\sqrt{b}}$ yields
			
			\begin{equation}
				\sum\limits_{\{i:\lambda_{\perp_{i}}=\lambda_1\}} \mu_i\frac{1}{2}\mathrm{e}^{-b(\lambda_{\parallel_i}-\lambda_{11})}=\sum\limits_{\{j:\widetilde{\lambda}_{\perp_{j}}=\lambda_1\}}\gamma_j\frac{1}{2}\mathrm{e}^{-b(\widetilde{\lambda}_{\parallel_j}-\lambda_{11})}.
			\end{equation}
			
			Take $b\to\infty$, the left hand side goes to a non-zero value while the right hand side goes to zero. This implies that $\mu_i=0$ for $\{i:\lambda_{\parallel_{i}}=\lambda_{11}\}$ $(\star\star\star)$.

			\subcase Only $M_1\equiv\emptyset$ or $N_1\equiv\emptyset$, say $N_1 \equiv\emptyset$, \eqref{eq:derirearrange} becomes 
			
			\begin{equation}\label{eq:Nempty}
				\begin{aligned}
					&\sum\limits_{\{i:\lambda_{\perp_{i}}=\lambda_1\}} \mu_i\frac{1}{2\sqrt{b}}\mathrm{e}^{-b\lambda_{\parallel_i}}+\sum\limits_{\{i:\lambda_{\perp_{i}}>\lambda_1\}}\mu_i\frac{1}{2\sqrt{b}}\mathrm{e}^{-b\lambda_{\parallel_i}}\\
					=
					&\sum\limits_{\{j:\widetilde{\lambda}_{\perp_{j}}=\lambda_1\}}\gamma_j\frac{1}{2\sqrt{b}}\mathrm{e}^{-b\widetilde{\lambda}_{\parallel_j}}+\sum\limits_{\{i:\lambda_{\perp_{i}}>\lambda_1\}} \mu_i\sqrt{b}
					(\lambda_{\perp_{i}} - \lambda_{1})
					\mathrm{e}^{-b\lambda_{\perp_{i}} }
					\int_{0}^{1} \mathrm{e}^{ -b( \lambda_{\parallel_i} - \lambda_{\perp_i}) x^{2}}dx
				\end{aligned}
			\end{equation}
			
			\begin{itemize}
				\item If $\lambda_{11}\neq\lambda_{21}$, say $\lambda_{11}<\lambda_{21}$.
				Dividing \eqref{eq:Nempty} by $\frac{\exp(-b\lambda_{11})}{\sqrt{b}}$  and taking $b\to\infty$, the left hand side goes to a non-zero value while the right hand side goes to zero. It implies that $\mu_i=0$ for $\{i:\lambda_{\parallel_{i}}=\lambda_{11}\}$ $(\star\star\star)$.
				
				\item If $\lambda_{11}=\lambda_{21}$
				
				Check whether $\lambda_{11}\in A_2$. 
				\begin{itemize}
					\item If $\lambda_{11}\in A_2$, we then have $\lambda_{11}\in B_3$. This is because $\lambda_{11}\in A_2$ implies that there exists $i$ such that $\lambda_{\parallel_i}=\lambda_{11}$. We also have $\lambda_{\perp_i}\geq\lambda_{11}$ and  $\lambda_{\perp_i}\geq\lambda_{\parallel_i}$. Combine all these three facts and we have $\lambda_{\perp_i}=\lambda_{\parallel_i}$. 
					Dividing \eqref{eq:Nempty} by $\exp(-b\lambda_{11})$ and taking $b\to\infty$, the left hand side goes to zero and the right hand side goes to infinity. This implies that $\mu_i=0$ ($\star\star\star$).
					
					\item If $\lambda_{11}\notin A_2$, we then have $\lambda_{11}\in A_1$ and $\lambda_{11}\in B_3$. 
					Dividing \eqref{eq:Nempty} by $\exp(-b\lambda_{11})$ and taking $b \to\infty$, the left hand side goes to zero while the right hand side goes to a positive value. This implies that $\mu_i=0$ for $\{i:\lambda_{\perp_{i}}=\lambda_{11}\}$ ($\star\star\star$).
				\end{itemize}
			\end{itemize}
			
			\subcase $M_1 \not\equiv \emptyset$ and $N_1 \not\equiv\emptyset$.
			\begin{itemize}
				\item If $\lambda_{11}\neq\lambda_{21}$, say $\lambda_{11}<\lambda_{21}$
				
				Dividing \eqref{eq:derirearrange} by $\frac{\exp(-b\lambda_{11})}{\sqrt{b}}$  and taking $b\to\infty$, the left hand side goes to a non-zero value or infinity while the right hand side goes to zero. It implies that $\mu_i=0$ for $\{i:\lambda_{\parallel_{i}}=\lambda_{11}\}$ ($\star\star\star$).
				
				\item If $\lambda_{11}=\lambda_{21}$, consider
				\begin{itemize}
					\item $\lambda_{11}\in A_2$ or $\lambda_{11}\in B_2$. Thus $\lambda_{11}\notin (A_2 \cap B_2)$ since there exist $i$ and $j$ that $\lambda_{\perp_{i}}=\lambda_{\parallel_{i}}=\widetilde{\lambda}_{\perp_{j}}=\widetilde{\lambda}_{\parallel_{j}}=\lambda_{11}$, which is contradictory with ($\lambda_{\parallel_{i}}\neq \widetilde{\lambda}_{\parallel_{j}} \lor \lambda_{\perp_{i}} \neq \widetilde{\lambda}_{\perp_{j}}$) for all $i \neq j$.
					
					Suppose $\lambda_{11}\in A_2$, then $\lambda_{\perp_{i}}=\lambda_{\parallel_{i}}$ for some $i$. Dividing \eqref{eq:derirearrange} by $\exp(-b\lambda_{11})$ and taking $b\to\infty$, the left hand side goes to zero while the right hand side goes to infinity. It implies that $\mu_i=0$ ($\star\star\star$).
					
					\item $\lambda_{11}\notin A_2$, $\lambda_{11}\notin B_2$, and ($\lambda_{11}\in A_1$ or $\lambda_{11}\in B_1$).
					
					Since $A_1\cap B_1=\emptyset$, suppose $\lambda_{11}\in A_1$ and then  $\lambda_{11}\in B_3$ on the right hand side only. 
					\begin{itemize}
						\item If $\lambda_{11}\notin A_3$
						
						Dividing \eqref{eq:derirearrange} by $\exp(-b\lambda_{11})$ and taking $b\to\infty$, the left hand side goes to zero while the right hand side goes to a positive value. It implies that $\mu_i=0$, for $\{i:\lambda_{\perp_{i}}=\lambda_{11}\}$ ($\star\star\star$).
						\item  If $\lambda_{11}\in A_3$
						
						Divide the equation \eqref{eq:derirearrange} by $\exp(-b\lambda_{11})$ and take derivative
						\begin{equation}\label{eq:lambda11inA3}
							\begin{aligned}
								&\sum\limits_{\{i:\lambda_{\perp_{i}}=\lambda_1, \lambda_{\parallel_{i}}=\lambda_{11}\}} -\mu_i\frac{1}{4b^{\frac{3}{2}}}+\sum\limits_{\{i:\lambda_{\perp_{i}}\geq\lambda_1,\lambda_{\parallel_{i}}>\lambda_{11}\}}\mu_i\mathrm{e}^{-b(\lambda_{\parallel_i}-\lambda_{11})}(-\frac{\lambda_{\parallel_i}-\lambda_{11}}{2\sqrt{b}}-\frac{1}{4b^{\frac{3}{2}}})\\
								&+\sum\limits_{\{j:\widetilde{\lambda}_{\perp_{j}}>\lambda_{1}\}} \gamma_j(\widetilde{\lambda}_{\perp_{j}} - \lambda_{1})\frac{1}{2\sqrt{b}}\mathrm{e}^{-b(\widetilde{\lambda}_{\parallel_j} - \lambda_{11})}\\
								&-\sum\limits_{\{j:\widetilde{\lambda}_{\perp_{j}}>\lambda_{11}\}} \gamma_j(\widetilde{\lambda}_{\perp_{j}} - \lambda_{1})\sqrt{b}
								(\widetilde{\lambda}_{\perp_{j}} - \lambda_{11})
								\mathrm{e}^{-b(\widetilde{\lambda}_{\perp_{j}} - \lambda_{11})}
								\int_{0}^{1} \mathrm{e}^{-b( \widetilde{\lambda}_{\parallel_j} - \widetilde{\lambda}_{\perp_j}) x^{2} }dx\\
								&=\sum\limits_{\{j:\widetilde{\lambda}_{\perp_{j}}\geq\lambda_1\}}\gamma_j\mathrm{e}^{-b(\widetilde{\lambda}_{\parallel_j}-\lambda_{11})}(-\frac{\widetilde{\lambda}_{\parallel_j}-\lambda_{11}}{2\sqrt{b}}-\frac{1}{4b^{\frac{3}{2}}})\\
								&+\sum\limits_{\{i:\lambda_{\perp_{i}}>\lambda_1\}} \mu_i(\lambda_{\perp_{i}} - \lambda_{1})\frac{1}{2\sqrt{b}}\mathrm{e}^{-b(\lambda_{\parallel_i} - \lambda_{11})}\\
								&-\sum\limits_{\{i:\lambda_{\perp_{i}}>\lambda_{11}\}} \mu_i(\lambda_{\perp_{i}} - \lambda_{1})\sqrt{b}
								(\lambda_{\perp_{i}} - \lambda_{11})
								\mathrm{e}^{-b(\lambda_{\perp_{i}} - \lambda_{11})}
								\int_{0}^{1} \mathrm{e}^{-b( \lambda_{\parallel_i} - \lambda_{\perp_i}) x^{2}}dx\\
							\end{aligned}
						\end{equation}
						Multiplying \eqref{eq:lambda11inA3} by $b^{\frac{3}{2}}$ and taking $b\to\infty$, the left hand side goes to a non-zero value while the right hand side goes to zero. It implies that $\mu_i=0$ for $\{i:\lambda_{\perp_{i}}=\lambda_1, \lambda_{\parallel_{i}}=\lambda_{11}\}$ ($\star\star\star$).
					\end{itemize}
					\item $\lambda_{11}\notin A_1$, $\lambda_{11}\notin A_2$, and $\lambda_{11}\in A_3$. $\lambda_{11}\notin B_1$, $\lambda_{11}\notin B_2$, and $\lambda_{11}\in B_3$.
					
					Multiply \eqref{eq:derirearrange} by $\sqrt{b}\exp(b\lambda_{11})$ and take derivative
					
					\begin{equation}
						\begin{aligned}
							&\sum\limits_{\{i:\lambda_{\perp_{i}}\geq\lambda_1\}}\mu_i\mathrm{e}^{-b(\lambda_{\parallel_i}-\lambda_{11})}(-\frac{\lambda_{\parallel_i}-\lambda_{11}}{2\sqrt{b}}-\frac{1}{4b^{\frac{3}{2}}})
							+\sum\limits_{\{j:\widetilde{\lambda}_{\perp_{j}}>\lambda_1\}} \gamma_j(\widetilde{\lambda}_{\perp_{j}} - \lambda_{1})\frac{1}{2\sqrt{b}}\mathrm{e}^{-b(\widetilde{\lambda}_{\parallel_j} - \lambda_{11})}\\
							-&\sum\limits_{\{j:\widetilde{\lambda}_{\perp_{j}}>\lambda_{11}\}} \gamma_j(\widetilde{\lambda}_{\perp_{j}} - \lambda_{1})\sqrt{b}
							(\widetilde{\lambda}_{\perp_{j}} - \lambda_{11})
							\mathrm{e}^{-b(\widetilde{\lambda}_{\perp_{j}} - \lambda_{11})}
							\int_{0}^{1} \mathrm{e}^{ -b( \widetilde{\lambda}_{\parallel_j} - \widetilde{\lambda}_{\perp_j}) x^{2}}dx\\
							=&\sum\limits_{\{j:\widetilde{\lambda}_{\perp_{j}}\geq\lambda_1\}}\gamma_j\mathrm{e}^{-b(\widetilde{\lambda}_{\parallel_j}-\lambda_{11})}(-\frac{\widetilde{\lambda}_{\parallel_j}-\lambda_{11}}{2\sqrt{b}}-\frac{1}{4b^{\frac{3}{2}}})
							+\sum\limits_{\{i:\lambda_{\perp_{i}}>\lambda_1\}} \mu_i(\lambda_{\perp_{i}} - \lambda_{1})\frac{1}{2\sqrt{b}}\mathrm{e}^{-b(\lambda_{\parallel_i} - \lambda_{11})}\\
							-&\sum\limits_{\{i:\lambda_{\perp_{i}}>\lambda_{11}\}} \mu_i(\lambda_{\perp_{i}} - \lambda_{1})\sqrt{b}
							(\lambda_{\perp_{i}} - \lambda_{11})
							\mathrm{e}^{-b(\lambda_{\perp_{i}} - \lambda_{11})}
							\int_{0}^{1} \mathrm{e}^{ -b( \lambda_{\parallel_i} - \lambda_{\perp_i}) x^{2} }dx\\
						\end{aligned}
					\end{equation}
					Multiply both sides with $\exp(-b\lambda_{11})$ and rearrange
					
					\begin{equation}\label{eq:subcase323}
						\begin{aligned}
							&\sum\limits_{\{i:\lambda_{\perp_{i}}=\lambda_1\}}\mu_i\mathrm{e}^{-b\lambda_{\parallel_i}}(\frac{\lambda_{\parallel_i}-\lambda_{11}}{2\sqrt{b}}+\frac{1}{4b^{\frac{3}{2}}})\\
							+&\sum\limits_{\{i:\lambda_{\perp_{i}}=\lambda_{11}\}}\mu_i\mathrm{e}^{-b\lambda_{\parallel_i}}(\frac{\lambda_{\parallel_i}-\lambda_{11}+\lambda_{\perp_{i}} - \lambda_{1}}{2\sqrt{b}}+\frac{1}{4b^{\frac{3}{2}}})
							+\sum\limits_{\{i:\lambda_{\perp_{i}}>\lambda_{11}\}}\mu_i\mathrm{e}^{-b\lambda_{\parallel_i}}(\frac{\lambda_{\parallel_i}-\lambda_{11}+\lambda_{\perp_{i}} - \lambda_{1}}{2\sqrt{b}}+\frac{1}{4b^{\frac{3}{2}}})\\
							+&\sum\limits_{\{j:\widetilde{\lambda}_{\perp_{j}}>\lambda_{11}\}} \gamma_j(\widetilde{\lambda}_{\perp_{j}} - \lambda_{1})\sqrt{b}
							(\widetilde{\lambda}_{\perp_{j}} - \lambda_{11})
							\mathrm{e}^{-b\widetilde{\lambda}_{\perp_{j}} }
							\int_{0}^{1} \mathrm{e}^{-b( \widetilde{\lambda}_{\parallel_j} - \widetilde{\lambda}_{\perp_j}) x^{2}}dx\\
							=&\sum\limits_{\{j:\widetilde{\lambda}_{\perp_{j}}=\lambda_1\}}\gamma_j\mathrm{e}^{-b\widetilde{\lambda}_{\parallel_j}}(\frac{\widetilde{\lambda}_{\parallel_j}-\lambda_{11}}{2\sqrt{b}}+\frac{1}{4b^{\frac{3}{2}}})\\
							+&\sum\limits_{\{j:\widetilde{\lambda}_{\perp_{j}}=\lambda_{11}\}}\gamma_j\mathrm{e}^{-b\widetilde{\lambda}_{\parallel_j}}(\frac{\widetilde{\lambda}_{\parallel_j}-\lambda_{11}+\widetilde{\lambda}_{\perp_{j}} - \lambda_{1}}{2\sqrt{b}}+\frac{1}{4b^{\frac{3}{2}}})
							+\sum\limits_{\{j:\widetilde{\lambda}_{\perp_{j}}>\lambda_{11}\}}\gamma_j\mathrm{e}^{-b\widetilde{\lambda}_{\parallel_j}}(\frac{\widetilde{\lambda}_{\parallel_j}-\lambda_{11}+\widetilde{\lambda}_{\perp_{j}} - \lambda_{1}}{2\sqrt{b}}+\frac{1}{4b^{\frac{3}{2}}})\\
							+&\sum\limits_{\{i:\lambda_{\perp_{i}}>\lambda_{11}\}} \mu_i(\lambda_{\perp_{i}} - \lambda_{1})\sqrt{b}
							(\lambda_{\perp_{i}} - \lambda_{11})
							\mathrm{e}^{-b\lambda_{\perp_{i}}}
							\int_{0}^{1} \mathrm{e}^{ -b( \lambda_{\parallel_i} - \lambda_{\perp_i}) x^{2}}dx
						\end{aligned}
					\end{equation}
					Denote $\lambda_{12}=\min(\{\lambda_{\parallel_{i}}:i=1\cdots n_1\}\cup\{\widetilde{\lambda}_{\perp_{j}}:\widetilde{\lambda}_{\perp_{j}}>\lambda_{11}\})$, the minimum of exponent on the left and $\lambda_{22}=\min(\{\widetilde{\lambda}_{\parallel_{j}}:j=1\cdots n_2\})\cup\{\lambda_{\perp_{i}}:\lambda_{\perp_{i}}>\lambda_{11}\})$, the minimum of exponent on the right. We also denote $A_{11}=\{\lambda_{\parallel_i}:\lambda_{\perp_{i}}\in\{\lambda_1,\lambda_{11}\}\}$, $A_{21}=\{\lambda_{\parallel_i}:\lambda_{\perp_{i}}>\lambda_{11}\}$, $A_{31}=\{\widetilde{\lambda}_{\perp_{j}}:\widetilde{\lambda}_{\perp_{j}}>\lambda_{11}\}$, $B_{11}=\{\widetilde{\lambda}_{\parallel_{j}}:\widetilde{\lambda}_{\perp_{j}}\in\{\lambda_1,\lambda_{11}\}\}$, $B_{21}=\{\widetilde{\lambda}_{\parallel_{j}}:\widetilde{\lambda}_{\perp_{j}}>\lambda_{11}\}$, $B_{31}=\{\lambda_{\perp_i}:\lambda_{\perp_{i}}>\lambda_{11}\}$, $M_2=\{i:\lambda_{\perp_{i}}>\lambda_{11}\}, N_2=\{j:\widetilde{\lambda}_{\perp_{j}}>\lambda_{11}\}$.
					
					\begin{itemize}
						\item If $M_2=N_2=\emptyset$, \eqref{eq:subcase323} becomes 
						\begin{equation}\label{eq:subcase3231}
							\begin{aligned}
								&\sum\limits_{\{i:\lambda_{\perp_{i}}=\lambda_1\}}\mu_i\mathrm{e}^{-b\lambda_{\parallel_i}}(\frac{\lambda_{\parallel_i}-\lambda_{11}}{2\sqrt{b}}+\frac{1}{4b^{\frac{3}{2}}})+\sum\limits_{\{i:\lambda_{\perp_{i}}=\lambda_{11}\}}\mu_i\mathrm{e}^{-b\lambda_{\parallel_i}}(\frac{\lambda_{\parallel_i}-\lambda_{11}+\lambda_{\perp_{i}} - \lambda_{1}}{2\sqrt{b}}+\frac{1}{4b^{\frac{3}{2}}})\\
								=&\sum\limits_{\{j:\widetilde{\lambda}_{\perp_{j}}=\lambda_1\}}\gamma_j\mathrm{e}^{-b\widetilde{\lambda}_{\parallel_j}}(\frac{\widetilde{\lambda}_{\parallel_j}-\lambda_{11}}{2\sqrt{b}}+\frac{1}{4b^{\frac{3}{2}}})+\sum\limits_{\{j:\widetilde{\lambda}_{\perp_{j}}=\lambda_{11}\}}\gamma_j\mathrm{e}^{-b\widetilde{\lambda}_{\parallel_j}}(\frac{\widetilde{\lambda}_{\parallel_j}-\lambda_{11}+\widetilde{\lambda}_{\perp_{j}} - \lambda_{1}}{2\sqrt{b}}+\frac{1}{4b^{\frac{3}{2}}})\\
							\end{aligned}
						\end{equation}
						\begin{itemize}
							\item If $\lambda_{12}\neq\lambda_{22}$, say $\lambda_{12}<\lambda_{22}$
							
							Dividing \eqref{eq:subcase3231} by $\frac{\exp(-b\lambda_{12})}{b^{\frac{3}{2}}}$ and taking $b\to\infty$, the left hand side goes to a non-zero value while the right hand side goes to zero. It implies that $\mu_i=0$ for $\{i:\lambda_{\parallel_{i}}=\lambda_{12}\}$ ($\star\star\star$).
							\item  	If $\lambda_{12}=\lambda_{22}$
							
							We have $\lambda_{12}\in \{\lambda_{\parallel_i}:\lambda_{\perp_{i}}=\lambda_1\}\cap\{\widetilde{\lambda}_{\parallel_{j}}:\widetilde{\lambda}_{\perp_{j}}=\lambda_{11}\}$ or $\lambda_{12}\in \{\lambda_{\parallel_i}:\lambda_{\perp_{i}}=\lambda_{11}\}\cap\{\widetilde{\lambda}_{\parallel_{j}}:\widetilde{\lambda}_{\perp_{j}}=\lambda_{1}\}$ since $\{\lambda_{\parallel_i}:\lambda_{\perp_{i}}=\lambda_1\}\cap\{\widetilde{\lambda}_{\perp_{j}}:\widetilde{\lambda}_{\perp_{j}}=\lambda_{1}\}=\{\lambda_{\parallel_i}:\lambda_{\perp_{i}}=\lambda_{11}\}\cap\{\widetilde{\lambda}_{\perp_{j}}:\widetilde{\lambda}_{\perp_{j}}=\lambda_{11}\}=\emptyset$.
							
							Suppose $\lambda_{12}\in \{\lambda_{\parallel_i}:\lambda_{\perp_{i}}=\lambda_1\}\cap\{\widetilde{\lambda}_{\parallel_{j}}:\widetilde{\lambda}_{\perp_{j}}=\lambda_{11}\}$ W.L.O.G. Thus, there exist  $i_1\in\{i:\lambda_{\perp_{i}}=\lambda_1\}$ and $j_1\in\{j:\widetilde{\lambda}_{\perp_{j}}=\lambda_{11}\}$ such that $\lambda_{\parallel_{i_1}}=\widetilde{\lambda}_{\parallel_{j_1}}=\lambda_{12}$
							
							Dividing \eqref{eq:subcase3231} $\exp(-\lambda_{12}b)/\sqrt{b}$ and taking $b\to\infty$, the left hand side goes to  $\mu_{i_1}\frac{\lambda_{\parallel_{i_1}}-\lambda_{11}}{2}$ and the right hand side goes to $\gamma_{j_1}\frac{\widetilde{\lambda}_{\parallel_{j_1}}-\lambda_{11}+\widetilde{\lambda}_{\perp_{j_1}}-\lambda_1}{2}$. It implies that $\mu_{i_1}(\lambda_{\parallel_{i_1}}-\lambda_{11}) =\gamma_{j_1}(\widetilde{\lambda}_{\parallel_{j_1}}-\lambda_{11}+\widetilde{\lambda}_{\perp_{j_1}}-\lambda_1)$. With this, \eqref{eq:subcase323} becomes 
							\begin{equation*}
								\begin{aligned}
									&\sum\limits_{\{i:\lambda_{\perp_{i}}=\lambda_1, i \neq i_1\}}\mu_i\mathrm{e}^{-b\lambda_{\parallel_i}}(\frac{\lambda_{\parallel_i}-\lambda_{11}}{2\sqrt{b}}+\frac{1}{4b^{\frac{3}{2}}})+\mu_{i_1}\mathrm{e}^{-b\lambda_{\parallel_{i_1}}}\frac{1}{4b^{\frac{3}{2}}}\\
									+&\sum\limits_{\{i:\lambda_{\perp_{i}}=\lambda_{11}\}}\mu_i\mathrm{e}^{-b\lambda_{\parallel_i}}(\frac{\lambda_{\parallel_i}-\lambda_{11}+\lambda_{\perp_{i}} - \lambda_{1}}{2\sqrt{b}}+\frac{1}{4b^{\frac{3}{2}}})\\
									=&\sum\limits_{\{j:\widetilde{\lambda}_{\perp_{j}}=\lambda_1\}}\gamma_j\mathrm{e}^{-b\widetilde{\lambda}_{\parallel_j}}(\frac{\widetilde{\lambda}_{\parallel_j}-\lambda_{11}}{2\sqrt{b}}+\frac{1}{4b^{\frac{3}{2}}})+\gamma_{j_1}\mathrm{e}^{-b\widetilde{\lambda}_{\parallel_{j_1}}}\frac{1}{4b^{\frac{3}{2}}}\\
									+&\sum\limits_{\{j:\widetilde{\lambda}_{\perp_{j}}=\lambda_{11}, j\neq j_1\}}\gamma_j\mathrm{e}^{-b\widetilde{\lambda}_{\parallel_j}}(\frac{\widetilde{\lambda}_{\parallel_j}-\lambda_{11}+\widetilde{\lambda}_{\perp_{j}} - \lambda_{1}}{2\sqrt{b}}+\frac{1}{4b^{\frac{3}{2}}})\\
								\end{aligned}
							\end{equation*}
							Dividing by $\frac{\exp(-b\lambda_{12})}{b^{\frac{3}{2}}}$ and taking $b\to\infty$, the left hand side goes to  $\frac{\mu_{i_1}}{4}$ and the right hand side goes to $\frac{\gamma_{j_1}}{4}$. This implies that $\mu_{i_1} =\gamma_{j_1}$, which is contradictory with $\mu_{i_1}(\lambda_{\parallel_{i_1}}-\lambda_{11}) =\gamma_{j_1}(\widetilde{\lambda}_{\parallel_{j_1}}-\lambda_{11}+\widetilde{\lambda}_{\perp_{j_1}}-\lambda_1)$. Thus $\lambda_{12}=\lambda_{22}$ does not happen.
						\end{itemize}
						\item If $M_1=\emptyset$ or $N_1=\emptyset$ only, say $N_2=\emptyset$
						
						\eqref{eq:subcase323} becomes 
						\begin{equation}\label{eq:subcase323121}
							\begin{aligned}
								&\sum\limits_{\{i:\lambda_{\perp_{i}}=\lambda_1\}}\mu_i\mathrm{e}^{-b\lambda_{\parallel_i}}(\frac{\lambda_{\parallel_i}-\lambda_{11}}{2\sqrt{b}}+\frac{1}{4b^{\frac{3}{2}}})+\sum\limits_{\{i:\lambda_{\perp_{i}}=\lambda_{11}\}}\mu_i\mathrm{e}^{-b\lambda_{\parallel_i}}(\frac{\lambda_{\parallel_i}-\lambda_{11}+\lambda_{\perp_{i}} - \lambda_{1}}{2\sqrt{b}}+\frac{1}{4b^{\frac{3}{2}}})\\
								+&\sum\limits_{\{i:\lambda_{\perp_{i}}>\lambda_{11}\}}\mu_i\mathrm{e}^{-b\lambda_{\parallel_i}}(\frac{\lambda_{\parallel_i}-\lambda_{11}+\lambda_{\perp_{i}} - \lambda_{1}}{2\sqrt{b}}+\frac{1}{4b^{\frac{3}{2}}})\\
								=&\sum\limits_{\{j:\widetilde{\lambda}_{\perp_{j}}=\lambda_1\}}\gamma_j\mathrm{e}^{-b\widetilde{\lambda}_{\parallel_j}}(\frac{\widetilde{\lambda}_{\parallel_j}-\lambda_{11}}{2\sqrt{b}}+\frac{1}{4b^{\frac{3}{2}}})+\sum\limits_{\{j:\widetilde{\lambda}_{\perp_{j}}=\lambda_{11}\}}\gamma_j\mathrm{e}^{-b\widetilde{\lambda}_{\parallel_j}}(\frac{\widetilde{\lambda}_{\parallel_j}-\lambda_{11}+\widetilde{\lambda}_{\perp_{j}} - \lambda_{1}}{2\sqrt{b}}+\frac{1}{4b^{\frac{3}{2}}})\\
								+&\sum\limits_{\{i:\lambda_{\perp_{i}}>\lambda_{11}\}} \mu_i(\lambda_{\perp_{i}} - \lambda_{1})\sqrt{b}
								(\lambda_{\perp_{i}} - \lambda_{11})
								\mathrm{e}^{-b\lambda_{\perp_{i}}}
								\int_{0}^{1} \mathrm{e}^{-b( \lambda_{\parallel_i} - \lambda_{\perp_i}) x^{2}}dx
							\end{aligned}
						\end{equation}
						\begin{itemize}
							\item If $\lambda_{12}\neq\lambda_{22}$, say $\lambda_{12}<\lambda_{22}$
							
							Dividing \eqref{eq:subcase323121} by $\frac{\exp(-b\lambda_{12})}{\sqrt{b}}$  and taking $b\to\infty$, the left hand side goes to a constant while the right hand side goes to zero. It implies that $\mu_i=0$ for $\{i:\lambda_{\parallel_{i}}=\lambda_{12}\}$ ($\star\star\star$).
							
							\item  If $\lambda_{12}=\lambda_{22}$
							
							Check whether $\lambda_{12}\in A_{21}$.
							
							\begin{itemize}
								\item If $\lambda_{12}\in A_{21}$
								
								We also have $\lambda_{12}\in B_{31}$. It implies that there exists $i$ such that $\lambda_{\perp_i}=\lambda_{\parallel_i}$. Dividing \eqref{eq:subcase323121} by $\exp(-b\lambda_{12})$ to and taking $b\to\infty$, the left hand side goes to zero while the right hand side goes to infinity. It implies $\mu_i=0$ ($\star\star\star$). 
								
								\item 	If $\lambda_{12}\notin A_{21}$
								
								We next see whether $\lambda_{12}\in B_{31}$.
								\begin{itemize}
									\item If $\lambda_{12}\in B_{31}$, we divide \eqref{eq:subcase323121} by $\exp(-b\lambda_{12})$ and take $b\to\infty$. The left hand side goes to zero while the right hand side goes to infinity. It implies that $\mu_i=0$ for $\{i:\lambda_{\perp_{i}}=\lambda_{12}\}$ ($\star\star\star$). 
									\item If $\lambda_{12}\notin B_{31}$, we use the same technique when $M_2=N_2=\emptyset$ and thus get the contradiction that this case does not exist.
								\end{itemize}
								
							\end{itemize}

						\end{itemize}
						\item If $M_2\neq\emptyset$ and $N_2\neq\emptyset$.
						
						\begin{itemize}
							\item 	If $\lambda_{12}\neq\lambda_{22}$, say $\lambda_{12}<\lambda_{22}$
							
							Dividing \eqref{eq:subcase323} by $\frac{\exp(-b\lambda_{12})}{\sqrt{b}}$ and taking $b\to\infty$, the left hand side goes to a non-zero value or infinity while the right hand side goes to zero. It implies that $\mu_i=0$ for $\{i:\lambda_{\parallel_{i}}=\lambda_{12}\}$ ($\star\star\star$).
							
							\item 	If $\lambda_{12}=\lambda_{22}$, consider
							
							\begin{itemize}
								\item $\lambda_{12}\in A_{21}$ or $\lambda_{12}\in B_{21}$
								
								This is the same with the previous case of $\lambda_{11}\in A_2$ or $\lambda_{11}\in B_2$. We omit the details.
								
								\item $\lambda_{12}\notin A_{21}$, $\lambda_{12}\notin B_{21}$, and ($\lambda_{12}\in A_{11}$ or $\lambda_{12}\in B_{11}$). 
								
								Here we claim that $\lambda_{12}\notin A_{11}\cap B_{11}$ and everything else is the same with the previous case of $\lambda_{11}\notin A_2$, $\lambda_{11}\notin B_2$, and ($\lambda_{11}\in A_1$ or $\lambda_{11}\in B_1$).
								
								This is because if $\lambda_{12}\in A_{11}\cap B_{11}$, then we can only have $\lambda_{12}\in\{\lambda_{\parallel_i}:\lambda_{\perp_{i}}=\lambda_1\}\cap\{\widetilde{\lambda}_{\parallel_{j}}:\widetilde{\lambda}_{\perp_{j}}=\lambda_{11}\}$ or $\lambda_{12}\in\{\lambda_{\parallel_i}:\lambda_{\perp_{i}}=\lambda_{11}\}\cap\{\widetilde{\lambda}_{\parallel_{j}}:\widetilde{\lambda}_{\perp_{j}}=\lambda_{1}\}$ since $\{\lambda_{\parallel_i}:\lambda_{\perp_{i}}=\lambda_1\}\cap\{\widetilde{\lambda}_{\parallel_{j}}:\widetilde{\lambda}_{\perp_{j}}=\lambda_{1}\}=\emptyset$ and $\{\widetilde{\lambda}_{\parallel_{j}}:\widetilde{\lambda}_{\perp_{j}}=\lambda_{11}\}\cap\{\lambda_{\parallel_i}:\lambda_{\perp_{i}}=\lambda_{11}\}=\emptyset$.
								
								Suppose $\lambda_{12}\in\{\lambda_{\parallel_i}:\lambda_{\perp_{i}}=\lambda_1\}\cap\{\widetilde{\lambda}_{\parallel_{j}}:\widetilde{\lambda}_{\perp_{j}}=\lambda_{11}\}$ W.L.O.G. Then there exist $i_1\in\{i:\lambda_{\perp_{i}}=\lambda_1\}$ and $j_1\in\{j:\widetilde{\lambda}_{\perp_{j}}=\lambda_{11}\}$ such that $\lambda_{\parallel_{i_1}}=\widetilde{\lambda}_{\parallel_{j_1}}=\lambda_{12}$.
								
								We divide \eqref{eq:subcase323} by $\exp(-b\lambda_{12})$ and take derivative; then multiply the result by $b^{\frac{3}{2}}$ and take $b\to\infty$. The left hand side goes to $-\frac{\mu_{i_1}(\lambda_{\parallel_{i_1}}-\lambda_{11})}{4}$ while the right hand side goes to $-\frac{\gamma_{j_1}(\widetilde{\lambda}_{\parallel_{j_1}}-\lambda_{11}+\widetilde{\lambda}_{\perp_{j_1}} - \lambda_{1})}{4}$, which implies that $\mu_{i_1}(\lambda_{\parallel_{i_1}}-\lambda_{11})=\gamma_{j_1}(\widetilde{\lambda}_{\parallel_{j_1}}-\lambda_{11}+\widetilde{\lambda}_{\perp_{j_1}} - \lambda_{1})$.
								Plugging back in \eqref{eq:subcase323} yields
								
								\begin{equation}\label{eq:case2long}
									\begin{aligned}
										&\sum\limits_{\{i:\lambda_{\perp_{i}}=\lambda_1,i\neq i_1\}}\mu_i\mathrm{e}^{-b\lambda_{\parallel_i}}(\frac{\lambda_{\parallel_i}-\lambda_{11}}{2\sqrt{b}}+\frac{1}{4b^{\frac{3}{2}}})+\mu_{i_1}\mathrm{e}^{-b\lambda_{\parallel_{i_1}}}(\frac{1}{4b^{\frac{3}{2}}})\\
										+&\sum\limits_{\{i:\lambda_{\perp_{i}}=\lambda_{11}\}}\mu_i\mathrm{e}^{-b\lambda_{\parallel_i}}(\frac{\lambda_{\parallel_i}-\lambda_{11}+\lambda_{\perp_{i}} - \lambda_{1}}{2\sqrt{b}}+\frac{1}{4b^{\frac{3}{2}}})\\
										+&\sum\limits_{\{i:\lambda_{\perp_{i}}>\lambda_{11}\}}\mu_i\mathrm{e}^{-b\lambda_{\parallel_i}}(\frac{\lambda_{\parallel_i}-\lambda_{11}+\lambda_{\perp_{i}} - \lambda_{1}}{2\sqrt{b}}+\frac{1}{4b^{\frac{3}{2}}})\\
										+&\sum\limits_{\{j:\widetilde{\lambda}_{\perp_{j}}>\lambda_{11}\}} \gamma_j(\widetilde{\lambda}_{\perp_{j}} - \lambda_{1})\sqrt{b}
										(\widetilde{\lambda}_{\perp_{j}} - \lambda_{11})
										\mathrm{e}^{-b\widetilde{\lambda}_{\perp_{j}}}
										\int_{0}^{1} \mathrm{e}^{-b( \widetilde{\lambda}_{\parallel_j} - \widetilde{\lambda}_{\perp_j}) x^{2} }dx\\
										=&\sum\limits_{\{j:\widetilde{\lambda}_{\perp_{j}}=\lambda_1\}}\gamma_j\mathrm{e}^{-b\widetilde{\lambda}_{\parallel_j}}(\frac{\widetilde{\lambda}_{\parallel_j}-\lambda_{11}}{2\sqrt{b}}+\frac{1}{4b^{\frac{3}{2}}})+\gamma_{j_1}\mathrm{e}^{-b\widetilde{\lambda}_{\parallel_{j_1}}}(\frac{1}{4b^{\frac{3}{2}}})\\
										+&\sum\limits_{\{j:\widetilde{\lambda}_{\perp_{j}}=\lambda_{11},j\neq j_1\}}\gamma_j\mathrm{e}^{-b\widetilde{\lambda}_{\parallel_j}}(\frac{\widetilde{\lambda}_{\parallel_j}-\lambda_{11}+\widetilde{\lambda}_{\perp_{j}} - \lambda_{1}}{2\sqrt{b}}+\frac{1}{4b^{\frac{3}{2}}})\\
										+&\sum\limits_{\{j:\widetilde{\lambda}_{\perp_{j}}>\lambda_{11}\}}\gamma_j\mathrm{e}^{-b\widetilde{\lambda}_{\parallel_j}}(\frac{\widetilde{\lambda}_{\parallel_j}-\lambda_{11}+\widetilde{\lambda}_{\perp_{j}} - \lambda_{1}}{2\sqrt{b}}+\frac{1}{4b^{\frac{3}{2}}})\\
										+&\sum\limits_{\{i:\lambda_{\perp_{i}}>\lambda_{11}\}} \mu_i(\lambda_{\perp_{i}} - \lambda_{1})\sqrt{b}
										(\lambda_{\perp_{i}} - \lambda_{11})
										\mathrm{e}^{-b\lambda_{\perp_{i}}}
										\int_{0}^{1} \mathrm{e}^{ -b( \lambda_{\parallel_i} - \lambda_{\perp_i}) x^{2} }dx
									\end{aligned}
								\end{equation}
								Again, we divide \eqref{eq:case2long} by $\exp(-b\lambda_{12})$ and take derivative; then multiply the result by $b^{\frac{5}{2}}$ and take $b\to\infty$. The left hand side goes to $-\frac{3\mu_{i_1}}{8}$ while the right hand side goes to $-\frac{3\gamma_{j_1}}{8}$, which implies that $\mu_{i_1}=\gamma_{j_1}$. This is contradictory with $\mu_{i_1}(\lambda_{\parallel_{i_1}}-\lambda_{11})=\gamma_{j_1}(\widetilde{\lambda}_{\parallel_{j_1}}-\lambda_{11}+\widetilde{\lambda}_{\perp_{j_1}} - \lambda_{1})$. Thus, this case does not exist.
								
								\item $\lambda_{11}\notin A_{1}$ and $\lambda_{11}\notin A_{21}$ and $\lambda_{11}\in A_{31}$. $\lambda_{11}\notin B_{11}$ and $\lambda_{11}\notin B_{21}$ and $\lambda_{11}\in B_{31}$.
								
								We multiply \eqref{eq:subcase323} by $\sqrt{b}\exp(b\lambda_{12})$ and take derivative. We then repeatedly follow the same procedures in the previous case of ``$\lambda_{11}\notin A_1$, $\lambda_{11}\notin A_2$, and $\lambda_{11}\in A_3$. $\lambda_{11}\notin B_1$, $\lambda_{11}\notin B_2$, and $\lambda_{11}\in B_3$" to finish the proof.
							\end{itemize}
						\end{itemize}
					\end{itemize}
				\end{itemize}
			\end{itemize}
			
		\end{mycases}
	\end{proof}

\end{document}


\centering
\resizebox{\linewidth}{!}
{
\begin{tikzpicture}

\def\ColorA{red!100!blue}
\def\ColorB{red!75!blue}
\def\ColorC{red!50!blue}
\def\DirectionColor{green!60!blue}
\definecolor{color1}{RGB}{55,126,184}
\definecolor{color2}{RGB}{215,26,28}
\definecolor{color3}{RGB}{77,175,74}
\definecolor{colorecic}{RGB}{146,100,51}
\definecolor{colorall}{RGB}{51,109,95}

\let\iwidth\relax
\let\iheight\relax
\newlength{\iwidth}
\newlength{\iheight}
\setlength{\iwidth}{0.095\textwidth}
\setlength{\iheight}{1.33\iwidth}
\setlength{\iwidth}{0.1\textwidth}

\tikzstyle{ellipsestyle} = [x radius=.07cm,y radius=0.4cm]
\tikzstyle{ellipsecolor} = [color=color2]

\begin{scope}
	\draw[ultra thick] (0,0) rectangle (5,5);

	\fill[ellipsecolor] (1,1) ellipse [ellipsestyle];
	\fill[ellipsecolor] (1,2) ellipse [ellipsestyle];
	\fill[ellipsecolor] (1,3) ellipse [ellipsestyle];
	\fill[ellipsecolor] (1,4) ellipse [ellipsestyle];

	\fill[ellipsecolor] (2,1) ellipse [ellipsestyle,rotate=-15];
	\fill[ellipsecolor] (2,2) ellipse [ellipsestyle,rotate=-15];
	\fill[ellipsecolor] (2,3) ellipse [ellipsestyle,rotate=-15];
	\fill[ellipsecolor] (2,4) ellipse [ellipsestyle,rotate=-15];

	\fill[ellipsecolor] (3,1) ellipse [ellipsestyle,rotate=-30];
	\fill[ellipsecolor] (3,2) ellipse [ellipsestyle,rotate=-30];
	\fill[ellipsecolor] (3,3) ellipse [ellipsestyle,rotate=-30];
	\fill[ellipsecolor] (3,4) ellipse [ellipsestyle,rotate=-30];

	\fill[ellipsecolor] (4,1) ellipse [ellipsestyle,rotate=-45];
	\fill[ellipsecolor] (4,2) ellipse [ellipsestyle,rotate=-45];
	\fill[ellipsecolor] (4,3) ellipse [ellipsestyle,rotate=-45];
	\fill[ellipsecolor] (4,4) ellipse [ellipsestyle,rotate=-45];
\end{scope}

\begin{scope} [yshift=0cm, xshift=6cm]
\draw[ultra thick] (0,0) rectangle (5,5);
\node at (1,1)
{\includegraphics [width=\iwidth, angle=0]{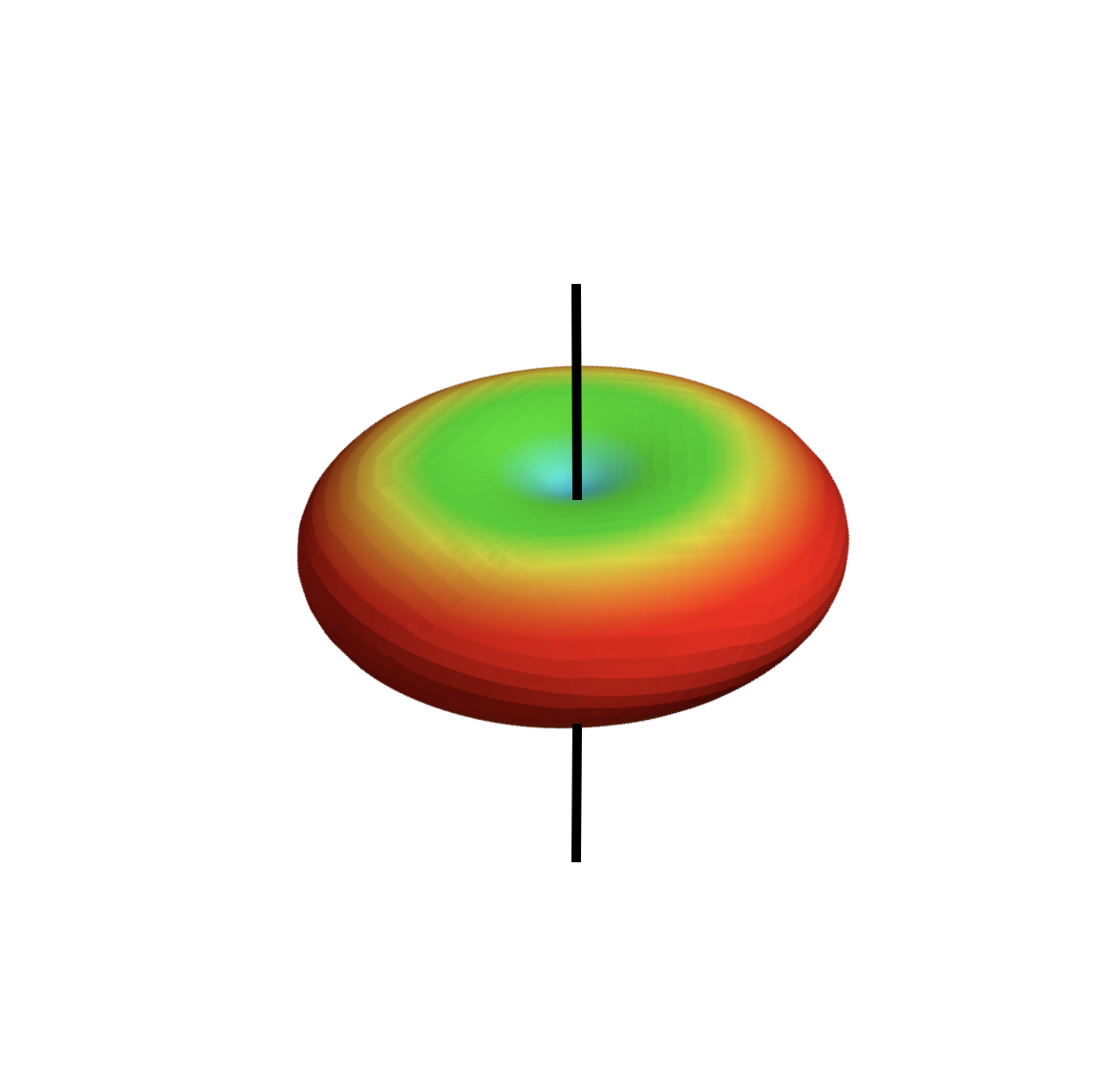}};
\node at (1,2)
{\includegraphics [width=\iwidth, angle=0]{signal_axis}};
\node at (1,3)
{\includegraphics [width=\iwidth, angle=0]{signal_axis}};
\node at (1,4)
{\includegraphics [width=\iwidth, angle=0]{signal_axis}};

\node at (2,1)
{\includegraphics [width=\iwidth, angle=-15]{signal_axis}};
\node at (2,2)
{\includegraphics [width=\iwidth, angle=-15]{signal_axis}};
\node at (2,3)
{\includegraphics [width=\iwidth, angle=-15]{signal_axis}};
\node at (2,4)
{\includegraphics [width=\iwidth, angle=-15]{signal_axis}};

\node at (3,1)
{\includegraphics [width=\iwidth, angle=-30]{signal_axis}};
\node at (3,2)
{\includegraphics [width=\iwidth, angle=-30]{signal_axis}};
\node at (3,3)
{\includegraphics [width=\iwidth, angle=-30]{signal_axis}};
\node at (3,4)
{\includegraphics [width=\iwidth, angle=-30]{signal_axis}};

\node at (4,1)
{\includegraphics [width=\iwidth, angle=-45]{signal_axis}};
\node at (4,2)
{\includegraphics [width=\iwidth, angle=-45]{signal_axis}};
\node at (4,3)
{\includegraphics [width=\iwidth, angle=-45]{signal_axis}};
\node at (4,4)
{\includegraphics [width=\iwidth, angle=-45]{signal_axis}};
\end{scope}

\begin{scope} [yshift=0cm, xshift=12cm]
\draw[ultra thick] (0,0) rectangle (5,5);
\node at (2.5,2.5)
{\includegraphics [width=4\iwidth, angle=-12,scale=0.5]{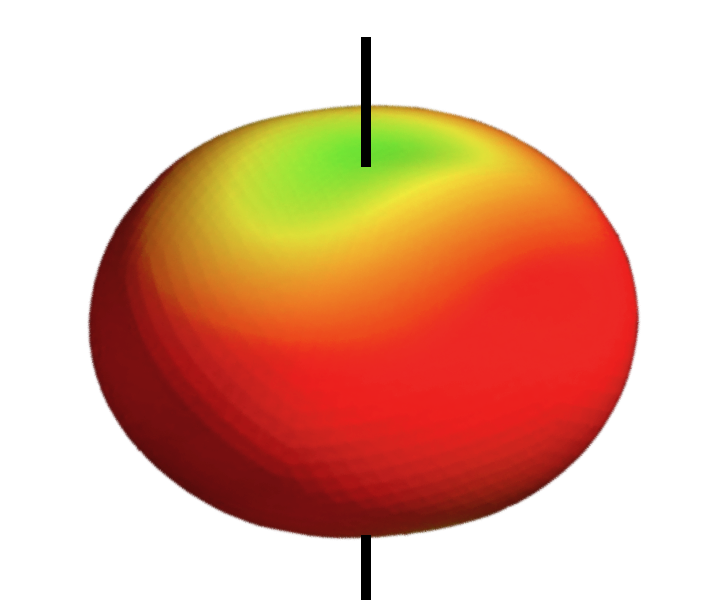}};
\end{scope}

\begin{scope} [yshift=0cm, xshift=18cm]
\draw[ultra thick] (0,0) rectangle (5,5);
\fill[ellipsecolor] (2.5,2.5) ellipse [x radius=.60cm,y radius=1.16cm,rotate=-12];
\end{scope}

\newcommand{\plotIC}[1]
{
  {    
    \begin{tikzpicture}[scale=#1,smooth]
    \draw [help lines, step=0.33] (0,0) grid (1,1);
    \draw[line width=2pt, color=color2] plot coordinates
    {
      (0,1)
      (1/6,0.6980)
      (2/6,0.5092)
      (3/6,0.3862)
      (4/6,0.3027)
      (5/6,0.2440)
      (6/6,0.2013)
    };
    \end{tikzpicture}
  }
}

\def\iscale{0.8}

\begin{scope} [yshift=-6cm, xshift= 6cm]
\draw[ultra thick] (0,0) rectangle (5,5);
\node at (1,1)
{\plotIC{\iscale}};
\node at (1,2)
{\plotIC{\iscale}};
\node at (1,3)
{\plotIC{\iscale}};
\node at (1,4)
{\plotIC{\iscale}};

\node at (2,1)
{\plotIC{\iscale}};
\node at (2,2)
{\plotIC{\iscale}};
\node at (2,3)
{\plotIC{\iscale}};
\node at (2,4)
{\plotIC{\iscale}};

\node at (3,1)
{\plotIC{\iscale}};
\node at (3,2)
{\plotIC{\iscale}};
\node at (3,3)
{\plotIC{\iscale}};
\node at (3,4)
{\plotIC{\iscale}};

\node at (4,1)
{\plotIC{\iscale}};
\node at (4,2)
{\plotIC{\iscale}};
\node at (4,3)
{\plotIC{\iscale}};
\node at (4,4)
{\plotIC{\iscale}};
\end{scope}

\begin{scope} [yshift=-6cm, xshift=12cm]
\draw[ultra thick] (0,0) rectangle (5,5);
\node at (2.5,2.5)
{\plotIC{4}};
\end{scope}

\begin{scope} [yshift=-6cm, xshift= 18cm]
\draw[ultra thick] (0,0) rectangle (5,5);
\draw[ultra thick] (0,0) rectangle (5,5);
\fill[ellipsecolor] (2.5,2.5) ellipse [x radius=0.21cm,y radius=1.2cm,rotate=-12];
\end{scope}

\node at (2.5,5.5) {Microstructure};
\node at (8.5,5.5) {Microscopic Signals};
\node at (14.5,5.5) {Voxel Signal};

\node at (8.5,-6.5) {Micro-level Spherical Means};
\node at (14.5,-6.5) {Voxel-level Spherical Mean};

\node at (5.5,2.5) {\Large{$=$}};
\node at (11.5,2.5) {\Large{$=$}};
\node at (17.5,2.5) {\Large{$\Rightarrow$}};
\node at (8.5,-0.5) {\rotatebox{270}{\Large{$\Rightarrow$}}};
\node at (14.5,-0.5) {\rotatebox{270}{\Large{$\Rightarrow$}}};
\node at (11.5,-3.5) {\Large{$=$}};
\node at (17.5,-3.5) {\Large{$\Rightarrow$}};


\node at (2.5,-0.5) {$\mu$FA = 0.8};
\node at (20.5,-0.5) {FA $<$ 0.8};
\node at (20.5,-6.5) {$\mu$FA = 0.8};

\end{tikzpicture}
}